\documentclass[pre,twocolumn,amsmath,amssymb]{revtex4-2}

\usepackage{amsthm}
\usepackage{bm}
\usepackage{xcolor}
\usepackage{tikz}
\usetikzlibrary{arrows.meta,decorations.markings}
\usepackage{hyperref}

\newtheorem{proposition}{Proposition}
\newtheorem{lemma}{Lemma}

\theoremstyle{remark}
\newtheorem{remark}{Remark}

\newcommand{\Qp}{\mathbb{Q}_p}
\newcommand{\Zp}{\mathbb{Z}_p}
\newcommand{\norm}[1]{\Vert #1\Vert_p}
\newcommand{\tphi}{\tilde\varphi}
\newcommand{\tpsi}{\tilde\psi}
\newcommand{\Dsig}{\boldsymbol{D}^{\sigma}}

\begin{document}

\title{$p$-Adic statistical field theory of fluctuation effects in hierarchical neural networks}

\author{W.~A. Z\'u\~niga-Galindo}
\email{wilson.zunigagalindo@utrgv.edu}
\affiliation{University of Texas Rio Grande Valley, School of Mathematical \& Statistical
Sciences, One West University Blvd, Brownsville, TX 78520, United States}

\date{Draft of \today}

\begin{abstract}
The Buice--Cowan field theory [PRE \textbf{75}, 051919 (2007)] describes fluctuations
and correlations of neural activity through a path integral whose mean-field limit is the
Wilson--Cowan equation and whose critical behavior belongs to the directed percolation
class. In its continuum limit the neurons are the points of a Euclidean space and
activity spreads by ordinary (local) diffusion. This description encodes neither the hierarchical
organization of cortical networks nor their small-world property. We formulate a version
of the theory in which the neurons are organized in an infinite, self-similar, tree-like
structure, described by the $p$-adic numbers. The master equation, the operator formulation and the
path integral carry over unchanged, and finite networks are exact restrictions of the
continuum theory, not approximations of it. The main change is in the linear response.
In the Euclidean theory every short-range connectivity kernel reduces to the same local
operator, the Laplacian. In the $p$-adic theory the connectivity kernel itself gives rise
to a nonlocal $p$-adic diffusion operator, the generator of a $p$-adic heat equation, and
different kernels give different operators. For couplings that decay by a fixed factor per level of the hierarchy this is
the Taibleson--Vladimirov operator, and the critical theory is an ultrametric Reggeon
field theory whose upper critical dimension is twice the decay exponent of the coupling.
Hence the rate at which the coupling decays up the hierarchy, and not the dimension of
space, decides whether mean-field exponents apply. We obtain a Ginzburg-type criterion,
loop-corrected hierarchical Wilson--Cowan equations, log-periodic corrections to scaling,
power-law correlations, and the counterparts of the Buice--Cowan results on the phase
transition and on self-organization. Exact simulations of the spike model on a tree
support the predictions. The avalanche-size exponent takes the mean-field value
three-halves when the coupling decays slowly, and decreases continuously toward the
directed-percolation value when it decays fast. The simulations also show that the
hierarchical network is a small-world network when the decay exponent is smaller than the
dimension: connections are highly clustered, while path lengths and spreading times
grow slowly, roughly as the logarithm of the number of neurons. The Buice--Cowan theory in
its local diffusion limit does not have this property. In a Euclidean space it requires
long-range coupling as an additional hypothesis, whereas in the $p$-adic setting
long-range coupling is the only possibility.
\end{abstract}

\maketitle

\section{Introduction}

There are two central hypotheses about the functioning of the brain. The first is that
the brain has a hierarchical organization~\cite{BullmoreSporns2009,HilgetagGoulas2020}. The
second, the \emph{critical brain hypothesis}, is that it operates near a phase transition,
where a delicate balance between ordered (synchronized) and disordered (random) states
occurs; the neuronal avalanches observed by Beggs and Plenz support this
view~\cite{BeggsPlenz2003}. Since deep neural networks (DNNs) are inspired by the brain,
both hypotheses are also relevant for
them~\cite{BornholdtRohl2003,Schoenholz2017,VockMeisel2025}.

Sporns and co-workers proposed a precise form of the hierarchy hypothesis: cortical
connections are organized in self-similar, fractal patterns across multiple nested
scales~\cite{Sporns2006,CostaSporns2005,SpornsTononiKotter2005}. This proposal provides a
specific geometric mechanism, nested and scale-free self-similarity, by which the brain
builds its structural and functional hierarchy. It has three aspects. (i) \emph{Nested
connectivity}: pathways link neuronal populations across scales, from whole brain regions
down to local minicolumns. (ii) \emph{Self-similarity}: the patterns of connections
repeat from one scale to the next. (iii) \emph{Small-world architecture}: cortical networks
combine a high local clustering of connections with short paths between distant
regions~\cite{WattsStrogatz1998,BassettBullmore2006}, which allows efficient communication
at a low wiring cost. As we explain below, nested self-similar connectivity is exactly the
structure encoded by the $p$-adic numbers. It was shown in Ref.~\cite{ZZ2023WC} that the
classical Wilson--Cowan model, in which the neurons are organized as the points of
$\mathbb R^N$ and interact through Gaussian-type kernels, is not compatible with the
small-world property. In this paper we examine the same question for the B-C theory.

Buice and Cowan (B-C)~\cite{BuiceCowan2007} gave a systematic statistical field theory for
the critical brain hypothesis. Starting from a Markovian ``effective spike'' model, they
constructed a Doi--Peliti path integral~\cite{Doi1976a,Peliti1985}. They showed that
its zero-loop (mean-field) limit is a Wilson--Cowan equation and that loops give
systematic corrections controlled by a Ginzburg-type criterion. They also argued that the
critical point lies in the directed percolation (DP) universality class~\cite{Janssen1981,Grassberger1982,CardySugar1980,Hinrichsen2000}.
In B-C the neurons live in $\mathbb{R}^N$ with $N=2$ or $3$. The hierarchy therefore enters,
at best, through the choice of kernel. A central problem is to develop new versions of B-C
statistical field theories compatible with hypotheses about the hierarchical organization
of the brain. This is the objective of this work; however, here, we take only a first step
toward it.

Another motivation is the correspondence between neural networks (NNs) and statistical
field theories (SFTs). This correspondence is a promising approach for understanding the
dynamics of DNNs~\cite{HeliasDahmen2020,Halverson2021,GrosvenorJefferson2022,ZG2023,ZG2024}.
Since DNNs are inspired by the structure and function of the biological brain, it is
natural to conjecture that SFTs of B-C type can play a central role in understanding how
large DNNs work. It is relevant to note that B-C developed such a correspondence for
cortical networks in the limit of infinitely many neurons, combined with a continuum limit
in which the neuron label becomes a position in space~\cite{BuiceCowan2007,BuiceCowan2009}. The correspondence between DNNs and lattice
SFTs, with a large but finite number of neurons, has been studied
intensively~\cite{RobertsYaida2022,Segadlo2022}. However, standard principles of
statistical mechanics suggest that understanding the dynamics of DNNs requires
understanding their thermodynamic limit. Based on that, the author has proposed that the
continuous SFTs developed by B-C should be taken as the basis for a correspondence between
DNNs and SFTs; see Ref.~\cite{ZG2026} and the references therein.

The present work supports this proposal. In the field-theoretic approaches to DNNs, the
natural large parameter is the width $n$ of the layers, or the number of neurons. At fixed
depth, the infinite-width limit is a Gaussian theory, and the non-Gaussian corrections come
in powers of $1/n$~\cite{RobertsYaida2022,Segadlo2022}. The $p$-adic B-C theory differs in
two ways. First, width and depth cannot be separated: a hierarchical network of depth $l$
in dimension $N$ has $p^{Nl}$ neurons, and its thermodynamic limit adds layers to the hierarchy at
the same time as it adds neurons, so the infinite-width limit involves infinitely many
layers. Second, the limiting theory is not Gaussian: its cubic couplings produce
fluctuations that are not small near criticality below the upper critical dimension
(Sec.~\ref{sec:RG}). The cutoffs of the
theory, the depth of the tree and the size of the network, are part of the architecture,
and a Ginzburg number plays the role of $1/n$. If the infinite-width limit is understood
as a network with one layer and infinitely many neurons, the $p$-adic SFTs are of a
completely different nature from the lattice SFTs of Ref.~\cite{RobertsYaida2022}. We
discuss this comparison in Sec.~\ref{sec:cutoffs}.

\paragraph*{Ultrametricity and $p$-adic numbers in physics.}
A distance is \emph{ultrametric} if it obeys the strong triangle inequality,
$d(x,z)\le\max\{d(x,y),d(y,z)\}$. A non-empty set endowed with an ultrametric is called an ultrametric
space. These spaces constitute the natural mathematical framework for modeling
hierarchical systems~\cite{RTV1986,KKZ2018}. In
physics they first appeared in the theory of disordered systems. Parisi and collaborators
showed that the pure states of the spin-glass phase of mean-field models, described by
replica symmetry breaking, are organized in an ultrametric
hierarchy~\cite{Mezard1984}. Frauenfelder and co-workers proposed that the conformational
substates of proteins are organized in a similar way, so that protein relaxation resembles
a random walk on a tree-like landscape of local energy minima~\cite{Frauenfelder1991}. The
$p$-adic numbers are the prototypical ultrametric spaces on which one can do analysis:
they carry a measure, a Fourier transform, and pseudodifferential operators~\cite{VVZ1994,Kochubei2001,AKS2010,ZunigaLNM2016,ZunigaTextbook2025}.
This made it possible to formulate replica symmetry breaking $p$-adically~\cite{ParisiSourlas2000},
and to describe relaxation in hierarchical energy landscapes by $p$-adic diffusion
equations~\cite{ABKO2002}. Since then, $p$-adic models have been used in statistical
mechanics, quantum field theory and string theory, and the theory of complex
systems~\cite{VVZ1994,KKZ2018}. $p$-Adic quantum field theory began in the 1980s under the
influence of Vladimirov and Volovich~\cite{VVZ1994}, and it has produced an extensive
literature~\cite{VVZ1994,KKZ2018}. Most of this work concerns formal models, whose connection
with measurable phenomena is indirect. To the author's knowledge, the $p$-adic B-C theory is
one of the first $p$-adic statistical field theories that is derived, without free modeling
choices, from the microscopic dynamics of a concrete system, here the master equation of a
spiking network, and whose predictions can be compared with simulations and, in principle,
with experiments.

\paragraph*{$p$-Adic statistical field theories and neural networks.}
In a series of papers, the author and collaborators have developed a correspondence between
NNs and SFTs in which the fields are defined on a $p$-adic space instead of a Euclidean
one~\cite{ZG2023,ZHZ2023,ZG2024ATMP,ZG2024,ZG2026}. In this framework, a field is a
real-valued function on the boundary of an infinite tree-like structure that provides the
architecture of a deep network. The correspondence was first established for
deep belief networks~\cite{ZG2023}: whereas a restricted Boltzmann machine corresponds to
a spin glass, a deep, hierarchical architecture corresponds to an ultrametric spin
glass, which is described naturally with $p$-adic numbers. Restricting the energy functional of a
$p$-adic SFT to a finite-dimensional space of test functions gives a discretization that
is exact, and that yields a new class of networks, the $p$-adic convolutional deep Boltzmann
machines~\cite{ZHZ2023,ZG2024ATMP}. Because of the weight sharing imposed by the symmetries
of the tree, these networks have $O(\mathcal N)$ parameters for $\mathcal N$ neurons, instead
of the $O(\mathcal N^2)$ of a fully connected machine. The same ideas were applied to
dynamical models: $p$-adic cellular neural networks, which are the continuum limits of
large hierarchical versions of the Chua--Yang networks~\cite{ZZ2023CNN}; hierarchical
Wilson--Cowan models~\cite{ZZ2023WC}; and continuous random networks, for which a dynamic
mean-field theory and a criticality condition were obtained~\cite{ZG2024,ZG2026}. The
exact discretization just mentioned is the mechanism used in Sec.~\ref{sec:geometry} of
the present paper.

A natural way to make a network intrinsically hierarchical is to index neurons by
$p$-adic numbers~\cite{VVZ1994,Kochubei2001,AKS2010,ZunigaLNM2016}. Working in dimension
one ($N=1$) for simplicity, the unit ball $\Zp$
is the boundary of a rooted tree in which each vertex has $p$ children. The ultrametric
distance $|x-y|_p$ records the depth of the last common ancestor of $x$ and $y$.

We use the following terminology throughout. The \emph{levels} of the tree play the role of
the \emph{layers} of the hierarchy, and we use the words \emph{level} and \emph{layer}
synonymously. A vertex at level $k$, together with all the neurons
that descend from it, is a ball of radius $p^{-k}$. We refer to it simply as a
\emph{ball}, or as a \emph{structure of level $k$}: the ball is built from the $p$
balls of level $k+1$ below it, just as higher structures in the brain are built from
lower ones. The \emph{hierarchical distance} between two neurons is
$|x-y|_p$: if it equals $p^{-k}$, the smallest structure containing both neurons is
of level $k$. In previous
work~\cite{ZG2023,ZG2024,ZG2026} it was observed that the B-C theory admits such a
formulation. One simply interprets the spatial integrals in Eqs.~(12)--(13) of
Ref.~\cite{BuiceCowan2007} as integrals over $\Qp^N$ with respect to the Haar measure.
The purpose of this paper is to work out what changes and what does not.

\paragraph*{Results and comparison with the B-C theory.}
Our starting point is the observation that the stochastic model of B-C, its operator
formulation and the resulting discrete action do not refer to any geometry. The geometry
enters only in the continuum limit, where B-C place the neurons at the points of
$\mathbb R^N$. We show that if the neurons are instead organized in a tree-like structure, the same limit
places them at the points of $\Qp^N$, and the sums over neurons become integrals with respect
to the Haar measure. In contrast with the Euclidean case, the finite network is not an
approximation of the continuum theory: the discrete action is exactly the restriction of
the continuum action to a finite-dimensional space of functions, a feature shared by
other $p$-adic statistical field theories~\cite{ZG2023,ZG2024,ZG2026,ZHZ2023}. Mean-field
theory is again
a Wilson--Cowan equation, now hierarchical, and the loop expansion of B-C carries over
formally unchanged.

The first substantial difference appears in the linear response. In the B-C theory, the
expansion of the Fourier transform of the connectivity kernel at small momenta produces a
local diffusion term, the Laplacian, the same for every short-range kernel. No such
expansion exists over $\Qp^N$. This does not mean that there is no $p$-adic diffusion. The
theory of $p$-adic heat equations is vast, and there are many more such equations than in
the Euclidean case~\cite{Kochubei2001,ZunigaLNM2016,KKZ2018}. What happens is that each
connectivity kernel gives rise, in a natural way, to its own nonlocal diffusion operator,
a $W$-operator, which is the generator of a $p$-adic heat equation. In the Euclidean theory the kernel is
reduced to a universal local operator; in the $p$-adic theory the kernel itself determines
the $p$-adic diffusion operator, a $W$-operator~\cite[Chap.~2]{ZunigaLNM2016}. For kernels
with compact support, the support of the kernel determines which neurons are truly
connected. If the connected neurons form a ball, only its total activity becomes marginal
at the critical point, and there are no scale-free correlations across the hierarchy. We also consider \emph{regional networks}, in which the neurons occupy a disjoint
union of finitely many (at least two) balls, as in the ultrametric transition networks of
Ref.~\cite{ZG2022Ultrametric}. Each ball represents a region of the cortex in which the
spike model holds with its own constant parameters, and the regions are coupled by
constant interregional connections. When all regions share the same decay rate and
firing-rate function, this is the B-C model with a block-structured connectivity.
Region-dependent parameters go beyond the B-C model, whose master equation has a single
firing-rate function and a single decay rate for all neurons. The criticality of a
regional network is decided by a finite matrix, and its critical behavior is of
mean-field type (Lemma~\ref{lem:regional}, Sec.~\ref{sec:Wop}). On a bounded network, the linear dynamics
is a Markov process exactly when the distance to criticality is nonnegative, and at the
critical point this process becomes conservative: branching exactly compensates decay
(Sec.~\ref{sec:master}). For isotropic
kernels that decay by a fixed factor per level of the hierarchy,
$w(\norm{x})\propto\norm{x}^{-N-\sigma}$ at large distances with $\sigma>0$, the operator is the
Taibleson--Vladimirov operator of order $\sigma$, which we show to be the unique isotropic
and homogeneous choice. The exponent $\sigma$ is a parameter of the connectivity, not of
the dynamics: each time one climbs one level of the hierarchy, the coupling between two
neurons is multiplied by $p^{-(N+\sigma)}$, and the total coupling that a neuron receives
from the neurons at that level is multiplied by $p^{-\sigma}$. Thus $\sigma$ measures how
fast the coupling decays with hierarchical distance, and it can be read off from the
connection weights. It is a new physical parameter with no counterpart in the B-C theory.
Throughout the paper $\sigma$ denotes this decay exponent.

The second difference concerns criticality. B-C argue that the critical point belongs to
the DP class, with upper critical dimension $4$, and they invoke the high connectivity of
cortex to justify mean-field exponents in dimension $2$ or $3$. In the $p$-adic theory the
effective action at criticality is an ultrametric Reggeon field theory whose upper
critical dimension is $N_c=2\sigma$. Whether mean-field exponents apply is therefore
decided by the decay exponent of the hierarchical coupling and not by the embedding
dimension. Unlike Euclidean DP with long-range spreading~\cite{Janssen1999}, there is no
competing local (Laplacian) term, so every $\sigma>0$ defines a distinct class. The Ginzburg-type
criterion of B-C becomes a statement about the level of the hierarchy that fluctuations
reach before they relax, and the discrete scale invariance of the hierarchical neuron
space, which is invariant only under the dilations $x\mapsto p^{-1}x$, produces
log-periodic corrections to scaling, which a Euclidean theory cannot produce. In the
mean-field regime the avalanche-size exponent is $3/2$, as in the B-C theory; for
$N<2\sigma$ we predict deviations that depend on $\sigma$, and we confirm them by direct
simulation of the spike model on a tree (Sec.~\ref{sec:simulations}).

The third difference concerns the small-world property. For a network with weighted
coupling we use two tests: the clustering and the path length of a sparse network whose
connections are drawn with probabilities proportional to the coupling, and the time that
activity needs to reach half of the neurons in the spike model. With both tests, the
hierarchical network is a small-world network when $\sigma<N$: the clustering is high,
and path lengths and spreading times grow slowly, roughly as the logarithm of the
number of neurons (Sec.~\ref{sec:smallworld}). For $\sigma>N$ it is not. The B-C theory
with the kernels of the classical model, Gaussian or exponential, which for practical
purposes have compact support, is not small-world: the spreading time grows in proportion
to the linear size of the network. This agrees with Ref.~\cite{ZZ2023WC}, where the
incompatibility is traced to the fact that $\mathbb R^N$ has no nontrivial compact
subgroups, while in $\Qp^N$ every ball is one. A finite Euclidean network can be made
small-world, by our tests, only by leaving the classical model: by imposing a coupling
with a power-law tail on a ring. In $\mathbb R^N$ long-range coupling is thus an additional
hypothesis, imposed on a theory whose generic limit is local diffusion. In $\Qp^N$ all
diffusion operators are nonlocal, long-range coupling is the only possibility, and it
comes with the nested structure of the hierarchy. The threshold $\sigma=N$ differs from the
threshold $\sigma=N/2$ for mean-field avalanches: the small-world property is necessary
for the exponent $3/2$ in our simulations, but it is not sufficient.

Finally, we give the $p$-adic counterparts of the phase transition and of the
self-organization mechanisms discussed in Ref.~\cite{BuiceCowan2009}. The linear spectrum
resolves exactly layer by layer. The zero-loop propagator is computed explicitly and has
power-law tails in hierarchical distance, so that correlations away from criticality are
not exponentially truncated in space, in contrast with the Euclidean prediction. Power
spectra of local averages scale as $\omega^{-(2-N/\sigma)}$ below a crossover frequency set
by the size of the averaging region. Standard self-organizing mechanisms, such as synaptic
depression, tune the distance to criticality but leave $\sigma$ unchanged. Throughout, we
distinguish results that are exact at zero-loop order from heuristic arguments and from
conjectures. In particular, the analysis below the upper critical dimension, which
requires the renormalization group, is not carried out here, which is why we regard this
work as a first step.

Table~\ref{tab:comparison} compares the Euclidean B-C theory with the $p$-adic theory of this
paper, item by item: the neuron space and finite networks, the mean-field equation and
the spreading operator, the propagator at large distance, the upper critical dimension and
the mean-field exponents, scale invariance, correlations and power spectra, the spreading
of activity and the small-world property, the avalanche-size exponent, and what
self-organization tunes.

\begin{table*}[t]
\caption{Comparison of the Euclidean B-C theory with the $p$-adic theory of this paper. Entries
of the $p$-adic column are zero-loop (mean-field) results unless stated otherwise.
$N$ is the dimension of the neuron space, $\sigma$ the decay exponent of the hierarchical
coupling [Eq.~\eqref{eq:perlevel}], and $\mu$ the distance to criticality.}
\label{tab:comparison}
\begin{ruledtabular}
\begin{tabular}{lll}
 & Euclidean B-C theory & $p$-Adic theory\\
\hline
Neuron space & $\mathbb R^N$ & a subset of $\Qp^N$, with a hierarchical geometry\\
Finite network & lattice; an approximation & finite rooted trees, $G_l^N$; an exact restriction (Prop.~\ref{prop:exact})\\
Mean-field equation & Wilson--Cowan & hierarchical Wilson--Cowan [Eq.~\eqref{eq:WC_delta}]\\
Spreading operator & local diffusion, $-D\nabla^2$ & nonlocal $p$-adic diffusion ($W$-operator) set by the kernel (Sec.~\ref{sec:Wop})\\
Propagator at large distance & Gaussian & power law, $\propto w(\norm{x})$ [Eq.~\eqref{eq:Delta_tail}]\\
Upper critical dimension & $4$ & $2\sigma$ (power counting)\\
Mean-field exponents & $\nu_\perp=\tfrac12$, $z=2$ & $\nu_\perp=1/\sigma$, $z=\sigma$\\
Scale invariance & continuous & discrete ($x\to px$); log-periodic corrections\\
Correlations for $\mu>0$ & exponentially truncated & power law [Eq.~\eqref{eq:C_tail}]\\
Spectrum of a local average & $\omega^{-2}$ & $\omega^{-(2-N/\sigma)}$ for $\omega\ll\omega_k$, with $\omega_k$ set by the electrode size [Eq.~\eqref{eq:LFP_spectrum}]\\
Spreading of activity ($\mu<0$) & constant velocity & constant layers per unit time for $\sigma<N$ (Fig.~\ref{fig:smallworld})\\
Small-world property & no (diffusion limit) & yes for $\sigma<N$ (Sec.~\ref{sec:smallworld})\\
Avalanche-size exponent & $3/2$ for $N\ge4$ & $3/2$ for $N\ge2\sigma$; smaller otherwise (Fig.~\ref{fig:avalanches})\\
Self-organization tunes & $\mu$ & $\mu$; $\sigma$ is unchanged\\
\end{tabular}
\end{ruledtabular}
\end{table*}

\paragraph*{Organization of the paper.}
The main physical results of this work were discussed in Ref.~\cite{ZGshort}. The present
paper is the companion paper to that publication: it contains the complete construction,
the proofs, and further results. Section~\ref{sec:prelim} collects the facts of $p$-adic analysis that we use.
Section~\ref{sec:model} constructs the $p$-adic effective spike model: the continuum limit of
a hierarchical network, the action, mean-field theory, and the free and interacting parts
of the action with the corresponding propagators. Section~\ref{sec:linear} studies the linear
response and the $W$-operators selected by the connectivity. Section~\ref{sec:RG} derives
the effective theory at criticality, its power counting and its mean-field exponents.
Section~\ref{sec:ginzburg} gives the Ginzburg-type criterion. Section~\ref{sec:loops} contains
the loop-corrected hierarchical Wilson--Cowan equations, and Sec.~\ref{sec:avalanches} the
avalanche exponents and their numerical simulation. Section~\ref{sec:smallworld} studies
the small-world property. Section~\ref{sec:phase} treats the phase transition, correlations,
power spectra, rhythms and self-organized criticality. Section~\ref{sec:discussion} discusses
open problems.

\paragraph*{Reading guide.}
The paper is addressed to physicists, neuroscientists and researchers in artificial intelligence.
Readers not familiar with field-theoretic methods for stochastic processes will find in
Appendix~\ref{app:primer} a self-contained introduction to the concepts used below: the
master equation, the Doi--Peliti formulation, the action, Feynman diagrams, and
universality. The appendix follows a simple branching process that contains all the
ingredients of the effective spike model, and it ends with a glossary.
Appendix~\ref{app:DP} introduces directed percolation and Reggeon field theory, the
universality class of the critical point. Appendix~\ref{app:padic} collects the basic facts of $p$-adic analysis, and
Appendix~\ref{app:integrals} the $p$-adic integrals used in the text.

\paragraph*{Mathematical format.}
In some parts of the paper we use the mathematical format of propositions, lemmas, and
proofs. This format gives precise control over several new calculations involving
$p$-adic heat equations and the asymptotics of the new $p$-adic integrals required in
perturbative calculations. It does not mean that we present a rigorous mathematical theory
of the B-C SFTs; that remains a very difficult open problem.

\section{$p$-adic preliminaries}\label{sec:prelim}

\begin{figure}[t]
\centering
\includegraphics[width=\columnwidth]{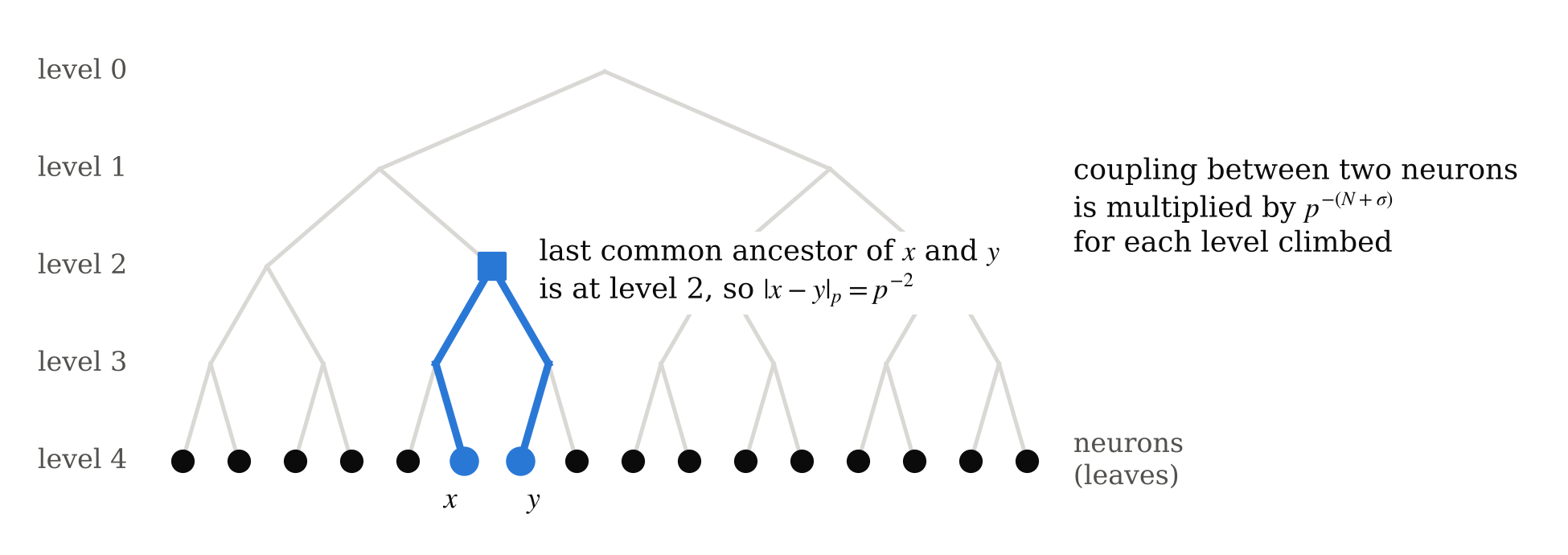}
\caption{Neurons organized in a tree-like structure ($p=2$, depth $4$: levels $0,\dots,4$). The $p$-adic
distance between two neurons is determined by the level of their last common ancestor.
With a kernel $w(\norm{x})\propto\norm{x}^{-N-\sigma}$, the exponent $\sigma$ measures how
fast coupling decays with hierarchical distance.}
\label{fig:tree}
\end{figure}

We collect the facts needed below; see
Refs.~\cite{VVZ1994,Kochubei2001,AKS2010,ZunigaLNM2016}. Definitions and further details
(the $p$-adic norm, balls, the Haar measure, test functions, the Fourier transform and
distributions) are given in Appendix~\ref{app:padic}. Throughout, $p$ denotes a fixed prime
number. Any nonzero $p$-adic number $x\in\Qp\smallsetminus\{0\}$ can be represented as a
series of the form $x=p^{\mathrm{ord}(x)}\sum_{j\ge0}x_jp^j$, with $x_j\in\{0,\dots,p-1\}$
and $x_0\neq0$; then $|x|_p=p^{-\mathrm{ord}(x)}$. The $p$-adic norm $|\cdot|_p$ is
ultrametric: $|x+y|_p\le\max\{|x|_p,|y|_p\}$. This inequality is stronger than the standard
triangle inequality, and it has strong implications for the geometry of $\Qp$. For
instance, two balls are either disjoint or one contains the other, which in turn implies
that the unit ball $\Zp=\{|x|_p\le1\}$ is an infinite rooted tree: the $p$-adic distance
between two points is determined by the level of their last common ancestor
(Fig.~\ref{fig:tree}). Throughout, $N$ denotes
the dimension of the neuron space ($d$ in Ref.~\cite{BuiceCowan2007}); we avoid $d$ so as
not to write $d^dx$. For $x\in\Qp^N$ set $\norm{x}=\max_i|x_i|_p$. Let $d^Nx$ denote the
Haar measure normalized by $\int_{\Zp^N}d^Nx=1$; then $d^N(ax)=|a|_p^N\,d^Nx$. The sphere
$S_k^N=\{\norm{x}=p^k\}$ has volume $(1-p^{-N})p^{kN}$. Hence, for radial
$F(x)=F(\norm{x})$,
\begin{equation}
\int_{\Qp^N}F(\norm{x})\,d^Nx=(1-p^{-N})\sum_{k\in\mathbb{Z}}p^{kN}F(p^k).
\label{eq:radial}
\end{equation}

We warn the reader that the integration theory cannot be reduced to
formula~\eqref{eq:radial}. Let $\chi_p(y)=e^{2\pi i\{y\}_p}$ be the standard additive
character (the $p$-adic analogue of the complex exponential function). For an integrable
function $g:\Qp^N\to\mathbb C$, its Fourier transform is defined as
$\hat g(\xi)=\int\chi_p(\xi\cdot x)g(x)\,d^Nx$. The basic identity is
\begin{equation}
\int_{\norm{x}\le p^k}\chi_p(\xi\cdot x)\,d^Nx=p^{kN}\,\Omega\!\left(p^k\norm{\xi}\right),
\label{eq:ballFT}
\end{equation}
where $\Omega$ is the indicator of $[0,1]$. The Fourier transform of a radial function
is radial.

The Taibleson--Vladimirov operator of order $\sigma>0$ is
\begin{equation}
\Dsig\phi(x)=\frac{1-p^{\sigma}}{1-p^{-\sigma-N}}\int_{\Qp^N}
\frac{\phi(x-y)-\phi(x)}{\norm{y}^{\sigma+N}}\,d^Ny,
\label{eq:Taibleson}
\end{equation}
with symbol $\norm{\xi}^{\sigma}$. It generates a Markov process on $\Qp^N$ (ultrametric
diffusion~\cite{ZunigaTextbook2025,Kochubei2001}). Radial convolution operators are
diagonalized by the $p$-adic wavelets~\cite{Kozyrev2002,AKS2010}
\begin{equation}
\psi_{\gamma nj}(x)=p^{-N\gamma/2}\,\chi_p\big(p^{-1}j\cdot(p^{\gamma}x-n)\big)\,
\Omega\big(\norm{p^{\gamma}x-n}\big),
\label{eq:wavelets}
\end{equation}
with $\gamma\in\mathbb{Z}$, $n\in(\Qp/\Zp)^N$ and
$j\in\{0,1,\dots,p-1\}^N\smallsetminus\{(0,\dots,0)\}$. The wavelet
$\psi_{\gamma nj}$ is supported on the ball of radius $p^{\gamma}$ centred at
$p^{-\gamma}n$, and the wavelets form an orthonormal basis of $L^2(\Qp^N)$. For radial $w\in L^1$,
\begin{equation}
w\star\psi_{\gamma nj}=\hat w\big(p^{1-\gamma}\big)\,\psi_{\gamma nj},
\label{eq:wavelet_eig}
\end{equation}
so the eigenvalue depends only on the level of the ball that supports the wavelet, not on its position
or on $j$; see also Ref.~\cite[Section~4.1]{ZG2022}.

\section{The $p$-adic effective spike model}\label{sec:model}

\subsection{Neuron geometry: from the discrete master equation to the continuum}\label{sec:geometry}

This subsection explains why the $p$-adic theory is a genuine version of the B-C theory and
not a modification of it. The main point is that the geometry of the neurons is not
fixed by the model. It enters only when the continuum limit is taken.

\paragraph*{The master equation has no geometry.}
The effective spike model of B-C [Sec.~II, Eq.~(1) of Ref.~\cite{BuiceCowan2007}] is
defined for an arbitrary finite set $\Lambda$ of neurons. Let $n_i$ be the number of
effective spikes of neuron $i\in\Lambda$, and $w_{ij}$ the weight with which neuron $j$
innervates neuron $i$. The probability $P(\mathbf n,t)$ of the configuration
$\mathbf n=(n_i)_{i\in\Lambda}$ obeys
\begin{multline}
\partial_tP(\mathbf n,t)=\sum_{i\in\Lambda}\Big\{\alpha\big[(n_i+1)P(\mathbf n_i^+,t)-n_iP(\mathbf n,t)\big]\\
+f\big(s_i(\mathbf n_i^-)\big)P(\mathbf n_i^-,t)-f\big(s_i(\mathbf n)\big)P(\mathbf n,t)\Big\},
\label{eq:master}
\end{multline}
where $s_i(\mathbf n)=\sum_{j}w_{ij}n_j+I$, and $\mathbf n_i^{\pm}$ differs from
$\mathbf n$ only in the $i$th entry, which is $n_i\pm1$, and $f$ is the firing-rate
function. Eq.~\eqref{eq:master} differs from Eq.~(1) of Ref.~\cite{BuiceCowan2007}, where the
gain term is $f(s_i(\mathbf n))P(\mathbf n_i^-,t)$, and from the corresponding Hamiltonian,
Eq.~(A11) there, in which $f$ stands to the left of $(1-\Phi_i^\dagger)$. In
Eq.~\eqref{eq:master} the firing rate is evaluated in the state before the jump,
$s_i(\mathbf n_i^-)=s_i(\mathbf n)-w_{ii}$, which ensures conservation of total probability
for any matrix $(w_{ij})$. The two forms coincide when $w_{ii}=0$; in the continuum limit
their difference vanishes (see the discussion after Eq.~\eqref{eq:wij}). Nothing in
Eq.~\eqref{eq:master} refers to a geometry. The set $\Lambda$ and the matrix $(w_{ij})$ are
arbitrary.

\paragraph*{The discrete action has no geometry.}
The Doi--Peliti construction [Appendices A and B of Ref.~\cite{BuiceCowan2007}] uses one
pair of creation and annihilation operators per neuron, $[\Phi_i,\Phi_j^\dagger]=\delta_{ij}$.
It is local in the neuron index and applies to any $\Lambda$. After the Doi shift and the
coherent-state representation, it gives the discrete action (see also Sec.~5.2 of
Ref.~\cite{ZG2023})
\begin{align}
S_\Lambda[\varphi,\tphi]=&\int_0^\infty\!dt\sum_{i\in\Lambda}\Big\{\tphi_i\,\partial_t\varphi_i+\alpha\tphi_i\varphi_i\nonumber\\
&\quad-\tphi_i\,f\Big(\sum_{j\in\Lambda}w_{ij}\big(\tphi_j\varphi_j+\varphi_j\big)+I\Big)\Big\}\nonumber\\
&-\sum_{i\in\Lambda}\bar n_i\,\tphi_i(0),
\label{eq:discrete_action}
\end{align}
where $f$ is the firing rate of the neurons and $\bar n_i\ge0$ specifies the initial
state. The interaction term is the coherent-state form of the Hamiltonian
$\sum_i(1-\Phi_i^\dagger)\,f_{\rm NO}\big(\sum_jw_{ij}\Phi_j^\dagger\Phi_j+I\big)$, with
$(1-\Phi_i^\dagger)$ to the left of $f$ (normal ordered as in Ref.~\cite{BuiceCowan2007}),
i.e.\ of Eq.~\eqref{eq:master}, followed by the Doi shift $\Phi_i^\dagger\to1+\tphi_i$. Thus
Eq.~\eqref{eq:discrete_action} agrees with the action of B-C [Eq.~(13) of
Ref.~\cite{BuiceCowan2007}, in discrete form], although the master equations differ.
Normal ordering the Hamiltonian (A11) of Ref.~\cite{BuiceCowan2007}, with $f$ to the left of
$(1-\Phi_i^\dagger)$, would produce additional terms proportional to $w_{ii}$, which do not
appear in Eq.~(13) there. The action of Ref.~\cite{BuiceCowan2007} is therefore the action
of the probability-conserving master equation~\eqref{eq:master}. As in Ref.~\cite{BuiceCowan2007}, the network starts in an uncorrelated state in
which the number of effective spikes of each neuron is Poisson distributed with mean
$\bar n_i$:
\begin{equation}
P(\mathbf n,0)=\prod_{i\in\Lambda}e^{-\bar n_i}\,\frac{\bar n_i^{\,n_i}}{n_i!},
\qquad \bar n_i=\langle n_i(0)\rangle .
\label{eq:Poisson_initial}
\end{equation}
In the operator representation this state is
$\exp\!\big(\sum_i\bar n_i(\Phi_i^\dagger-1)\big)|0\rangle$. After the Doi shift it
produces the initial-state term $-\sum_i\bar n_i\tphi_i(0)$ in the action. Other
(correlated) initial states give other initial-state terms~\cite{BuiceCowan2007}. The
geometry is still absent.

\paragraph*{The continuum limit chooses a geometry.}
To pass to a field theory, one must (i) embed the index set in a space $X$ with a measure
$dx$, (ii) let $\Lambda$ fill $X$, and (iii) replace the sums in
Eq.~\eqref{eq:discrete_action} by integrals over $X$. B-C take $\Lambda$ to be a square
lattice $a\mathbb Z^N$ with spacing $a$ and $w_{ij}=w(|x_i-x_j|)$. The sums become Riemann
sums, $\sum_ia^NF(x_i)\to\int_{\mathbb R^N}F(x)\,d^Nx$ as $a\to0$, and the neurons become the
points of $\mathbb R^N$. This is one particular choice. B-C themselves note that $N$ should be
$2$ or $3$, ``or possibly in between to allow for a fractal lattice'', and that $N$ reflects
how many neighbors a neuron has~\cite{BuiceCowan2007}. The Euclidean geometry is an
assumption made at the continuum step, and it is precisely the assumption that
suppresses the hierarchical organization of the network.

\begin{figure*}[t]
\centering
\includegraphics[width=0.92\textwidth]{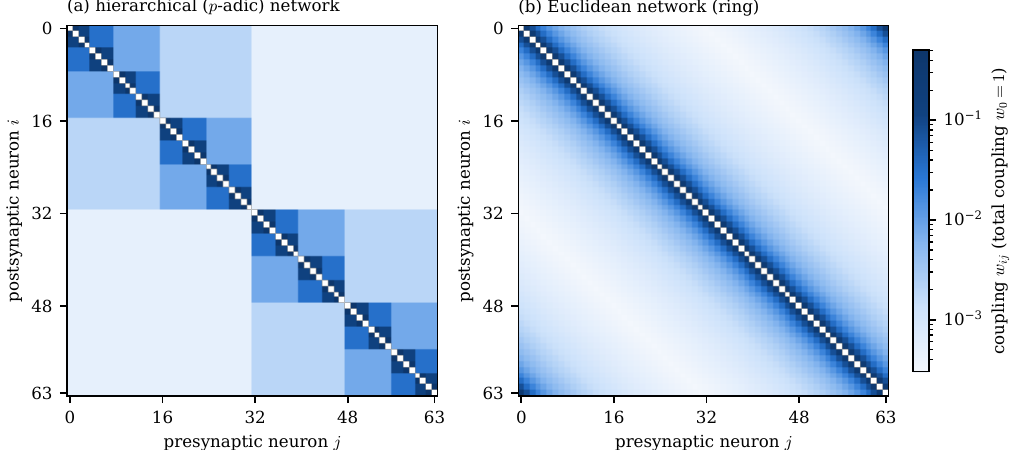}
\caption{Connection matrices $w_{ij}$ of two networks of $64$ neurons, with $\sigma=1$
(logarithmic color scale, common to both panels; the diagonal is excluded). In both
networks the proportionality constant is fixed by the condition that the total coupling
received by a neuron is $w_0=\sum_{j\neq i}w_{ij}=1$. (a) The hierarchical network $G_l$ with
$p=2$, $l=6$, $N=1$ and $w_{ij}=C_p\,|i-j|_p^{-(1+\sigma)}$, where $C_p=1/2016$. The
neurons are ordered along the boundary of the tree, so that every ball is a block of
consecutive indices; the nested blocks are the balls of levels $5,4,\dots,1$. (b) A
Euclidean network: a ring with $w_{ij}=C_E\,|i-j|^{-(1+\sigma)}$, where $|i-j|$ is the
distance along the ring and $C_E\simeq0.310$. The hierarchical matrix is constant on blocks, which
is the property used in Proposition~\ref{prop:exact}.}
\label{fig:matrices}
\end{figure*}

\paragraph*{Hierarchically organized neurons.}
Instead, take the neurons to be organized hierarchically. For $l\ge1$ let $G_l$ be the set
of numbers of the form
\begin{equation}
i=i_0+i_1p+\dots+i_{l-1}p^{l-1},
\end{equation}
where the $i_k$ are $p$-adic digits, i.e.\ elements of $\{0,1,\dots,p-1\}$, and let $G_l^N$
be the Cartesian product of $N$ copies of $G_l$, so that the cardinality of $G_l^N$ is
$p^{Nl}$~\cite{ZG2023}. The elements of $G_l^N$ are representatives of the quotient
$\Zp^N/p^l\Zp^N$, and
\begin{equation}
\Zp^N=\bigsqcup_{i\in G_l^N}\big(i+p^l\Zp^N\big).
\label{eq:Gl_union}
\end{equation}
Geometrically, $G_l$ is a rooted tree with $l+1$ layers in which each vertex has $p$
children: the truncations $i\bmod p^k$, $k=0,\dots,l$, are its vertices, and the elements
of $G_l$ are its leaves (layer $l$). The layer $k=0$ is the root of the tree; see
Fig.~\ref{fig:tree}. The set $G_l^N$ can be visualized as a forest of $N$
identical rooted trees; a neuron is an $N$-tuple of leaves, one from each tree. Each
neuron $i\in G_l^N$ is identified with the ball $B_{-l}^N(i)=i+p^l\Zp^N$ of radius $p^{-l}$ (we use the notation
$B_r^N(a)$ of Appendix~\ref{app:padic} for the ball of radius $p^r$ with center $a$). For
$N=1$, two neurons $i,j$ share their first $k$ digits exactly when $|i-j|_p=p^{-k}$, so the
ultrametric distance measures how far up the tree one must climb to reach their common
ancestor (Fig.~\ref{fig:matrices}). For $N>1$, $\norm{i-j}=p^{-k}$ means that $k$ is the
number of leading digits shared in every coordinate, since
$\norm{x}=p^{-\mathrm{ord}(x)}$ with $\mathrm{ord}(x)=\min_i\mathrm{ord}(x_i)$
(Appendix~\ref{app:padic}). B-C's homogeneity hypothesis, that $w_{ij}$
depends only on the distance between $i$ and $j$~\cite{BuiceCowan2007}, becomes
\begin{equation}
w_{ij}=w(\norm{i-j})\qquad(i\neq j),
\label{eq:wij_hier}
\end{equation}
for a radial kernel $w\in L^1$, and it is invariant under the translations of the finite
group $G_l^N$. For the self-coupling of the smallest structure we set
\begin{equation}
w_{ij}=p^{Nl}\int_{B_{-l}^N(j)}w(\norm{x_i-y})\,d^Ny ,
\label{eq:wij}
\end{equation}
which agrees with Eq.~\eqref{eq:wij_hier} for $i\neq j$ (by ultrametricity
$\norm{x_i-y}=\norm{i-j}$ for $y\in B_{-l}^N(j)$). For $i=j$ it defines $w_{ii}$ as the average
coupling inside $B_{-l}^N(i)$. Excluding self-coupling, as in Ref.~\cite{ZG2023}, changes the
theory by a term of order $\int_{\norm{y}\le p^{-l}}w\,d^Ny$, which vanishes as
$l\to\infty$.

\paragraph*{The continuum limit of the hierarchical network.}
Let $\mathcal D^l\subset\mathcal D(\Zp^N)$ be the space of functions constant on the balls
$B_{-l}^N(i)$, $i\in G_l^N$, supported in the $N$-dimensional unit ball $\Zp^N$. Then
$\dim\mathcal D^l=p^{Nl}$, and
\begin{equation}
\mathcal D^l\hookrightarrow\mathcal D^{l+1}\hookrightarrow\mathcal D(\Zp^N)=\bigcup_l\mathcal D^l .
\end{equation}
Here, $\mathcal D(\Zp^N)$ denotes the space of locally constant functions with support in
the $N$-dimensional unit ball.
Each element of $G_l^N$ represents the smallest structure of the network, a population whose
size is proportional to its volume $p^{-Nl}$. This is the same interpretation as B-C's
effective spikes, which count active neurons in a small region. We therefore relate the
discrete variables to densities by
\begin{align}
\varphi_i&=p^{-Nl}\varphi(x_i), & \tphi_i&=\tphi(x_i),\nonumber\\
\bar n_i&=p^{-Nl}\bar n(x_i), &&
\label{eq:densities}
\end{align}
with $\varphi,\tphi,\bar n\in\mathcal D^l$ and $x_i\in B_{-l}^N(i)$. This is the analogue of the
factors $a^N$ in the Euclidean limit. The firing-rate function must be scaled in the same
way. In the master equation~\eqref{eq:master} and in the action~\eqref{eq:discrete_action}
of the network $G_l^N$ we take, for all neurons, the firing-rate function
\begin{equation}
f_{(l)}=p^{-Nl}f ,
\label{eq:f_scaling}
\end{equation}
where $f$ is the firing-rate function of the continuum theory: the rate at which effective
spikes are produced in the ball $B_{-l}^N(i)$ is the rate per unit volume, $f$, times the volume
$p^{-Nl}$ of the ball. The function $f_{(l)}$ depends on the depth $l$ of the network but
not on the neuron. At fixed $l$ the factor $p^{-Nl}$ can be absorbed into $f$; it matters
only for $l\to\infty$.

\begin{proposition}[Exact discretization]\label{prop:exact}
Let $w$ be radial and integrable, and let $S$ be the action~\eqref{eq:action} below with
$\Qp^N$ replaced by $\Zp^N$. For fields $\varphi(\cdot,t),\tphi(\cdot,t),\bar n\in\mathcal D^l$
and the identifications~\eqref{eq:wij}, \eqref{eq:densities} and~\eqref{eq:f_scaling},
\begin{equation}
S_{G_l^N}[\varphi_i,\tphi_i]=S[\varphi,\tphi]\big|_{\mathcal D^l}.
\label{eq:exact_restriction}
\end{equation}
\end{proposition}
\begin{proof}
For $g\in\mathcal D^l$, $\int_{\Zp^N}g\,d^Nx=p^{-Nl}\sum_{i\in G_l^N}g(x_i)$ exactly. This turns
the terms $\tphi_i\partial_t\varphi_i$, $\alpha\tphi_i\varphi_i$ and $\bar n_i\tphi_i(0)$
into the corresponding integrals. The product $\tphi\varphi$ lies in $\mathcal D^l$, and
translation invariance of the Haar measure gives $w\star\mathcal D^l\subset\mathcal D^l$ with
$(w\star g)(x_i)=\sum_jp^{-Nl}w_{ij}\,g(x_j)$. Hence the argument of $f$ in
Eq.~\eqref{eq:discrete_action} equals $\big(w\star(\tphi\varphi+\varphi)+I\big)(x_i)$, and
$\sum_i\tphi_i\,f_{(l)}(\cdot)=p^{-Nl}\sum_i\tphi(x_i)\,f(\cdot)=\int\tphi\,f(\cdot)\,d^Nx$.
\end{proof}

Proposition~\ref{prop:exact} is not specific to the present model. The fact that the
discrete theory is an exact restriction of the continuum theory to a finite-dimensional
space of test functions is a common feature of $p$-adic statistical field theories; see
Refs.~\cite{ZG2023,ZG2024,ZG2026,ZHZ2023}. We include the statement and its proof here
because the rest of the paper depends on it: it is the core of the construction. For a
hierarchically
organized network, the passage to the continuum replaces the sums over $G_l^N$ by
integrals with respect to the Haar measure of $\Zp^N$. The neurons become the points of
$\Zp^N$ (for $N=1$, the leaves of an infinite rooted tree). To reach $\Qp^N$, use
$p^{-l}\Zp^N/p^l\Zp^N$ in place of $\Zp^N/p^l\Zp^N$. For $l\ge1$,
\begin{equation}
p^{-l}\Zp^N=\bigsqcup_{i\in\widetilde G_l^N}\big(i+p^l\Zp^N\big),
\end{equation}
where $\widetilde G_l$ is the set of numbers of the form
$i=i_{-l}p^{-l}+\dots+i_0+i_1p+\dots+i_{l-1}p^{l-1}$, with $p$-adic digits $i_k$, and
$\widetilde G_l^N$ is the Cartesian product of $N$ copies of $\widetilde G_l$, of
cardinality $p^{2Nl}$. Multiplying $\widetilde G_l$ by $p^l$ one obtains the finite rooted
tree $G_{2l}$. The finite network is therefore of the same kind as before, with depth
$2l$, and the balls $p^{-l}\Zp^N$ exhaust $\Qp^N$ as $l\to\infty$. Note that the points of
$\Qp^N$ are not the leaves of a rooted tree: since the balls grow without bound, there is
no root. The points of $\Qp^N$ are organized in a tree-like structure, which is the union
of the rooted trees of the balls $p^{-l}\Zp^N$.

In the Euclidean setting, the thermodynamic limit (infinite volume at fixed spacing) and
the continuum limit (spacing tending to zero at fixed volume) are logically independent,
and a lattice of spacing $a$ approximates the continuum only as $a\to0$. In
Ref.~\cite{BuiceCowan2007} the two limits are taken together and not distinguished. In the
$p$-adic setting they are a single limit. On $\Zp^N$, the network $G_l^N$ has $p^{Nl}$
neurons and spacing $p^{-l}$, so $l\to\infty$ sends the number of neurons to infinity and
the spacing to zero simultaneously, at fixed volume. The exhaustion by the balls
$p^{-l}\Zp^N$ makes the volume grow and the spacing shrink with the same parameter $l$.
Moreover, by Proposition~\ref{prop:exact}, every step of this limit is exact.

In general, the neuron space is a subset of $\Qp^N$, determined by the support of $w$. Its
geometry can be very general, but it is always hierarchical. In general one may consider
a network topology consisting of a finite number of clusters, each of them a finite
rooted tree, with connections between the clusters. In the simulations of
Secs.~\ref{sec:simulations} and~\ref{sec:smallworld} we use a single finite rooted tree,
which is acceptable and convenient.

A hierarchical network is not a tree. A tree is a particular architecture: a graph whose
edges are the connections. Hierarchical architectures are tree-like structures, topologies
made of trees, such as a single rooted tree ($\Zp^N$), a union of rooted trees with no root
($\Qp^N$), or a finite number of connected clusters. In them the trees organize the neurons;
they are not the connections. The neurons are the leaves, the internal vertices are
modules (balls), and the couplings~\eqref{eq:wij_hier} join every pair of neurons, with a
strength fixed by the level of their nearest common ancestor
[Fig.~\ref{fig:matrices}(a)]. In particular, the model is not directed percolation on a
Cayley tree (Bethe lattice). There, the units sit on all the vertices, activity spreads
along the edges, and the number of sites within graph distance $d$ grows exponentially in
$d$. That lattice is effectively infinite-dimensional, and DP on it is mean-field. Here
the ball of radius $p^{k}$ has Haar volume $p^{kN}$, so the neuron space has finite
dimension $N$. This is why the upper critical dimension $N_c=2\sigma$ of Sec.~\ref{sec:RG}
is finite, and why fluctuation effects appear for $N<2\sigma$.

There is also a structural
difference from the Euclidean case. On $a\mathbb Z^N$ the lattice action agrees with the
continuum action only as $a\to0$, because Riemann sums are approximations and convolution
does not preserve piecewise-constant functions. On $G_l^N$ the identity
\eqref{eq:exact_restriction} holds \emph{for every $l$}. The discrete theory is the
continuum theory restricted to the finite-dimensional space $\mathcal D^l$.

\subsection{The continuum action}

By the results of Sec.~\ref{sec:geometry}, in the limit $l\to\infty$, the discrete action
$S_{G_l^N}[\varphi_i,\tphi_i]$ tends to the continuous action of the $p$-adic effective
spike model
\begin{align}
S[\varphi,\tphi]=&\int_{\Qp^N}\!d^Nx\!\int_0^\infty\!dt\,\Big\{\tphi\,\partial_t\varphi+\alpha\tphi\varphi\nonumber\\
&\qquad-\tphi\,f\big(w\star(\tphi\varphi+\varphi)+I\big)\Big\}\nonumber\\
&-\int_{\Qp^N}d^Nx\,\bar n(x)\,\tphi(x,0),
\label{eq:action}
\end{align}
with $(w\star g)(x)=\int_{\Qp^N}w(\norm{x-y})\,g(y)\,d^Ny$. Here $\bar n(x)\ge0$ is the
\emph{initial density of effective spikes}. The initial state is uncorrelated, and the
number of effective spikes in a small ball $B$ is Poisson distributed with mean
$\int_B\bar n(x)\,d^Nx$. It is the continuum limit of the discrete means $\bar n_i$ of
Eq.~\eqref{eq:Poisson_initial}, via $\bar n_i=p^{-Nl}\bar n(x_i)$ [Eq.~\eqref{eq:densities}].
In particular $\langle\varphi(x,0)\rangle=\bar n(x)$, so $\bar n$ is the initial condition
of the mean field (Sec.~\ref{sec:meanfield}). The generating functional, Feynman rules,
effective action and loop expansion of Secs.~III--IV of Ref.~\cite{BuiceCowan2007} are
formally identical.

\subsection{Mean-field theory: the Wilson--Cowan equation}\label{sec:meanfield}

As in Ref.~\cite{BuiceCowan2007}, mean-field theory is the zero-loop
approximation of the theory, i.e.\ the lowest order of the steepest-descent evaluation of
the path integral. In field theory this order is usually called \emph{tree level}, since it
corresponds to Feynman diagrams without closed loops; we avoid this name, because in this
paper ``tree'' always refers to the hierarchical organization of the neurons. At this order the moments are the solutions of the classical equations
of motion $\delta S/\delta\tphi=0$, $\delta S/\delta\varphi=0$, and all loop corrections
are neglected. We write $a(x,t)=\langle\varphi(x,t)\rangle$ and
$\tilde a(x,t)=\langle\tphi(x,t)\rangle$ for the zero-loop means.

The initial-state term of Eq.~\eqref{eq:action} is not integrated over time. To treat it on
the same footing as the other terms, we write it as a source localized at $t=0$:
\begin{align}
&-\int_{\Qp^N}d^Nx\,\bar n(x)\,\tphi(x,0)\nonumber\\
&\quad=-\int_{\Qp^N}\!d^Nx\!\int_0^\infty\!dt\;\bar n(x)\,\delta(t)\,\tphi(x,t).
\label{eq:initial_source}
\end{align}
Here the time integral runs over $[0,\infty)$, as in all the actions of this paper, and
$\delta(t)$ is understood as a unit mass at the endpoint $t=0$, i.e.\
$\int_0^\infty\delta(t)\,g(t)\,dt=g(0)$.
The equation $\delta S/\delta\varphi=0$ is linear and homogeneous in $\tphi$. With the
final condition inherited from the Doi shift, its solution is $\tilde a=0$, reflecting
conservation of probability [cf.\ Eq.~(5) of Ref.~\cite{BuiceCowan2007}]. Setting
$\tphi=\tilde a=0$ in $\delta S/\delta\tphi(x,t)=0$ kills every term in which the
derivative does not act on the explicit prefactor $\tphi$ in front of $f$. The result is
\begin{equation}
\partial_t a(x,t)=-\alpha\,a(x,t)+f\big((w\star a)(x,t)+I\big)+\bar n(x)\,\delta(t),
\label{eq:WC_delta}
\end{equation}
with $a(x,t)=0$ for $t<0$.
Equation~\eqref{eq:WC_delta} is the Wilson--Cowan equation of the effective spike model,
written with the initial state as a source. In the present setting the convolution is taken
over $\Qp^N$ with respect to the Haar measure, and we call it the \emph{hierarchical
Wilson--Cowan equation}; deterministic equations of this type were studied in
Ref.~\cite{ZZ2023WC}.

\paragraph*{The term $\bar n\,\delta(t)$.}
The source $\bar n(x)\delta(t)$ encodes the initial state. Since $a=0$ for $t<0$,
integrating Eq.~\eqref{eq:WC_delta} over a small interval around $t=0$ gives
that $a$ jumps from $0$ to $\bar n(x)$ at $t=0$, and for $t>0$ the source vanishes.
Therefore, for $t\ge0$, Eq.~\eqref{eq:WC_delta} is equivalent to the Cauchy problem
\begin{equation}
\partial_t a=-\alpha a+f(w\star a+I)\quad(t>0),\qquad a(x,0)=\bar n(x),
\label{eq:WC}
\end{equation}
which is the form given in Eq.~(44) of Ref.~\cite{BuiceCowan2007}, with the initial
condition $a(x,0)=\bar n(x)$, and the form used in the rest of the paper. The condition $a=0$ for $t<0$ requires
$f(I)=0$. From now on, we assume that $f(0)=0$ and $I=0$. In
Ref.~\cite{BuiceCowan2007} the external input $I$ is kept general and $f$ is expanded
about $I$. The condition $f(I)=0$ is not stated there; it is only implicit. Section~IV~C
of Ref.~\cite{BuiceCowan2007} considers ``perturbations around a homogeneous equilibrium
solution'' and speaks of ``the absorbing state'', and Sec.~V describes the transition
``from the absorbing (nonfluctuating) fully quiescent state to a spontaneously active
state''. A quiescent absorbing state exists only if $f(I)=0$.

\paragraph*{Mean field as an approximation of the exact equation for the mean.}
Equivalently, taking the expectation of the operator form of the master equation gives
the exact equation
$\partial_t\langle\varphi\rangle=-\alpha\langle\varphi\rangle+\langle f(w\star\varphi+I)\rangle$
for $t>0$ [Eqs.~(111)--(112) of Ref.~\cite{BuiceCowan2009}]. This is not a closed equation
for the mean $\langle\varphi\rangle$: since $f$ is nonlinear, the term
$\langle f(w\star\varphi+I)\rangle$ depends on all the higher moments of $\varphi$, whose
equations involve still higher moments, and so on
(Appendix~\ref{app:master}). Mean-field theory is the approximation
\begin{equation}
\langle f(w\star\varphi+I)\rangle\approx f(w\star\langle\varphi\rangle+I),
\end{equation}
in which the average of $f$ is replaced by $f$ of the average. It amounts to assuming
that all moments factorize into products of the mean, i.e.\ that fluctuations are
neglected, and it turns the exact equation into a closed equation for the mean, which is
Eq.~\eqref{eq:WC} [Eq.~(113) of Ref.~\cite{BuiceCowan2009}]. The loop expansion of
Secs.~\ref{sec:split} and~\ref{sec:loops} replaces this uncontrolled approximation by a
systematic expansion: the one-loop correction adds to the right-hand side of
Eq.~\eqref{eq:WC} a term coupling $a$ to the two-point correlation [Eq.~\eqref{eq:WC1}].

\paragraph*{Discrete version.}
On the finite network $G_l^N$, with the identifications~\eqref{eq:densities} and~\eqref{eq:f_scaling},
Eq.~\eqref{eq:WC_delta} becomes the system
$\dot a_i=-\alpha a_i+f\big(p^{-Nl}\sum_jw_{ij}a_j+I\big)+p^{Nl}\bar n_i\,\delta(t)$,
$i\in G_l^N$, for the densities $a_i=p^{Nl}\langle n_i\rangle$. The space
$\mathcal D^l$ is invariant under the convolution with $w$ and under the pointwise action
of $f$, so this system is exactly the restriction of Eq.~\eqref{eq:WC_delta} to
$\mathcal D^l$: if $\bar n\in\mathcal D^l$, then $a(\cdot,t)\in\mathcal D^l$ for all $t$.
Existence, uniqueness and positivity of the solutions for integrable kernels were
established in Ref.~\cite{ZZ2023WC}. For general initial data the discrete solutions
approximate the continuous one on every finite time interval, with an error controlled
by the error in the initial data. The rigorous construction of path integrals for
actions involving $W$-operators, as limits of finite-dimensional ones, was carried out
in Ref.~\cite{ZG2022RMP}.

\subsection{Free and interacting actions}\label{sec:split}

Reference~\cite{BuiceCowan2007} splits the action into a free part $S_{\rm free}$ and an interacting
part $S_i$ [their Eqs.~(14)--(15)] but gives the two parts only in words. Since the
propagators used below depend on this choice, we state it explicitly. There are two
distinct splits: one about the vacuum $\varphi=\tphi=0$ (used for the perturbative
expansion of Sec.~III of Ref.~\cite{BuiceCowan2007}), and one about the mean field (used
for the loop expansion of Sec.~IV). They coincide only when the mean field vanishes,
$a\equiv0$, i.e.\ when the network is at the absorbing state $\varphi=0$, which requires
$f(I)=0$ [Eq.~\eqref{eq:absorbing}].

\paragraph*{The firing-rate function.}
The firing-rate function $f:\mathbb R\to\mathbb R$ is a function with a convergent Taylor
expansion on the whole real line, nonnegative for $s\ge0$, and satisfying $f(0)=0$,
$f'(0)>0$ and $f''(0)<0$. Nonnegativity is required only for $s\ge0$: since $w\ge0$ and
$I=0$, the input of a neuron is never negative (see Remark~\ref{rem:activation}). A
prototype of such a function is
\begin{equation}
f(s)=f_{\max}\big(1-e^{-s/s_0}\big),\qquad f_{\max},s_0>0 .
\label{eq:f_choice}
\end{equation}
This function is real analytic on the whole real line, so the Taylor
expansion~\eqref{eq:taylor} below converges for every argument and the vertex expansion is
well defined. It also provides the smoothness
that B-C require for the renormalization argument [Sec.~V of Ref.~\cite{BuiceCowan2007}].
Its derivatives at the origin are $f^{(n)}(0)=(-1)^{n+1}f_{\max}/s_0^{\,n}$. In particular,
\begin{equation}
f(0)=0,\quad f'(0)=\frac{f_{\max}}{s_0}>0,\quad f''(0)=-\frac{f_{\max}}{s_0^{2}}<0 .
\label{eq:f_derivatives}
\end{equation}
These are exactly the conditions used in the critical analysis: the condition $f(0)=0$
gives an absorbing state [Eq.~\eqref{eq:absorbing}]; $f'(0)>0$ gives branching and
propagation; and $f''(0)<0$ (saturation) gives the relative minus sign between the cubic vertices of
Eq.~\eqref{eq:RFT}, which B-C identify as necessary for the DP fixed point. With these
values, the distance to criticality is $\mu=\alpha-f'(0)\,\hat w(0)=\alpha-f_{\max}\hat w(0)/s_0$, and B-C's
rescaling of the fields [Eq.~(58) of Ref.~\cite{BuiceCowan2007}] involves the factor
\begin{equation*}
\Big(\frac{2f'(0)}{\hat w(0)\,|f''(0)|}\Big)^{1/2}=\Big(\frac{2s_0}{\hat w(0)}\Big)^{1/2}.
\end{equation*}

\begin{remark}[Firing-rate function versus activation function]\label{rem:activation}
In neural-network and rate models, an \emph{activation function} (a sigmoid, $\tanh$,
ReLU, \dots) maps a unit's (or node's) input to its output (or state), determined by the
inputs connected to the node and their weights. That output can take negative values;
e.g., $\tanh$ represents deviations from a baseline rate in the symmetric range $(-1,1)$.
The function $f$ in the effective spike model differs in three ways.
\begin{enumerate}
\item \emph{It is a transition rate, not a state.} In the master
  equation~\eqref{eq:master}, $f(s_i)$ is the probability per unit time that neuron $i$
  emits another effective spike. It has units of inverse time and must be nonnegative
  wherever it is evaluated. The state of the neuron is the spike count $n_i$, not $f$.
\item \emph{It becomes an activation function only in mean field.} In the Wilson--Cowan
  equation~\eqref{eq:WC}, obtained at zero-loop order, $f$ plays the role of a deterministic
  activation function acting on the mean input $w\star a+I$. Beyond zero-loop order, $f$ enters
  through all its derivatives, as the vertices $V_{n,m}$ of Eq.~\eqref{eq:vertex}, and it
  fixes the statistics of the fluctuations, not only the mean response.
\item \emph{Criticality constrains it at a point.} The requirements $f(I)=0$, $f'(I)>0$ and
  $f''(I)<0$ concern the behavior of $f$ at the operating point, not its global shape. The
  hyperbolic tangent fails them. Nonnegativity forces $I=0$, where $\tanh''(0)=0$. The cubic
  aggregation vertex then vanishes at zero-loop order and the leading saturating term is
  quartic ($\tanh'''(0)=-2$), which is the setting of tricritical directed percolation
  rather than DP.
\end{enumerate}
Nonnegativity and analyticity together have a consequence. A real analytic function with
$f(0)=0$ and $f'(0)>0$ is negative for small $s<0$, so the inputs must satisfy $s\ge0$. We
therefore take $w\ge0$ and $I=0$. As B-C note, inhibition cannot be distinguished from the
decay rate $\alpha$ in the effective spike model~\cite{BuiceCowan2007}, so net inhibition is
represented through $\alpha$. Models with negative weights require a nonnegative extension
of $f$ to $s<0$ (e.g.\ $f=0$ there), which is not analytic at $s=0$.
\end{remark}

\paragraph*{Expansion about the vacuum.}
Expand the firing-rate function about $I=0$:
\begin{equation}
f\big(w\star(\tphi\varphi+\varphi)\big)=\sum_{n\ge1}\frac{f^{(n)}(0)}{n!}
\Big(w\star(\tphi\varphi)+w\star\varphi\Big)^{n}.
\label{eq:taylor}
\end{equation}
The sum starts at $n=1$ because $f(0)=0$.
The \emph{free action} is the part of Eq.~\eqref{eq:action} bilinear in $\varphi$ and $\tphi$:
\begin{align}
S_{\rm free}[\varphi,\tphi]&=\int_{\Qp^N}\!d^Nx\!\int_0^\infty\!dt\;\tphi\,(\partial_t+\alpha)\,\varphi\nonumber\\
&\quad-f'(0)\!\int_{\Qp^N}\!d^Nx\!\int_0^\infty\!dt\;\tphi\,(w\star\varphi)\nonumber\\
&\equiv\int_{\Qp^N}\!d^Nx\!\int_0^\infty\!dt\;\tphi\,L\,\varphi ,
\label{eq:Sf}
\end{align}
with
\begin{equation}
L=\partial_t+\alpha-f'(0)\,w\star .
\label{eq:L}
\end{equation}
\emph{The derivation of Eq.~\eqref{eq:Sf} uses $f(0)=0$.} The zeroth-order term of the
expansion~\eqref{eq:taylor}, $f(0)\int\tphi$, vanishes [Eq.~\eqref{eq:f_derivatives}], and
$f'(0)=f_{\max}/s_0$ for the prototype~\eqref{eq:f_choice}. The condition $f(0)=0$ is
also what makes the vacuum a solution: the classical equations of motion are
$\delta S/\delta\varphi=0$ and $\delta S/\delta\tphi=0$ (Sec.~\ref{sec:meanfield}); for
$\tphi=0$ they reduce to Eq.~\eqref{eq:WC_delta}, and in the absence of initial spikes
($\bar n=0$) the vacuum $\varphi=\tphi=0$ solves them if and only if $f(0)=0$. Then the
action splits as $S=S_{\rm free}+S_i$, with $S_{\rm free}$ the purely bilinear
form~\eqref{eq:Sf}, and $S_i$ contains no term linear in $\tphi$ other than the initial
state. If $f(0)\neq0$, the action contains in addition the source $-f(0)\int\tphi$. The vacuum
is then not stationary, and Eq.~\eqref{eq:Sf} is no longer the free action for
fluctuations about a solution. This source cannot be absorbed into a normalization
constant: after the Doi shift every term of Eq.~\eqref{eq:action} contains at least one
factor of $\tphi$, so $S[\varphi,0]=0$ and $Z[0,0]=1$ (conservation of probability), and a
term linear in $\tphi$ shifts the mean activity. In that case one must expand about the
mean field instead (see below).

The \emph{interacting action} is the remainder, $S_i=S-S_{\rm free}$. Using the binomial theorem in
Eq.~\eqref{eq:taylor},
\begin{align}
S_i[\varphi,\tphi]=&\;{\sum_{n\ge1}\sum_{m=0}^{n}}{}^{'}\;V_{n,m}[\varphi,\tphi]
 -\int_{\Qp^N}d^Nx\,\bar n(x)\,\tphi(x,0),
\label{eq:Si}\\
V_{n,m}=&-\frac{f^{(n)}(0)}{n!}\binom{n}{m}\nonumber\\
&\times\int_{\Qp^N}\!d^Nx\!\int_0^\infty\!dt\;\tphi\,\big(w\star(\tphi\varphi)\big)^{m}
 \big(w\star\varphi\big)^{n-m}.
\label{eq:vertex}
\end{align}
The prime on the sum means that the term $(n,m)=(1,0)$ is omitted, since it is already
contained in $S_{\rm free}$. The vertex $V_{n,m}$ has $n$ incoming legs (factors of $\varphi$) and
$m+1$ outgoing legs (factors of $\tphi$). This is the vertex structure described in
Ref.~\cite{BuiceCowan2007}, where ``the $n=1$, $m=0$ term has already been incorporated
into the propagator.'' The only term of Eq.~\eqref{eq:Si} that is a source rather than an interaction is the
initial-state term $-\int\bar n\,\tphi(\cdot,0)$, which has no time integral and only
outgoing legs (the bold dots in the figures of Ref.~\cite{BuiceCowan2007}). The term
$n=0$ of the expansion, $-f(0)\int\!d^Nx\!\int_0^\infty\!dt\,\tphi$, would be a second
source describing spontaneous firing (see above); it is absent because $f(0)=0$.

\paragraph*{Fourier conventions on $\Qp^N$.}
We use $\hat g(\xi)=\int_{\Qp^N}\chi_p(\xi\cdot x)\,g(x)\,d^Nx$, with the Haar measure
normalized as in Sec.~\ref{sec:prelim}. This measure is self-dual, so the inverse is
$g(x)=\int_{\Qp^N}\chi_p(-\xi\cdot x)\,\hat g(\xi)\,d^N\xi$ and the convolution theorem
reads $\widehat{w\star g}=\hat w\,\hat g$. The Dirac distribution $\delta(x)$ on $\Qp^N$
has $\hat\delta=1$. For a real radial kernel $w$, $\hat w$ is real and radial, because
$w$ is invariant under $x\mapsto-x$.

\paragraph*{Zero-loop propagator of $S_{\rm free}$.}
The propagator $\Delta(x,t)=\langle\varphi(x,t)\tphi(0,0)\rangle_{S_{\rm free}}$ is the Green's
function of $L$: $L\Delta=\delta(x)\,\delta(t)$ [Eq.~(15) of Ref.~\cite{BuiceCowan2007}].
In Fourier variables $\partial_t\hat\Delta=-\lambda_0(\xi)\hat\Delta+\delta(t)$. Hence
\begin{equation}
\hat\Delta(\xi,t)=\theta(t)\,e^{-\lambda_0(\xi)t},\qquad
\lambda_0(\xi)=\alpha-f'(0)\,\hat w(\xi),
\label{eq:Delta0}
\end{equation}
where $\theta$ denotes the Heaviside step function, $\theta(t)=1$ for $t>0$ and
$\theta(t)=0$ for $t<0$, with the convention $\theta(0)=0$. This convention is the one of
Ref.~\cite{BuiceCowan2007}. It follows from normal ordering of the operators in the
Hamiltonian, i.e.\ it is the It\^o prescription, and it implies that the propagator does
not connect a field to itself at equal times.

\paragraph*{A note on the ``response field''.}
The name of $\tphi$ can be misread. The field $\tphi$ is not the response of the network: it
marks the point where a stimulus is applied, and $\varphi$ the point where the response is
measured. Injecting extra spikes at rate $h(x',t')$ adds the term $\int h\,\tphi$ to $-S$,
so $h$ enters as a source of $\tphi$, and
$\delta\langle\varphi(x,t)\rangle/\delta h(x',t')=\langle\varphi(x,t)\tphi(x',t')\rangle=\Delta(x-x',t-t')$,
which vanishes unless $t>t'$. In particular, $\langle\varphi(x',t')\tphi(x,t)\rangle=0$ for
$t>t'$, since an average in which $\tphi$ carries the latest time vanishes.

\begin{figure*}[t]
\centering
\includegraphics[width=0.92\textwidth]{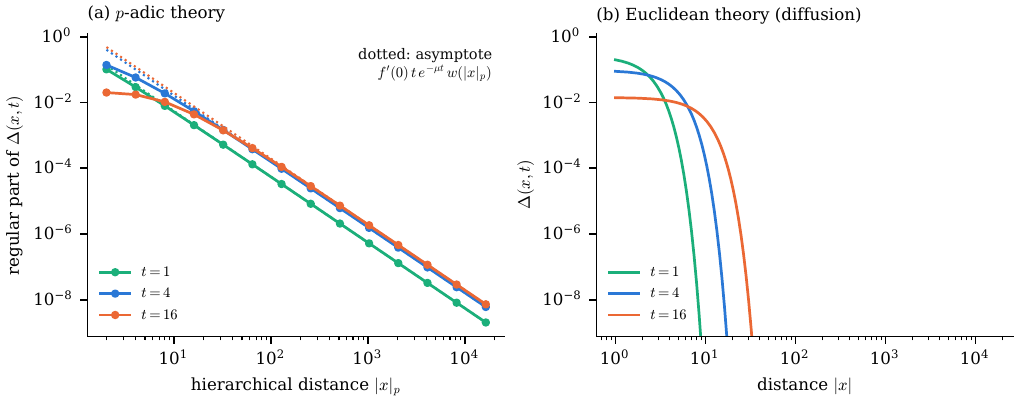}
\caption{Zero-loop propagator at three times. (a) Regular part $e^{-\alpha t}K_t(x)$ of the
$p$-adic propagator~\eqref{eq:Delta_decomp}, computed from the explicit
formula~\eqref{eq:Kt_explicit} of Appendix~\ref{app:integrals}, for $p=2$, $N=1$, a kernel that is constant for
$|x|_p\le1$ and decays as $|x|_p^{-(1+\sigma)}$ with $\sigma=1$ beyond, $\alpha=1$ and
$f'(0)w_0=0.9$ (so $\mu=0.1$). Dotted lines: the asymptote~\eqref{eq:Delta_tail}. The
computation is for the continuum theory on $\Qp$, not for a finite tree: the series
of Eq.~\eqref{eq:Kt_explicit} are summed over $300$ levels, which is numerically exact.
The neuron space $\Qp$ is the disjoint union of infinitely many unit balls; it is not
compact, and it is not a regional network in the sense of Lemma~\ref{lem:regional}, which
requires finitely many regions. Restricting it to the ball of radius $2^{14}$ centered at the origin,
$\{|x|_2\le2^{14}\}=2^{-14}\mathbb Z_2$, which is compact and is the disjoint union of
the $2^{14}$ unit balls $i+\mathbb Z_2$, $i\in2^{-14}\mathbb Z_2/\mathbb Z_2$,
with the same $\alpha$ and $f'(0)$, changes the plotted values by less than $2\times10^{-5}$
and only cuts off the tail at the size of the network.
The points are the values at $|x|_p=2^k$, $k=1,\dots,14$, i.e.\ at the $14$ levels of
the hierarchy above the unit ball $\Zp$, on which the kernel is constant. (b) The
Euclidean propagator of the B-C theory on the real line,
$e^{-\mu t}(4\pi Dt)^{-1/2}e^{-x^2/4Dt}$, with the same $\mu$ and $D=1$, for
$1\le|x|\le2^{14}$. In both panels the unit of distance is the short-distance scale of
the model (the radius of the unit ball in (a), the length $\sqrt{D/\alpha}$ in (b)),
and a ball of radius $p^k$ has volume $p^k$ in $\Qp$ (radii are powers of $p$), while an
interval of radius $r$ has length $2r$ in $\mathbb R$, so the two
horizontal scales measure comparable numbers of neurons. The $p$-adic propagator has a
power-law tail at every time; the Euclidean one is cut off at the diffusion length
$\sqrt{4Dt}$.}
\label{fig:propagator}
\end{figure*}

\paragraph*{The propagator in position space.}
Since $w$ is integrable, the Riemann--Lebesgue theorem implies that $\hat w(\xi)\to0$ as
$\norm{\xi}\to\infty$, so $\lambda_0(\xi)\to\alpha$. Hence, for each fixed $t>0$,
$e^{-\lambda_0(\xi)t}\to e^{-\alpha t}\neq0$ as $\norm{\xi}\to\infty$: as a function of
$\xi$, $e^{-\lambda_0(\xi)t}$ tends to a nonzero constant and is therefore not integrable
over $\Qp^N$, and the inverse Fourier transform must be taken in the sense of
distributions:
\begin{equation}
\Delta(x,t)=\theta(t)\,\mathcal F^{-1}_{\xi\to x}\big(e^{-\lambda_0(\xi)t}\big),
\label{eq:Delta0x}
\end{equation}
where $\mathcal F$ is the Fourier transform on the space $\mathcal D'(\Qp^N)$ of
distributions on $\Qp^N$. Writing
$e^{-\lambda_0(\xi)t}=e^{-\alpha t}\big[1+\big(e^{f'(0)t\,\hat w(\xi)}-1\big)\big]$ and
using $\mathcal F^{-1}(\hat w^{\,n})=w^{\star n}$ gives
\begin{align}
\Delta(x,t)&=\theta(t)\,e^{-\alpha t}\big[\delta(x)+K_t(x)\big],\nonumber\\
K_t&=\sum_{n\ge1}\frac{\big(f'(0)\,t\big)^n}{n!}\,w^{\star n}.
\label{eq:Delta_decomp}
\end{align}
The series converges in $L^1(\Qp^N)$, since $\Vert w^{\star n}\Vert_1\le\Vert w\Vert_1^n$,
so $K_t$ is an integrable function. The two terms of $\Delta(x,t)$ in
Eq.~\eqref{eq:Delta_decomp} have a direct interpretation. The
$\delta$ term is the injected spike decaying in place at rate $\alpha$. $K_t$ sums the
chains of $n$ synaptic transmissions. Because $e^{f'(0)t\hat w}-1$ is radial, $K_t$ can be
computed explicitly as a sum over spheres (Proposition~\ref{prop:propagator} of
Appendix~\ref{app:integrals}). At large hierarchical distance, the propagator is proportional to the
kernel $w$: it decays with the distance exactly as the coupling between neurons does,
\begin{equation}
\Delta(x,t)\simeq\theta(t)\,f'(0)\,t\,e^{-\lambda_0(0)t}\,w(\norm{x}),\qquad\norm{x}\to\infty,
\label{eq:Delta_tail}
\end{equation}
for nonnegative, radially nonincreasing $w$ (Proposition~\ref{prop:propagator}). The $p$-adic
numbers appear in Eqs.~\eqref{eq:Delta0}--\eqref{eq:Delta_tail} only through the fact that the
convolution and the Fourier transform are taken over $\Qp^N$. The decomposition~\eqref{eq:Delta_decomp}
holds in the Euclidean theory as well; what is specific to $\Qp^N$ is the explicit
radial evaluation and the exact tail~\eqref{eq:Delta_tail}. Figure~\ref{fig:propagator}
compares the two propagators. The comparison already suggests why the B-C theory in its
diffusion limit is not compatible with the small-world property: its propagator is cut
off at the diffusion length, so that a spike has no influence on distant neurons at short
times, whereas the $p$-adic propagator reaches all hierarchical distances at every time. A
power-law tail is necessary for the small-world property but not sufficient; the question
is examined in Sec.~\ref{sec:smallworld}.

\paragraph*{Ultraviolet behavior.}
The $\delta$ term in Eq.~\eqref{eq:Delta_decomp} controls the ultraviolet behavior. A product of
propagators at coinciding points, such as $\Delta(x-y,\tau)^2$ in a bubble diagram,
contains $\delta^2$ and is not defined. Correspondingly, the momentum integral
$\int e^{-\lambda_0(q)\tau}e^{-\lambda_0(\eta-q)\tau}\,d^Nq$ diverges, because its integrand
tends to $e^{-2\alpha\tau}$ as $\norm{q}\to\infty$. Such diagrams are finite only when the
vertices supply factors of $\hat w$, as in the loop terms of B-C, where every vertex acts
through $w\star$, or with an ultraviolet cutoff. The finite network $G_l^N$ provides the
cutoff naturally: on $\mathcal D^l$ the distribution $\delta$ is replaced by
$p^{Nl}\,\Omega(p^l\norm{x})$, and momenta are restricted to $\norm{\xi}\le p^{l}$. This makes
precise B-C's remark that the integrals are ``well behaved in the ultraviolet because of the
weight function, if not by a neural lattice cutoff''~\cite{BuiceCowan2007}.

\paragraph*{Expansion about the mean field.}
For the loop expansion, Ref.~\cite{BuiceCowan2007} shifts the fields by their means,
$\varphi\to a+\varphi$, $\tphi\to\tilde a+\tphi$, with $a(x,t)=\langle\varphi(x,t)\rangle$
and $\tilde a=0$. After the shift, $\varphi$ and $\tphi$ denote the fluctuations about the
mean. The action then takes the form $S=S_0[a]+S_F[a;\varphi,\tphi]$ [Eqs.~(33)--(35) of
Ref.~\cite{BuiceCowan2007}], where $S_0$ is the action evaluated at the mean and $S_F$,
the action of the fluctuations, contains all the terms that depend on $\varphi$ and
$\tphi$. Let
\begin{equation}
s(x,t)=(w\star a)(x,t)+I .
\label{eq:s}
\end{equation}
The argument of $f$ becomes $s+w\star\varphi+w\star(a\tphi)+w\star(\tphi\varphi)$.
We expand $f$ in a Taylor series about $s$ (not about $I$), multiply by $-\tphi$, and
collect the terms of $S_F$ according to their degree in the fluctuation fields $\varphi$
and $\tphi$:
\begin{enumerate}
\item The term linear in $\tphi$ collects the drift and the initial-state
  term, Eq.~\eqref{eq:initial_source}:
  $\int\!d^Nx\!\int_0^\infty\!dt\;\tphi\,\big[\partial_ta+\alpha a-f(s)-\bar n\,\delta(t)\big]$. It vanishes
  at zero-loop order, because $a$ then solves Eq.~\eqref{eq:WC_delta}. In this way the
  initial-state term is absorbed into the equation that determines $a$: it cancels
  against the other terms of the bracket and does not remain in the shifted action.
  After the shift, $\bar n$ no longer appears in the action as a source; it enters only
  through $a$, i.e.\ through $s=w\star a+I$.
\item The terms of second degree, which we denote by $S_F^{(2)}$, are
\begin{align}
S_F^{(2)}=&\int\!d^Nx\!\int_0^\infty\!dt\;\Big\{\tphi\,(\partial_t+\alpha)\varphi
 -f'(s)\,\tphi\,(w\star\varphi)\Big\}\nonumber\\
&-\int\!d^Nx\!\int_0^\infty\!dt\;f'(s)\,\tphi\,w\star(a\tphi).
\label{eq:SF2}
\end{align}
The first line defines the response propagator of the shifted theory,
$\Delta_a(x,t;x',t')=\langle\varphi(x,t)\tphi(x',t')\rangle$ [Eq.~(47) of
Ref.~\cite{BuiceCowan2007}]:
\begin{multline}
\partial_t\Delta_a(x,t;x',t')=-\alpha\Delta_a(x,t;x',t')+\delta(x-x')\,\delta(t-t')\\
+f'\big(s(x,t)\big)\,(w\star\Delta_a)(x,t;x',t'),
\label{eq:Delta_a}
\end{multline}
where $w\star$ acts on the first spatial argument. When $a$ is not
homogeneous, i.e.\ when $a(x,t)$ depends on the position $x$, the coefficient
$f'(s(x,t))$ depends on $x$ and $\Delta_a$ is not translation invariant: it is not a
function of $x-x'$ alone, which is why it is written as a function of both points. Equation~\eqref{eq:Delta_a}
contains no term $\bar n\,\delta(t)$, for two reasons. First, the initial-state term has
already been absorbed into the equation for the mean [item 1]. Second, $\Delta_a$ is a
\emph{response function}: it is the solution of the linearization of
Eq.~\eqref{eq:WC_delta} about $a$, driven by a unit impulse $\delta(x-x')\delta(t-t')$
applied at $(x',t')$. The initial state is part of the background $a$ about which one
linearizes, not of the perturbation. The inhomogeneous term $\bar n\,\delta(t)$ of
Eq.~\eqref{eq:WC_delta} drops out of the linearization because it does not depend on $a$.
The second line of Eq.~\eqref{eq:SF2} has two factors of $\tphi$ and no $\varphi$. It
is not part of the propagator. Following Ref.~\cite{BuiceCowan2007}, we treat it as the $(2,0)$ vertex
\begin{equation}
-f'(s(x))\,\tphi(x)\int_{\Qp^N}w(\norm{x-y})\,a(y)\,\tphi(y)\,d^Ny ,
\label{eq:Gamma20}
\end{equation}
which is Eq.~(54) of B-C; it generates the zero-loop two-point correlation [their
Eq.~(55)] and the source term
in Eq.~\eqref{eq:C} below.
\item Cubic and higher terms, such as $-f'(s)\,\tphi\,w\star(\tphi\varphi)$ and
  $-\tfrac12f''(s)\,\tphi\,(w\star\varphi)^2$, are the vertices of the loop expansion. The
  latter produces the one-loop term in Eq.~\eqref{eq:WC1}.
\end{enumerate}

\paragraph*{Homogeneous states and the absorbing state.}
For a homogeneous stationary mean field $a(x,t)=a^*$ we have $s=\hat w(0)a^*+I$, where
$a^*$ solves $\alpha a^*=f(\hat w(0)a^*+I)$. Eq.~\eqref{eq:Delta_a} then gives
\begin{align}
\hat\Delta_{a^*}(\xi,t)&=\theta(t)\,e^{-\lambda_{a^*}(\xi)t},\nonumber\\
\lambda_{a^*}(\xi)&=\alpha-f'\big(\hat w(0)a^*+I\big)\,\hat w(\xi).
\label{eq:Delta_astar}
\end{align}
In general $\lambda_{a^*}\neq\lambda_0$: the propagator of the shifted theory is not the
propagator~\eqref{eq:Delta0} of the free action $S_{\rm free}$ of Eq.~\eqref{eq:Sf}, which was
obtained by expanding about the vacuum. The two propagators coincide when $a^*=0$. However, $a^*=0$ is a stationary state
only if
\begin{equation}
f(I)=0 ,
\label{eq:absorbing}
\end{equation}
in which case there is no constant source and $\varphi=0$ is an absorbing state. We
recall that an absorbing state is a state that the dynamics can reach but cannot leave:
if no neuron has an effective spike and $f(I)=0$, no spike is ever created, and the
network remains silent forever. If
$f(I)>0$, the constant source $-f(I)\int\!d^Nx\!\int_0^\infty\!dt\,\tphi$ drives spontaneous
activity, there is no absorbing state, and the
DP-type analysis of Secs.~\ref{sec:RG}--\ref{sec:avalanches} does not apply directly.
Condition~\eqref{eq:absorbing} is implicit in the critical analysis of
Ref.~\cite{BuiceCowan2007}, which treats ``the absorbing (nonfluctuating) fully quiescent
state''. We impose it explicitly whenever we discuss criticality.

\section{Linear response: $p$-adic diffusion}\label{sec:linear}

We linearize about a homogeneous stationary state $a^*$. By Sec.~\ref{sec:split}, the
relevant propagator is that of the shifted theory, Eq.~\eqref{eq:Delta_astar}. For the
critical analysis below we take the absorbing state $a^*=0$, which requires $f(I)=0$
[Eq.~\eqref{eq:absorbing}]; for $I=0$ this is the condition $f(0)=0$. Then
\begin{equation}
\hat\Delta(\xi,t)=\theta(t)\,e^{-\lambda(\xi)t},\qquad
\lambda(\xi)=\lambda_0(\xi)=\alpha-f'(0)\,\hat w(\xi),
\label{eq:propagator}
\end{equation}
and Eq.~\eqref{eq:propagator} is also the zero-loop propagator~\eqref{eq:Delta0} of the free
action $S_{\rm free}$.

In real space the same result is obtained by separation of variables, in any dimension
$N$. The linearized equation is
\begin{equation}
\partial_tu(x,t)+\alpha\,u(x,t)-f'(0)\,(w\star u)(x,t)=0 ,
\label{eq:linearized}
\end{equation}
which is $Lu=0$ with the operator $L$ of Eq.~\eqref{eq:L}. We look for
solutions of the form $u(x,t)=\psi_{\gamma nj}(x)\,T(t)$, where $\psi_{\gamma nj}$ is the
wavelet defined in Eq.~\eqref{eq:wavelets}. By Eq.~\eqref{eq:wavelet_eig},
\begin{equation}
\partial_tT(t)+\alpha\,T(t)-f'(0)\,\hat w\big(p^{1-\gamma}\big)\,T(t)=0 ,
\end{equation}
so that $T(t)=T(0)\,e^{-\lambda(p^{1-\gamma})t}$, where $\lambda(p^{1-\gamma})$ denotes the
value of $\lambda(\xi)$ for $\norm{\xi}=p^{1-\gamma}$. Since the wavelets form an
orthonormal basis of $L^2(\Qp^N)$, the solution of the Cauchy problem for
Eq.~\eqref{eq:linearized} with initial datum $u(x,0)=u_0(x)\in L^2(\Qp^N)$ is
\begin{equation}
u(x,t)=\sum_{\gamma,n,j}c_{\gamma nj}\,e^{-\lambda(p^{1-\gamma})t}\,\psi_{\gamma nj}(x),
\qquad t\ge0,
\label{eq:Cauchy_wavelets}
\end{equation}
with $c_{\gamma nj}=\int_{\Qp^N}u_0(x)\,\overline{\psi_{\gamma nj}(x)}\,d^Nx$. The
linear response thus decomposes into independent \emph{hierarchical modes}, one for each
ball, with relaxation rate set by the level of the ball.

\subsection{The operator selected by the connectivity}\label{sec:Wop}

Recall that $\mu=\alpha-f'(0)\,\hat w(0)$ is the bare distance to criticality (B-C denote
this bare value by $\mu_0$ and reserve $\mu$ for the renormalized parameter), and introduce the operator
\begin{equation}
\boldsymbol W\phi(x)=f'(0)\!\int_{\Qp^N}\!\big(\phi(x)-\phi(x-y)\big)\,w(\norm{y})\,d^Ny
\label{eq:Wop}
\end{equation}
for $\phi\in\mathcal D(\Qp^N)$. This operator is pseudodifferential with radial symbol $A_w(\xi)=A_w(\norm{\xi})$:
\begin{equation}
\boldsymbol W\phi(x)=\mathcal F^{-1}_{\xi\to x}\big(A_w(\norm{\xi})\,\mathcal F_{x\to\xi}\phi\big),
\label{eq:Wpseudo}
\end{equation}
where
\begin{align}
A_w(\xi)&=f'(0)\big(\hat w(0)-\hat w(\xi)\big)\nonumber\\
&=f'(0)\!\int_{\Qp^N}w(\norm{y})\,\big(1-\chi_p(\xi\cdot y)\big)\,d^Ny .
\label{eq:Wsymbol}
\end{align}
With these definitions, the relaxation rate $\lambda(\xi)=\lambda_0(\xi)$ of
Eqs.~\eqref{eq:propagator} and~\eqref{eq:Delta0} can be written exactly, for every
$\xi\in\Qp^N$, as
\begin{equation}
\lambda(\xi)=\mu+A_w(\xi).
\label{eq:lambdaW}
\end{equation}
Operators of this form, and the corresponding heat equations
$\partial_tu+\boldsymbol Wu=0$, are the $W$-operators studied in
Refs.~\cite{ZunigaLNM2016,ZG2022Ultrametric}. Their symbols are nonnegative and radial, and the heat semigroups are
Markovian. The free part of the action is therefore
$\tphi(\partial_t+\mu+\boldsymbol W)\varphi$ \emph{with no approximation}. The linear
dynamics of the $p$-adic B-C theory is the heat equation determined by the synaptic kernel
$w$, not by an independent choice of ``Laplacian''.

The operator $\boldsymbol W$ of Eq.~\eqref{eq:Wop} is also defined over $\mathbb R^N$,
replacing $\norm{y}$ by the Euclidean norm $|y|$, and its symbol is given by
Eq.~\eqref{eq:Wsymbol} with $\chi_p(\xi\cdot y)$ replaced by $e^{iq\cdot y}$. Then
$\lambda(q)=\mu+A_w(q)$ holds exactly, and at $\mu=0$ the linearized equation
\begin{equation}
\partial_tu(x,t)+\boldsymbol Wu(x,t)=0,\qquad x\in\mathbb R^N,\ t\ge0,
\label{eq:nonlocal_RN}
\end{equation}
is a nonlocal diffusion equation. B-C did not consider this equation; instead, they expand
$\hat w(q)=\hat w(0)-\tfrac12w_2q^2+\dots$. If $w$ has a finite second moment
$w_2=\int x_1^2\,w(|x|)\,d^Nx$, then $\hat w(q)=\hat w(0)-\tfrac12w_2|q|^2+o(|q|^2)$, and
the symbol of $\boldsymbol W$ is $A_w(q)=D|q|^2+o(|q|^2)$, with $D=\tfrac12f'(0)\,w_2$.
Since $|q|^2$ is the symbol of $-\nabla^2$, $\boldsymbol W$ and $-D\nabla^2$ have the same
symbol to leading order as $q\to0$; equivalently, expanding $\phi(x-y)$ in
Eq.~\eqref{eq:Wop} to second order gives $\boldsymbol W\phi=-D\nabla^2\phi+\dots$ for
$\phi$ that varies slowly on the range of $w$. In this sense the long-wavelength limit of
$\boldsymbol W$ is a diffusion operator, with diffusion constant proportional to the
second moment of $w$ [B-C's diffusion length is $l_d^2=D/\mu$, Eq.~(52) of
Ref.~\cite{BuiceCowan2007}]. Analyticity of $\hat w$ at the origin, which for $w\ge0$
holds only if $w$ decays exponentially, is not needed. B-C assume short-range connectivity and use the moments $w_2,w_4,\dots$
[Eqs.~(50)--(52) of Ref.~\cite{BuiceCowan2007}]. Kernels with power-law tails,
$w(x)\sim|x|^{-N-\sigma}$, never have an analytic Fourier transform: $\hat w(0)-\hat w(q)$
contains a term proportional to $|q|^{\sigma}$ ($|q|^{\sigma}\log|q|$ for even integer
$\sigma$). For $\sigma>2$ the second moment is finite, the term $q^2$ dominates,
$|q|^{\sigma}$ is an irrelevant correction, and the critical behavior is that of DP. For
$\sigma<2$ the second moment is infinite, $\hat w(0)-\hat w(q)\propto|q|^{\sigma}$, the
nonlocal term is relevant, and the critical behavior is that of DP with long-range
spreading~\cite{Janssen1999,HinrichsenHoward1999}. The dividing line between local and
nonlocal diffusion is thus $\sigma=2$, the finiteness of the second moment, not
analyticity. There is
no such mechanism over $\Qp^N$, because a nonconstant locally constant function has no
Taylor expansion. There are infinitely many inequivalent heat equations on $\Qp^N$, and
each connectivity kernel selects one. The infrared behavior therefore depends on how
$A_w(\xi)$ behaves as $\norm{\xi}\to0$. We distinguish the cases
below.

\paragraph*{Networks on a single ball.}
Suppose that $w$ is supported in the ball $B_r^N=p^{-r}\Zp^N$, with $r$ the smallest possible
integer. The support of $w$ determines the neurons that are really connected:
$w(\norm{x-y})=0$ whenever $x$ and $y$ lie in distinct balls of radius $p^{r}$, so the
neurons outside a ball of radius $p^{r}$ do not interact with those inside it. We
therefore take the network to be the ball $B_r^N$, and discard the neurons outside it. A
function on $B_r^N$ is identified with a function on $\Qp^N$ by extending it by zero outside
$B_r^N$, so that $L^\rho(B_r^N)\subset L^\rho(\Qp^N)$ for $\rho=1,2$. This is possible because of the topology of
$\Qp^N$: $B_r^N$ is open and compact, so the extension by zero preserves local constancy,
and $B_r^N$ is an additive subgroup, so $B_r^N+\operatorname{supp}w\subset B_r^N$ and the
dynamics never leaves $B_r^N$. In this way networks with different supports can be treated
within the same space $\Qp^N$. The network on a ball is the case $K=B_r^N$ of
Sec.~\ref{sec:master}. In Fourier variables, $\hat w(\xi)=\hat w(0)$ for
$\norm{\xi}\le p^{-r}$, since $\chi_p(\xi\cdot x)=1$ on the support of $w$, so $A_w(\xi)=0$
and $\lambda(\xi)=\mu$ on a whole ball of momenta. For a function $\phi$ supported in
$B_r^N$, $\hat\phi$ is constant on this ball and equal to $\int\phi$. Hence the only mode of
the network with $A_w=0$ is its total activity, which is conserved at $\mu=0$
(Sec.~\ref{sec:master}). All the other modes, the wavelets supported in $B_r^N$, are damped,
since $A_w$ is bounded below by a positive constant on $\norm{\xi}>p^{-r}$. There is no
diverging correlation length, and the critical behavior is that of a single well-mixed
population (mean field), which becomes sharp only in the limit of infinitely many neurons
in the ball. A connected hierarchical network extending over several balls requires
connections between them: constant couplings between regions (Lemma~\ref{lem:regional}
below), or tails that extend over all levels of the hierarchy (Lemma~\ref{lem:tail}).

\paragraph*{Regional networks.}
A more realistic model couples the balls. Suppose that the neurons occupy a disjoint
union of finitely many balls, at least two,
\begin{equation}
K=\bigsqcup_{a\in A}B_{l_a}^N(a),\qquad B_{l_a}^N(a)=a+p^{-l_a}\Zp^N ,
\label{eq:K_union}
\end{equation}
with $A$ a finite index set of cardinality $|A|\geq2$ as in
Ref.~\cite{ZG2022Ultrametric}. Then $K$ is compact and open. The case of
infinitely many balls is different. For instance, $\Qp^N$ is the disjoint union of the
unit balls $a+\Zp^N$, with $a$ running over $\Qp^N/\Zp^N$; this neuron space is not compact,
the matrix $M$ below becomes infinite, and Lemma~\ref{lem:regional} does not apply. This
is the continuum theory, treated separately. We abbreviate $B_a=B_{l_a}^N(a)$, the ball of radius
$p^{l_a}$ with center $a$. Each ball $B_a$ represents a \emph{region} of the
cortex in which the spike model holds with constant parameters: a decay rate $\alpha_a$, a
firing-rate function $f_a$ with $f_a(0)=0$, $f_a'(0)>0$ and $f_a''(0)<0$, and a radial,
integrable and continuous connectivity $w_a(\norm{x-y})\ge0$ between neurons $x,y\in B_a$.
Neurons in different regions are coupled with constant strength: $w(x,y)=c_{ab}\ge0$ for
$x\in B_a$, $y\in B_b$, $a\neq b$. The linearized dynamics is
\begin{equation}
\partial_tu(x,t)=-\alpha_a\,u(x,t)+f_a'(0)\!\int_Kw(x,y)\,u(y,t)\,d^Ny
\label{eq:lin_K}
\end{equation}
for $x\in B_a$, where $w(x,y)=w_a(\norm{x-y})$ if $x,y\in B_a$. Transitions between regions
are branching events whose offspring land in another region: the elementary events are
those of Eq.~\eqref{eq:app_reactions}, with a block-structured connectivity $K_{ij}$. For
$\alpha_a=\alpha$ and $f_a=f$, the model is the B-C model with block-structured
connectivity. Region-dependent parameters extend the B-C master
equation~\eqref{eq:master}; the derivation of the Doi--Peliti action goes through
unchanged with indexed rates. A translation-invariant radial kernel restricted to $K$ is
a particular regional network (Appendix~\ref{app:regional}). Set
$\omega_a=\int_{B_{l_a}^N}w_a(\norm{z})\,d^Nz$, let $\hat w_a$ be the Fourier transform
of $w_a1_{B_{l_a}^N}$, and define the matrix
\begin{equation}
M_{aa}=-\alpha_a+f_a'(0)\,\omega_a,\qquad
M_{ab}=f_a'(0)\,\mathrm{vol}(B_a)\,c_{ab},
\label{eq:M}
\end{equation}
for $a\neq b$.

\begin{lemma}[Regional networks]\label{lem:regional}\leavevmode
\begin{enumerate}
\item Every wavelet $\psi_{\gamma nj}$ supported in $B_a$ is an eigenfunction of the
  right-hand side of Eq.~\eqref{eq:lin_K}, with eigenvalue
  $-\alpha_a+f_a'(0)\,\hat w_a(p^{1-\gamma})\le M_{aa}$.
\item The regional activities $U_a(t)=\int_{B_a}u(y,t)\,d^Ny$ satisfy
  $dU_a/dt=\sum_bM_{ab}U_b$.
\item The leading growth rate of Eq.~\eqref{eq:lin_K} is the spectral abscissa $s(M)$,
  i.e.\ the largest real part of the eigenvalues of $M$. The network is critical if and
  only if $s(M)=0$.
\end{enumerate}
\end{lemma}
\begin{proof}
See Appendices~\ref{app:ultra}--\ref{app:spectrum}.
\end{proof}

\begin{figure*}[t]
\centering
\includegraphics[width=0.92\textwidth]{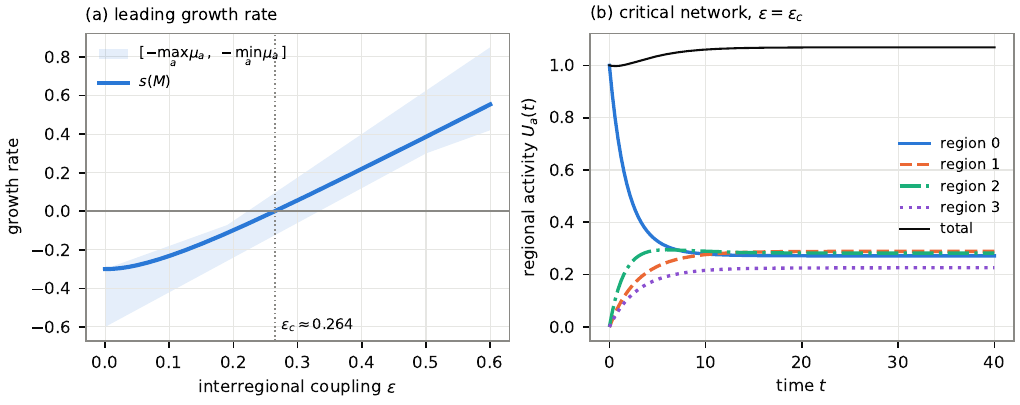}
\caption{A regional network (Lemma~\ref{lem:regional}) with four regions,
$K=\bigsqcup_{a=0}^{3}B_{-2}^1(a)$, $B_{-2}^1(a)=a+4\mathbb Z_2$ ($p=2$, $N=1$). Inside each region the kernel is
constant, so the wavelet modes of region $a$ relax at rate $\alpha_a$. The parameters are
$\alpha_a=(1,1.2,0.8,1.5)$, $f_a'(0)=(1,1.5,0.8,1.2)$ and $f_a'(0)\,\omega_a=(0.6,0.7,0.5,0.9)$,
and the interregional couplings are $c_{ab}=\varepsilon\,d_{ab}^{-2}$, with
$d_{ab}=|a-b|_2\in\{1/2,1\}$. (a) The leading growth rate $s(M)$ as a function of
$\varepsilon$, and the band $[-\max_a\mu_a,-\min_a\mu_a]$ of Appendix~\ref{app:regional}.
The network is critical at $\varepsilon_c\approx0.264$, where
$\mu_a\approx(0.003,-0.095,-0.017,0.124)$: the $\mu_a$ are not equal, so the critical
dynamics is not conservative. (b) Regional activities $U_a(t)$ at $\varepsilon_c$, with
all activity initially in region $0$. They converge to the Perron eigenvector of $M$ and
then stay constant. The total activity also tends to a constant, $1.07$, different from
its initial value $1$.}
\label{fig:regional}
\end{figure*}

With finitely many regions, the critical point $s(M)=0$ is that of a finite network of
well-mixed regions. The modes inside each region have growth rates at most
$M_{aa}\le s(M)$, generically strictly smaller, so at criticality they remain damped.
There is no diverging correlation length inside a region, and the critical behavior is of
mean-field (branching-process) type (Fig.~\ref{fig:regional}). When $c_{ab}=0$, the regions are independent networks
of the kind discussed above. Scale-free critical behavior across levels requires coupling across infinitely many
levels of the hierarchy, as for the power-law tails of the next lemma.

\begin{lemma}[Hierarchical power-law tails]\label{lem:tail}
Suppose $w$ is radial and integrable, and $w(x)=C_w\norm{x}^{-N-\sigma}$ for
$\norm{x}>p^{r}$, with $C_w>0$ and $\sigma>0$. Then, for all $\norm{\xi}\le p^{-r}$,
\begin{equation}
\hat w(0)-\hat w(\xi)=c_\sigma\,\norm{\xi}^{\sigma},\qquad
c_\sigma=C_w\,\frac{1-p^{-\sigma-N}}{p^{\sigma}-1}.
\label{eq:tail}
\end{equation}
Equivalently, $A_w(\xi)=\kappa_W\norm{\xi}^\sigma$ with
$\kappa_W=f'(0)\,c_\sigma$, i.e.\ $\boldsymbol W$ and $\kappa_W\Dsig$ have the same symbol on the
ball $\norm{\xi}\le p^{-r}$.
\end{lemma}
\begin{proof}
See Appendix~\ref{app:prooftail}.
\end{proof}

The hypothesis of Lemma~\ref{lem:tail} has a direct hierarchical meaning. On
$\Qp^N$, $\norm{x}$ takes the values $p^j$, and a power law in $\norm{x}$ is the same as
\emph{geometric decay per level of the tree}:
\begin{equation}
\frac{w(p^{j+1})}{w(p^{j})}=p^{-(N+\sigma)}\qquad (j\ge r).
\label{eq:perlevel}
\end{equation}
The coupling between two neurons is attenuated by the same factor each time their
common ancestor lies one layer closer to the root. The next lemma shows that this self-similarity,
together with isotropy, is what singles out the Taibleson--Vladimirov operator.

In the next lemma we use the following notation. $\boldsymbol A$ denotes a
pseudodifferential operator of the form
\begin{equation*}
\boldsymbol A\phi=\mathcal F^{-1}_{\xi\to x}\big(\hat A(\xi)\,\mathcal F_{x\to\xi}\phi\big),
\qquad \phi\in\mathcal D(\Qp^N),
\end{equation*}
whose symbol $\hat A:\Qp^N\smallsetminus\{0\}\to\mathbb C$ is a continuous function; such
an operator commutes with translations. $\Dsig$ denotes the Taibleson--Vladimirov operator
of order $\sigma>0$, Eq.~\eqref{eq:Taibleson}, i.e.\ the pseudodifferential operator with
symbol $\norm{\xi}^{\sigma}$. $GL_N(\Zp)$ denotes the group of $N\times N$ matrices $k$ with
entries in $\Zp$ and $|\det k|_p=1$; it acts on $\Qp^N$ by $\xi\mapsto k\xi$, and
$\norm{k\xi}=\norm{\xi}$ for every $k\in GL_N(\Zp)$. Indeed, since the entries of $k$
are in $\Zp$, the strong triangle inequality gives $\norm{k\xi}\le\norm{\xi}$; since
$|\det k|_p=1$, the entries of $k^{-1}=(\det k)^{-1}\operatorname{adj}(k)$ are also in
$\Zp$, so $\norm{\xi}=\norm{k^{-1}k\xi}\le\norm{k\xi}$.

\begin{lemma}[Uniqueness of $\Dsig$]\label{lem:unique}
Let $\boldsymbol A$ be a pseudodifferential operator with symbol $\hat A(\xi)$, as above.
Assume:
\begin{enumerate}
\item \emph{Isotropy:} $\hat A(k\xi)=\hat A(\xi)$ for all $k\in GL_N(\Zp)$. For the
  kernel operators~\eqref{eq:Wop} this is the statement that $w$ depends only on
  $\norm{y}$.
\item \emph{Homogeneity of degree $\sigma$:} $\hat A(a\xi)=|a|_p^{\sigma}\hat A(\xi)$ for all
  $a\in\Qp^{\times}$. Equivalently, $\boldsymbol A[\phi(a\,\cdot)]=|a|_p^\sigma(\boldsymbol A\phi)(a\,\cdot)$.
\end{enumerate}
Then $\hat A(\xi)=\hat A(1)\,\norm{\xi}^\sigma$, where $1$ denotes any vector of norm one,
i.e.\ $\boldsymbol A=\hat A(1)\,\Dsig$.
\end{lemma}
\begin{proof}
The group $GL_N(\Zp)$ acts transitively on each sphere $\{\norm{\xi}=p^k\}$, so by (1),
$\hat A(\xi)=g(\norm{\xi})$ for a function $g$ on $p^{\mathbb Z}$. Taking $a=p^{-k}$ in (2)
gives $g(p^k)=p^{k\sigma}g(1)$.
\end{proof}

\subsection{Isotropy, self-similarity and universality}\label{sec:isotropy}

The two hypotheses of Lemma~\ref{lem:unique} are idealizations of features of cortical
hierarchies, and each mirrors a familiar Euclidean symmetry.

\paragraph*{Isotropy: no preferred branch.}
Each $g\in GL_N(\Zp)$ maps every ball onto a ball of the same radius and preserves
inclusions, so it acts as an automorphism of the tree: it fixes the balls centered at the
origin and permutes, consistently at every level, the branches below each vertex. It also
preserves the hierarchical distance. Invariance
under this group means that the coupling between two neurons depends only on their
hierarchical distance, i.e.\ on the level of their last common ancestor, and not on the
ball (structure) of the hierarchy to which they belong. Physically, all the balls of a
given level are equivalent: the pattern of connections inside a ball of radius $p^{k}$, and
between that ball and the rest of the network, is the same for every ball of radius
$p^{k}$, so no subtree is wired differently from its siblings. For $N\ge2$, isotropy also
means that the $N$ ``directions'' of branching are equivalent; the anisotropic case is
discussed in Remark~\ref{rem:elliptic}. For $N=1$, condition (1) is automatic
($GL_1(\Zp)=\Zp^{\times}$), and homogeneity alone forces $\Dsig$. Isotropy is the
hierarchical counterpart of rotational invariance in $\mathbb R^N$, which makes a
Euclidean kernel depend only on $|x-y|$.

\paragraph*{Homogeneity: self-similarity across levels.}
Multiplication by $a\in\Qp^{\times}$ is a change of hierarchical level. For example,
$x\mapsto p^{-1}x$ maps the ball $B_k^N$ of radius $p^{k}$ onto the ball $B_{k+1}^N$, i.e.\
it moves one level up the tree. Homogeneity of degree $\sigma$ says that the dynamics looks
the same at every level, up to a rescaling of the rates: the modes supported on balls of
radius $p^{k}$ (the wavelets $\psi_{\gamma nj}$ with $\gamma=k$, for which
$\norm{\xi}=p^{1-k}$) relax at a rate $p^{-k\sigma}$ times that of the modes supported on
balls of radius $1$. Indeed, $\psi_{\gamma nj}$ is an eigenfunction of $\Dsig$ with
eigenvalue $\norm{\xi}^{\sigma}=p^{(1-\gamma)\sigma}$, as in Eq.~\eqref{eq:wavelet_eig},
and in the solution~\eqref{eq:Cauchy_wavelets} of the Cauchy problem it decays as
$e^{-\lambda(p^{1-\gamma})t}$, with $\lambda(p^{1-\gamma})=\mu+\kappa_Wp^{(1-\gamma)\sigma}$
for $A_w=\kappa_W\norm{\xi}^{\sigma}$. Physically, the connectivity is self-similar: climbing one level of
the hierarchy always attenuates the coupling by the same factor $p^{-(N+\sigma)}$
[Eq.~\eqref{eq:perlevel}]. This is aspect (ii) of the hypothesis of Sporns and co-workers
recalled in the Introduction: the patterns of connections repeat from one scale to the
next. Equivalently, the hierarchy has no intrinsic scale apart from the microscopic
cutoff.

\paragraph*{What Lemma~\ref{lem:unique} says.}
Together, ``no preferred branch'' and ``no preferred level'' leave a single operator up to
a constant, $\kappa_W\Dsig$, whose only parameter is the per-level attenuation $\sigma$.
In $\mathbb R^N$, isotropy and homogeneity alone also leave a one-parameter family, the
fractional Laplacians $(-\nabla^2)^{\sigma/2}$, whose symbols are $|q|^{\sigma}$; locality,
or a finite second moment of the kernel, then fixes $\sigma=2$, i.e.\ the
Laplacian~\cite{Stein1970,Kwasnicki2017}. Over $\Qp^N$ there is no locality
condition to impose, and $\sigma$ remains a physical parameter of the connectivity.

\paragraph*{Why it matters for criticality.}
A real connectome is not exactly self-similar, and homogeneity is an idealization. It is
relevant because at a critical point only the large-scale behavior matters: by power
counting, the deviations of an asymptotically homogeneous kernel from the homogeneous one
are irrelevant (see the classes below). For this reason $\kappa_W\Dsig$ is not the literal connectivity, but the
representative of a class of kernels, as we now explain.

\paragraph*{Universality classes of $W$-operators.}
What matters for critical behavior is the infrared behavior of $A_w$. Using the power counting of Sec.~\ref{sec:RG} (momentum
dimension $1$ for $\norm{\xi}$), we distinguish three classes.
\begin{enumerate}
\item \emph{Asymptotically homogeneous kernels:}
  $w(p^j)=C_wp^{-j(N+\sigma)}\big(1+O(p^{-j\delta})\big)$ as $j\to\infty$, with $\delta>0$.
  This condition imposes an important restriction on the symbol of the operator. By
  Eq.~\eqref{eq:Wsymbol},
  $A_w(\xi)=f'(0)\int_{\Qp^N}w(\norm{y})\big(1-\chi_p(\xi\cdot y)\big)\,d^Ny$. Let
  $\norm{\xi}=p^{m}$, and split the integral over the spheres $S_j^N$, on each of which $w$
  is constant. The factor $1-\chi_p(\xi\cdot y)$ vanishes for $\norm{y}\le p^{-m}$, since
  then $|\xi\cdot y|_p\le1$; by Eq.~\eqref{eq:ballFT},
  $\int_{S_j^N}\chi_p(\xi\cdot y)\,d^Ny$ equals $-p^{-mN}$ for $j=1-m$ and $0$ for
  $j>1-m$. Together with the volumes of the spheres [Eq.~\eqref{eq:radial}], this gives
  the exact formula
  \begin{align}
  A_w(\xi)=f'(0)\Big[&(1-p^{-N})\sum_{j\ge1-m}p^{jN}w(p^j)\nonumber\\
  &+p^{-mN}w(p^{1-m})\Big],\qquad \norm{\xi}=p^{m}.
  \label{eq:symbol_spheres}
  \end{align}
  The symbol is thus a sum over the levels $j\ge1-m$, i.e.\ over the hierarchical
  distances larger than $\norm{\xi}^{-1}$. Inserting the asymptotic form of $w$, the sum
  is geometric, and as $\norm{\xi}\to0$ (i.e.\ $m\to-\infty$),
  \begin{equation}
  A_w(\xi)=\kappa_W\norm{\xi}^{\sigma}\big(1+O(\norm{\xi}^{\delta})\big),
  \qquad \norm{\xi}\to0 ,
  \end{equation}
  with $\kappa_W=f'(0)\,c_\sigma$ as in Lemma~\ref{lem:tail}: the leading terms reproduce
  $c_\sigma$, and the corrections $O(p^{-j\delta})$, summed over $j\ge1-m$, are of relative
  order $p^{(m-1)\delta}\propto\norm{\xi}^{\delta}$. The correction has degree $\sigma+\delta>\sigma$ and is irrelevant by the power
  counting of Sec.~\ref{sec:RG}. All such kernels are expected to share the critical behavior of
  $\kappa_W\Dsig$. In this sense $\Dsig$ represents a \emph{universality class labelled by
  $\sigma$}, as $\nabla^2$ represents all short-range kernels in the Euclidean theory.
  The difference is that $\sigma$ is not fixed by analyticity. It is a physical
  parameter of the connectivity, the per-level attenuation in Eq.~\eqref{eq:perlevel}.
\item \emph{Slowly varying corrections:} $w(p^j)=C_wp^{-j(N+\sigma)}\ell(j)$ with $\ell$
  slowly varying (e.g., a power of $j$). Then
  $A_w(\xi)\sim \kappa_W\norm{\xi}^\sigma\ell(-\log_p\norm{\xi})$, and we
  expect the exponents of the $\Dsig$ class with logarithmic corrections.
\item \emph{Faster than any power:} if $w(p^j)p^{j(N+\sigma)}\to0$ for every
  $\sigma>0$, there is no scale-invariant infrared limit. Compact support is the
  extreme case (see the discussion before Lemma~\ref{lem:regional}).
\end{enumerate}
The statements above are exact at zero-loop order. That the corrections in class 1 remain
irrelevant beyond zero-loop order is expected from power counting but not proved here.

\begin{remark}[Anisotropic hierarchies]\label{rem:elliptic}
For $N\ge2$, B-C's homogeneity assumption requires only translation invariance, so the
kernel need not be radial. Dropping isotropy but keeping homogeneity allows other
operators. An example is the class of pseudodifferential operators with symbol
$|\mathfrak f(\xi)|_p^{\beta}$, where $\mathfrak f$ is an elliptic homogeneous polynomial
of degree $m$~\cite{ZunigaLNM2016}. Ellipticity gives
$C_1\norm{\xi}^m\le|\mathfrak f(\xi)|_p\le C_2\norm{\xi}^m$, so these operators have the
same power counting as $\Dsig$ with $\sigma=m\beta$, and $N_c=2\sigma$ is unchanged. They
model hierarchies in which the $N$ ``directions'' of branching are not equivalent.
Whether they fall in the same universality class as $\Dsig$ is an open question.
\end{remark}

\paragraph*{Summary.}
The long-distance linear dynamics is the heat equation of the $W$-operator selected by
the connectivity. For isotropic, self-similar connectivity (and, we expect, for every
asymptotically homogeneous kernel) this is ultrametric diffusion generated by
$\kappa_W\Dsig$~\cite{ABKO2002,Kochubei2001,ZunigaLNM2016}. Its Green function is the transition
density of a Markov process on the tree, and it decays as a power of $\norm{x}$ rather
than as a Gaussian. The exponent $\sigma$ is, in principle, measurable from connectomic
data as a function of hierarchical distance.

\subsection{Master equation and criticality on networks supported on a single ball}\label{sec:master}

The condition $\mu\geq0$ has a probabilistic meaning: it is the condition
under which the linear dynamics on a network supported on a single ball, or on a
regional network, is a Markov process.
Here, we discuss the simplest case, in which the neurons occupy a ball $K=B_m^N$ containing the support of $w$. The general case,
in which $K$ is a finite disjoint union of balls, as in Ref.~\cite{ZG2022Ultrametric}, is
treated in Appendix~\ref{app:regional}. We assume that $w$ is continuous.

Since $K$ is an additive subgroup, $K+\operatorname{supp}w\subset K$, so a solution of
Eq.~\eqref{eq:linearized} with initial datum supported in $K$ stays supported in $K$ for
all $t\ge0$. Moreover, $\int_Kw(\norm{x-y})\,d^Ny=\hat w(0)$ for every $x\in K$, and
Eq.~\eqref{eq:linearized} restricted to $K$ takes the form of a master equation,
\begin{equation}
\partial_tu(x,t)=\int_K\big\{j(x\,|\,y)\,u(y,t)-j(y\,|\,x)\,u(x,t)\big\}\,d^Ny
\label{eq:masterK}
\end{equation}
for $x\in K$, $t\ge0$, with transition density rates
\begin{align*}
j(x\,|\,y)&=f'(0)\,w(\norm{x-y}),\\
j(y\,|\,x)&=\frac{\alpha}{\hat w(0)}\,w(\norm{x-y}).
\end{align*}
The condition $j(x\,|\,y)\le j(y\,|\,x)$ for all $x,y\in K$ (Hypothesis~1 of
Ref.~\cite{ZG2022Ultrametric}) is equivalent to $f'(0)\hat w(0)\le\alpha$, i.e.\
$\mu\ge0$. Under this condition, by Theorem~3.1 of Ref.~\cite{ZG2022Ultrametric} (stated
there for $N=1$ for simplicity; the results extend to arbitrary dimension $N$),
Eq.~\eqref{eq:masterK} generates a Feller semigroup on the space of continuous functions
defined on $K$, $C(K,\mathbb R)$, and the solution of the Cauchy problem is
$u(x,t)=\int_Ku_0(y)\,p_t(x,dy)$, where $p_t$ is the transition function of a Markov
process with state space $K$. The condition is also necessary: since
$\boldsymbol W1_K=0$ on $K$, the solution with $u_0=1_K$ is $e^{-\mu t}1_K$, so
$p_t(x,K)=e^{-\mu t}$, which exceeds $1$ when $\mu<0$. Three cases follow.
\begin{itemize}
\item $\mu=0$: the process is conservative. The dynamics is the heat equation
  $\partial_tu+\boldsymbol Wu=0$, a random walk on the hierarchy with no loss of
  activity: the rate $f'(0)\hat w(0)$ at which active neurons trigger new activity equals
  the rate $\alpha$ at which activity decays (branching ratio $m=f'(0)\hat w(0)/\alpha=1$),
  so the total activity $\int_Ku(x,t)\,d^Nx$ is conserved.
\item $\mu>0$: the walk is killed at rate $\mu$, i.e.\ it is sent to a cemetery state
  outside $K$, which is an absorbing state~\cite{ZG2022Ultrametric}; in the present
  context it corresponds to the quiescent state.
\item $\mu<0$: there is no Markov interpretation. $u$ is the mean density of a
  supercritical branching random walk and grows like $e^{|\mu|t}$; in this regime the
  linearization ceases to be valid, and the nonlinear terms of $f$, which saturate the
  growth, must be kept already at the level of mean-field theory.
\end{itemize}
Thus the critical point $\mu=0$ is exactly the point at which the linear dynamics on a
ball $K$ containing the support of $w$ becomes a conservative Markov process. When $K$ is
a disjoint union of balls, the killing rate is constant on each ball, and a conservative
dynamics is always critical (Appendix~\ref{app:regional}).

\paragraph*{Comparison with $\mathbb R^N$.}
Over $\mathbb R^N$ the situation is different. A bounded set $K$ is not a subgroup, and a
solution with initial datum supported in $K$ is, for every $t>0$, supported on all of
$\mathbb R^N$ (for $w\ge0$ positive near the origin). A bounded region of cortex must
therefore be modeled by restricting the integral to $K$. The resulting master equation has
the symmetric rates $f'(0)\,w(|x-y|)$ and the killing rate
$\alpha-f'(0)\int_Kw(|x-y|)\,d^Ny$, which depends on $x$ and grows near the boundary of
$K$. The linear dynamics is then not translation invariant, and at $\mu=0$ it is not
conservative. Over $\Qp^N$, a ball has no boundary in this sense, and on a disjoint union
of balls the killing rate is constant on each ball.

\section{Criticality: ultrametric Reggeon field theory}\label{sec:RG}

\paragraph*{Physical picture.}
We consider the network without external drive, $f(I)=0$
[Eq.~\eqref{eq:absorbing}], so that the silent state $a\equiv0$, where
$a=\langle\varphi\rangle$ is the mean activity (Sec.~\ref{sec:meanfield}), is
\emph{absorbing}: once all neurons are silent, they stay silent. The absorbing state is an idealization. Here
``silent'' means that there are no effective spikes, the activity that propagates through
the network, not that there is no electrical activity at all. Real cortex is never
silent: even at rest it receives weak input and fires spontaneously. This corresponds to
a small source $f(I)>0$, which turns the absorbing state into a stationary state of low
activity, $a^*\simeq f(I)/\mu$ (Sec.~\ref{sec:corr}). Resting cortical activity is
described by this weakly driven state near $\mu=0$, in which spikes produced by the input
start avalanches that spread and die out; the absorbing state is its limit as $I\to0$.
A
single spike triggers on average $m=f'(0)\hat w(0)/\alpha$ further
spikes (the branching ratio), and $\mu=\alpha(1-m)$. For $\mu>0$
($m<1$) every avalanche of activity dies out, and the network returns
to the silent state. For $\mu<0$ ($m>1$) activity can sustain itself.
Its growth is stopped by the \emph{saturation} of $f$: the firing rate
is bounded, $f\le f_{\max}$, so the response grows less than
proportionally to the input, and $f''(0)<0$
[Eq.~\eqref{eq:f_derivatives}]. The network then settles in an active
state of density $\rho>0$. The value $\mu=0$ separates the two phases.
This is a nonequilibrium phase transition into an absorbing state. When
the neurons occupy Euclidean space $\mathbb R^N$, as in B-C, it belongs
to the universality class of directed percolation
(Appendix~\ref{app:DP}). Here the neurons occupy $\Qp^N$; the
transition is of the same type, but its critical behavior is governed
by the hierarchical connectivity, through the operator
$\boldsymbol W$. At $\mu=0$ avalanches of all sizes occur, and their
statistics follow power laws. Near $\mu=0$ the activity is small and
slowly varying, while its correlation length and time diverge.
Mean-field theory (the Wilson--Cowan equation) neglects the fluctuations
of this sparse activity, and the question is whether they change the
scaling laws. To answer it we expand the action about
$\varphi=\tphi=0$, which, since $a\equiv0$ and $\tilde a=0$, is the expansion about the
mean field (Sec.~\ref{sec:split}), in powers of the fields, and keep only the
terms that matter at large scales. The quadratic terms,
$\tphi(\partial_t+\mu+\boldsymbol W)\varphi$, describe the linear
dynamics of Sec.~\ref{sec:linear}: propagation of activity through the
hierarchy and its decay. The cubic terms describe the two elementary
processes that the linear dynamics misses. \emph{Branching},
$\tphi^2\varphi$: the spikes triggered by one active neuron are
correlated, which is the source of fluctuations; their strength is
proportional to the activity itself (multiplicative noise).
\emph{Aggregation}, or saturation, $\tphi\varphi^2$: two active inputs
to the same neuron produce less than twice the response, which limits
the growth for $\mu<0$ and fixes $\rho$. Quartic and higher terms are
irrelevant near the upper critical dimension (power counting below).
The resulting action, Eq.~\eqref{eq:RFT}, is Reggeon field theory with
the Laplacian replaced by the $W$-operator. It describes both phases:
the absorbing phase, in which $\varphi=0$ is stable, and the active
phase, in which its classical equation has the stable solution
$\rho\propto|\mu|$. The rest of this section rescales the fields so
that the two cubic couplings are equal, which exhibits the
rapidity-reversal symmetry, and uses power counting to find when the
fluctuations generated by branching change the mean-field exponents:
this happens for $N<N_c=2\sigma$.

\paragraph*{The effective action.}
We follow B-C [Sec.~V of Ref.~\cite{BuiceCowan2007}]. Among the vertices of the
interacting action (Sec.~\ref{sec:split}), the relevant ones near the critical point are
the cubic vertices $-f'(0)\,\tphi\,w\star(\tphi\varphi)$ (branching) and
$-\tfrac12f''(0)\,\tphi\,(w\star\varphi)^2$ (aggregation, or saturation). For the slowly
varying fields that dominate at criticality, $w\star$ may be replaced by $\hat w(0)$ in
these vertices; the difference contains the operator $\boldsymbol W$ and is irrelevant.
This gives $-f'(0)\hat w(0)\,\tphi^2\varphi+\tfrac12|f''(0)|\hat w(0)^2\,\tphi\varphi^2$.
B-C then rescale the fields [Eq.~(58) of Ref.~\cite{BuiceCowan2007}],
\begin{equation}
\tpsi=c\,\tphi,\qquad \psi=c^{-1}\varphi,\qquad
c=\Big(\frac{2f'(0)}{\hat w(0)\,|f''(0)|}\Big)^{1/2}.
\label{eq:BCrescaling}
\end{equation}
The quadratic part of the action depends only on the product $\tphi\varphi=\tpsi\psi$ and
is unchanged, while the two cubic coefficients become equal in magnitude:
$f'(0)\hat w(0)/c=\tfrac12|f''(0)|\hat w(0)^2c=g/2$, with
\begin{equation}
g=\big(2f'(0)\,|f''(0)|\,\hat w(0)^3\big)^{1/2}.
\label{eq:g_bare}
\end{equation}
Keeping the relevant interactions, the result for a general connectivity kernel is
\begin{equation}
\begin{aligned}
S_{\rm eff}^{W}[\psi,\tpsi]=\int\!d^Nx\!\int_0^\infty\!dt\,\Big\{&\tpsi\big(\partial_t+\mu+\boldsymbol W\big)\psi\\
&+\frac{g}{2}\big(\tpsi\psi^2-\tpsi^2\psi\big)\Big\},
\end{aligned}
\label{eq:RFT}
\end{equation}
with $\boldsymbol W$ the $W$-operator~\eqref{eq:Wop} selected by the connectivity, $\mu$ the
distance to criticality and $g$ the coupling~\eqref{eq:g_bare}. The relative minus sign
between the cubic terms comes from $f''(0)<0$. Eqs.~\eqref{eq:BCrescaling} and
\eqref{eq:g_bare} give the bare values of $\mu$ and $g$; beyond zero-loop order both are
modified by fluctuations (renormalized), as in B-C. Their Eq.~(59) writes the cubic term as
$g(\varphi^2\tphi-\tphi^2\varphi)$, with a coupling $g$ for which they give no explicit
expression; their rescaling, Eq.~(58) of Ref.~\cite{BuiceCowan2007}, gives it the bare value
$g/2$, with our $g$ of Eq.~\eqref{eq:g_bare}. We keep the notation $\boldsymbol W$ throughout: the kinetic operator is not an independent
modeling choice, it is fixed by $w$. For the scaling analysis we assume from now on that
$\boldsymbol W$ is in the class of asymptotically homogeneous kernels
(Sec.~\ref{sec:isotropy}). This is the class in which scale-free critical behavior occurs.
Kernels with compact support couple only finitely many levels, and their critical behavior
is of mean-field (branching-process) type (Lemma~\ref{lem:regional}); more generally,
kernels that decay faster than any power have no scale-invariant infrared limit (class~3
of Sec.~\ref{sec:isotropy}). For an asymptotically homogeneous kernel,
\begin{equation}
A_w(\xi)=\kappa_W\norm{\xi}^{\sigma}\big(1+O(\norm{\xi}^{\delta})\big),
\qquad \norm{\xi}\to0,
\label{eq:Wasym}
\end{equation}
with $\kappa_W>0$, $\sigma>0$ and $\delta>0$. The Taibleson--Vladimirov operator
$\kappa_W\Dsig$ is the exactly homogeneous representative of this class
(Lemma~\ref{lem:unique}). It corresponds to isotropic connectivity with per-level
attenuation $p^{-(N+\sigma)}$. Only the leading term of Eq.~\eqref{eq:Wasym} enters the power
counting below; the correction is irrelevant.

\paragraph*{Rapidity reversal.}
The action~\eqref{eq:RFT} is invariant under the transformation
\begin{equation}
\psi(x,t)\to-\tpsi(x,-t),\qquad \tpsi(x,t)\to-\psi(x,-t),
\label{eq:rapidity}
\end{equation}
called \emph{rapidity reversal}, which exchanges the density field $\psi$ and the response
field $\tpsi$ and reverses time. Indeed, after an integration by parts in $t$ the quadratic
term is invariant because $\boldsymbol W$ is self-adjoint ($w$ is radial), and the two
cubic terms are exchanged because their coefficients have equal magnitude, which is what
the rescaling~\eqref{eq:BCrescaling} achieves. As in B-C [Sec.~V of
Ref.~\cite{BuiceCowan2007}], this is an emergent symmetry of the critical theory; the
initial-state term of the full action breaks it. The symmetry relates two order
parameters of the absorbing-state transition. In the active phase ($\mu<0$),
\begin{equation}
\rho=\lim_{t\to\infty}\langle\psi(x,t)\rangle\propto(-\mu)^{\beta},\qquad
P_\infty\propto(-\mu)^{\beta'},
\label{eq:betas}
\end{equation}
where $\rho$ is the stationary density of activity, starting from a fully active network,
and $P_\infty$ is the probability that the activity started by a single seed (one
spike) never dies out. The exponents $\beta$ and $\beta'$ are defined by
Eq.~\eqref{eq:betas}; they are the exponents $\beta$ and $\beta'$ of Eq.~(60) of
Ref.~\cite{BuiceCowan2007}. B-C state these definitions, the mean-field values
$\beta=\beta'=1$ and the relation $\beta=\beta'$ without derivation: all of this is standard
directed percolation material, taken from the literature (Ref.~\cite{Hinrichsen2000},
which is Ref.~[9] of B-C). We derive the mean-field values in Appendix~\ref{app:MF} and
justify $\beta=\beta'$ below. The density is a one-point function of $\psi$, while the
survival probability of a seed is a one-point function of $\tpsi$ (the response to a
localized initial perturbation). Rapidity reversal maps one onto the other, so
$\beta=\beta'$, as in DP~\cite{Hinrichsen2000,Tauber2014}. The same argument gives equal
exponents for the decay in time of the density and of the survival probability at
$\mu=0$ (Appendix~\ref{app:RFT}).

\paragraph*{Power counting.}
Assign momentum dimension $1$ to $\norm{\xi}$, and write $[t]=-z$ for the scaling
dimension of time, where $z$ is the dynamic exponent. In the quadratic part of
Eq.~\eqref{eq:RFT}, $\partial_t$ and $\boldsymbol W$ (whose symbol is
$\kappa_W\norm{\xi}^{\sigma}$ at small $\norm{\xi}$) are added, so they have the same
dimension; hence, at zero-loop order, $z=\sigma$, the order of the Taibleson--Vladimirov
operator. The quadratic part is dimensionless if $[\psi]+[\tpsi]=N$, and rapidity
reversal gives $[\psi]=[\tpsi]=N/2$. Requiring the cubic term
$\int d^Nx\,dt\,\frac g2\tpsi\psi^2$ to be dimensionless gives
$-N-z+3N/2+[g]=0$, where $-N-z$ is the dimension of the measure $d^Nx\,dt$; thus
\begin{equation}
[g]=\sigma-\frac N2 .
\end{equation}
Equation~\eqref{eq:RFT} keeps only the cubic vertices. The interacting action of
Sec.~\ref{sec:split} also contains quartic and higher vertices: by
Eq.~\eqref{eq:taylor}, the term $n=2$ contains $-f''(0)\,\tphi\,(w\star(\tphi\varphi))(w\star\varphi)$,
of type $\tphi^2\varphi^2$, and the term $n=3$ contains $-\tfrac16f'''(0)\,\tphi\,(w\star\varphi)^3$,
of type $\tphi\varphi^3$. For a quartic coupling $u$ the same count gives
$-N-z+4N/2+[u]=0$, i.e.\ $[u]=\sigma-N$, which is negative near $N=2\sigma$. These vertices
are therefore irrelevant near the upper critical dimension, and dropping them in
Eq.~\eqref{eq:RFT} is justified there. For $N$ well below $2\sigma$ (in particular for
$N\le\sigma$, where $[u]\ge0$), that they remain irrelevant at the non-Gaussian fixed point
is an assumption, the same one made for Euclidean directed percolation below $d=4$, and it
is not proved here. The cubic coupling is relevant for $N<2\sigma$, so
\begin{equation}
N_c=2\sigma .
\end{equation}
Equivalently, in terms of the spectral dimension $d_s=2N/\sigma$ of the $p$-adic diffusion,
Eq.~\eqref{eq:ds} of Sec.~\ref{sec:phase}, mean-field theory holds for $d_s\ge4$, as in the
Euclidean theory with local coupling.
More generally, consider a term
$\int\!d^Nx\!\int_0^\infty\!dt\,\tpsi^{\,b}\psi^{c-1}\,\boldsymbol D^{a}\psi$, where
$\boldsymbol D^{a}$ is the Taibleson--Vladimirov operator~\eqref{eq:Taibleson} of order
$a\ge0$ (with $\boldsymbol D^{0}$ the identity), i.e.\ the operator with symbol
$\norm{\xi}^{a}$; here $a$ is real, not necessarily an even integer. The dimension
of the integral is $-N-z+a+(b+c)N/2$, and its coupling has the opposite dimension. With
$z=\sigma$, the coupling is relevant or marginal iff
\begin{equation}
N\le\frac{2(\sigma-a)}{b+c-2}.
\label{eq:powercount}
\end{equation}
For $\sigma=2$ this is Eq.~(132) of Ref.~\cite{BuiceCowan2009}. For $a=0$, cubic vertices
give $N_c=2\sigma$ and quartic vertices give $N_c=\sigma$, so quartic and higher
couplings are irrelevant near $N_c=2\sigma$. There is no local
gradient term available over $\Qp^N$. A symbol term $\norm{\xi}^{\sigma'}$ with
$\sigma'>\sigma$, for example from a faster-decaying component of the kernel, is
irrelevant. Therefore, unlike
Euclidean L\'evy-flight DP, where $\nabla^2$ takes over for $\sigma\ge2$~\cite{Janssen1999},
\emph{every} $\sigma>0$ defines a distinct theory. In particular the $p$-adic line
($N=1$) is below its upper critical dimension whenever $\sigma>1/2$.

\paragraph*{Mean-field exponents.}
At zero-loop order, the stationary density in the active phase ($\mu<0$) is
$\rho=2|\mu|/g$, and a mode of momentum $\xi$ relaxes at the rate
$\lambda(\xi)=\mu+A_w(\xi)\simeq\mu+\kappa_W\norm{\xi}^\sigma$. These two facts give
\begin{equation}
\beta=\beta'=1,\qquad \nu_\parallel=1,\qquad \nu_\perp=\frac1\sigma,\qquad z=\sigma .
\label{eq:MF}
\end{equation}
Compare the Euclidean values $(1,1,1,\tfrac12,2)$ used by B-C. The exponents of the total
activity ($\beta$, $\beta'$, $\nu_\parallel$) coincide with those of B-C; the spatial ones
($\nu_\perp$, $z$) carry the decay exponent $\sigma$, and coincide with them for
$\sigma=2$. The derivation is given in Appendix~\ref{app:MF}.

\paragraph*{Interpretation for cortex.}
In the Euclidean theory $N_c=4$, so for $N=2,3$ fluctuations dominate sufficiently close
to the critical point, and the exponents are not those of mean-field theory. B-C argued
indirectly that cortex should nevertheless display mean-field DP exponents: since each
neuron receives input from very many others over long distances, the fluctuation
correction in their Ginzburg criterion [Eq.~(52) of Ref.~\cite{BuiceCowan2007}] carries a
small prefactor, so mean-field behavior should be observed over a wide range of $\mu$ and
fail only in a narrow region around $\mu=0$. In this argument mean-field theory is a good
approximation, not an exact result.

In the $p$-adic theory the argument becomes the sharp statement $\sigma\le N/2$: for
$N\ge2\sigma$ the mean-field exponents are exact asymptotically, arbitrarily close to the
critical point, and for $N<2\sigma$ they eventually fail, however small the fluctuation
region. A simple Ginzburg estimate, derived in Appendix~\ref{app:MFginzburg}, shows where
the inequality comes from. In the active phase $\rho=2|\mu|/g$, and the noise of DP is
multiplicative, with strength of order $g\rho$. Averaged over a correlation volume $\xi_\perp^N$ and a correlation time
$\xi_\parallel$, the relative fluctuations are
\begin{equation}
\frac{\langle\delta\rho^2\rangle}{\rho^2}\simeq\frac{g\,\xi_\parallel}{\rho\,\xi_\perp^{N}}
\simeq g^2\,|\mu|^{N/\sigma-2},
\label{eq:ginzburg_heur}
\end{equation}
by Eq.~\eqref{eq:MF} ($\xi_\parallel=1/|\mu|$, $\xi_\perp=(\kappa_W/|\mu|)^{1/\sigma}$).
This ratio vanishes as $\mu\to0$ exactly when $N>2\sigma$, the condition found by power
counting and by the one-loop integral of Sec.~\ref{sec:ginzburg}.

The inequality has a direct reading in terms of connectivity. Let $W_k$ be the total
coupling that a neuron receives from the neurons at hierarchical distance $p^{k}$
[Eq.~\eqref{eq:Wk}]. For the power-law kernel, $W_{k+1}/W_k=p^{-\sigma}$: each level up
the hierarchy contributes $p^{\sigma}$ times less total input than the level below. Since
each vertex of the tree has $p^N$ children, the condition $\sigma\le N/2$ reads
\begin{equation*}
\frac{W_{k+1}}{W_k}=p^{-\sigma}\ \ge\ p^{-N/2}=\frac{1}{\sqrt{p^{N}}} :
\end{equation*}
mean-field theory holds when the input from successive levels of the hierarchy falls off
no faster than one over the square root of the branching number. If distant levels still
contribute strongly (small $\sigma$), each neuron averages over a large and growing
population, and fluctuations average out. If the coupling is concentrated on nearby levels
(large $\sigma$), fluctuations are not averaged away, and the exponents depart from mean
field; the simulations of Sec.~\ref{sec:simulations} show this trend in the avalanche-size
exponent. Whether hierarchical cortex is mean-field is therefore a measurable question
about how fast the synaptic input to a neuron decays with hierarchical distance, not about
the dimension of physical space.

\paragraph*{Beyond mean field.}
For $N<2\sigma$ the cubic coupling is relevant, fluctuations dominate sufficiently close to
the critical point, and the exponents differ from Eq.~\eqref{eq:MF}. Computing them
requires a renormalization-group analysis of Eq.~\eqref{eq:RFT}, which we do not attempt
here. In this respect we follow B-C, who do not compute the exponents either: they
identify their effective action with Reggeon field theory and take its critical behavior,
directed percolation, from the literature [Sec.~V of Ref.~\cite{BuiceCowan2007}]. That
step is not available for Eq.~\eqref{eq:RFT}, whose critical behavior below $N_c=2\sigma$
is, to our knowledge, not known. The renormalization group over the $p$-adics is well
developed, from Dyson's hierarchical model~\cite{Dyson1969,BleherSinai1973} and the
$p$-adic renormalization group~\cite{LernerMissarov1989} to rigorous constructions with
anomalous dimensions~\cite{AbdesselamChandraGuadagni2013} and the $p$-adic $O(n)$
models~\cite{Gubser2017}; these provide the natural framework for this step. At present,
the simulations of Sec.~\ref{sec:simulations} give numerical estimates of the
avalanche-size exponent below $N_c$. Like B-C, we assume that the relative minus sign
between the cubic terms of Eq.~\eqref{eq:RFT}, which reflects the saturation of $f$,
persists beyond zero-loop order.

\section{A $p$-adic Ginzburg criterion}\label{sec:ginzburg}

\paragraph*{Origin of the criterion.}
The B-C criterion [Eqs.~(50)--(53) of Ref.~\cite{BuiceCowan2007}] compares the zero-loop
and one-loop contributions to the vertex $\Gamma^{(1,1)}$ (Fig.~8 of
Ref.~\cite{BuiceCowan2007}).

\paragraph*{The vertex $\Gamma^{(1,1)}$.}
Let $\Gamma[a,\tilde a]$ be the effective action of B-C [Eqs.~(36)--(41) of
Ref.~\cite{BuiceCowan2007}]: the Legendre transform of $W=\ln Z$ with respect to the sources,
regarded as a functional of the means $a=\langle\varphi\rangle$, $\tilde a=\langle\tphi\rangle$. Its
stationary points are the exact means, and its derivatives are the proper
(one-particle-irreducible) vertices. At the quiescent state we set
\begin{equation}
\Gamma^{(1,1)}(x,t;x',t')=\frac{\delta^2\Gamma}{\delta\tilde a(x,t)\,\delta a(x',t')}\bigg|_{a=\tilde a=0}.
\label{eq:Gamma11}
\end{equation}
It is the inverse of the full propagator:
\begin{align*}
&\int\Gamma^{(1,1)}(x,t;y,s)\,\langle\varphi(y,s)\tphi(x',t')\rangle\,d^Ny\,ds\\
&\qquad=\delta(x-x')\,\delta(t-t').
\end{align*}
At zero loops $\Gamma=S$, and $\Gamma^{(1,1)}$ is the kernel of the operator
$\partial_t+\alpha-f'(0)\,w\star$ of the free action. By translation invariance in space and
time, $\Gamma^{(1,1)}$ depends only on $(x-x',t-t')$. With
$\hat F(\xi,\omega)=\int\chi_p(\xi\cdot x)\,e^{i\omega t}F(x,t)\,d^Nx\,dt$ and
$\alpha-f'(0)\hat w(\xi)=\mu+A_w(\xi)$ [Eq.~\eqref{eq:Wsymbol}], one has, to one loop,
\begin{align}
\hat\Gamma^{(1,1)}(\xi,\omega)&=-i\omega+\mu+A_w(\xi)-\Sigma(\xi,\omega),\nonumber\\
\Sigma(\xi,\omega)&=f''(0)f'(0)\nonumber\\
&\times\int_{\Qp^N}
\frac{\hat w(\eta)^2\,\hat w(\xi-\eta)\,d^N\eta}{-i\omega+2\mu+A_w(\eta)+A_w(\xi-\eta)} .
\label{eq:selfenergy}
\end{align}
The self-energy $\Sigma$ is the loop of Fig.~8 of Ref.~\cite{BuiceCowan2007}. It is built from
the branching vertex $f'(0)\,\tphi\,w\star(\tphi\varphi)$ and the merging vertex
$\tfrac12f''(0)\,\tphi\,(w\star\varphi)^2$, joined by two propagators
$\theta(t)e^{-(\mu+A_w)t}$; the time integral gives the denominator, and the two ways of joining
the legs cancel the factor $\tfrac12$. Define
\begin{align}
\mu_R(\mu)&:=\hat\Gamma^{(1,1)}(\xi=0,\omega=0)=\mu-f''(0)f'(0)\,J(\mu),\nonumber\\
J(\mu)&=\int_{\Qp^N}\frac{\hat w(\xi)^3}{2\big(\mu+A_w(\xi)\big)}\,d^N\xi ,
\label{eq:BC50}
\end{align}
to one loop. The bare parameter $\mu$ is the decay rate of the uniform mode in mean-field
theory; $\mu_R$ is its decay rate in the full theory, i.e.\ the inverse of the static response
of the network to a uniform stimulus. Equivalently, $\mu_R=1/\langle S\rangle$, where
$\langle S\rangle=\int dt\int d^Nx\,\langle\varphi(x,t)\tphi(0,0)\rangle=1/\hat\Gamma^{(1,1)}(0,0)$ is the
mean total number of spikes triggered by one injected spike (the mean avalanche size),
fluctuations included; in mean-field theory $\langle S\rangle=1/\mu$. Thus $\mu_R$ is the
renormalized distance to criticality, and the critical point is where it vanishes. There are no factors $(2\pi)^{-N}$, because the
Fourier transform uses the character $\chi_p$ and the self-dual Haar measure
(Appendix~\ref{app:padic}).

B-C's Eq.~(50) compares the two terms of Eq.~\eqref{eq:BC50}. Its symbol $\gg$ compares
magnitudes: for a saturating $f$, $f''(0)<0$, and the one-loop term, taken with its sign, has
the opposite sign to $\mu$. This is why $|f''(0)|$ appears below, as in B-C's Eq.~(52); the sign
of $f''(0)$ fixes the direction of the shift of the critical point. Taken literally at zero
momentum, B-C's condition $|\mu|\gg|f''(0)f'(0)J(\mu)|$ fails as $\mu\to0$ whenever $J(0)$ is
finite ($N>\sigma$, see below), even for $N>2\sigma$, where mean-field exponents are exact. It
detects the shift of the critical point, not a breakdown of mean-field theory. The criterion
must therefore be applied to the distance $\mu-\mu_c$ from the true critical point, as below;
B-C do this implicitly by keeping only the infrared singular part of the integral (their
Eq.~(51)). B-C write the prefactor as
$\tfrac12f''f'$, omitting the factor $2$ from the two pairings (as they omit the combinatorial
factor in their Eq.~(25)). We checked the factor on a single neuron with birth rate
$f'(0)\,n+\tfrac12f''(0)\,n(n-1)$ and death rate $\alpha n$. There, the one-loop self-energy
$f''(0)f'(0)/(-i\omega+2\mu)$ predicts the decay rate $\mu-f''(0)f'(0)/\mu+O(f''(0)^2)$ of the
mean, and the leading eigenvalue of the master equation agrees; B-C's prefactor would halve
the correction.

For $N>\sigma$ the value $J(0)$ is finite. Near $\xi=0$, by Eq.~\eqref{eq:Wasym}, the integrand
behaves as $\hat w(0)^3/(2\kappa_W\norm{\xi}^{\sigma})$, and summing over spheres gives
$\sum_{k\le0}p^{k(N-\sigma)}$, which converges if and only if $N>\sigma$. At intermediate
$\norm{\xi}$, $A_w>0$ by Lemma~\ref{lem:I2}(1) below. At large $\norm{\xi}$,
$A_w(\xi)\to f'(0)\hat w(0)>0$, and $\int|\hat w|^3<\infty$ if, for instance, $w\in L^1\cap L^2$,
since then $|\hat w|^3\le\Vert\hat w\Vert_\infty|\hat w|^2$; alternatively, the cutoff of
Sec.~\ref{sec:cutoffs} makes this region finite.

The critical point is the value of the bare parameter $\mu=\alpha-f'(0)\hat w(0)$ at which the
infinite network on $\Qp^N$, with fluctuations included, undergoes the absorbing-state
transition. Physically, $\mu$ is set by $\alpha$, $f'(0)$, and $w$. In mean-field theory, this
happens at $\mu=0$. With fluctuations, to one loop, it happens at $\mu_c=f''(0)f'(0)J(0)$. On
a finite network, there is no sharp transition (Sec.~\ref{sec:cutoffs}), so the statement
refers to $\Qp^N$. Equivalently, $\mu_c$ is defined by $\mu_R(\mu_c)=0$, i.e.
$\mu_c=f''(0)f'(0)\,J(\mu_c)\simeq f''(0)f'(0)\,J(0)$; the last step is valid to one loop, since
$J(\mu)-J(0)\to0$ as $\mu\to0$ for $N>\sigma$. For a saturating $f$, $f''(0)<0$ and $\mu_c<0$,
i.e.\ $\alpha<f'(0)\hat w(0)$ at the transition: the network needs more excitation than
mean-field theory predicts. Fluctuations lower the mean firing rate, as Jensen's inequality
requires for a concave $f$. The shift is non-universal, like the shift of the critical
temperature in equilibrium theory, and says nothing about the approach to the transition.
That is controlled by
\begin{equation*}
\mu_R\simeq(\mu-\mu_c)\big[1-f''(0)f'(0)\,J'(\mu)\big].
\end{equation*}
Mean-field theory is reliable when the one-loop term in the bracket is small,
$c\,|J'(\mu)|\ll1$ with $c=|f''(0)|f'(0)$, i.e.
\begin{equation}
\frac{|f''(0)|\,f'(0)}{2}\int_{\Qp^N}\frac{\hat w(\xi)^3}{\big(\mu+A_w(\xi)\big)^2}\,d^N\xi\ll1 .
\end{equation}
The singular dependence on $\mu$ comes from small $\norm{\xi}$. By
Eqs.~\eqref{eq:Wsymbol} and~\eqref{eq:Wasym}, $\hat w(\xi)=\hat w(0)-A_w(\xi)/f'(0)$ and
$A_w(\xi)=O(\norm{\xi}^{\sigma})$, so
$\hat w(\xi)^3=\hat w(0)^3\big(1+O(\norm{\xi}^{\sigma})\big)$. The correction contributes
at most a multiple of $\int A_w/(\mu+A_w)^2\,d^N\xi$, which is smaller than the leading
term by a factor $O\big(\mu^{\min(1,\,2-N/\sigma)}\big)$ (times $|\ln\mu|$ if $N=\sigma$).
Replacing $\hat w(\xi)^3$ by $\hat w(0)^3$ therefore does not change the leading singular
behavior, and by Eq.~\eqref{eq:g_bare} the criterion becomes
\begin{equation}
\frac{g^2}{4}\,I_2(\mu)\ll1,\qquad
I_2(\mu)=\int_{\norm{\xi}\le p^{\Lambda}}\frac{d^N\xi}{\big(\mu+A_w(\xi)\big)^2},
\label{eq:I2def}
\end{equation}
with $\Lambda\in\mathbb Z$, and $g$ as in Eq.~\eqref{eq:g_bare}. For $N\le\sigma$, $J(0)$ diverges and there is no shift to
subtract; the direct ratio $c\,J(\mu)/\mu$ in Eq.~\eqref{eq:BC50} then behaves as
$\mu^{N/\sigma-2}$, the same power as below.

\paragraph*{The cutoff.}
Since $w$ is radial, $A_w(\xi)$ depends only on $\norm{\xi}$. Write $A_w(p^k)$ for its
value on the sphere $S_k^N=\{\norm{\xi}=p^k\}$. By Lemma~\ref{lem:radial} of
Appendix~\ref{app:integrals},
\begin{equation}
I_2(\mu)=(1-p^{-N})\sum_{k=-\infty}^{\Lambda}\frac{p^{kN}}{\big(\mu+A_w(p^k)\big)^2}.
\label{eq:I2sum}
\end{equation}
The ultraviolet limit of Eq.~\eqref{eq:I2sum} is controlled by the terms involving larger
positive powers of $p^{k}$ (associated with small balls $p^{k}\Zp^N$), while the infrared
limit is controlled by powers of $p^{k}$, with $k\to-\infty$. The cutoff is needed because $A_w$ is bounded:
$0\le A_w(\xi)\le2f'(0)\hat w(0)$, and $A_w(\xi)\to f'(0)\hat w(0)$ as
$\norm{\xi}\to\infty$ by the Riemann--Lebesgue lemma. Without a cutoff the sum would
diverge at large $k$. This divergence is created by the replacement
$\hat w(\xi)^3\to\hat w(0)^3$, since $\hat w(\xi)\to0$; it is not a property of the model.
The natural values of $\Lambda$ are discussed in Sec.~\ref{sec:cutoffs}.

\begin{lemma}[Infrared asymptotics of $I_2$]\label{lem:I2}
Let $w\ge0$ be radial and integrable, and $f'(0)>0$. Assume Eq.~\eqref{eq:Wasym} in the
form $A_w(\xi)=\kappa_W\norm{\xi}^{\sigma}\big(1+\varepsilon(\xi)\big)$, with
$|\varepsilon(\xi)|\le C\norm{\xi}^{\delta}$ for $\norm{\xi}\le p^{-r}$, for some
$r\in\mathbb Z$ and $C,\delta>0$. Fix $\Lambda\in\mathbb Z$. Then:
\begin{enumerate}
\item $A_w(\xi)>0$ for every $\xi\neq0$.
\item $I_2(0)<\infty$ if and only if $N>2\sigma$.
\item If $N<2\sigma$, then, as $\mu\to0^+$,
\begin{align}
I_2(\mu)={}&(1-p^{-N})\,\kappa_W^{-N/\sigma}\,\mu^{N/\sigma-2}\nonumber\\
&\times\Big[\Phi(s)+O\big((\mu/\kappa_W)^{\delta/\sigma}\big)+O\big(\mu^{2-N/\sigma}\big)\Big],
\label{eq:I2}
\end{align}
where $\mu/\kappa_W=p^{\sigma s}$, $s\in\mathbb R$, and
\begin{equation}
\Phi(s)=\sum_{k\in\mathbb{Z}}\frac{p^{(k-s)N}}{\big(1+p^{\sigma(k-s)}\big)^2}.
\label{eq:Phi}
\end{equation}
If $N+\delta=2\sigma$, the first error term carries an extra factor $|\ln\mu|$; if
$N+\delta>2\sigma$, it is absorbed into the second. The constant in the second error term
depends on $\Lambda$.
\end{enumerate}
\end{lemma}

The proof is given in Appendix~\ref{app:integrals}. Three remarks:
\begin{itemize}
\item No hypothesis beyond Eq.~\eqref{eq:Wasym} is needed. In particular, positivity of
  $A_w$ away from the origin is a consequence of Eq.~\eqref{eq:Wasym}, not an extra
  assumption. If only $\varepsilon(\xi)\to0$ is known, Eq.~\eqref{eq:I2} still holds with
  an error $o(1)$ inside the bracket.
\item The limit $\mu\to0$ is taken at fixed $\Lambda$: the constant in the second error
  term grows like $p^{\Lambda N}$. The power law~\eqref{eq:Wasym} is used only in the
  infrared.
\item For the power-law tails of Lemma~\ref{lem:tail}, $\varepsilon\equiv0$ on
  $\norm{\xi}\le p^{-r}$, and the first error term is absent. In the Euclidean case a
  power-law tail always produces correction terms.
\end{itemize}

The function $\Phi$ is periodic in $s=\sigma^{-1}\log_p(\mu/\kappa_W)$, with period $1$.
Define the ultrametric diffusion length
\begin{equation}
\ell_W=\Big(\frac{\kappa_W}{\mu}\Big)^{1/\sigma},\qquad k^*=\log_p\ell_W .
\end{equation}
Here $\mu=\alpha-f'(0)\hat w(0)$ is the bare distance to criticality (Sec.~\ref{sec:RG}). Near
the shifted critical point it should be read as $\mu-\mu_c$, which is $\simeq\mu_R$ in the
region where the criterion holds; the two differ by the shift $\mu_c=O(g^2)$.
By Eqs.~\eqref{eq:I2def} and~\eqref{eq:I2}, the mean-field criterion reads
\begin{equation}
\frac{1-p^{-N}}{4}\,\frac{g^2}{\kappa_W^2}\,\ell_W^{\,2\sigma-N}\,\Phi(s)\ll1 .
\label{eq:ginzburg}
\end{equation}
Recall that the coupling $g$ of Eq.~(59) of Ref.~\cite{BuiceCowan2007}, for which B-C give no
explicit expression, has the bare value $g/2$, with our $g$ of Eq.~\eqref{eq:g_bare}; this
follows from their rescaling, Eq.~(58) of Ref.~\cite{BuiceCowan2007} (Sec.~\ref{sec:RG}).

\paragraph*{Physical meaning of the criterion.}
The left-hand side of Eq.~\eqref{eq:ginzburg} is the one-loop term in the bracket of
$\mu_R\simeq(\mu-\mu_c)[1-f''(0)f'(0)J'(\mu)]$, i.e.\ the relative size of the fluctuation
correction to the response of the network. Since
$\langle S\rangle=1/\mu_R\simeq(\mu-\mu_c)^{-1}[1+f''(0)f'(0)J'(\mu)]$, it is also the
relative correction to the mean avalanche size: mean-field theory is reliable when
fluctuations change $\langle S\rangle$ by a small fraction. Its two
factors have a direct reading. The coupling $g^2\propto f'(0)\,|f''(0)|$ measures the strength
of the spike-count noise that survives in the dynamics of the mean: it comes from the branching
of activity, $f'(0)$, and from the saturation of the firing rate, $f''(0)$, and it vanishes for
a linear $f$, for which mean-field theory is exact. The factor $\ell_W^{\,2\sigma-N}$ measures how
much of this noise is averaged out. A perturbation of the quiescent state lives a time of order
$\ell_W^{\sigma}/\kappa_W=1/\mu$ and, during that time, spreads over a ball of radius $\ell_W$,
which contains a number of neurons proportional to $\ell_W^{N}$; its noise is averaged over that
ball and that time. Indeed, since $\ell_W^{\,2\sigma-N}/\kappa_W^2=\kappa_W^{-N/\sigma}\mu^{N/\sigma-2}$, the
left-hand side of Eq.~\eqref{eq:ginzburg} is, up to the factor $\Phi(s)$, the relative fluctuation
of the activity in a correlation volume during a correlation time, Eq.~\eqref{eq:ginzburg_heur}
of Sec.~\ref{sec:RG}. Mean-field theory is reliable when these fluctuations are small: when each
neuron, through the connections it receives, averages over many independent sources of noise
before the perturbation relaxes. As $\mu\to0$, the ball of radius $\ell_W$ climbs the hierarchy,
up to level $k^*=\log_p\ell_W$. If $N>2\sigma$, the number of neurons it contains grows faster
than the noise accumulates, and the criterion improves; this is the hierarchical form of the
condition $W_{k+1}/W_k=p^{-\sigma}\ge p^{-N/2}$ of Sec.~\ref{sec:RG}. If $N<2\sigma$, the
connectivity decays too fast with hierarchical distance, the effective number of independent
inputs does not keep up, and fluctuations dominate sufficiently close to the critical point. On a
finite network the ball cannot exceed the network itself, which leads to the Ginzburg number of
Sec.~\ref{sec:cutoffs}.

Two features are new:
\begin{enumerate}
\item \emph{Hierarchical reading.} $k^*$ is the level of the tree that a
  fluctuation reaches before relaxing. Mean-field theory fails when fluctuations
  propagate high enough in the hierarchy, i.e.\ when small balls start to act
  coherently with balls that are far away in the hierarchical distance.
\item \emph{Log-periodicity.} The factor $\Phi(s)$ reflects discrete scale invariance under
  $x\to px$: observables near criticality carry modulations that are periodic functions of
  $\ln\mu$ with period $\sigma\ln p$, i.e.\ invariant under $\mu\mapsto p^{\sigma}\mu$. The Fourier series of $\Phi$, derived in
  Appendix~\ref{app:integrals}, gives the size of the modulation. With $a=N/\sigma$ and
  $\omega_n=2\pi n/(\sigma\ln p)$,
  \begin{equation}
  \Phi(s)=\frac{1}{\sigma\ln p}\sum_{n\in\mathbb Z}\Gamma(a-i\omega_n)\,
  \Gamma(2-a+i\omega_n)\,e^{-2\pi ins},
  \label{eq:Phi_Fourier}
  \end{equation}
  where $\Gamma$ denotes Euler's Gamma function, not the vertex function $\Gamma^{(1,1)}$ of
  Eqs.~\eqref{eq:Gamma11}--\eqref{eq:selfenergy}.
  The mean value of $\Phi$ is $(1-a)\pi/\big(\sigma\ln p\,\sin(\pi a)\big)$, equal to
  $1/(\sigma\ln p)$ for $a=1$. The oscillating part decreases like
  $|\omega_1|e^{-\pi|\omega_1|}$, with $\omega_1=2\pi/(\sigma\ln p)$, so it is exponentially
  small unless $\sigma\ln p$ is large. For $N=\sigma=1$, the relative amplitude of the
  oscillation of $\Phi$ is about $5\times10^{-11}$ for $p=2$, $2\times10^{-3}$ for $p=7$,
  $0.07$ for $p=31$, and $0.24$ for $p=101$. The value for $p=7$ is of the same order as
  the amplitude found in the avalanche simulations of Sec.~\ref{sec:avalanches}.
  Log-periodicity is thus a qualitative signature of a $p$-adic (discrete-hierarchical)
  architecture that a Euclidean theory cannot produce, but it is quantitatively
  observable only when $\sigma\ln p$ is large.
\end{enumerate}

\subsection{Natural cutoffs}\label{sec:cutoffs}

Two scales enter the criterion~\eqref{eq:ginzburg}. The ultraviolet cutoff $p^{\Lambda}$
controls non-universal quantities: the shift $\mu_c$ of the critical point (defined after
Eq.~\eqref{eq:BC50}) and the amplitude of the $O(\mu^{2-N/\sigma})$ correction in Eq.~\eqref{eq:I2}. The infrared
scale is the size $L$ (diameter) of the network ($L=p^{r}$, $r\in\mathbb Z$). The universal content of this section, namely
$N_c=2\sigma$, the power $\mu^{N/\sigma-2}$, and the function $\Phi$, does not depend on
either, provided the scale $\ell_W$ at which the criterion is evaluated lies many levels of
the hierarchy away from both: $p^{-\Lambda}\ll\ell_W\ll L$. Since the terms of $\Phi(s)$ decay
geometrically away from the level $k=-\log_p\ell_W$, the cutoffs then change these results
only by relative amounts of order $p^{-(2\sigma-N)(\Lambda+\log_p\ell_W)}$ and $(\ell_W/L)^N$. This
scale separation is
the real assumption behind any Ginzburg criterion; B-C make it implicitly when they
extract the infrared singular behavior of their Eq.~(51). In the $p$-adic model both
cutoffs are natural, and both are exact.

\paragraph*{The network cutoff.}
Fields on the finite network $G_l^N$ belong to $\mathcal D^l$: they are supported in
$\Zp^N$ and constant on the balls of radius $p^{-l}$. By Appendix~\ref{app:padic}, their
Fourier transforms are supported in $\norm{\xi}\le p^{l}$ and constant on the balls of
radius $1$. Hence $\Lambda=l$. The momenta form the finite set
$p^{-l}\Zp^N/\Zp^N$, one class per neuron: $\xi=0$, and $(1-p^{-N})p^{kN}$ classes on
each sphere $\norm{\xi}=p^k$, $1\le k\le l$. By Proposition~\ref{prop:exact} this is not
an approximation. Suppose the network keeps only the connections inside $\Zp^N$, i.e.,
it uses $w_{\mathrm{in}}=w\,1_{\Zp^N}$, and set
$\mu_l=\alpha-f'(0)\hat w_{\mathrm{in}}(0)$. The characters $\chi_p(\xi\cdot x)$,
$\xi\in p^{-l}\Zp^N$, are then exact eigenfunctions of the linearized dynamics on $\Zp^N$,
with decay rates $\mu_l+A_{w_{\mathrm{in}}}(\xi)$, and the finite-network version of
Eq.~\eqref{eq:I2sum} is
\begin{equation}
I_2^{(l)}(\mu_l)=\frac{1}{\mu_l^2}+(1-p^{-N})\sum_{k=1}^{l}
\frac{p^{kN}}{\big(\mu_l+A_{w_{\mathrm{in}}}(p^k)\big)^2}.
\label{eq:I2finite}
\end{equation}
For a power law inside the network, $w_{\mathrm{in}}(x)=C_w\norm{x}^{-N-\sigma}$ for
$p^{1-l}\le\norm{x}\le1$, Eq.~\eqref{eq:symbol_spheres} gives, for $1\le k\le l$,
\begin{equation}
A_{w_{\mathrm{in}}}(p^k)=\kappa_Wp^{k\sigma}-c_\infty,\qquad
c_\infty=f'(0)\,C_w\,\frac{1-p^{-N}}{p^{\sigma}-1},
\end{equation}
where $c_\infty<\kappa_Wp^{\sigma}$ is the coupling lost to the outside of $\Zp^N$. Three
consequences follow:
\begin{itemize}
\item The finite network is critical at $\mu_l=0$, i.e.\
  $\alpha=f'(0)\hat w_{\mathrm{in}}(0)$. Only the uniform mode becomes critical: every mode
  with $\xi\neq0$ keeps a decay rate bounded below,
  $\mu_l+A_{w_{\mathrm{in}}}(\xi)\ge A_{w_{\mathrm{in}}}(p)=\kappa_Wp^{\sigma}-c_\infty>0$ (a spectral
  gap), and relaxes in a finite time even at $\mu_l=0$.
\item The scaling form~\eqref{eq:I2} holds only while the relevant momentum
  $\norm{\xi}\approx\ell_W^{-1}$ lies between $p$ and $p^l$, i.e., in the \emph{scaling window}
  $p^{-l}\ll\ell_W\ll1$ (equivalently, $\kappa_W\ll\mu_l\ll\kappa_Wp^{l\sigma}$): the diffusion length must be large compared with the size of a
  neuron (a leaf of the tree) and small compared with the size of the network. When
  $\ell_W$ becomes comparable to the size of the network, the zero mode $1/\mu_l^2$
  dominates. This is finite-size rounding: on a finite network an absorbing-state
  transition is never sharp. The depth $l$ is the natural finite-size parameter.
\item On a network whose neuron space is the ball $B_r^N=p^{-r}\Zp^N$, of diameter $L=p^{r}$
  (here $r=0$, $L=1$), the left-hand side
  of~\eqref{eq:ginzburg} is largest at $\ell_W\approx L$. Mean-field theory therefore
  fails somewhere inside the scaling window if and only if the Ginzburg number
  \begin{equation}
  \mathrm{Gi}=\frac{g^2}{\kappa_W^2}\,L^{2\sigma-N}
  \label{eq:Gi}
  \end{equation}
  is of order one or larger. If $\mathrm{Gi}\ll1$, mean-field theory holds throughout the
  window, and the approach to $\mu_l=0$ is governed by the fluctuations of the zero mode,
  i.e.\ by the finite size of the network. On $\Qp^N$ ($L=\infty$) and for $N<2\sigma$,
  mean-field theory fails sufficiently close to criticality, whatever the value of $g$.
\end{itemize}
The exhaustion $p^{-l}\Zp^N\uparrow\Qp^N$ of Sec.~\ref{sec:geometry} removes both cutoffs
at once: the momenta run over $p^{1-l}\le\norm{\xi}\le p^{l}$.

\paragraph*{The kernel cutoff.}
In $\Qp^N$ the Fourier transform exchanges the scale of local constancy with the radius
of the support (Appendix~\ref{app:padic}). A radial $w$ is constant on each sphere, so if
it is constant on the ball $B_{-m}^N$ it is constant on all balls of radius $p^{-m}$, and
then $\hat w(\xi)=0$ for $\norm{\xi}>p^{m}$. In this case the integrand of the exact
criterion vanishes for $\norm{\xi}>p^{m}$, the cutoff $\Lambda=m$ is supplied by $w$
itself, and $A_w(\xi)=f'(0)\hat w(0)$ exactly for $\norm{\xi}>p^m$. This makes precise
B-C's remark that the integrals are ``well behaved in the ultraviolet because of the
weight function, if not by a neural lattice cutoff''~\cite{BuiceCowan2007}. A Euclidean
kernel cannot do this: a compactly supported $\hat w$ forces $w$ to be analytic, hence not
compactly supported. Biologically, $p^{-m}$ is the size of the smallest population that the
connectivity distinguishes. The dual statement concerns the support. If $w$ is supported
in a ball of radius $p^{r}$, then $A_w(\xi)=0$ for $\norm{\xi}\le p^{-r}$. In an
ultrametric space this is literal: each neuron interacts only inside its own ball of
radius $p^r$, and the network splits into disconnected pieces. All the modes with
$\norm{\xi}\le p^{-r}$ become massless together, $I_2\sim\mu^{-2}$, and each piece behaves
as a finite network, with no scaling regime. Eq.~\eqref{eq:Wasym} excludes this case
(Lemma~\ref{lem:I2}): a power-law criterion requires connections at all hierarchical
distances.

\paragraph*{Comparison with field theories of artificial neural networks.}
In the field-theoretic approaches to DNNs, the natural large parameter is the width $n$
of the layers, or the number of
neurons~\cite{HeliasDahmen2020,Halverson2021,GrosvenorJefferson2022,RobertsYaida2022,Segadlo2022}.
At fixed depth, the infinite-width limit is a Gaussian theory, and the non-Gaussian
corrections come in powers of $1/n$; in Ref.~\cite{RobertsYaida2022} the effective
expansion parameter is the depth-to-width ratio. Other works regularize the perturbative
integrals with a cutoff in input space and obtain the Wilsonian running of the couplings
by requiring that physical predictions do not depend on
it~\cite{Halverson2021,ErbinLahocheSamary2021}. The $p$-adic setting differs in two ways.
First, width and depth cannot be separated. The network $G_l^N$ has depth $l$ and
$p^{Nl}$ neurons, and the thermodynamic limit $l\to\infty$ (Sec.~\ref{sec:geometry}) adds
layers to the hierarchy at the same time as it adds neurons: the infinite-width limit
involves infinitely many layers. Second, the limiting theory is not Gaussian. The
continuum action has the cubic couplings of Eq.~\eqref{eq:RFT}, and for $N<2\sigma$ the
fluctuations they produce are not small near criticality. The cutoffs $\Lambda=l$ and $L$
are part of the architecture, and the Ginzburg number~\eqref{eq:Gi} plays the role that
$1/n$, or the depth-to-width ratio, plays in the works above. In addition, the fluctuations
here come from the stochastic spiking dynamics, not from the random initialization of the
weights. If the infinite-width limit is understood as a network with one layer and
infinitely many neurons, the $p$-adic SFTs are of a completely different nature from the
lattice SFTs of Ref.~\cite{RobertsYaida2022}. This supports the proposal of
Ref.~\cite{ZG2026} that continuous SFTs of B-C type, rather than lattice SFTs with a
large but finite number of neurons, should be the basis of the correspondence between
DNNs and SFTs.

\section{Loop-corrected hierarchical Wilson--Cowan equations}\label{sec:loops}

B-C's one-loop equations [Eqs.~(48)--(49) and (54)--(57) of Ref.~\cite{BuiceCowan2007}] carry over. Let
$s(x,t)=(w\star a)(x,t)+I$ and let $C(x,y,t)$ be the equal-time connected correlation of
$\varphi$ (a factorial cumulant of $n$). At one loop,
\begin{widetext}
\begin{align}
\partial_t a(x)=&-\alpha a(x)+f(s(x))\nonumber\\
&+\tfrac12 f''(s(x))\int_{\Qp^N}\int_{\Qp^N} w(\norm{x-y})\,w(\norm{x-z})\,C(y,z)\,d^Ny\,d^Nz,
\label{eq:WC1}\\[2pt]
\partial_tC(x,y)=&-2\alpha C(x,y)+f'(s(x))\,(w\star_1C)(x,y)+f'(s(y))\,(w\star_2C)(x,y)\nonumber\\
&+f'(s(x))\,w(\norm{x-y})\,a(y)+f'(s(y))\,w(\norm{x-y})\,a(x),
\label{eq:C}
\end{align}
\end{widetext}
where $\star_i$ denotes convolution over $\Qp^N$ in the $i$th argument, e.g.
$(w\star_1C)(x,y)=\int_{\Qp^N}w(\norm{x-u})\,C(u,y)\,d^Nu$. The source term in
Eq.~\eqref{eq:C} comes from the $(2,0)$ vertex~\eqref{eq:Gamma20}, and the one-loop term in
Eq.~\eqref{eq:WC1} from the $-\tfrac12f''(s)\,\tphi\,(w\star\varphi)^2$ vertex of
Sec.~\ref{sec:split}. Equations~\eqref{eq:WC1}--\eqref{eq:C} coincide with the generalized activity
equations~(3.23)--(3.24) of Ref.~\cite{BuiceCowanChow2010}, written for a radial kernel on
$\Qp^N$.

Two consequences:
\begin{itemize}
\item \emph{Correlation matrices between balls.} Expanding $C$ in the $p$-adic wavelets $\psi_{\gamma nj}$,
  $C=\sum C_{(\gamma nj),(\gamma'n'j')}\psi_{\gamma nj}\otimes\psi_{\gamma'n'j'}$, gives
  correlations between balls at arbitrary levels. These objects can be compared
  directly with functional-connectivity matrices.
\item \emph{Mathematical program.} Well-posedness of Eqs.~\eqref{eq:WC1}--\eqref{eq:C}
  on $\Zp^N$, convergence of level-$l$ truncations, positivity of $a$ for saturating
  $f$, and stability of hierarchical patterns. B-C remark that $N$ controls pattern
  formation; here patterns are indexed by balls, and the question becomes which balls
  synchronize.
\end{itemize}

\section{Avalanches}\label{sec:avalanches}

An \emph{avalanche} is the cascade of activity triggered by a single spike in an otherwise
quiescent network. Two conditions make the background quiescent: $\bar n=0$, i.e.\ there are no
spikes initially, and $f(0)=0$, i.e.\ there is no spontaneous firing, so that the quiescent state
is absorbing. The seed spike may make other neurons fire, each of these may trigger further
spikes, and so on, until the activity dies out and the network returns to the absorbing state.
An avalanche is thus a transient excursion from the quiescent state; it changes the activity
but not the parameters $\alpha$, $f'(0)$ and $w$, and hence not the distance $\mu$ to
criticality. Avalanches are a probe, not a mechanism: their statistics, for example the
mean size $\langle S\rangle=1/\mu_R$ and the cutoff of the size distribution, reveal how
close the network is to the critical point, but they do not change $\mu$. Only a slow,
activity-dependent plasticity acting on the parameters, i.e.\ a change of the synaptic
strengths driven by the activity itself, such as synaptic depression, can bring the
network there (self-organized criticality, Sec.~\ref{sec:SOC}). The \emph{size} $S$ of the avalanche is the total number of
spikes, and its \emph{duration} $T$ is the time until extinction. In the language of directed
percolation, an avalanche is a cluster grown from a single seed
(Appendices~\ref{app:DPmodel}--\ref{app:DPobs}), and its statistics are controlled by the
distance to the DP transition:
\begin{itemize}
\item in the absorbing phase ($\mu>0$), a spike has on average fewer than one descendant;
  avalanches are finite, and $P(S)$ is cut off exponentially beyond a cutoff size $S_c$:
  $P(S)\simeq S^{-\tau_S}G(S/S_c)$, where $G(u)$ tends to a constant for $u\ll1$ and decays
  exponentially for $u\gg1$. For $N\le N_c$,
  $S_c\sim\mu^{-(\nu_\parallel+N\nu_\perp-\beta)}$ (hyperscaling); in mean-field theory,
  $N\ge N_c$, $S_c\sim\mu^{-2}$, the value of this formula at $N=N_c=2\sigma$;
\item in the active phase ($\mu<0$), a finite fraction $P_\infty\sim|\mu|^{\beta'}$ of the
  avalanches never stops;
\item at the critical point, the branching ratio is one, the cutoff size $S_c$ diverges, and the
  distributions become power laws, $P(S)\sim S^{-\tau_S}$ and $P(T)\sim T^{-\tau_T}$; the mean
  avalanche size $\langle S\rangle=1/\mu_R$ (Sec.~\ref{sec:ginzburg}) diverges.
\end{itemize}
Only at the critical point are avalanches scale-free: in the absorbing phase the cutoff
$S_c$ sets a characteristic size, and in the active phase some avalanches never stop.
Scale-free avalanches are thus the signature of a network poised at the DP transition, and
their exponents are universal: they are fixed by the DP exponents $\beta$, $\beta'$, $\nu_\parallel$
and $\nu_\perp$. This is why B-C interpreted the avalanches observed in cortical
slices~\cite{BeggsPlenz2003} as evidence of a DP transition, and why $\tau_S$ tests the
universality class.

Following B-C [Eqs.~(60)--(64)], the avalanche-size distribution $P(S)\sim S^{-\tau_S}$ is
obtained from the path integral with $\bar n=0$ and a single seed. In terms of the
survival exponent $\delta=\beta'/\nu_\parallel$ and the initial-slip exponent $\vartheta$,
$\tau_S=1+\delta/(1+\vartheta+\delta)$~\cite{Munoz1999}. Below $N_c$, hyperscaling gives
$\vartheta=(N\nu_\perp-\beta-\beta')/\nu_\parallel$, and then
\begin{equation}
\tau_S=1+\frac{\beta'}{\nu_\parallel+N\nu_\perp-\beta}.
\end{equation}

\begin{itemize}
\item For $N\ge2\sigma$ (mean field), $\vartheta=0$ and $\delta=1$, so $\tau_S=3/2$ and the
  duration exponent is $2$. The $p$-adic theory reproduces the Beggs--Plenz
  exponent~\cite{BeggsPlenz2003}, as B-C's does. At $N=2\sigma$ the hyperscaling expression
  also gives $1+1/(1+2-1)=3/2$, consistently.
\item For $N<2\sigma$, the exponents acquire $O(\varepsilon)$ corrections that depend on
  $\sigma$, and $\tau_S\neq3/2$. This is confirmed by the simulations of
  Sec.~\ref{sec:simulations}.
\end{itemize}

\subsection{Numerical simulations}\label{sec:simulations}

The statement that $\tau_S$ departs from $3/2$ for $N<2\sigma$ rests on power counting. We
tested it by simulating the effective spike model directly.

\paragraph*{Model and method.}
The network is the tree $G_l$ with $p=2$, $N=1$ and depth $l=40$ ($2^{40}$ neurons, so
that the boundary of the network is not reached on the simulated time scales, except
marginally for the smallest $\sigma$). Two neurons $i\neq j$ have a last common ancestor
in the tree; the neurons below it form the smallest ball containing both. If this ball has
$2^k$ neurons ($k=1,\dots,l$), i.e.\ if the common ancestor lies $k$ levels above the
leaves, then $|i-j|_2=2^{k-l}$. The coupling between them is
$w\propto2^{-k(1+\sigma)}\propto|i-j|_2^{-(1+\sigma)}$ [this is Eq.~\eqref{eq:perlevel}
with $N=1$], with no self-coupling,
normalized to total coupling $w_0=1$. Each effective spike decays at rate $\alpha=1$, and
neuron $i$ emits a spike at rate $f(s_i)$ with the firing rate function
Eq.~\eqref{eq:f_choice}, $s_0=0.25$. The control parameter is
$b=f'(0)w_0/\alpha$; mean-field theory places the critical point at $b=1$, i.e.\ $\mu=0$.
The simulation is exact in continuous time: since $f(s)\le f'(0)s$, each spike attempts
to trigger a spike at rate $b$ at a neuron chosen with probability $w_{ij}/w_0$, and the
attempt is accepted with probability $f(s_i)/[f'(0)s_i]$. An avalanche starts from a single
spike and ends when no spike is left. Its size $S$ is the total number of spikes.

For each $\sigma$ the critical value $b_c$ was located by bisection, requiring that the
mean number of spikes $\mathcal N(t)$ be a pure power law of $t$ up to $t=10^3$. At $b_c$ we
generated $2\times10^5$ avalanches, followed up to $t=10^3$. We measured the exponents
$\delta$ and $\vartheta$ of the survival probability $P(t)\propto t^{-\delta}$ and of
$\mathcal N(t)\propto t^{\vartheta}$ for $30\le t\le10^3$, and the exponent $\tau_S$ by a
maximum-likelihood fit of the size distribution for $10\le S\le2000$. As a control we
ran the same program on a Euclidean network, a chain with nearest-neighbor coupling, and
on the model without saturation (a pure branching process). The exact exponents are known
in both cases.

\begin{table*}[t]
\caption{Simulation results at the critical point. $b_c$: critical value of
$b=f'(0)w_0/\alpha$ (mean-field value: $1$). $\delta$, $\vartheta$: survival and growth
exponents. $\tau_S$ (fit): maximum-likelihood fit of the size distribution; statistical
errors are below $0.005$. $\tau_S$ (scaling): from $\tau_S=1+\delta/(1+\vartheta+\delta)$.
In parentheses, the exact or known values of $(\delta,\vartheta,\tau_S)$ for the two
controls: the critical branching process and DP in one dimension. The two controls show that the fitted $\tau_S$ is systematically too large by about
$0.03$.}
\label{tab:simulations}
\begin{ruledtabular}
\begin{tabular}{lccccc}
 & $b_c$ & $\delta$ & $\vartheta$ & $\tau_S$ (fit) & $\tau_S$ (scaling)\\
\hline
Pure branching (exact: $1$, $0$, $3/2$) & 1.000 & 1.02 & -0.06 & 1.53 & 1.52\\
Tree, $\sigma=0.25$ & 1.106 & 0.99 & +0.03 & 1.52 & 1.49\\
Tree, $\sigma=0.5$ & 1.371 & 0.90 & -0.01 & 1.46 & 1.48\\
Tree, $\sigma=0.75$ & 1.739 & 0.65 & +0.16 & 1.39 & 1.36\\
Tree, $\sigma=1$ & 2.169 & 0.49 & +0.24 & 1.32 & 1.28\\
Tree, $\sigma=1.5$ & 3.157 & 0.28 & +0.32 & 1.23 & 1.17\\
Euclidean chain (DP: $0.159$, $0.314$, $1.108$) & 4.848 & 0.16 & +0.31 & 1.14 & 1.11\\
\end{tabular}
\end{ruledtabular}
\end{table*}

\begin{figure*}[t]
\centering
\includegraphics[width=0.92\textwidth]{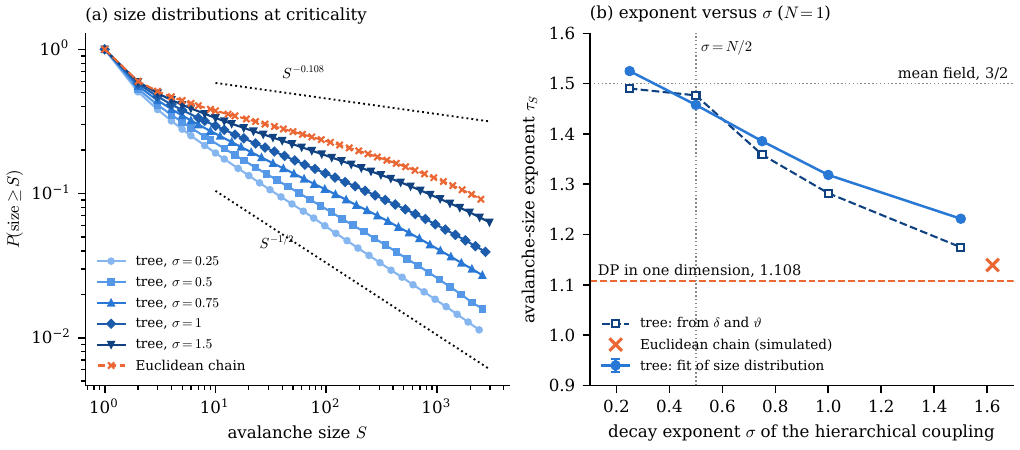}
\caption{Avalanches at the critical point. (a) Probability that an avalanche has size at
least $S$, for the hierarchical network with five values of $\sigma$ and for a Euclidean
chain with nearest-neighbor coupling. Dotted lines: the mean-field law $S^{-1/2}$
($\tau_S=3/2$) and the law $S^{-0.108}$ of directed percolation in one dimension. (b)
Avalanche-size exponent $\tau_S$ against $\sigma$. Filled circles: fit of the size
distribution. Open squares: from the measured $\delta$ and $\vartheta$. Cross: the Euclidean chain
with nearest-neighbor coupling (simulated, fit of the size distribution); it has no
parameter $\sigma$ and is drawn to the right of the tree data. Dashed horizontal line: the
value $\tau_S=1.108$ of directed percolation in one dimension~\cite{Munoz1999}, not a
simulation. The vertical line
marks $\sigma=N/2$, where $N=N_c$. Parameters as in Table~\ref{tab:simulations}.}
\label{fig:avalanches}
\end{figure*}

\paragraph*{Methods.}
The simulations follow standard methods for absorbing-state
transitions~\cite{Hinrichsen2000}. The continuous-time dynamics is simulated exactly, by
an event-driven (Gillespie-type) algorithm~\cite{Gillespie1976} combined with
thinning~\cite{LewisShedler1979}, as described above, so that the saturation of $f$ is
treated without approximation. Each run starts from a single spike, as in the spreading
method of Grassberger and de la Torre~\cite{GrassbergerTorre1979}, and the critical point is
located by requiring that $\mathcal N(t)$ be a pure power law. The exponents $\delta$ and
$\vartheta$ are obtained by least-squares fits on logarithmic scales; $\tau_S$ is obtained by
maximum likelihood for a power law truncated to the fitting window~\cite{Clauset2009}, with
bootstrap errors, and independently from the scaling relation
$\tau_S=1+\delta/(1+\vartheta+\delta)$~\cite{Munoz1999}. The two controls, with known exponents,
calibrate the systematic bias of the fit. The simulation code, the data and the scripts that produce the figures are available at
\url{https://wazuniga.github.io/padic-BC/}. We
used Claude (Anthropic; model \mbox{claude-opus-5-5}) to help implement the numerical
simulation code of this section and of Sec.~\ref{sec:smallworld}, for some well-known
techniques in statistical
physics~\cite{Hinrichsen2000,Gillespie1976,LewisShedler1979,GrassbergerTorre1979}. The control
was carried out as discussed above.

\paragraph*{Results.}
Table~\ref{tab:simulations} and Fig.~\ref{fig:avalanches} show three things.
\begin{enumerate}
\item \emph{The controls are reproduced.} The chain gives $\delta\simeq0.16$ and
  $\vartheta\simeq0.31$, the values of directed percolation in one dimension. The pure
  branching process gives $\delta\simeq1$ and $\vartheta\simeq0$. In both cases the fitted
  $\tau_S$ exceeds the exact value by about $0.03$, which we take as the systematic error
  of the fit in this window.
\item \emph{Mean-field behavior for $\sigma\le N/2$.} For $\sigma=0.25$ the exponents are
  those of the branching process, $\tau_S\simeq3/2$. At $\sigma=0.5=N/2$ they are close to
  the mean-field values, with small deviations of the size expected from logarithmic
  corrections at the upper critical dimension.
\item \emph{Non-mean-field behavior for $\sigma>N/2$.} For $\sigma=0.75$, $1$ and $1.5$,
  $\tau_S$ decreases steadily, from about $1.36$--$1.39$ to about $1.18$--$1.23$, and
  $\vartheta$ becomes positive. The exponents vary continuously with $\sigma$ and move
  toward the short-range values as $\sigma$ grows.
\end{enumerate}
These results agree with $N_c=2\sigma$, and they show that the hierarchical decay exponent
$\sigma$ controls the avalanche statistics. They are numerical evidence, not precise
determinations: the runs are limited to $t\le10^3$, the critical points carry an
uncertainty of order $10^{-3}$, and the two estimates of $\tau_S$ differ by up to $0.06$.
The simulation also shows that the critical point is shifted well above its mean-field
value, $b_c>1$, as the one-loop estimate of Appendix~\ref{app:feynman} indicates
qualitatively. The shift is large here because a single spike already produces an
input comparable to $s_0$ at its closest neighbors. We also simulated, with the same rules and the same
procedure, a Euclidean chain with power-law coupling $w(r)\propto r^{-1-\sigma}$, whose
critical point is directed percolation with long-range
spreading~\cite{Janssen1999,HinrichsenHoward1999}. For $\sigma\le1.5$ its avalanche
exponents coincide with those of the hierarchical network within numerical accuracy
(the fitted values of $\tau_S$ differ by at most $0.005$ for $\sigma\le1$ and by $0.02$ at
$\sigma=1.5$), while the critical points differ. Whether the two networks differ at
larger $\sigma$, where the Euclidean theory crosses over to ordinary directed
percolation, is an open question.

\paragraph*{Comparison with the Euclidean model.}
The chain is more than a control: it is the Euclidean counterpart of the tree, with the
same number of dimensions ($N=1$), the same local rules and the same total coupling. The
two networks behave differently for a simple reason. On the chain a spike excites only
its nearest neighbors, so the activity spreads as a compact patch and new spikes keep
landing on neurons that already receive input. This crowding is what the saturation of
$f$ penalizes, and it drives the exponents away from their mean-field values:
$\tau_S\simeq1.11$, far from $3/2$. On the tree with small $\sigma$ the coupling decays
slowly with the hierarchical distance, most new spikes land far from the active neurons,
crowding is rare, and the avalanches are those of a branching process. As $\sigma$ grows
the activity stays inside small balls, crowding returns, and $\tau_S$ falls toward the
Euclidean value. Hence in the Euclidean theory with short-range coupling the dimension
of space decides whether the avalanches are mean-field, and the Beggs--Plenz exponent
$3/2$ requires $N\ge4$; in the $p$-adic theory the decay exponent $\sigma$ decides, and
a network with $N=1$ gives $3/2$ when $\sigma\le1/2$. The shift of the critical
point shows the same ordering: $b_c$ grows from $1.11$ at $\sigma=0.25$ to $3.16$ at
$\sigma=1.5$, and is largest for the chain, $b_c=4.85$. Mean-field theory is therefore a
better guide for a hierarchical network with slowly decaying coupling than for the
Euclidean chain. Section~\ref{sec:smallworld} relates this difference to the
small-world property. This comparison is limited to one tree ($p=2$, $N=1$) and one value of
$s_0$; other values of $p$ and $N$ were not simulated.

Thus an estimate of $\sigma$ from connectivity data, from how the coupling decays with
hierarchical distance, predicts whether the avalanche exponents are mean-field
($\sigma\le N/2$) or not ($\sigma>N/2$); the size of the departure is, at present, known
only from simulations.

\section{The small-world property}\label{sec:smallworld}

A network has the small-world property if it combines a high clustering of connections
with short paths: the mean number of steps between two nodes grows as the logarithm of
the number of nodes, as in a random graph, and not as a power of it, as in a
lattice~\cite{WattsStrogatz1998}. Cortical networks have this
property~\cite{BassettBullmore2006}. In this section we ask whether the networks described
by the $p$-adic theory and by the B-C theory have it.

\subsection{Two tests for a network with weighted coupling}

In the effective spike model every neuron is coupled to every other neuron, with a
weight $w$ that depends on the distance. Path lengths in the usual sense are then
trivially equal to one, so the definition has to be adapted. We use two tests.
\begin{enumerate}
\item \emph{Sparse network.} Each neuron draws $k$ connections, and the probability
  of a connection to a given neuron is proportional to $w$. On the resulting network we
  measure the clustering coefficient $C$ (the probability that two neighbors of a neuron
  are connected to each other) and the mean path length $\ell$.
\item \emph{Spreading time.} We simulate the effective spike model itself, with the full
  weighted coupling, in the active phase. One spike is placed in a network of
  $n$ neurons, and $T_{1/2}(n)$ is the mean time at which half of the neurons have fired at
  least once. Runs in which the activity dies out are discarded.
\end{enumerate}
The network is small-world if $C$ stays finite while $\ell$ and $T_{1/2}$ grow as $\ln n$,
or as a power of $\ln n$.

\subsection{A simple estimate}

Consider the kernel~\eqref{eq:perlevel}, $w\propto\norm{x}^{-N-\sigma}$, and a ball of level
$m$, which has $p^{mN}$ neurons. The total coupling of one neuron to the neurons outside
the ball is proportional to $p^{-m\sigma}$. The coupling of the whole ball to its exterior
is therefore proportional to
\begin{equation}
p^{mN}\,p^{-m\sigma}=p^{m(N-\sigma)}.
\label{eq:escape}
\end{equation}
If $\sigma<N$ this grows with $m$: the larger the structure, the more strongly it is
connected to the rest of the network, at every level. Once a ball is active, its activity
passes to the next level in a time that does not grow with $m$, and the time to reach
$n=p^{lN}$ neurons is at most of order $l\propto\ln n$. If $\sigma>N$ the
quantity~\eqref{eq:escape} decreases with $m$: large structures are almost isolated,
activity must wait a time of order $p^{m(\sigma-N)}$ to leave a ball of level $m$, and the
spreading time grows as the power $n^{(\sigma-N)/N}$. The clustering is high in both
cases, since most of the coupling of a neuron stays inside the smallest balls that
contain it. This estimate gives the threshold
\begin{equation}
\sigma<N
\label{eq:sw_threshold}
\end{equation}
for the small-world property. It is the same threshold that separates the two regimes of
long-range percolation on a hierarchical lattice~\cite{KovalMeesterTrapman2012} and on a
Euclidean lattice~\cite{BenjaminiBerger2001,Biskup2004}. It is also the condition under
which the random walk generated by the Taibleson--Vladimirov operator is transient, since
$\int\norm{k}^{-\sigma}d^Nk$ converges at $k=0$ exactly when $\sigma<N$. (Heuristic.)

\subsection{Numerical results}

\begin{figure*}[t]
\centering
\includegraphics[width=0.92\textwidth]{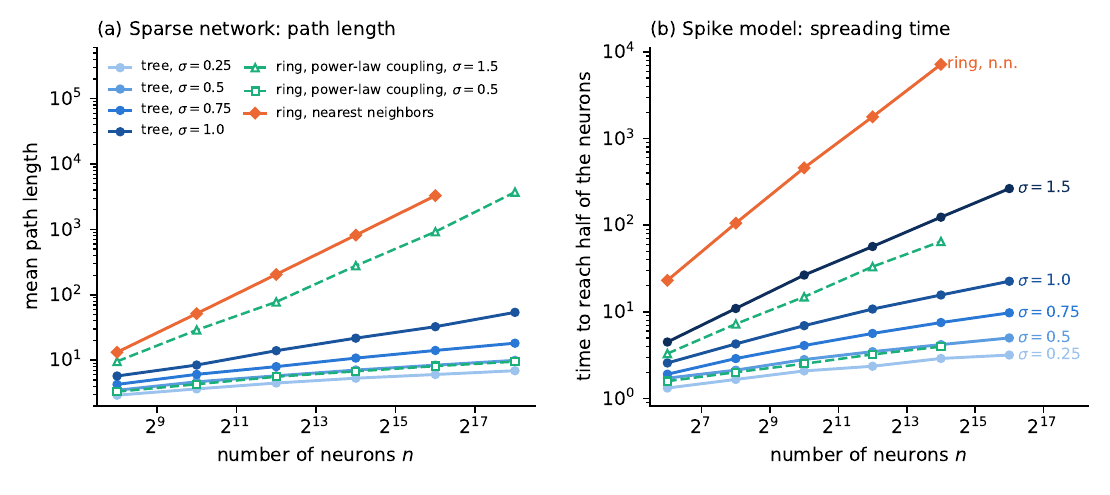}
\caption{Small-world tests. (a) Mean path length of sparse networks with $n$ neurons, in
which each neuron draws $k=6$ connections with probabilities proportional to the
coupling. Filled circles: tree with $p=2$, $N=1$ (the network with $\sigma=1.5$ is not
connected and is not shown). Open symbols: Euclidean ring with power-law coupling.
Diamonds: Euclidean ring in which each neuron is connected to its five nearest neighbors
on each side. (b) Spike model in the active phase, $b=6$: mean time at which half of
the $n$ neurons have fired at least once, for the same networks; here the short-range
ring has nearest-neighbor coupling. Error bars are standard errors and are mostly smaller
than the symbols.}
\label{fig:smallworld}
\end{figure*}

\begin{table}[b]
\caption{Small-world tests for $N=1$. $C$ and $\ell$: clustering coefficient and mean
path length of the sparse network with $n=2^{16}$ neurons. $T_{1/2}$: spreading time of
the spike model with $n=2^{12}$ and $n=2^{14}$ neurons, in units of $1/\alpha$. The sparse tree network with $\sigma=1.5$ is not connected. A random
graph of the same size and degree has $C\approx10^{-4}$.}
\label{tab:smallworld}
\begin{ruledtabular}
\begin{tabular}{lcccc}
 & $C$ & $\ell$ & $T_{1/2}(2^{12})$ & $T_{1/2}(2^{14})$\\
\hline
Tree, $\sigma=0.25$ & 0.08 & 6.0 & 2.4 & 2.9\\
Tree, $\sigma=0.5$ & 0.18 & 8.4 & 3.5 & 4.2\\
Tree, $\sigma=0.75$ & 0.27 & 14.1 & 5.6 & 7.5\\
Tree, $\sigma=1$ & 0.34 & 32.6 & 10.8 & 15.7\\
Tree, $\sigma=1.5$ & 0.45 & --- & 57 & 124\\
Ring, power law, $\sigma=0.5$ & 0.17 & 8.1 & 3.2 & 4.0\\
Ring, power law, $\sigma=1.5$ & 0.40 & 921 & 33 & 65\\
Ring, short range & 0.67 & 3277 & 1783 & 7186\\
\end{tabular}
\end{ruledtabular}
\end{table}

We took $p=2$ and $N=1$, as in Sec.~\ref{sec:simulations}, with finite trees of
$n=2^{l}$ neurons, and compared them with Euclidean rings of the same size. The sparse
networks have $k=6$ and $n$ up to $2^{18}$. The spreading times were measured with the
simulator of Sec.~\ref{sec:simulations}, with $s_0=0.25$ and $b=6$, which is above the
critical point of every network considered, with $5$ to $37$ surviving runs per point. For the
Euclidean ring with power-law coupling, $w(r)\propto r^{-1-\sigma}$ with $r$ the distance
along the ring, we wrote a separate program with the same rules. The results are in
Fig.~\ref{fig:smallworld} and Table~\ref{tab:smallworld}.

\paragraph*{Hierarchical network.}
For $\sigma=0.25$ and $\sigma=0.5$ the path length and the spreading time increase by
an approximately constant amount each time the network is enlarged by a factor of four:
they grow as $\ln n$. The clustering coefficient stays between $0.08$ and $0.18$, of order
a thousand times that of a random graph. These networks are small-world. For $\sigma=0.75$
the growth is faster than $\ln n$ but slower than a power of $n$. The case $\sigma=1=N$ is
marginal: over the range studied, $\ell$ and $T_{1/2}$ grow approximately as $n^{0.3}$. For
$\sigma=1.5$ the sparse network falls apart into small pieces (the largest contains less
than $1\%$ of the neurons at $n=2^{16}$), and $T_{1/2}$ grows as $n^{0.55}$, close to the
estimate $n^{\sigma-N}=n^{0.5}$. A tree is therefore not automatically a small-world
network: the property holds for $\sigma<N$, in agreement with Eq.~\eqref{eq:sw_threshold}.

\paragraph*{Euclidean networks.}
On the ring with short-range coupling both $\ell$ and $T_{1/2}$ are proportional to $n$:
activity advances as a front with constant velocity, as in the B-C theory. At $n=2^{14}$
the spreading time is $7186$, against $4.2$ on the tree with $\sigma=0.5$. The ring
with power-law coupling behaves like the tree with the same exponent. For $\sigma=0.5$
the two networks have almost the same clustering, path length and spreading time. For
$\sigma=1.5$ neither is small-world.

\subsection{Comparison with the B-C theory}

The master equation and the action of B-C are written for a general weight function, so
the formalism does not exclude a small-world network. The restriction enters in the
continuum analysis of the critical point. There the connectivity is assumed to be short
range, its Fourier transform is expanded as $\hat w(q)\simeq w_0-\tfrac12w_2q^2$, which gives the
diffusion term $D\nabla^2$, and the Ginzburg criterion uses the moments $w_2,w_4,\dots$
[Eqs.~(50)--(52) of Ref.~\cite{BuiceCowan2007}]. These hypotheses exclude the small-world
property. A kernel with power-law tail $|x|^{-N-\sigma}$ produces a small-world network
only for $\sigma<N$~\cite{BenjaminiBerger2001,Biskup2004}, and then its moment of order $N$
is infinite. A finite fourth moment forces $\sigma>4>N$ for $N\le3$. (A finite second
moment alone does not suffice when $N=3$: a tail with $2<\sigma<3$ has a finite second
moment and gives a small-world network, while the critical behavior remains that of DP.)
Under the hypotheses of B-C, activity passes only to nearby neurons, the time to reach $n$
neurons grows as the power $n^{1/N}$, and the network is not small-world. This is the case
represented by our short-range ring.

By our tests, which concern finite networks, the property is recovered within the B-C
formalism if the hypothesis is dropped and the
kernel is taken to have a power-law tail: $w$ bounded for $|x|<1$ and
$w(x)\propto|x|^{-N-\sigma}$ for $|x|\ge1$, with $\sigma<\min(N,2)$. The cutoff at short
distances is necessary, because the pure power law is not integrable at the origin. With
it, $w$ is integrable for every $\sigma>0$, although its second moment is infinite for
$\sigma\le2$. Integrability is what the mean-field theory requires. The Wilson--Cowan
model on a general locally compact group was studied in Ref.~\cite{ZZ2023WC} under the
hypothesis that the kernels are integrable. The existence proof uses only the inequality
$\Vert w\star a\Vert_\infty\le\Vert w\Vert_1\Vert a\Vert_\infty$, so that it holds on
$\mathbb R^N$ as well: for an integrable kernel and a Lipschitz firing-rate function the
Wilson--Cowan equation has a unique global solution in $L^\infty$. Hence the
long-range Euclidean model is well defined at the mean-field level. In our simulations
the neurons are on a lattice, $r\ge1$, and the kernel is summable. The diffusion term is
then replaced by the fractional Laplacian $(-\nabla^2)^{\sigma/2}$, and the critical
behavior is that of directed percolation with long-range
spreading~\cite{Janssen1999,HinrichsenHoward1999}, whose upper critical dimension is also
$2\sigma$. Our simulations of the ring with power-law coupling show that this network is
small-world for $\sigma=0.5$.

The conclusion is the following. The B-C theory in its diffusion limit describes
networks that are not small-world. The small-world property requires long-range
coupling. In the Euclidean setting this coupling must be imposed as an additional
hypothesis, on a theory whose generic limit is local diffusion. In the $p$-adic setting
all diffusion operators are nonlocal (Sec.~\ref{sec:Wop}), so long-range coupling is the only possibility,
and the property holds for $\sigma<N$. What
distinguishes the $p$-adic theory from a finite Euclidean network with a power-law
coupling is the hierarchy: nested balls, discrete scale invariance and
log-periodic corrections, none of which the Euclidean kernel has.

\paragraph*{Relation to the compactness argument of Ref.~\cite{ZZ2023WC}.}
Reference~\cite{ZZ2023WC} treats the small-world property of the Wilson--Cowan model in
the continuum, with a different and stronger criterion: the interaction between any two
groups of neurons must be non-negligible, $w(x)\ge\varepsilon>0$ for all $x$. The kernels
must also be integrable, since this is the hypothesis under which the Wilson--Cowan
equations are well posed on a general group. The two conditions together force the group
to be compact. The group $(\mathbb R^N,+)$ has no nontrivial compact subgroups, which is
a consequence of the Archimedean axiom. Hence the classical model, whose Gaussian or
exponential kernels have compact support for practical purposes, is incompatible with
the small-world property, and by this criterion so is every integrable kernel on
$\mathbb R^N$, including one with a power-law tail. In $\Qp^N$ the balls are compact
subgroups. On a ball every positive radial kernel satisfies the criterion, and the theory
restricted to a ball is again a theory of the same type (Proposition~\ref{prop:exact}).

The two approaches answer different questions and are consistent. The criterion of
Ref.~\cite{ZZ2023WC} concerns the continuum model on the whole space. Our tests concern
finite networks of increasing size, and they separate the kernels on a ball further:
path lengths and spreading times grow logarithmically only for $\sigma<N$. The Euclidean
ring with power-law coupling used in our simulations does not contradict the compactness
argument. A ring is a compact group, but it is not a subgroup of $\mathbb R^N$: it is
obtained by imposing periodic boundary conditions, and the power-law coupling is a
further hypothesis foreign to the classical model. In the $p$-adic setting the space
does not have to be changed: every finite network is a ball, and every ball is a compact
subgroup.

\paragraph*{Small-world property and avalanches.}
The threshold $\sigma=N$ is not the threshold $\sigma=N/2$ of Sec.~\ref{sec:avalanches}. There
are three regimes. For $\sigma\le N/2$ the network is small-world and the avalanche
exponents are mean-field. For $N/2<\sigma<N$ the network is still small-world, but the
avalanche exponents are not mean-field: at $\sigma=0.75$ we found $\tau_S\simeq1.37$
(Table~\ref{tab:simulations}). For $\sigma>N$ the network is not small-world and the
exponents approach the short-range values. In our simulations the small-world property is
thus necessary for the exponent $3/2$, but not sufficient, and the avalanche exponents
alone cannot be used as a test of it.

\paragraph*{Relation to fast transition modes.}
In ultrametric models of transition networks, the hierarchical organization of the states
produces fast transition modes that are absent in an ordinary Markov
chain~\cite{ZG2022Ultrametric}. The logarithmic spreading time found here for $\sigma<N$
is an effect of the same kind, since activity leaves each ball at a rate that does not
decrease with its size. We have not made this connection quantitative.

\paragraph*{Limitations.}
The tests were made for one prime, $N=1$, one value of $b$ and network sizes up to
$2^{18}$. The laws quoted above describe the range studied; in particular, the data do not
distinguish $\ln n$ from a power of $\ln n$ for $\sigma$ close to $N$.

\section{Phase transition and self-organization}\label{sec:phase}

The review~\cite{BuiceCowan2009} develops the physical consequences of the B-C theory:
the neocortical phase transition (Sec.~10 there), and large avalanches, rhythms,
correlations, power spectra and self-organized criticality (Sec.~12 there). In this
section we give the $p$-adic counterparts. We label each statement as \emph{exact}
(it follows from the $p$-adic setting at zero-loop order, with no further approximation) or
\emph{heuristic}. Equation
numbers of the form ``R(\dots)'' refer to Ref.~\cite{BuiceCowan2009}.

\subsection{The control parameter, layer by layer}

Let
\begin{equation}
W_k=\int_{\norm{x}=p^k}w(\norm{x})\,d^Nx=(1-p^{-N})\,p^{kN}\,w(p^k),
\label{eq:Wk}
\end{equation}
$k\in\mathbb Z$,
be the total coupling that a neuron receives from the neurons at hierarchical distance
$p^k$. Then $w_0=\hat w(0)=\sum_kW_k$, and the control parameter of the transition
[R(137)] is
\begin{equation}
\mu=\alpha-f'(0)\sum_{k\in\mathbb Z}W_k .
\label{eq:mu_layers}
\end{equation}
The balance condition $\mu=0$ constrains only the total, so the balance between
branching ($f'(0)\hat w(0)$) and decay ($\alpha$) can be distributed across layers in many
ways. (Exact.) If negative weights are allowed, so that excitation within small balls is
compensated by inhibition at larger hierarchical distances, $f$ must be extended to $s<0$
as in Remark~\ref{rem:activation}.

\subsection{Level-resolved linear spectrum}\label{sec:levels}

\begin{proposition}\label{prop:levels}
Let $w$ be radial and integrable, and let $\norm{\xi}=p^{1-\gamma}$, i.e.\ $\xi$ labels
the wavelets supported on balls of radius $p^{\gamma}$. Then
\begin{equation}
\hat w(\xi)=\sum_{j\le\gamma-1}W_j-\frac{W_\gamma}{p^N-1},
\label{eq:what_levels}
\end{equation}
and the linearized growth rate of the level-$\gamma$ modes is
\begin{equation}
\varrho_\gamma=-\lambda(p^{1-\gamma})=-\alpha+f'(0)\Big[\sum_{j<\gamma}W_j-\frac{W_\gamma}{p^N-1}\Big].
\label{eq:rho_gamma}
\end{equation}
\end{proposition}
\begin{proof}
Using Eq.~\eqref{eq:ballFT}, for $\norm{\xi}=p^m$,
$\int_{\norm{x}=p^j}\chi_p(\xi\cdot x)\,d^Nx$ equals $(1-p^{-N})p^{jN}$ if $j\le-m$,
equals $-p^{-mN}$ if $j=1-m$, and vanishes otherwise. Sum over spheres and set $m=1-\gamma$.
\end{proof}

\begin{figure*}[t]
\centering
\includegraphics[width=0.92\textwidth]{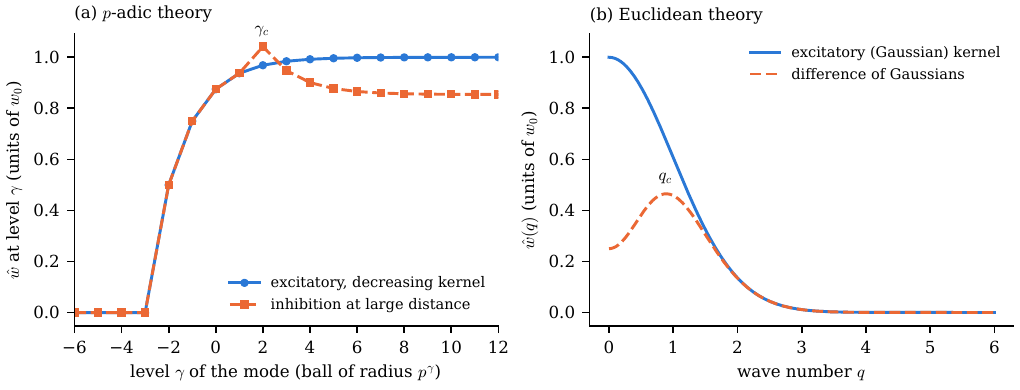}
\caption{Spectrum of the linearized dynamics (Sec.~\ref{sec:levels}). Linearizing the
Wilson--Cowan equation about the quiescent state gives $\partial_tu=-\alpha u+f'(0)\,w\star u$,
whose eigenfunctions are the wavelets; a mode supported on balls of radius $p^\gamma$ grows
at the rate $\varrho_\gamma=-\alpha+f'(0)\hat w(p^{1-\gamma})$ [Eq.~\eqref{eq:rho_gamma}], so
the first instability occurs where $\hat w$ is largest. (a) $p$-adic theory: $\hat w$ at each level
$\gamma$, from Eq.~\eqref{eq:what_levels}, for $p=2$, $N=1$. Solid: a nonnegative kernel,
constant for $|x|_p\le p^{-3}$ and decaying as $|x|_p^{-2}$ beyond; $\hat w$ increases
monotonically to its value $w_0$ at the homogeneous mode ($\gamma\to\infty$). Dashed: the
same kernel with inhibitory (negative) coupling for $|x|_p\ge p^{2}$; the maximum is
at a finite level $\gamma_c$. (b) Euclidean theory: $\hat w(q)$ for a Gaussian kernel and for a
difference of Gaussians, whose maximum is at a nonzero wave number $q_c$. In both theories
a pattern-forming instability requires inhibition.}
\label{fig:spectrum}
\end{figure*}

A mode supported on a ball of radius $p^\gamma$ averages to zero over that ball and is
constant on each of the $p^N$ balls of radius $p^{\gamma-1}$ into which it splits (its
children in the tree); these are the sub-balls meant below. It receives the full coupling
from inside these sub-balls (levels $j<\gamma$). The coupling between different sub-balls (level $\gamma$)
enters with weight $-1/(p^N-1)$. For power-law tails, Eq.~\eqref{eq:what_levels}
reproduces Lemma~\ref{lem:tail}. Since
$\hat w(p^{-\gamma})-\hat w(p^{1-\gamma})=(p^NW_\gamma-W_{\gamma+1})/(p^N-1)$ and
$W_{\gamma+1}/W_\gamma=p^N\,w(p^{\gamma+1})/w(p^{\gamma})$, $\hat w$ increases towards
coarser levels exactly when $w$ decreases with hierarchical distance
(Fig.~\ref{fig:spectrum}). (Exact.)

\paragraph*{Pattern-forming instabilities and large avalanches.}
If $w\ge0$, then $|\hat w(\xi)|\le\hat w(0)=w_0$, so the homogeneous DP transition always
comes first, even for non-monotone $w$. A pattern-forming instability requires inhibition
(negative coupling at some hierarchical distances, with the extension of $f$ of
Remark~\ref{rem:activation}), as in the Euclidean theory, where the band-pass kernel is a
difference of Gaussians. The first instability then occurs at the level $\gamma_c$ that
maximizes the bracket in Eq.~\eqref{eq:rho_gamma}, and all wavelets of that level, one set
per ball of radius $p^{\gamma_c}$, become unstable simultaneously, a degeneracy much larger
than that of the Euclidean ring $|q|=q_c$. When inhibition is blocked pharmacologically in
cortical cultures (e.g.\ by picrotoxin), the avalanche-size distribution develops a second
peak at large sizes, and the review attributes this peak to such an instability (its
Sec.~12.5). In the $p$-adic theory the peak should sit at sizes proportional to
$p^{N\gamma_c}$, and several destabilized levels give several peaks at geometrically spaced
sizes. (Exact for the linear instability; the avalanche interpretation is heuristic.)

\subsection{Propagator, scaling and exponents}

Below criticality the review's propagator R(134) is a Gaussian times $e^{-\Delta t}$.
In the $p$-adic theory,
\begin{align}
G(x,t)&=\theta(t)\,e^{-\mu t}\,Z_t(x),\nonumber\\
Z_t(x)&:=Z_t^{W}(x)=\int_{\Qp^N}\chi_p(-\xi\cdot x)\,e^{-tA_w(\xi)}\,d^N\xi ,
\label{eq:G_heat}
\end{align}
where $Z_t=Z_t^W$ is the heat kernel of the $W$-operator of Sec.~\ref{sec:Wop}, understood,
as in Eq.~\eqref{eq:Delta0x}, as an inverse Fourier transform in $\mathcal D'(\Qp^N)$. It
depends on the whole connectivity kernel $w$ through $A_w$, not only on $\sigma$.
Since $A_w$ is bounded [$A_w(\xi)\to f'(0)w_0$ as
$\norm{\xi}\to\infty$], $Z_t$ is not a function. By Eq.~\eqref{eq:Delta_decomp},
\begin{equation}
Z_t=e^{-tf'(0)w_0}\big[\delta+K_t\big],
\end{equation}
and by Proposition~\ref{prop:propagator}, for nonnegative, radially nonincreasing $w$,
\begin{equation}
G(x,t)\simeq\theta(t)\,f'(0)\,t\,e^{-\mu t}\,w(\norm{x}),\qquad \norm{x}\to\infty .
\label{eq:G_tail}
\end{equation}
For kernels with the power-law tail of Lemma~\ref{lem:tail},
$G(x,t)\simeq\theta(t)\,f'(0)C_w\,t\,e^{-\mu t}\norm{x}^{-N-\sigma}$. The propagator therefore
has \emph{power-law} tails in hierarchical distance for every $\mu$, not Gaussian ones.
(Exact.)

The representative $\kappa_W\Dsig$ of the universality class has an unbounded symbol.
Its heat kernel, which we denote by $Z^{(\sigma)}_t$ (it depends only on $\sigma$ and
$\kappa_W$), is a genuine function, with no $\delta$ part. In Ref.~\cite{BGPW2014}, $\Dsig$ is the
Taibleson operator $T_\sigma$ (Definition~5.16 there, with the same norm
$\norm{x}=\max_i|x_i|_p$), and it is shown there to be an isotropic Laplacian on the
ultrametric space $(\Qp^N,\norm{\cdot})$. Theorem~5.17 of that reference gives the
two-sided estimate
\begin{equation}
c\,F_t(x)\le Z^{(\sigma)}_t(x)\le C\,F_t(x),
\label{eq:Z_estimate}
\end{equation}
with
\begin{equation*}
F_t(x)=\kappa_Wt\,\big((\kappa_Wt)^{1/\sigma}+\norm{x}\big)^{-N-\sigma},
\end{equation*}
for all $t>0$ and $x\in\Qp^N$, where $0<c\le C$ depend only on $p$, $N$, and $\sigma$.
The upper bound was obtained earlier in Ref.~\cite{RodriguezZuniga2008} (see also
Ref.~\cite[Chap.~4]{Kochubei2001} for $N=1$). The factors of $\kappa_W$ follow from
$Z^{(\sigma)}_t=p_\sigma(\kappa_Wt,\cdot)$, where $p_\sigma$ is the heat kernel of
$T_\sigma$. Explicit admissible values of $c$ and $C$ are derived in
Appendix~\ref{app:heatkernel}, Eq.~\eqref{eq:cC_explicit}.

\paragraph*{Spectral dimension.}
Let $X_t$ be the Markov process generated by $\kappa_W\Dsig$ (the $p$-adic diffusion),
started at $X_0=y$. The function $u(x,t)=Z^{(\sigma)}_t(x-y)$ solves the heat equation
$\partial_tu+\kappa_W\Dsig u=0$ with initial condition $u(x,0)=\delta(x-y)$, and it is the
probability density of $X_t$. Its value at the starting point, $Z^{(\sigma)}_t(0)$, is the
\emph{return probability} (density) of the diffusion; by translation invariance it does not
depend on $y$. For a ball $B$ of radius $r\le(\kappa_Wt)^{1/\sigma}$ centered at $y$, the
probability that the walker is back in $B$ at time $t$ is
$P(X_t\in B)\asymp Z^{(\sigma)}_t(0)\,r^N$, since $F_t(x)\asymp F_t(0)$ for
$\norm{x}\le(\kappa_Wt)^{1/\sigma}$.

The \emph{spectral dimension} $d_s$ of a diffusion is defined by the decay of its return
probability at large times: if $h_t$ denotes its heat kernel, $h_t(0)\asymp t^{-d_s/2}$. For Brownian motion in $\mathbb R^N$
with diffusion constant $D$, $h_t(0)=(4\pi Dt)^{-N/2}$, so $d_s=N$: the spectral dimension
coincides with the dimension of space. On fractals, graphs and ultrametric spaces the two
can differ. The name refers to the spectrum of the generator. Writing
$h_t(0)=\int e^{-tA(\xi)}\,d^N\xi=\int_0^\infty e^{-t\lambda}\,d\mathcal N(\lambda)$, where
$A$ is the symbol of the generator and $\mathcal N(\lambda)=\mathrm{vol}\{\xi:A(\xi)\le\lambda\}$
counts the modes with relaxation rate at most $\lambda$, the decay $t^{-d_s/2}$ corresponds
to $\mathcal N(\lambda)\asymp\lambda^{d_s/2}$ as $\lambda\to0$: a Weyl law in which $d_s$
plays the role of the dimension. For the $p$-adic diffusion generated by $\kappa_W\Dsig$,
$\mathcal N(\lambda)=\mathrm{vol}\{\xi:\kappa_W\norm{\xi}^\sigma\le\lambda\}\asymp(\lambda/\kappa_W)^{N/\sigma}$
(compare the spectral distribution function of Ref.~\cite{BGPW2014}, Sec.~5.3.2).
Directly, at $x=0$, Eq.~\eqref{eq:Z_estimate} gives
\begin{equation}
Z^{(\sigma)}_t(0)\asymp(\kappa_Wt)^{-N/\sigma}=(\kappa_Wt)^{-d_s/2},\qquad
d_s=\frac{2N}{\sigma},
\label{eq:ds}
\end{equation}
so the spectral dimension of the $p$-adic diffusion generated by $\kappa_W\Dsig$ is $d_s=2N/\sigma$.
The neuron space has dimension $N$, the Haar volume of a ball of radius $r$ being $r^N$, and
the hierarchy enters only through $\sigma$. The mean-field condition $N\ge2\sigma$ of
Sec.~\ref{sec:RG} is exactly $d_s\ge4$, the same criterion as in the Euclidean theory with
local coupling, where $d_s=N$ and the condition reads $N\ge4$. Likewise, the
transience criterion $\sigma<N$ of the random walk generated by $\Dsig$, which is also the
small-world threshold~\eqref{eq:sw_threshold} of Sec.~\ref{sec:smallworld}, is $d_s>2$, as for
Euclidean diffusion. Thus the hierarchy does not change the dimension of the
neuron space; it changes the spectral dimension of the diffusion that the connectivity
supports, and this is what decides whether fluctuations are relevant.
For the $W$-operator of the B-C theory, Eq.~\eqref{eq:Z_estimate} is expected to describe
the regular part $K_t$ in the scaling regime of large $t$ and $\norm{x}$, where only the
small-$\xi$ behavior $A_w\simeq\kappa_W\norm{\xi}^\sigma$ matters. Its
tail agrees in form with Eq.~\eqref{eq:G_tail}. Proving this is an open problem; see
Remark~\ref{rem:Kt_open}.

\begin{remark}[Open problem: heat-kernel estimate for the regular part $K_t$]\label{rem:Kt_open}
Let $A_w$ be radial and satisfy Eq.~\eqref{eq:Wasym}, and write
$Z_t=e^{-tf'(0)w_0}\big[\delta+K_t\big]$. The $\delta$ part carries the factor
$e^{-tf'(0)w_0}$ and is negligible at large $t$. The problem is to show that the regular part
$e^{-tf'(0)w_0}K_t$ obeys the estimate~\eqref{eq:Z_estimate} in the scaling regime: there
are $t_0>0$ and constants $0<c'\le C'$ such that
\begin{equation}
c'\,F_t(x)\le e^{-tf'(0)w_0}K_t(x)\le C'\,F_t(x),\qquad t\ge t_0,\ x\ne0,
\label{eq:Kt_open}
\end{equation}
and, more precisely, that $e^{-tf'(0)w_0}K_t(x)/Z^{(\sigma)}_t(x)\to1$ as $t\to\infty$,
uniformly in $x\ne0$. Theorem~5.17 of Ref.~\cite{BGPW2014} does not apply directly: $A_w$
is bounded, so the $W$-operator is not of the form $\psi(T_1)$ with $\psi$ a bijection of
$[0,\infty)$ (Theorem~5.18 of Ref.~\cite{BGPW2014}), and $Z_t$ has a $\delta$ part. A possible
route is the series of Appendix~\ref{app:heatkernel}. Because $A_w$ is radial, the
same summation by parts gives, for $x\ne0$,
\begin{equation}
Z_t(x)=\sum_{k\in\mathbb Z}p^{kN}\,\Omega(p^k\norm{x})\,
\big(e^{-tA_k}-e^{-tA_{k+1}}\big),
\end{equation}
where $A_k$ is the value of $A_w$ on the sphere $\norm{\xi}=p^k$. The terms are nonnegative when
$A_k$ is nondecreasing in $k$. For large $t$, the dominant terms have $tA_k=O(1)$, hence
small $p^k$, where Eq.~\eqref{eq:Wasym} gives $A_k=\kappa_Wp^{k\sigma}(1+O(p^{k\delta}))$, and
the bounds of Appendix~\ref{app:heatkernel} should carry over. The remaining work is to control
the terms with large $p^k$, where $A_k$ saturates at $f'(0)w_0$, uniformly in $\norm{x}$.
\end{remark}

The review's critical scaling form R(143) holds with $z=\sigma$, but only under the
discrete dilations $x\mapsto p^{-1}x$, $t\mapsto p^{\sigma}t$. Under them
$Z^{(\sigma)}_{p^\sigma t}(p^{-1}x)=p^{-N}Z^{(\sigma)}_t(x)$ exactly, and the B-C propagator
inherits this covariance asymptotically in the scaling regime. The scaling function therefore carries a
periodic dependence on $\sigma^{-1}\log_p t$, the same discrete scale invariance found in
Sec.~\ref{sec:ginzburg}.

With the mean-field values~\eqref{eq:MF}, the exponent used in the review,
$\vartheta=N/z-2\beta/\nu_\parallel$ [R(136)], where it is denoted $\theta$, becomes $\vartheta=N/\sigma-2$. The scaling
relation $\beta=\tfrac12\nu_\parallel(N/z-\vartheta)$ gives $\beta=1$, as it should. The
susceptibility R(136) is
$\chi=\int_0^\infty dt\,\hat G(0,t)=1/\mu$, unchanged: all quantities of the homogeneous
($\xi=0$) mode coincide with their Euclidean counterparts. (Exact at zero-loop order.)

\paragraph*{Active phase.}
The coherent-state description of the active phase and the time-reversal argument
[R(138)--R(140)] do not depend on the geometry and carry over unchanged, with
$a\propto\mu^{\beta}$. The ballistic front of R(141) does not. For $\mu<0$, combining the
growth $e^{|\mu|t}$ with the power-law tail of Eq.~\eqref{eq:Z_estimate}, the activity
reaches the saturation level at $\log_p\norm{x}\simeq|\mu|t/\big((N+\sigma)\ln p\big)$: it
climbs the hierarchy at a constant number of levels per unit time, so recruiting
$\mathcal N$ neurons takes a time of order $\ln\mathcal N$ rather than $\mathcal N^{1/N}$, as
for Fisher--KPP fronts with fractional diffusion~\cite{CabreRoquejoffre2013}. This
heuristic estimate is the small-world spreading studied, with simulations, in
Sec.~\ref{sec:smallworld}.

\subsection{Correlations}\label{sec:corr}

The review's correlation lengths and susceptibility (Sec.~12.6 there) are computed in
the absorbing state perturbed by a weak background input. In this subsection we therefore
relax the assumption $I=0$ made in Sec.~\ref{sec:meanfield}. The reason is that with $I=0$
and $\mu>0$ the only stationary state is the absorbing one, $a^*=0$, in which there is no
activity and hence no stationary correlations to compute. A weak constant input $I>0$
produces a small spontaneous firing rate $f(I)>0$, which maintains a stationary state
with low activity, and the correlations of this state are what the review computes. We
take $f(I)>0$ small, so
that the homogeneous state is slightly subcritical with $a^*\simeq f(I)/\mu$. The
absorbing state is recovered in the limit $I\to0$.

In the stationary, homogeneous state, $a(x,t)=a^*$ for all $x$ and $t$, so the input is the
same at every neuron, $s(x)=(w\star a^*)(x)+I=\hat w(0)\,a^*+I=:s^*$, and, by translation
invariance, $C(x,y)=C(x-y)$. Since $w$ is radial, both convolutions in Eq.~\eqref{eq:C}
reduce to $(w\star C)(x-y)$; for instance,
$(w\star_2C)(x,y)=\int w(\norm{y-u})\,C(x-u)\,d^Nu=(w\star C)(x-y)$. Setting $\partial_tC=0$,
Eq.~\eqref{eq:C} becomes
\begin{equation*}
0=-2\alpha\,C+2f'(s^*)\,w\star C+2f'(s^*)\,a^*\,w ,
\end{equation*}
and its Fourier transform in $x-y$ gives
\begin{equation}
\hat C(\xi)=\frac{f'(s^*)\,a^*\,\hat w(\xi)}{\alpha-f'(s^*)\,\hat w(\xi)}
=\frac{f'(s^*)\,a^*\,\hat w(\xi)}{\lambda(\xi)},
\label{eq:C_stat}
\end{equation}
with $\lambda(\xi)=\mu+A_w(\xi)$.
Here $\mu=\alpha-f'(s^*)\,\hat w(0)$, and $A_w(\xi)=f'(s^*)\big(\hat w(0)-\hat w(\xi)\big)$ is the
symbol~\eqref{eq:Wsymbol} with $f'(0)$ replaced by $f'(s^*)$.
The value of the inverse Fourier transform of a radial function at $\norm{x}=p^k$ depends
only on its restriction to $\norm{\xi}\le p^{1-k}$. On that ball,
$A_w(\xi)\simeq\kappa_W\norm{\xi}^\sigma\le p^\sigma\mu\,(\xi_\perp/\norm{x})^\sigma$, so the
expansion parameter $A_w/\mu$ is small precisely when $\norm{x}\gg\xi_\perp$. Expanding
Eq.~\eqref{eq:C_stat} to first order in $A_w/\mu$, and using
$\hat w(\xi)=\hat w(0)-A_w(\xi)/f'(s^*)$ and $\mu+f'(s^*)\,w_0=\alpha$, gives
$\hat C(\xi)\simeq\big(f'(s^*)a^*/\mu\big)\big[\hat w(0)-\alpha A_w(\xi)/(f'(s^*)\mu)\big]$.
For $x\neq0$ the constant term does not contribute, and the inverse Fourier transform of
$-A_w$ is $f'(s^*)\,w(\norm{x})$; hence
\begin{equation}
C(x)\simeq\frac{\alpha f'(s^*)\,a^*}{\mu^2}\;w(\norm{x})
\label{eq:C_tail}
\end{equation}
for $\norm{x}\gg\xi_\perp=(\kappa_W/\mu)^{1/\sigma}$.
At $\mu=0$, $\hat C\propto\norm{\xi}^{-\sigma}$ at small $\xi$, and
$C(x)\propto\norm{x}^{\sigma-N}$ for $N>\sigma$. The review does not give the critical form of $C$. Its Euclidean
counterpart follows from the denominator $\mu+\tfrac12Dq^2$ of R(150) (where the momentum
is denoted $p$): $\hat C\propto q^{-2}$
and $C(r)\propto r^{2-N}$ for $N>2$, the Ornstein--Zernike form~\cite{OrnsteinZernike1914,Tauber2014},
which our result reproduces at $\sigma=2$. Three consequences follow.
\begin{itemize}
\item Away from criticality, spatial correlations follow the connectivity and remain
  power laws in hierarchical distance. They cross over from $\norm{x}^{\sigma-N}$ to
  $\norm{x}^{-N-\sigma}$ at $\xi_\perp$, but they are \emph{not} exponentially truncated.
  Temporal correlations are still cut off by $e^{-\mu t}$, with $\xi_\parallel=1/\mu$.
\item The review predicts that power laws become exponentially truncated as one moves away
  from criticality [its Sec.~12.12.2]. The $p$-adic theory predicts exponential
  truncation in time only.
\item The correlation length is a level of the hierarchy,
  $k_\xi=\log_p\xi_\perp=\sigma^{-1}\log_p(\kappa_W/\mu)$, the same as $k^*$ in
  Sec.~\ref{sec:ginzburg}. It grows only logarithmically as $\mu\to0$.
\end{itemize}
(Exact to leading order in $A_w/\mu$.)

\subsection{Power spectra of local field potentials}

The review derives the $1/\omega^2$ spectrum of local field potentials (LFP) and of the
electroencephalogram (EEG) from the propagator of the homogeneous mode, i.e.\ at zero
momentum [R(150)--R(152); there the momentum is denoted $p$, which in this paper is the
prime]. An electrode averages activity over a region. In the
$p$-adic model we take that region to be a ball $B$ of radius $p^k$. The Fourier
transform of its indicator is exactly $p^{kN}\,\Omega(p^k\norm{\xi})$, so the ball
average retains precisely the modes with $\norm{\xi}\le p^{-k}$, the levels coarser than
the electrode, with no leakage.

\paragraph*{Gaussian approximation.}
We treat the fluctuation $\delta\varphi=\varphi-a^*$ about the stationary state of
Sec.~\ref{sec:corr} as a Gaussian random field. This is the linear noise approximation of
van Kampen's system-size expansion~\cite{vanKampen2007}: the action is expanded about $a^*$
and only the terms quadratic in $\delta\varphi$ are kept, so the fluctuations obey a linear
equation driven by noise whose covariance is evaluated at $a^*$, and are therefore Gaussian.
The neglected cubic terms are the vertices whose one-loop effect is measured by the Ginzburg
criterion of Sec.~\ref{sec:ginzburg}, so the approximation is controlled where that
criterion holds: for $N>2\sigma$ up to the critical point, and for $N<2\sigma$ when $\mu$ is
not too small. In addition, the electrode signal is an average over many neurons; when the
ball $B$ contains many correlation volumes (radius of $B$ much larger than $\xi_\perp$), it
is a sum of many weakly dependent contributions and is close to Gaussian by the central
limit theorem. Close to $\mu=0$ with $N<2\sigma$ the fluctuations are not Gaussian, and the
results below do not apply.

\paragraph*{Power spectrum.}
In this approximation each Fourier mode of $\delta\varphi$ obeys an Ornstein--Uhlenbeck
equation,
\begin{equation*}
\partial_t\delta\hat\varphi(\xi,t)=-\lambda(\xi)\,\delta\hat\varphi(\xi,t)+\hat\eta(\xi,t),
\end{equation*}
with $\lambda(\xi)=\mu+A_w(\xi)$ as in Eq.~\eqref{eq:C_stat}. The noise $\hat\eta$ is
uncorrelated in time, and its covariance in momentum space is
$\hat Q(\xi)=2f'(s^*)\,a^*\hat w(\xi)$, the source term of Eq.~\eqref{eq:C}; indeed, the
stationary condition $2\lambda\hat C=\hat Q$ reproduces Eq.~\eqref{eq:C_stat}. At small
$\xi$, $\hat Q(\xi)\simeq2f'(s^*)\,a^*w_0$ is approximately constant: the noise is white in
space as well as in time, and the $\xi$-dependence of the spectrum comes only from
$\lambda(\xi)$. Each mode relaxes at its own rate, so the two-time correlation is
$\hat C(\xi,\tau)=\hat C(\xi)\,e^{-\lambda(\xi)|\tau|}$, and its Fourier transform in $\tau$ is
$\hat Q(\xi)/\big(\lambda(\xi)^2+\omega^2\big)$, which is flat for $\omega\ll\lambda(\xi)$ and
decays as $\omega^{-2}$ for $\omega\gg\lambda(\xi)$; integrating it over $\omega$ returns the
equal-time correlation~\eqref{eq:C_stat}. The electrode records the ball average
$X_B(t)=\mathrm{vol}(B)^{-1}\int_B\delta\varphi(x,t)\,d^Nx$. Its power spectrum is
\begin{align*}
S_B(\omega)&=\int_{\mathbb R}e^{i\omega\tau}\,\langle X_B(t+\tau)X_B(t)\rangle\,d\tau\\
&\propto\int_{\norm{\xi}\le p^{-k}}\frac{\hat Q(\xi)}{\lambda(\xi)^2+\omega^2}\,d^N\xi ,
\end{align*}
the sum of the spectra of the modes retained by the ball. At $\mu=0$ (in practice, for
$\omega\gg\mu$), summing over spheres gives
\begin{align}
S_B(\omega)&\propto\sum_{j\le-k}\frac{p^{jN}}{\kappa_W^2p^{2j\sigma}+\omega^2}\nonumber\\
&\propto
\begin{cases}\omega^{-(2-N/\sigma)}, & \omega\ll\omega_k,\\[2pt]
\omega^{-2}, & \omega\gg\omega_k,\end{cases}
\label{eq:LFP_spectrum}
\end{align}
with $\omega_k=\kappa_Wp^{-k\sigma}$,
for $N<2\sigma$, with log-periodic ripple. The high-frequency regime reproduces the
review's $1/\omega^2$ law. Below the crossover $\omega_k$, set by the electrode's level,
the exponent $2-N/\sigma$ depends on the hierarchy; $1/f$ noise corresponds to
$\sigma=N$. For $N>2\sigma$ the low-frequency spectrum saturates. For $\mu>0$ it saturates
below $\omega\sim\mu$, as in R(152). (Exact within the Gaussian approximation.)

\subsection{Rhythms and finite size}\label{sec:rhythms}

In the review (Sec.~12.5.1 and Figs.~27--29 there), the periods of large-scale rhythms are
related to the number $\mathcal N$ of neurons recruited. In the power-law class at $\mu=0$, the
relaxation rate of the level-$\gamma$ modes is
$\lambda(p^{1-\gamma})=\kappa_Wp^{(1-\gamma)\sigma}\propto p^{-\gamma\sigma}$.
\begin{itemize}
\item \emph{Geometric progression.} The characteristic time scales of successive levels
  form a geometric progression with ratio $p^{\sigma}$. The EEG frequency bands form a
  geometric progression with ratio close to $e$~\cite{PenttonenBuzsaki2003} (Fig.~28 of
  Ref.~\cite{BuiceCowan2009}).
\item \emph{Scaling with size.} A ball of level $\gamma$ contains $\mathcal N\propto p^{N\gamma}$
  neurons, so $f\propto \mathcal N^{-\sigma/N}$. Figure~29 of Ref.~\cite{BuiceCowan2009} is fitted
  by $f\propto \mathcal N^{-1/2}$, which corresponds to $\sigma=N/2$, i.e.\ $N=N_c$. This is
  marginal mean-field behavior, consistent with B-C's claim that cortex is mean-field.
\item \emph{Caveat.} These are relaxation rates of overdamped modes, not oscillation
  frequencies. Turning them into rhythms requires a second variable at each level:
  inhibition (Sec.~12.5.2 of Ref.~\cite{BuiceCowan2009}), refractoriness, or synaptic
  depression (Sec.~\ref{sec:SOC} below).
\end{itemize}

(Heuristic and speculative.)

\subsection{Self-organized criticality}\label{sec:SOC}

In the review (Sec.~12.9 there), the homogeneous equilibrium and its stability are
$\alpha\varphi_0=f(w_0\varphi_0+h_0)$ and $\mu=\alpha-f'(w_0\varphi_0+h_0)\,w_0$ [R(148)--R(149)]; in
our notation, $\varphi_0=a^*$ and $h_0=I$. The control parameters are $w_0=\int w$ and the
background input $I$. Candidate mechanisms are synaptic depression, spike-timing-dependent
plasticity, thalamocortical control of $I$, and stochastic resonance.

\paragraph*{Plasticity.}
In neuroscience, \emph{synaptic plasticity} is the ability of synapses to change their
strength in response to the activity of the neurons they connect~\cite{AbbottNelson2000}.
It acts on several time scales. Short-term plasticity, such as synaptic depression, lasts
from milliseconds to seconds: repeated firing temporarily depletes the resources
(neurotransmitter vesicles) that a synapse can release, and weakens it until they recover.
Long-term plasticity lasts from minutes to much longer. It includes Hebbian rules, in which
a synapse is strengthened when the two neurons it connects are active together, and
spike-timing-dependent plasticity, in which the sign of the change depends on the order of
the pre- and postsynaptic spikes. Homeostatic plasticity slowly rescales synaptic strengths
to keep firing rates within a range. In the model, plasticity is a slow dynamics of the
connectivity $w$, and hence of $\mu$, driven by the activity $a$ or by the correlations $C$;
the mechanisms below are of this kind.

\paragraph*{What must be tuned.}
By Eq.~\eqref{eq:mu_layers}, criticality ($\mu=0$) is a single scalar condition on the
layer profile $\{W_k\}$. It is not sufficient on its own. Every level must also remain
stable, $\varrho_\gamma\le0$ in Eq.~\eqref{eq:rho_gamma}. Otherwise a hierarchical pattern
instability precedes the DP transition.

\paragraph*{Synaptic depression.}
Section~\ref{sec:avalanches} argues that only a slow, activity-dependent plasticity can bring
the network to the critical point. The standard example is short-term synaptic depression:
Levina, Herrmann and Geisel~\cite{Levina2007} showed that networks with depressing synapses
self-organize to criticality, with the avalanche exponent $3/2$, over a wide range of
parameters. The mechanism is a negative feedback on $\mu$: high activity depletes the synaptic
resources and weakens the effective coupling, which raises $\mu$, and low activity lets them
recover, which lowers $\mu$. We ask what this mechanism does in a hierarchical network. Let
$u(x,t)\in[0,1]$ be the fraction of synaptic resources available at the outgoing synapses of
neuron $x$; each spike uses a fraction $v\in(0,1]$ of them, and they recover toward $1$ with time
constant $\tau_R$. In mean-field theory, with $a(x,t)$ the mean activity of
Sec.~\ref{sec:meanfield}, the dynamics is $\partial_ta=-\alpha a+f(s)$, with
\begin{align}
s(x,t)&=\int_{\Qp^N}w(\norm{x-y})\,u(y,t)\,a(y,t)\,d^Ny+I,\nonumber\\
\partial_tu(x,t)&=\frac{1-u(x,t)}{\tau_R}-v\,u(x,t)\,a(x,t).
\label{eq:depression}
\end{align}
The second equation is nonlinear: the depletion rate is proportional to the product $ua$.
In the homogeneous state $u^*=1/(1+v\tau_Ra^*)$, so depression rescales the connectivity
\emph{uniformly across all layers}: $w_{\rm eff}=u^*w$. It tunes $\mu$ and leaves the
shape of $\{W_k\}$, hence $\sigma$ and the universality class, unchanged. Linear
stability is decided level by level through the $2\times2$ matrix
\begin{equation}
J(\xi)=\begin{pmatrix}-\alpha+f'(s^*)\,u^*\hat w(\xi) & f'(s^*)\,a^*\hat w(\xi)\\[2pt]
-v\,u^* & -\big(\tau_R^{-1}+v\,a^*\big)\end{pmatrix},
\label{eq:Jacobian}
\end{equation}
in which $s^*=u^*\hat w(0)a^*+I$ is now the stationary input and $\hat w(\xi)$ takes one value per level [Eq.~\eqref{eq:what_levels}]. A Hopf
instability, if present, appears first at the level maximizing $\hat w$. This is one
mechanism for level-specific oscillations (Sec.~\ref{sec:rhythms}). (Exact within mean
field.)

\paragraph*{Correlation-based plasticity.}
By Eq.~\eqref{eq:C_tail}, $C\propto w$ at large hierarchical distance. Any rule of the
form $\partial_tw=\eta\,C-\varepsilon\,w$ therefore has a growth rate
$\eta\alpha f'(\hat w(0)a^*+I)\,a^*/\mu^2-\varepsilon$ that is the same for every layer, and it also
preserves $\sigma$. On its own such a Hebbian rule is unstable: the fixed point
$\mu^*=\big(\eta\alpha f'(\hat w(0)a^*+I)\,a^*/\varepsilon\big)^{1/2}$ repels, because growth of $w$ lowers $\mu$ and
increases $C$. It must be combined with depression or homeostasis. (Exact to leading
order.)

\paragraph*{Conclusion.}
In the $p$-adic B-C theory, the standard self-organizing mechanisms tune the distance to
criticality $\mu$ but leave the hierarchy exponent $\sigma$ untouched. This matters because
$\mu$ and $\sigma$ control different things: $\mu$ decides how close the network is to the
critical point, while $\sigma$ decides which critical behavior it shows there, mean-field
avalanches ($\tau_S=3/2$) for $\sigma\le N/2$ and anomalous ones for $\sigma>N/2$.
Plasticity can therefore explain why cortex is near criticality, but not which exponents it
displays: these are fixed by the architecture. Experiments that alter plasticity should
change the distance to criticality, and hence the cutoff and the mean avalanche size, but
not the critical exponents. Changing $\sigma$ requires a mechanism that depends explicitly
on the level, such as a wiring cost or conduction delay that grows with hierarchical
distance. Whether $\sigma$ is tuned dynamically by such a mechanism, or fixed by
development, is an open problem (Sec.~\ref{sec:discussion}).

\section{Discussion and open problems}\label{sec:discussion}

\paragraph*{The $p$-adic theory as an extension of the B-C picture.}
Buice and Cowan argued, using field-theoretic methods, that their effective spike model
with an absorbing state has a phase transition in the directed-percolation universality
class, and that the avalanche-size exponent takes its mean-field value $3/2$, the value
observed by Beggs and Plenz~\cite{BuiceCowan2007,BeggsPlenz2003}, when fluctuations are
negligible: exactly for $N\ge4$, the upper critical dimension of DP, and approximately in
cortex, by their Ginzburg argument (Sec.~\ref{sec:RG}). In the $p$-adic model the
upper critical dimension is $N_c=2\sigma$, and the exponent $3/2$ holds for $N\ge2\sigma$.
Since the Taibleson--Vladimirov operator is defined for every $\sigma>0$, the $p$-adic theory
gives a one-parameter family of ultrametric DP classes, with $N_c$ ranging over $(0,\infty)$.
At $\sigma=2$ its power counting and mean-field exponents coincide with those of the B-C
model. In this sense the $p$-adic theory extends the B-C picture, and the $3/2$ law no longer
requires $N\ge4$: on a hierarchical network it holds even for $N=1$ when $\sigma\le1/2$.
In Euclidean L\'evy-flight DP, by contrast, $N_c=2\sigma$ only for $\sigma<2$, since
$\nabla^2$ takes over for $\sigma\ge2$~\cite{Janssen1999} (Sec.~\ref{sec:RG}).
More generally, the $p$-adic B-C theory provides a framework to study criticality in large
deep neural networks for which a thermodynamic-continuous limit is available: in
hierarchical architectures this limit exists and is exact, and the criterion
$\sigma\le N/2$ and the Ginzburg number then decide whether the critical behavior of the
network is mean-field.

\paragraph*{General neuron spaces.}
The construction of Sec.~\ref{sec:model} does not depend on the geometry of the neuron
space. Let $\Omega$ be a neuron space with a measure $\mu$, and let $w(x,y)$ be a
connectivity kernel on $\Omega\times\Omega$. Replacing $\int_{\Qp^N}d^Nx$ by
$\int_\Omega d\mu(x)$ and $w\star g$ by $\int_\Omega w(x,y)\,g(y)\,d\mu(y)$ in
Eq.~\eqref{eq:action}, one obtains the action of the effective spike model on $\Omega$. It
is the limit of the discrete actions $S_\Lambda$ of finite networks $\Lambda$ that fill up
$\Omega$. Its mean-field theory is the Wilson--Cowan equation
\begin{equation}
\begin{aligned}
\partial_ta(x,t)={}&-\alpha\,a(x,t)+\bar n(x)\,\delta(t)\\
&+f\Big(\int_\Omega w(x,y)\,a(y,t)\,d\mu(y)+I\Big),
\end{aligned}
\end{equation}
and the generating functional and the loop expansion of Ref.~\cite{BuiceCowan2007} carry
over formally. In the original formulation of B-C, $\Omega=\mathbb R^N$ and
$w=w(|x-y|)$. Here $\Omega$ is $\Qp^N$, or a finite disjoint union of balls of $\Qp^N$, and
$w=w(\norm{x-y})$. In the $p$-adic theory the hierarchy hypothesis of Sporns and
co-workers~\cite{Sporns2006,CostaSporns2005,SpornsTononiKotter2005} is implemented
naturally, because the $p$-adic numbers are organized in a tree-like structure
(Sec.~\ref{sec:geometry}). The study of more general architectures, such as
\begin{equation}
\Omega=\mathbb R^{N_0}\times\mathbb Q_{p_1}^{N_1}\times\dots\times\mathbb Q_{p_k}^{N_k},
\end{equation}
which combine Euclidean and hierarchical directions, is an open problem. Cellular neural
networks are bio-inspired by the Wilson--Cowan equation, and they have been applied to
several practical tasks, such as image processing; see Ref.~\cite{ZZ2023CNN} and the
references therein.

\paragraph*{Not DP on a Cayley tree.}
The tree enters the two models in different ways. In DP on a Cayley tree (Bethe lattice),
the units sit on all the vertices of the tree, internal ones included, and activity spreads
only along its edges. Distance is graph distance, and the number of units within distance
$d$ grows exponentially in $d$; the lattice is effectively infinite-dimensional, and DP on
it is mean-field. In the $p$-adic model the tree is not the graph along which activity
spreads. Only its leaves are neurons, and the internal vertices label the modules (balls) of
the hierarchy. Every pair of neurons is coupled directly, with a strength fixed by the level
of their last common ancestor, and a ball of radius $r$ contains a number of neurons
proportional to $r^N$, a power of the radius. The neuron space therefore has finite
dimension $N$, and the theory has a finite upper critical dimension $N_c=2\sigma$, like
Dyson's hierarchical model~\cite{Dyson1969,BleherSinai1973}. The simulations of
Sec.~\ref{sec:simulations} show branching-process avalanches at small $\sigma$ and
departures from them as $\sigma$ grows.

\paragraph*{Relation to Ref.~\cite{ZG2024}.}
That work treats Gaussian-noise-driven continuous networks (Martin--Siggia--Rose type).
There, criticality is defined through the largest Lyapunov exponent in a double-copy
system, on both real and $p$-adic neuron spaces. The present theory is Poissonian
(Doi--Peliti) and has an absorbing state, and its criticality is DP-like. Comparing the
two notions of criticality on the same $p$-adic architecture is a natural next step:
when do the edge of chaos and the absorbing-state transition coincide?

\paragraph*{Disorder and Griffiths phases.}
Hierarchical networks with quenched structural heterogeneity display extended
critical (Griffiths) regions~\cite{MorettiMunoz2013}. The $p$-adic model is
translation-invariant under $\Zp^N$ and has a sharp transition. Level-dependent
quenched disorder in $w$ or $\alpha$ interpolates between the two pictures. B-C's remark
that spatiotemporal disorder preserves DP~\cite{Hinrichsen2000} should be revisited in
the ultrametric setting.

\paragraph*{Finite size.}
B-C leave finite-size effects open. On $G_l^N$ the tree depth $l$ is the natural finite-size
parameter, and scaling forms in $p^{l}$ follow from Sec.~\ref{sec:RG}.

\paragraph*{Refractoriness.}
The two- and three-state Cowan actions (B-C Appendix C) transfer directly to $\Qp^N$.
Whether refractoriness drives the model toward an ultrametric version of dynamic isotropic
percolation is open.

\paragraph*{Pathology.}
In B-C's reading, seizures correspond to $\hat w(0)$ exceeding threshold. Hierarchically,
one can ask at which level $\gamma$ the effective rate $\lambda(p^{1-\gamma})$ first changes
sign. Instability confined to a small ball versus instability of the coarsest modes offers a
caricature of focal versus generalized seizures.

\paragraph*{Summary of open technical tasks.}
\begin{enumerate}
\item Rigorous $l\to\infty$ limit of the Doi--Peliti integral (Sec.~\ref{sec:model}).
\item Rigorous proof of the mean-field critical behavior of regional networks
  (Lemma~\ref{lem:regional}).
\item Critical exponents for $N<2\sigma$, using the $p$-adic renormalization
  group~\cite{LernerMissarov1989,AbdesselamChandraGuadagni2013} (Sec.~\ref{sec:RG}).
\item Amplitude of log-periodic terms in observables other than $I_2$, such as avalanche
  statistics (Sec.~\ref{sec:ginzburg}).
\item Analysis of Eqs.~\eqref{eq:WC1}--\eqref{eq:C} (Sec.~\ref{sec:loops}).
\item Rigorous version of the logarithmic spreading law of the active phase
  (Sec.~\ref{sec:phase}), and nonlinear selection of hierarchical patterns (Sec.~\ref{sec:phase}).
\item Whether $\sigma$ is fixed by development or tuned by a level-dependent
  self-organizing mechanism, e.g.\ toward the marginal value $\sigma=N/2$
  (Sec.~\ref{sec:SOC}).
\item Fluctuation effects in regional networks (Lemma~\ref{lem:regional} and
  Appendix~\ref{app:regional} treat only the linear dynamics).
\end{enumerate}

\begin{acknowledgments}
During the preparation of this work, the author used Claude (Anthropic; model
\mbox{claude-opus-5-5}) to improve the clarity and readability of the language of select
sections. After using this tool, the author reviewed and edited the content as needed and
took full responsibility for the publication. The author was partially supported by the
Debnath Endowment.
\end{acknowledgments}

\appendix
\section{Field theory of branching processes: a primer}\label{app:primer}

This appendix explains the field-theoretic tools used in the main text. It is written for
readers from neuroscience or machine learning who are not familiar with quantum field
theory (QFT). We work with a simple branching process on
a network that contains all the ingredients of the effective spike model of
Sec.~\ref{sec:model}. At each step we indicate the corresponding objects of the main text.
Standard references are
Refs.~\cite{Doi1976a,Peliti1985,Cardy1996notes,TauberHowardVollmayrLee2005,Tauber2014,Hinrichsen2000}.

\subsection{Core concepts}\label{app:concepts}

\paragraph*{What a branching process is.}
A branching process is a model of a population in which each individual (a ``particle''),
independently of the others, disappears or produces new individuals at random times. The
classical example is the Galton--Watson process of population genetics~\cite{Harris1963}.
In our setting a particle is an \emph{effective spike}, i.e.\ a spike that still
influences the network~\cite{BuiceCowan2007}, and particles live on the nodes $i$ of a
network (neurons, or small groups of neurons). We use a single branching process
throughout this appendix. Its dynamics consist of three kinds of elementary random
events:
\begin{align}
&\text{decay:} && A_i\to\varnothing && \text{rate } \alpha,\nonumber\\
&\text{branching:} && A_j\to A_j+A_i && \text{rate } \beta K_{ij},\nonumber\\
&\text{saturation:} && A_i+A_i\to A_i && \text{rate } \lambda .
\label{eq:app_reactions}
\end{align}
Here $A_i$ denotes one effective spike at node $i$, and the rates are per particle (for saturation, the
total rate at node $i$ is $\lambda n_i(n_i-1)$).
\begin{itemize}
\item \emph{Decay.} An effective spike stops influencing the network. Each spike does so at
  rate $\alpha$, so its influence lasts on average $1/\alpha$.
\item \emph{Branching.} An effective spike at node $j$ causes a new spike at node $i$. For
  each spike at $j$ this happens at rate $\beta K_{ij}$, where $K_{ij}\ge0$ is the strength of
  the connection from $j$ to $i$. This is the only event that increases activity.
\item \emph{Saturation.} When two effective spikes are present at the same node, one of
  them is removed, at total rate $\lambda n_i(n_i-1)$. This limits growth.
\end{itemize}
With decay and branching only, one obtains the classical, linear branching process: each
spike independently has on average $m=\beta\sum_iK_{ij}/\alpha$ offspring during its
lifetime, and in the supercritical regime ($m>1$) activity grows exponentially without
bound. Saturation is the simplest way to keep activity bounded, as real firing rates are.

\paragraph*{Relation to the effective spike model.}
The effective spike model of Sec.~\ref{sec:model} has two kinds of events: the decay of an
effective spike at rate $\alpha$, and the emission of a new spike at neuron $i$ at rate
$f(s_i)$, where $s_i=\sum_jw_{ij}n_j$ is the input from the other neurons. Expanding $f$ for
small input, with $f(0)=0$ [Eq.~\eqref{eq:f_derivatives}],
\begin{equation}
f(s_i)\simeq f'(0)\,s_i+\tfrac12f''(0)\,s_i^2 .
\label{eq:app_fexpansion}
\end{equation}
The linear term makes the emission rate at $i$ proportional to the number of spikes $n_j$
at the presynaptic neurons (i.e., the neurons $j$ that send connections to neuron $i$,
those with $w_{ij}\neq0$). This is exactly branching, with $\beta K_{ij}=f'(0)w_{ij}$. The
quadratic term is negative, since $f''(0)<0$, so the emission rate grows more slowly than
linearly when many spikes are present. This is the effect of saturation. In the effective
spike model it acts on the total input $s_i$ rather than on pairs of spikes at one node;
this difference does not change the large-scale behavior (Appendix~\ref{app:DP}). The
process~\eqref{eq:app_reactions} is therefore a simplified version of the effective spike
model with the same three ingredients: decay, branching and saturation. We use it because
every step below can be written out explicitly.

\paragraph*{Absorbing state and criticality.}
If there are no particles, nothing can happen: the empty network is an \emph{absorbing
state}. Whether activity survives depends on the \emph{branching ratio}
$m=\beta\sum_iK_{ij}/\alpha$, the mean number of offspring of one particle during its
lifetime. For $m<1$ (subcritical) activity dies out quickly. For $m>1$ (supercritical) it
can grow and persist. The \emph{critical point} $m=1$ separates the two phases. There,
activity neither dies out quickly nor explodes, perturbations spread over all scales,
and statistical quantities follow power laws. This is the non-equilibrium phase transition
behind the critical-brain hypothesis. In machine learning, the analogous
``edge of chaos'' separates layers that damp and layers that amplify signals in deep
networks~\cite{Poole2016,Schoenholz2017}. In the main text the distance to criticality
is $\mu=\alpha-f'(0)\hat w(0)$, and $\mu=0$ corresponds to $m=1$.

\paragraph*{Why fluctuations matter.}
The simplest description of the process is a deterministic \emph{rate equation} for the
mean activity, derived below; see Eq.~\eqref{eq:app_rate}. It ignores
fluctuations. The number of events in a short time is random, with a variance
proportional to the number of particles. Relative fluctuations are therefore large when
activity is low, which is exactly the situation near an absorbing state. A systematic
theory of fluctuations is needed, and field theory provides one.

\subsection{The master equation}\label{app:master}

Let $n_i$ be the number of particles at node $i$, let $\mathbf n=(n_i)$ be the state (or
configuration) of the network, which varies with time, and let $P(\mathbf n,t)$ be the
probability of configuration $\mathbf n$ at time $t$. The dynamics is modeled by a continuous-time
Markov chain, so $P$ obeys a \emph{master equation}: its rate of change is the probability
flowing in from other configurations minus the probability flowing out. For the events
\eqref{eq:app_reactions},
\begin{align}
\partial_tP(\mathbf n,t)=&\sum_i\alpha\big[(n_i+1)P(\mathbf n_i^+)-n_iP(\mathbf n)\big]\nonumber\\
&+\sum_{i,j}\beta K_{ij}\big[n_jP(\mathbf n_i^-)-n_jP(\mathbf n)\big]\nonumber\\
&+\sum_i\lambda\big[(n_i+1)n_iP(\mathbf n_i^+)-n_i(n_i-1)P(\mathbf n)\big],
\label{eq:app_master}
\end{align}
where $\mathbf n_i^{\pm}$ differs from $\mathbf n$ only by $n_i\to n_i\pm1$, and we take
$K_{ii}=0$. The master equation of the effective spike model, Eq.~\eqref{eq:master}, has the
same structure, with the creation rate $f(s_i)$ in place of the branching and saturation
terms; by Eq.~\eqref{eq:app_fexpansion} the two agree to second order in the input.

Equation~\eqref{eq:app_master} is exact but is a system of infinitely many coupled linear
equations. Multiplying it by $n_i$ and summing over $\mathbf n$ gives
\begin{equation}
\frac{d\langle n_i\rangle}{dt}=-\alpha\langle n_i\rangle+\beta\sum_jK_{ij}\langle n_j\rangle
-\lambda\langle n_i(n_i-1)\rangle .
\label{eq:app_hierarchy}
\end{equation}
The mean couples to the second moment, the second moment to the third, and so on. This is
the \emph{closure problem}. Mean-field theory closes the hierarchy by the approximation
$\langle n_i(n_i-1)\rangle\approx\langle n_i\rangle^2$, which neglects fluctuations. Writing
$a_i=\langle n_i\rangle$, one obtains the \emph{rate equation}
\begin{equation}
\frac{da_i}{dt}=-\alpha a_i+\beta\sum_jK_{ij}a_j-\lambda a_i^2 .
\label{eq:app_rate}
\end{equation}

\paragraph*{Master equation versus rate equation.}
The master equation~\eqref{eq:app_master} is an
\emph{exact} equation for the \emph{probability distribution} $P(\mathbf n,t)$ of all
configurations. It is linear in $P$, but it involves infinitely many unknowns (one for
each configuration), and it contains the full statistics: means, variances, correlations,
avalanche distributions. The rate equation~\eqref{eq:app_rate} is an \emph{approximate}
equation for the \emph{means} $a_i$ only. It is a closed, finite system of deterministic
ordinary differential equations, but it is nonlinear and it contains no information about
fluctuations. It is the analogue for this process of the Wilson--Cowan
equation~\eqref{eq:WC} of the effective spike model. The field theory replaces the ad hoc
closure leading from Eq.~\eqref{eq:app_hierarchy} to Eq.~\eqref{eq:app_rate} by a
controlled expansion, in which the rate equation is the lowest order
(Appendix~\ref{app:feynman}).

\subsection{The Doi--Peliti formulation}\label{app:doi}

The Doi--Peliti method~\cite{Doi1976a,Peliti1985} rewrites the master equation as a linear
evolution equation that can be turned into a path integral.

\paragraph*{Occupation-number states.}
At each node introduce a vector $|n\rangle$ for each $n=0,1,2,\dots$, and two linear
operators $a$ and $a^\dagger$ (``annihilation'' and ``creation'') acting as
\begin{equation}
a^\dagger|n\rangle=|n+1\rangle,\qquad a|n\rangle=n\,|n-1\rangle .
\label{eq:app_ladder}
\end{equation}
Then $a^\dagger a|n\rangle=n|n\rangle$: the operator $a^\dagger a$ counts particles. One
checks $aa^\dagger-a^\dagger a=1$. These are the same algebraic rules as for the quantum
harmonic oscillator, but nothing quantum is involved; they are bookkeeping devices. For a network, each node $i$ carries its own pair of operators $a_i,a_i^\dagger$, which
act on the number of particles at that node; operators associated with different nodes
commute, so that $a_ia_j^\dagger-a_j^\dagger a_i=\delta_{ij}$.

\paragraph*{The state vector.}
The probability distribution is encoded in
$|\Psi(t)\rangle=\sum_{\mathbf n}P(\mathbf n,t)\,|\mathbf n\rangle$, with
$|\mathbf n\rangle=\prod_i(a_i^\dagger)^{n_i}|0\rangle$. The master
equation~\eqref{eq:app_master} becomes
\begin{equation}
\partial_t|\Psi(t)\rangle=-H\,|\Psi(t)\rangle ,
\label{eq:app_schrodinger}
\end{equation}
with the \emph{Liouvillian} (``Hamiltonian'') $H$ obtained from a simple rule: an event
that removes $k$ particles and puts back $l$ particles at rate $r$ contributes
$-r\,[(a^\dagger)^l-(a^\dagger)^k]\,a^k$. For Eq.~\eqref{eq:app_reactions},
\begin{align}
H=&\sum_i\alpha\,(a_i^\dagger-1)\,a_i
-\sum_{i,j}\beta K_{ij}\,(a_i^\dagger-1)\,a_j^\dagger a_j\nonumber\\
&+\sum_i\lambda\,(a_i^\dagger-1)\,a_i^\dagger a_i^2 .
\label{eq:app_H}
\end{align}
For example, applying $\alpha(a_i-a_i^\dagger a_i)$ to $|\mathbf n\rangle$ gives
$\alpha n_i(|\mathbf n_i^-\rangle-|\mathbf n\rangle)$, which reproduces the decay terms of
Eq.~\eqref{eq:app_master}.

\paragraph*{Averages and the Doi shift.}
Averages are computed with the ``projection'' vector
$\langle\mathcal P|=\langle0|\prod_ie^{a_i}$, which satisfies
$\langle\mathcal P|\mathbf n\rangle=1$ for every $\mathbf n$. Then
$\langle\mathcal P|\Psi\rangle=\sum_{\mathbf n}P(\mathbf n)=1$, and
$\langle n_i\rangle=\langle\mathcal P|a_i|\Psi\rangle$. Since
$\langle\mathcal P|a_i^\dagger=\langle\mathcal P|$, every term of $H$ vanishes when
$a_i^\dagger$ is replaced by $1$. This is conservation of probability. It suggests the
\emph{Doi shift} $a_i^\dagger=1+\tilde a_i^\dagger$, which moves the projection vector
into the operators. After the shift, every term of $H$ contains at least one factor
$\tilde a^\dagger$. This is the origin of the property $S[\varphi,0]=0$ used in
Sec.~\ref{sec:split}.

\paragraph*{Initial condition.}
An initial state in which each $n_i$ is independently Poisson distributed with mean
$\bar n_i$ is $|\Psi(0)\rangle=\prod_ie^{\bar n_i(a_i^\dagger-1)}|0\rangle$. After the
Doi shift it becomes $\prod_ie^{\bar n_i\tilde a_i^\dagger}|0\rangle$. This produces the
initial-state term $-\sum_i\bar n_i\tphi_i(0)$ of Eq.~\eqref{eq:discrete_action}.

\subsection{The action}\label{app:action}

\paragraph*{From operators to a path integral.}
The formal solution of Eq.~\eqref{eq:app_schrodinger} is
$|\Psi(t)\rangle=e^{-Ht}|\Psi(0)\rangle$. Divide $[0,t]$ into many small steps and insert,
at each step, a resolution of the identity in terms of \emph{coherent states}, the
eigenvectors of $a_i$: $a_i|\varphi\rangle=\varphi_i|\varphi\rangle$. In the limit of
infinitely many steps, the matrix elements of $e^{-Ht}$ become an integral over the
trajectories of two sets of variables: $\varphi_i(t)$, the eigenvalue of $a_i$, and
$\tphi_i(t)$, the variable associated with $\tilde a_i^\dagger$. The result is the
generating functional
\begin{equation}
Z[J,\tilde J]=\int\mathcal D\varphi\,\mathcal D\tphi\;
e^{-S[\varphi,\tphi]+\int_0^\infty\!dt\sum_i(J_i\tphi_i+\tilde J_i\varphi_i)} ,
\label{eq:app_Z}
\end{equation}
whose derivatives with respect to the ``sources'' $J,\tilde J$ at $J=\tilde J=0$ give all
moments. The \emph{action} is obtained by replacing, in the normal-ordered operator $H$,
$a_i\to\varphi_i$ and $\tilde a_i^\dagger\to\tphi_i$:
\begin{equation}
S=\int_0^\infty\!dt\sum_i\big[\tphi_i\,\partial_t\varphi_i+H(\tphi_i,\varphi_i)\big]
-\sum_i\bar n_i\,\tphi_i(0).
\label{eq:app_S_general}
\end{equation}
The derivation is standard~\cite{Cardy1996notes,TauberHowardVollmayrLee2005,Tauber2014}.
The only technical point is that, after the Doi shift, $\tphi$ is not the complex
conjugate of $\varphi$. The two are treated as independent variables, with $\tphi$
integrated along the imaginary axis.

\paragraph*{The action of the branching process.}
For Eq.~\eqref{eq:app_H}, after the shift and in the continuum (sums over nodes become
integrals), and assuming that $K_{ij}$ depends only on the difference $i-j$, so that
$K_{ij}\to K(x-y)$ and the sums $\sum_jK_{ij}(\cdot)_j$ become convolutions
$(K\star\cdot)(x)$,
\begin{align}
S=\int_0^\infty\!dt\!\int\!d^Nx\,\Big\{&\tphi\,\partial_t\varphi+\alpha\tphi\varphi
-\beta\,\tphi\,K\star\varphi\nonumber\\
&-\beta\,\tphi\,K\star(\tphi\varphi)+\lambda\tphi\varphi^2+\lambda\tphi^2\varphi^2\Big\}\nonumber\\
&-\int\!d^Nx\,\bar n\,\tphi(x,0).
\label{eq:app_S}
\end{align}
For a general kernel $K(x,y)$, not translation invariant, $K\star\varphi$ and
$K\star(\tphi\varphi)$ are replaced by the integrals $\int K(x,y)\,\varphi(y,t)\,d^Ny$ and
$\int K(x,y)\,\tphi(y,t)\varphi(y,t)\,d^Ny$.
Compare with the B-C action, Eq.~\eqref{eq:action}, expanded to second order in $f$:
$-\tphi f\big(w\star(\tphi\varphi+\varphi)\big)\simeq-f'(0)\,\tphi\,w\star\varphi
-f'(0)\,\tphi\,w\star(\tphi\varphi)-\tfrac12f''(0)\,\tphi\,(w\star\varphi)^2+\dots$.
The first two terms are exactly the branching terms of Eq.~\eqref{eq:app_S} with
$\beta K=f'(0)w$. The third is a (nonlocal) saturation term with positive coefficient
$\tfrac12|f''(0)|$, playing the role of $\lambda\tphi\varphi^2$.

\paragraph*{Meaning of the two fields.}
The field $\varphi$ is the ``density'' field: $\langle\varphi(x,t)\rangle=\langle n(x,t)\rangle$
exactly, and higher moments of $\varphi$ are the factorial moments of $n$ (e.g.\
$\langle\varphi^2\rangle=\langle n(n-1)\rangle$ at a single node). The field $\tphi$ is the
\emph{response field}. Adding a source $h\,\tphi$ to the action means injecting particles
at rate $h$, and correlations with $\tphi$ measure the response to such injections. The
name should not be read literally: $\tphi$ sits at the earlier point, where the injection
acts, and $\varphi$ at the later point, where the response is measured (see the note after
Eq.~\eqref{eq:Delta0}).

\paragraph*{Connection with stochastic differential equations.}
Readers familiar with stochastic differential equations (SDEs) may find the following
useful. If the action is truncated at second order in $\tphi$, it takes the form
$S=\int\tphi[\partial_t\varphi-F(\varphi)]-\tfrac12\int\tphi\,Q(\varphi)\,\tphi$. This
is the Martin--Siggia--Rose--Janssen--De Dominicis (MSRJD) action~\cite{MSR1973,Janssen1976,DeDominicis1976}
of the Langevin equation
\begin{equation}
\partial_t\varphi=F(\varphi)+\eta,\qquad
F(\varphi)=-\alpha\varphi+\beta K\star\varphi-\lambda\varphi^2,
\label{eq:app_langevin}
\end{equation}
with Gaussian noise $\eta$ whose covariance $Q(\varphi)$ is read off from the terms
$\beta\tphi K\star(\tphi\varphi)$ and $\lambda\tphi^2\varphi^2$. The noise covariance is
proportional to $\varphi$, so it vanishes in the absorbing state: fluctuations are
multiplicative. The response field plays the role of a Lagrange multiplier enforcing the
equation of motion. (The saturation term contributes a negative piece to $Q$; this
known subtlety of the Doi--Peliti Langevin representation does not affect the
perturbative expansion below.)

\paragraph*{Mean field from the action.}
The classical equations of motion, $\delta S/\delta\tphi=0$ and $\delta S/\delta\varphi=0$,
admit the solution $\tphi=0$, and then $\delta S/\delta\tphi=0$ gives
\begin{equation}
\partial_t\varphi=-\alpha\varphi+\beta K\star\varphi-\lambda\varphi^2+\bar n\,\delta(t),
\label{eq:app_mf}
\end{equation}
which is the rate equation~\eqref{eq:app_rate} with its initial condition. For the B-C action
the same computation gives the Wilson--Cowan equation~\eqref{eq:WC_delta}
(Sec.~\ref{sec:meanfield}). Mean-field theory is the lowest order of a systematic
expansion, to which we now turn.

\subsection{Perturbation theory and Feynman diagrams}\label{app:feynman}

\paragraph*{Free part and propagator.}
Split the action into a \emph{free} (bilinear) part and an \emph{interacting} part,
$S=S_0+S_{\rm int}$, with
\begin{equation}
S_0=\int_0^\infty\!dt\!\int\!d^Nx\;\tphi\,\big(\partial_t+\alpha-\beta K\star\big)\varphi .
\end{equation}
With $S_0$ alone, the path integral is Gaussian and can be computed exactly. Its only
nontrivial moment is the \emph{propagator}
$G(x,t)=\langle\varphi(x,t)\tphi(0,0)\rangle_0$. It is the Green function of the
linearized dynamics, i.e.\ the mean response at $(x,t)$ to one particle injected at
$(0,0)$. In Fourier variables,
\begin{equation}
\hat G(q,t)=\theta(t)\,e^{-\mu(q)t},\qquad \mu(q)=\alpha-\beta\hat K(q),
\label{eq:app_G}
\end{equation}
where $\theta$ is the Heaviside function with $\theta(0)=0$ (causality: a response cannot
precede its cause). The rate $\mu(0)=\alpha-\beta\hat K(0)$ is the distance to criticality;
it vanishes at $m=1$. In the Euclidean case with a short-range kernel,
$\hat K(q)\simeq\hat K(0)-\tfrac12K_2q^2$, so $\mu(q)\simeq\mu+Dq^2$: particles diffuse. In
the $p$-adic case, $\mu(q)=\mu+A_w(q)$ with the $W$-operator of
Sec.~\ref{sec:Wop}.

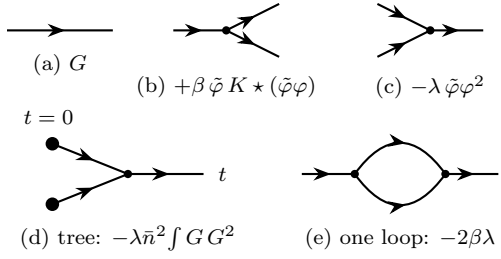
\begin{figure}[t]
\centering
\begin{tikzpicture}[
  line/.style={thick,decoration={markings,mark=at position 0.55 with {\arrow{Stealth}}},postaction={decorate}},
  dot/.style={circle,fill=black,inner sep=1.8pt},
  vtx/.style={circle,fill=black,inner sep=1.1pt},
  lab/.style={font=\footnotesize}]
\draw[line] (0,0) -- (1.4,0);
\node[lab] at (0.7,-0.45) {(a) $G$};
\begin{scope}[xshift=2.2cm]
\draw[line] (0,0) -- (0.7,0);
\draw[line] (0.7,0) -- (1.4,0.35);
\draw[line] (0.7,0) -- (1.4,-0.35);
\node[vtx] at (0.7,0) {};
\node[lab] at (0.7,-0.75) {(b) $+\beta\,\tphi\,K\star(\tphi\varphi)$};
\end{scope}
\begin{scope}[xshift=4.9cm]
\draw[line] (0,0.35) -- (0.7,0);
\draw[line] (0,-0.35) -- (0.7,0);
\draw[line] (0.7,0) -- (1.4,0);
\node[vtx] at (0.7,0) {};
\node[lab] at (0.7,-0.75) {(c) $-\lambda\,\tphi\varphi^2$};
\end{scope}
\begin{scope}[yshift=-1.9cm,xshift=0.6cm]
\node[dot] (d1) at (0,0.4) {};
\node[dot] (d2) at (0,-0.4) {};
\draw[line] (0,0.4) -- (1.0,0);
\draw[line] (0,-0.4) -- (1.0,0);
\draw[line] (1.0,0) -- (2.0,0);
\node[vtx] at (1.0,0) {};
\node[lab] at (-0.05,0.75) {$t=0$};
\node[lab] at (2.25,0) {$t$};
\node[lab] at (1.0,-0.85) {(d) tree: $-\lambda\bar n^2\!\int G\,G^2$};
\end{scope}
\begin{scope}[yshift=-1.9cm,xshift=3.9cm]
\draw[line] (0,0) -- (0.7,0);
\draw[line] (0.7,0) .. controls (1.1,0.55) and (1.5,0.55) .. (1.9,0);
\draw[line] (0.7,0) .. controls (1.1,-0.55) and (1.5,-0.55) .. (1.9,0);
\draw[line] (1.9,0) -- (2.6,0);
\node[vtx] at (0.7,0) {};
\node[vtx] at (1.9,0) {};
\node[lab] at (1.3,-0.85) {(e) one loop: $-2\beta\lambda$};
\end{scope}
\end{tikzpicture}
\caption{Feynman diagrams for the branching process of Appendix~\ref{app:primer}. Time
runs from left to right. (a) The propagator $G$: the mean response to one injected
particle. (b) Branching vertex: one particle becomes two. (c) Saturation vertex: two
particles become one. (d) A zero-loop diagram contributing to the mean activity: two particles
from the initial state (dots) merge; its value is the second term of
Eq.~\eqref{eq:app_tree}. (e) The one-loop correction to the propagator,
Eq.~\eqref{eq:app_oneloop}: a fluctuation in which a particle splits into two that
later merge again.}
\label{fig:diagrams}
\end{figure}
\paragraph*{Wick's theorem and diagrams.}
Expand $e^{-S_{\rm int}}$ in powers of the couplings. Each term is a Gaussian average of a
product of fields, and Gaussian averages are sums over all ways of pairing each $\varphi$
with a $\tphi$ (\emph{Wick's theorem}), each pair contributing a propagator. These sums are
organized by \emph{Feynman diagrams} (Fig.~\ref{fig:diagrams}):
\begin{itemize}
\item a \emph{line} is a propagator $G$, drawn from an earlier to a later time;
\item a \emph{vertex} is a term of $-S_{\rm int}$, with one incoming line for each
  $\varphi$ and one outgoing line for each $\tphi$; its position and time are integrated;
\item an \emph{initial-state dot} is a factor $\bar n$ with one outgoing line at $t=0$.
\end{itemize}
In the branching process, $+\beta\tphi K\star(\tphi\varphi)$ is a vertex with one incoming
and two outgoing lines (one particle becomes two), and $-\lambda\tphi\varphi^2$ has two
incoming lines and one outgoing line (two particles become one). The factor $\theta(0)=0$
forbids lines that start and end at the same time, so every diagram respects causality.

\paragraph*{Zero-loop diagrams give mean-field theory.}
The \emph{loop expansion} orders diagrams by their number of loops. Diagrams without closed
loops form the zero-loop order. In field theory they are usually called \emph{tree
diagrams}, and this order \emph{tree level}; as in Sec.~\ref{sec:meanfield}, we avoid these
names, because in this paper ``tree'' refers to the hierarchical organization of the
neurons. Zero-loop order is mean-field theory, and each loop adds one order of fluctuation
corrections. The sum of the zero-loop diagrams is exactly the solution of the classical
equation~\eqref{eq:app_mf}. For a single node with $\beta=0$,
$\mu=\alpha$, the mean has the exact zero-loop expansion
\begin{align}
a(t)&=\bar n\,e^{-\mu t}-\lambda\bar n^{\,2}\int_0^t\!ds\,e^{-\mu(t-s)}e^{-2\mu s}+\dots\nonumber\\
&=\bar n\,e^{-\mu t}\Big[1-\frac{\lambda\bar n}{\mu}\big(1-e^{-\mu t}\big)+\dots\Big],
\label{eq:app_tree}
\end{align}
where the second term is the diagram in which two lines from initial-state dots merge at
one saturation vertex. The series is the expansion of
$a(t)=\bar n e^{-\mu t}/[1+(\lambda\bar n/\mu)(1-e^{-\mu t})]$, the exact solution of
$\dot a=-\mu a-\lambda a^2$.

\paragraph*{Loops are fluctuation corrections.}
Diagrams with closed loops describe fluctuations. The simplest one is a line that splits
into two at a branching vertex, after which the two lines merge again at a saturation
vertex. It corrects the propagator. Collecting the combinatorial factor (two ways to join
the two lines),
\begin{align}
\hat G^{-1}(q,\omega)=&-i\omega+\mu(q)\nonumber\\
&+2\beta\lambda\int\!\frac{d^Nk}{-i\omega+\mu(k)+\mu(q-k)}+\dots
\label{eq:app_oneloop}
\end{align}
(for a local kernel, $\hat K\simeq\hat K(0)$ at the vertex; in the Euclidean case
$d^Nk$ stands for $d^Nk/(2\pi)^N$, in the $p$-adic case for the Haar measure). At $q=\omega=0$ the
effective decay rate is $\mu_R=\mu+\beta\lambda\int d^Nk/\mu(k)$. It is larger than
$\mu$, so fluctuations shift the critical point to a larger branching rate than predicted
by mean-field theory. The size of the loop relative to the zero-loop terms is controlled by
\begin{equation}
u(\mu)=\beta\lambda\int\frac{d^Nk}{\big(\mu+Dk^2\big)^2}\;\propto\;\mu^{N/2-2}.
\label{eq:app_ginzburg}
\end{equation}
For $N>4$ this stays finite as $\mu\to0$, so mean-field theory is correct near the critical
point. For $N<4$ it diverges, and fluctuations dominate. This is the Ginzburg-type criterion
of Ref.~\cite{BuiceCowan2007} and of Sec.~\ref{sec:ginzburg}. In the $p$-adic theory,
$Dk^2$ is replaced by $\kappa_W\norm{k}^\sigma$, and the same argument gives
$u\propto\mu^{N/\sigma-2}$, i.e.\ the upper critical dimension $N_c=2\sigma$. Adding the
loop corrections order by order is how the closure problem of
Eq.~\eqref{eq:app_hierarchy} is solved in a controlled way.

\subsection{Universality}\label{app:universality}

\paragraph*{Coarse-graining and relevance.}
Near a critical point, correlations extend over lengths and times much larger than the
microscopic ones. The large-scale behavior can then be studied by coarse-graining:
integrating out fluctuations on short scales and rescaling. This is the renormalization
group (RG). Under this procedure each coupling in the action grows, shrinks or stays
constant. Growing couplings are \emph{relevant}, shrinking ones \emph{irrelevant}, and the
others \emph{marginal}. Irrelevant couplings do not affect the large-scale behavior.
Consequently, microscopically different models that share the same relevant couplings have
the same critical exponents. They belong to the same \emph{universality class}.

\paragraph*{Power counting.}
Whether a coupling is relevant can be read off from dimensions. With diffusive scaling,
$[x]=-1$ and $[t]=-2$ in momentum units. After a rescaling of $\varphi$ and $\tphi$ that
makes the branching and saturation couplings equal (both $=g/2$), $[\varphi]=[\tphi]=N/2$
and $[g]=2-N/2$. The cubic couplings are therefore relevant for $N<4$, while the quartic
term $\lambda\tphi^2\varphi^2$, with dimension $2-N$, is irrelevant near $N=4$. The
remaining action,
\begin{align}
S_{\rm RFT}=\int_0^\infty\!dt\!\int\!d^Nx\,\Big\{&\tpsi\big(\partial_t+\mu-D\nabla^2\big)\psi\nonumber\\
&+\frac g2\big(\tpsi\psi^2-\tpsi^2\psi\big)\Big\},
\label{eq:app_RFT}
\end{align}
is Reggeon field theory~\cite{CardySugar1980}, discussed in detail in
Appendix~\ref{app:DP}. Its critical behavior defines the
\emph{directed percolation} (DP) universality class~\cite{Janssen1981,Grassberger1982,Hinrichsen2000}.
The action is invariant under the \emph{rapidity reversal}
$\psi(x,t)\leftrightarrow-\tpsi(x,-t)$. The $p$-adic counterpart, with $-D\nabla^2$
replaced by $\boldsymbol W$ and $[t]=-z$ with $z=\sigma$, is Eq.~\eqref{eq:RFT} of the main text.

\paragraph*{The DP conjecture.}
Janssen and Grassberger conjectured~\cite{Janssen1981,Grassberger1982} that a stochastic
process belongs to the DP class if (i) it has a unique absorbing state, (ii) its activity is
described by a single non-negative order parameter, (iii) its interactions are short-ranged,
and (iv) it has no additional symmetries, conservation laws or quenched disorder. The
effective spike model with $f(0)=0$, $f'(0)>0$ and $f''(0)<0$ satisfies (i), (ii) and (iv).
Condition (iii) is where the geometry enters. In the $p$-adic theory, interactions are
long-ranged in the sense of hierarchical distance, and the class is modified to the
ultrametric DP class labelled by $\sigma$ (Sec.~\ref{sec:RG}).

\paragraph*{Critical exponents.}
The DP class is characterized by the exponents
\begin{equation}
\rho\propto(-\mu)^{\beta},\quad \xi_\perp\propto|\mu|^{-\nu_\perp},\quad
\xi_\parallel\propto|\mu|^{-\nu_\parallel},\quad z=\nu_\parallel/\nu_\perp ,
\end{equation}
for the stationary density of active sites, the spatial correlation length and the
temporal correlation length. Rapidity reversal implies that the survival probability of
activity started from a single seed has the same exponent $\beta'=\beta$. Mean-field values
are $\beta=1$, $\nu_\perp=\tfrac12$, $\nu_\parallel=1$, $z=2$. Numerical estimates in low
dimensions are (Table~2 of Ref.~\cite{Hinrichsen2000}; the digits in parentheses are the
uncertainty in the last digits):
\begin{center}
\begin{tabular}{lcccc}
\hline\hline
$N$ & $\beta$ & $\nu_\perp$ & $\nu_\parallel$ & $z$\\
\hline
1 & 0.276486(8) & 1.096854(4) & 1.733847(6) & 1.580745(10)\\
2 & 0.584(4) & 0.734(4) & 1.295(6) & 1.76(3)\\
3 & 0.81(1) & 0.581(5) & 1.105(5) & 1.90(1)\\
$\ge4$ & 1 & 1/2 & 1 & 2\\
\hline\hline
\end{tabular}
\end{center}

\paragraph*{Avalanches.}
In the mean-field regime, the process started from one particle is a critical
Galton--Watson branching process. The total number of particles produced (the avalanche
size $S$) and the duration $T$ then have the distributions $P(S)\propto S^{-3/2}$ and
$P(T)\propto T^{-2}$~\cite{Harris1963}. The exponent $3/2$ is the one measured by Beggs and
Plenz in cortical slices~\cite{BeggsPlenz2003}, and its $p$-adic counterpart is discussed in
Sec.~\ref{sec:avalanches}.

\section{Directed percolation and Reggeon field theory}\label{app:DP}

This appendix explains the universality class to which the critical point of the
effective spike model belongs. Directed percolation (DP) is a simple lattice model. Reggeon
field theory (RFT) is the field theory that describes its large-scale behavior; see
Refs.~\cite{Hinrichsen2000,Tauber2014,TauberHowardVollmayrLee2005}.

\subsection{Directed percolation as a model of spreading}\label{app:DPmodel}

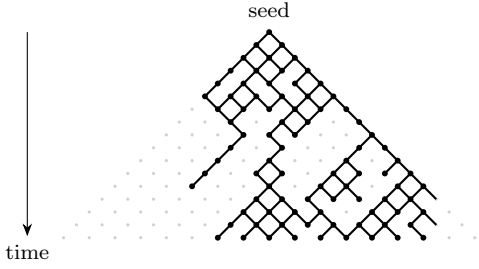
\begin{figure}[t]
\centering
\begin{tikzpicture}
\fill[black!18] (0.00,0.00) circle (0.6pt);
\fill[black!18] (-0.17,-0.17) circle (0.6pt);
\fill[black!18] (0.17,-0.17) circle (0.6pt);
\fill[black!18] (-0.34,-0.34) circle (0.6pt);
\fill[black!18] (0.00,-0.34) circle (0.6pt);
\fill[black!18] (0.34,-0.34) circle (0.6pt);
\fill[black!18] (-0.51,-0.51) circle (0.6pt);
\fill[black!18] (-0.17,-0.51) circle (0.6pt);
\fill[black!18] (0.17,-0.51) circle (0.6pt);
\fill[black!18] (0.51,-0.51) circle (0.6pt);
\fill[black!18] (-0.68,-0.68) circle (0.6pt);
\fill[black!18] (-0.34,-0.68) circle (0.6pt);
\fill[black!18] (0.00,-0.68) circle (0.6pt);
\fill[black!18] (0.34,-0.68) circle (0.6pt);
\fill[black!18] (0.68,-0.68) circle (0.6pt);
\fill[black!18] (-0.85,-0.85) circle (0.6pt);
\fill[black!18] (-0.51,-0.85) circle (0.6pt);
\fill[black!18] (-0.17,-0.85) circle (0.6pt);
\fill[black!18] (0.17,-0.85) circle (0.6pt);
\fill[black!18] (0.51,-0.85) circle (0.6pt);
\fill[black!18] (0.85,-0.85) circle (0.6pt);
\fill[black!18] (-1.02,-1.02) circle (0.6pt);
\fill[black!18] (-0.68,-1.02) circle (0.6pt);
\fill[black!18] (-0.34,-1.02) circle (0.6pt);
\fill[black!18] (0.00,-1.02) circle (0.6pt);
\fill[black!18] (0.34,-1.02) circle (0.6pt);
\fill[black!18] (0.68,-1.02) circle (0.6pt);
\fill[black!18] (1.02,-1.02) circle (0.6pt);
\fill[black!18] (-1.19,-1.19) circle (0.6pt);
\fill[black!18] (-0.85,-1.19) circle (0.6pt);
\fill[black!18] (-0.51,-1.19) circle (0.6pt);
\fill[black!18] (-0.17,-1.19) circle (0.6pt);
\fill[black!18] (0.17,-1.19) circle (0.6pt);
\fill[black!18] (0.51,-1.19) circle (0.6pt);
\fill[black!18] (0.85,-1.19) circle (0.6pt);
\fill[black!18] (1.19,-1.19) circle (0.6pt);
\fill[black!18] (-1.36,-1.36) circle (0.6pt);
\fill[black!18] (-1.02,-1.36) circle (0.6pt);
\fill[black!18] (-0.68,-1.36) circle (0.6pt);
\fill[black!18] (-0.34,-1.36) circle (0.6pt);
\fill[black!18] (0.00,-1.36) circle (0.6pt);
\fill[black!18] (0.34,-1.36) circle (0.6pt);
\fill[black!18] (0.68,-1.36) circle (0.6pt);
\fill[black!18] (1.02,-1.36) circle (0.6pt);
\fill[black!18] (1.36,-1.36) circle (0.6pt);
\fill[black!18] (-1.53,-1.53) circle (0.6pt);
\fill[black!18] (-1.19,-1.53) circle (0.6pt);
\fill[black!18] (-0.85,-1.53) circle (0.6pt);
\fill[black!18] (-0.51,-1.53) circle (0.6pt);
\fill[black!18] (-0.17,-1.53) circle (0.6pt);
\fill[black!18] (0.17,-1.53) circle (0.6pt);
\fill[black!18] (0.51,-1.53) circle (0.6pt);
\fill[black!18] (0.85,-1.53) circle (0.6pt);
\fill[black!18] (1.19,-1.53) circle (0.6pt);
\fill[black!18] (1.53,-1.53) circle (0.6pt);
\fill[black!18] (-1.70,-1.70) circle (0.6pt);
\fill[black!18] (-1.36,-1.70) circle (0.6pt);
\fill[black!18] (-1.02,-1.70) circle (0.6pt);
\fill[black!18] (-0.68,-1.70) circle (0.6pt);
\fill[black!18] (-0.34,-1.70) circle (0.6pt);
\fill[black!18] (0.00,-1.70) circle (0.6pt);
\fill[black!18] (0.34,-1.70) circle (0.6pt);
\fill[black!18] (0.68,-1.70) circle (0.6pt);
\fill[black!18] (1.02,-1.70) circle (0.6pt);
\fill[black!18] (1.36,-1.70) circle (0.6pt);
\fill[black!18] (1.70,-1.70) circle (0.6pt);
\fill[black!18] (-1.87,-1.87) circle (0.6pt);
\fill[black!18] (-1.53,-1.87) circle (0.6pt);
\fill[black!18] (-1.19,-1.87) circle (0.6pt);
\fill[black!18] (-0.85,-1.87) circle (0.6pt);
\fill[black!18] (-0.51,-1.87) circle (0.6pt);
\fill[black!18] (-0.17,-1.87) circle (0.6pt);
\fill[black!18] (0.17,-1.87) circle (0.6pt);
\fill[black!18] (0.51,-1.87) circle (0.6pt);
\fill[black!18] (0.85,-1.87) circle (0.6pt);
\fill[black!18] (1.19,-1.87) circle (0.6pt);
\fill[black!18] (1.53,-1.87) circle (0.6pt);
\fill[black!18] (1.87,-1.87) circle (0.6pt);
\fill[black!18] (-2.04,-2.04) circle (0.6pt);
\fill[black!18] (-1.70,-2.04) circle (0.6pt);
\fill[black!18] (-1.36,-2.04) circle (0.6pt);
\fill[black!18] (-1.02,-2.04) circle (0.6pt);
\fill[black!18] (-0.68,-2.04) circle (0.6pt);
\fill[black!18] (-0.34,-2.04) circle (0.6pt);
\fill[black!18] (0.00,-2.04) circle (0.6pt);
\fill[black!18] (0.34,-2.04) circle (0.6pt);
\fill[black!18] (0.68,-2.04) circle (0.6pt);
\fill[black!18] (1.02,-2.04) circle (0.6pt);
\fill[black!18] (1.36,-2.04) circle (0.6pt);
\fill[black!18] (1.70,-2.04) circle (0.6pt);
\fill[black!18] (2.04,-2.04) circle (0.6pt);
\fill[black!18] (-2.21,-2.21) circle (0.6pt);
\fill[black!18] (-1.87,-2.21) circle (0.6pt);
\fill[black!18] (-1.53,-2.21) circle (0.6pt);
\fill[black!18] (-1.19,-2.21) circle (0.6pt);
\fill[black!18] (-0.85,-2.21) circle (0.6pt);
\fill[black!18] (-0.51,-2.21) circle (0.6pt);
\fill[black!18] (-0.17,-2.21) circle (0.6pt);
\fill[black!18] (0.17,-2.21) circle (0.6pt);
\fill[black!18] (0.51,-2.21) circle (0.6pt);
\fill[black!18] (0.85,-2.21) circle (0.6pt);
\fill[black!18] (1.19,-2.21) circle (0.6pt);
\fill[black!18] (1.53,-2.21) circle (0.6pt);
\fill[black!18] (1.87,-2.21) circle (0.6pt);
\fill[black!18] (2.21,-2.21) circle (0.6pt);
\fill[black!18] (-2.38,-2.38) circle (0.6pt);
\fill[black!18] (-2.04,-2.38) circle (0.6pt);
\fill[black!18] (-1.70,-2.38) circle (0.6pt);
\fill[black!18] (-1.36,-2.38) circle (0.6pt);
\fill[black!18] (-1.02,-2.38) circle (0.6pt);
\fill[black!18] (-0.68,-2.38) circle (0.6pt);
\fill[black!18] (-0.34,-2.38) circle (0.6pt);
\fill[black!18] (0.00,-2.38) circle (0.6pt);
\fill[black!18] (0.34,-2.38) circle (0.6pt);
\fill[black!18] (0.68,-2.38) circle (0.6pt);
\fill[black!18] (1.02,-2.38) circle (0.6pt);
\fill[black!18] (1.36,-2.38) circle (0.6pt);
\fill[black!18] (1.70,-2.38) circle (0.6pt);
\fill[black!18] (2.04,-2.38) circle (0.6pt);
\fill[black!18] (2.38,-2.38) circle (0.6pt);
\fill[black!18] (-2.55,-2.55) circle (0.6pt);
\fill[black!18] (-2.21,-2.55) circle (0.6pt);
\fill[black!18] (-1.87,-2.55) circle (0.6pt);
\fill[black!18] (-1.53,-2.55) circle (0.6pt);
\fill[black!18] (-1.19,-2.55) circle (0.6pt);
\fill[black!18] (-0.85,-2.55) circle (0.6pt);
\fill[black!18] (-0.51,-2.55) circle (0.6pt);
\fill[black!18] (-0.17,-2.55) circle (0.6pt);
\fill[black!18] (0.17,-2.55) circle (0.6pt);
\fill[black!18] (0.51,-2.55) circle (0.6pt);
\fill[black!18] (0.85,-2.55) circle (0.6pt);
\fill[black!18] (1.19,-2.55) circle (0.6pt);
\fill[black!18] (1.53,-2.55) circle (0.6pt);
\fill[black!18] (1.87,-2.55) circle (0.6pt);
\fill[black!18] (2.21,-2.55) circle (0.6pt);
\fill[black!18] (2.55,-2.55) circle (0.6pt);
\fill[black!18] (-2.72,-2.72) circle (0.6pt);
\fill[black!18] (-2.38,-2.72) circle (0.6pt);
\fill[black!18] (-2.04,-2.72) circle (0.6pt);
\fill[black!18] (-1.70,-2.72) circle (0.6pt);
\fill[black!18] (-1.36,-2.72) circle (0.6pt);
\fill[black!18] (-1.02,-2.72) circle (0.6pt);
\fill[black!18] (-0.68,-2.72) circle (0.6pt);
\fill[black!18] (-0.34,-2.72) circle (0.6pt);
\fill[black!18] (0.00,-2.72) circle (0.6pt);
\fill[black!18] (0.34,-2.72) circle (0.6pt);
\fill[black!18] (0.68,-2.72) circle (0.6pt);
\fill[black!18] (1.02,-2.72) circle (0.6pt);
\fill[black!18] (1.36,-2.72) circle (0.6pt);
\fill[black!18] (1.70,-2.72) circle (0.6pt);
\fill[black!18] (2.04,-2.72) circle (0.6pt);
\fill[black!18] (2.38,-2.72) circle (0.6pt);
\fill[black!18] (2.72,-2.72) circle (0.6pt);
\draw[thick] (0.00,0.00) -- (-0.17,-0.17);
\draw[thick] (0.00,0.00) -- (0.17,-0.17);
\draw[thick] (0.17,-0.17) -- (0.00,-0.34);
\draw[thick] (0.17,-0.17) -- (0.34,-0.34);
\draw[thick] (-0.17,-0.17) -- (-0.34,-0.34);
\draw[thick] (-0.17,-0.17) -- (0.00,-0.34);
\draw[thick] (0.00,-0.34) -- (-0.17,-0.51);
\draw[thick] (0.00,-0.34) -- (0.17,-0.51);
\draw[thick] (0.34,-0.34) -- (0.17,-0.51);
\draw[thick] (0.34,-0.34) -- (0.51,-0.51);
\draw[thick] (-0.34,-0.34) -- (-0.51,-0.51);
\draw[thick] (-0.34,-0.34) -- (-0.17,-0.51);
\draw[thick] (0.17,-0.51) -- (0.00,-0.68);
\draw[thick] (0.51,-0.51) -- (0.34,-0.68);
\draw[thick] (0.51,-0.51) -- (0.68,-0.68);
\draw[thick] (-0.51,-0.51) -- (-0.68,-0.68);
\draw[thick] (-0.17,-0.51) -- (-0.34,-0.68);
\draw[thick] (-0.17,-0.51) -- (0.00,-0.68);
\draw[thick] (0.00,-0.68) -- (0.17,-0.85);
\draw[thick] (0.34,-0.68) -- (0.51,-0.85);
\draw[thick] (0.68,-0.68) -- (0.51,-0.85);
\draw[thick] (0.68,-0.68) -- (0.85,-0.85);
\draw[thick] (-0.68,-0.68) -- (-0.85,-0.85);
\draw[thick] (-0.34,-0.68) -- (-0.51,-0.85);
\draw[thick] (-0.34,-0.68) -- (-0.17,-0.85);
\draw[thick] (0.17,-0.85) -- (0.00,-1.02);
\draw[thick] (0.17,-0.85) -- (0.34,-1.02);
\draw[thick] (0.51,-0.85) -- (0.34,-1.02);
\draw[thick] (0.51,-0.85) -- (0.68,-1.02);
\draw[thick] (0.85,-0.85) -- (0.68,-1.02);
\draw[thick] (0.85,-0.85) -- (1.02,-1.02);
\draw[thick] (-0.85,-0.85) -- (-0.68,-1.02);
\draw[thick] (-0.51,-0.85) -- (-0.68,-1.02);
\draw[thick] (-0.51,-0.85) -- (-0.34,-1.02);
\draw[thick] (-0.17,-0.85) -- (-0.34,-1.02);
\draw[thick] (-0.17,-0.85) -- (0.00,-1.02);
\draw[thick] (0.34,-1.02) -- (0.17,-1.19);
\draw[thick] (0.34,-1.02) -- (0.51,-1.19);
\draw[thick] (0.68,-1.02) -- (0.51,-1.19);
\draw[thick] (1.02,-1.02) -- (1.19,-1.19);
\draw[thick] (-0.68,-1.02) -- (-0.51,-1.19);
\draw[thick] (-0.34,-1.02) -- (-0.51,-1.19);
\draw[thick] (0.17,-1.19) -- (0.00,-1.36);
\draw[thick] (0.17,-1.19) -- (0.34,-1.36);
\draw[thick] (0.51,-1.19) -- (0.34,-1.36);
\draw[thick] (-0.51,-1.19) -- (-0.34,-1.36);
\draw[thick] (1.19,-1.19) -- (1.36,-1.36);
\draw[thick] (0.00,-1.36) -- (0.17,-1.53);
\draw[thick] (1.36,-1.36) -- (1.19,-1.53);
\draw[thick] (1.36,-1.36) -- (1.53,-1.53);
\draw[thick] (-0.34,-1.36) -- (-0.51,-1.53);
\draw[thick] (0.17,-1.53) -- (0.00,-1.70);
\draw[thick] (-0.51,-1.53) -- (-0.68,-1.70);
\draw[thick] (1.53,-1.53) -- (1.70,-1.70);
\draw[thick] (1.19,-1.53) -- (1.02,-1.70);
\draw[thick] (0.00,-1.70) -- (-0.17,-1.87);
\draw[thick] (0.00,-1.70) -- (0.17,-1.87);
\draw[thick] (1.70,-1.70) -- (1.53,-1.87);
\draw[thick] (1.70,-1.70) -- (1.87,-1.87);
\draw[thick] (-0.68,-1.70) -- (-0.85,-1.87);
\draw[thick] (1.02,-1.70) -- (0.85,-1.87);
\draw[thick] (1.02,-1.70) -- (1.19,-1.87);
\draw[thick] (0.17,-1.87) -- (0.00,-2.04);
\draw[thick] (0.85,-1.87) -- (0.68,-2.04);
\draw[thick] (0.85,-1.87) -- (1.02,-2.04);
\draw[thick] (1.19,-1.87) -- (1.02,-2.04);
\draw[thick] (1.87,-1.87) -- (1.70,-2.04);
\draw[thick] (1.87,-1.87) -- (2.04,-2.04);
\draw[thick] (-0.85,-1.87) -- (-1.02,-2.04);
\draw[thick] (-0.17,-1.87) -- (0.00,-2.04);
\draw[thick] (0.00,-2.04) -- (-0.17,-2.21);
\draw[thick] (0.00,-2.04) -- (0.17,-2.21);
\draw[thick] (0.68,-2.04) -- (0.51,-2.21);
\draw[thick] (0.68,-2.04) -- (0.85,-2.21);
\draw[thick] (1.02,-2.04) -- (0.85,-2.21);
\draw[thick] (1.02,-2.04) -- (1.19,-2.21);
\draw[thick] (1.70,-2.04) -- (1.53,-2.21);
\draw[thick] (1.70,-2.04) -- (1.87,-2.21);
\draw[thick] (2.04,-2.04) -- (1.87,-2.21);
\draw[thick] (2.04,-2.04) -- (2.21,-2.21);
\draw[thick] (0.17,-2.21) -- (0.00,-2.38);
\draw[thick] (0.17,-2.21) -- (0.34,-2.38);
\draw[thick] (0.51,-2.21) -- (0.68,-2.38);
\draw[thick] (1.53,-2.21) -- (1.36,-2.38);
\draw[thick] (1.53,-2.21) -- (1.70,-2.38);
\draw[thick] (1.87,-2.21) -- (1.70,-2.38);
\draw[thick] (1.87,-2.21) -- (2.04,-2.38);
\draw[thick] (-0.17,-2.21) -- (-0.34,-2.38);
\draw[thick] (-0.17,-2.21) -- (0.00,-2.38);
\draw[thick] (0.00,-2.38) -- (-0.17,-2.55);
\draw[thick] (0.00,-2.38) -- (0.17,-2.55);
\draw[thick] (0.34,-2.38) -- (0.17,-2.55);
\draw[thick] (0.34,-2.38) -- (0.51,-2.55);
\draw[thick] (0.68,-2.38) -- (0.51,-2.55);
\draw[thick] (0.68,-2.38) -- (0.85,-2.55);
\draw[thick] (1.36,-2.38) -- (1.19,-2.55);
\draw[thick] (1.36,-2.38) -- (1.53,-2.55);
\draw[thick] (1.70,-2.38) -- (1.53,-2.55);
\draw[thick] (2.04,-2.38) -- (1.87,-2.55);
\draw[thick] (2.04,-2.38) -- (2.21,-2.55);
\draw[thick] (-0.34,-2.38) -- (-0.51,-2.55);
\draw[thick] (-0.34,-2.38) -- (-0.17,-2.55);
\draw[thick] (0.17,-2.55) -- (0.00,-2.72);
\draw[thick] (0.17,-2.55) -- (0.34,-2.72);
\draw[thick] (0.85,-2.55) -- (0.68,-2.72);
\draw[thick] (0.85,-2.55) -- (1.02,-2.72);
\draw[thick] (1.19,-2.55) -- (1.02,-2.72);
\draw[thick] (1.19,-2.55) -- (1.36,-2.72);
\draw[thick] (1.53,-2.55) -- (1.36,-2.72);
\draw[thick] (1.53,-2.55) -- (1.70,-2.72);
\draw[thick] (1.87,-2.55) -- (2.04,-2.72);
\draw[thick] (-0.51,-2.55) -- (-0.68,-2.72);
\draw[thick] (-0.17,-2.55) -- (-0.34,-2.72);
\draw[thick] (-0.17,-2.55) -- (0.00,-2.72);
\fill (-0.34,-2.38) circle (1.1pt);
\fill (0.51,-1.19) circle (1.1pt);
\fill (0.68,-1.02) circle (1.1pt);
\fill (0.85,-2.21) circle (1.1pt);
\fill (0.51,-2.21) circle (1.1pt);
\fill (-0.51,-1.53) circle (1.1pt);
\fill (0.34,-0.34) circle (1.1pt);
\fill (-0.51,-2.55) circle (1.1pt);
\fill (0.00,-2.38) circle (1.1pt);
\fill (1.70,-2.04) circle (1.1pt);
\fill (0.17,-1.53) circle (1.1pt);
\fill (1.87,-1.87) circle (1.1pt);
\fill (0.34,-1.36) circle (1.1pt);
\fill (-0.68,-0.68) circle (1.1pt);
\fill (0.17,-2.55) circle (1.1pt);
\fill (-0.68,-1.70) circle (1.1pt);
\fill (-0.34,-2.72) circle (1.1pt);
\fill (-0.17,-2.21) circle (1.1pt);
\fill (2.04,-2.04) circle (1.1pt);
\fill (-0.85,-0.85) circle (1.1pt);
\fill (0.51,-2.55) circle (1.1pt);
\fill (-1.02,-2.04) circle (1.1pt);
\fill (-0.85,-1.87) circle (1.1pt);
\fill (1.36,-1.36) circle (1.1pt);
\fill (0.34,-0.68) circle (1.1pt);
\fill (0.00,-0.68) circle (1.1pt);
\fill (0.00,-2.72) circle (1.1pt);
\fill (1.70,-2.38) circle (1.1pt);
\fill (0.17,-1.87) circle (1.1pt);
\fill (1.87,-2.21) circle (1.1pt);
\fill (1.02,-1.70) circle (1.1pt);
\fill (-0.34,-1.02) circle (1.1pt);
\fill (-0.17,-2.55) circle (1.1pt);
\fill (2.04,-2.38) circle (1.1pt);
\fill (0.85,-0.85) circle (1.1pt);
\fill (0.00,0.00) circle (1.1pt);
\fill (0.00,-1.02) circle (1.1pt);
\fill (1.70,-2.72) circle (1.1pt);
\fill (0.17,-2.21) circle (1.1pt);
\fill (1.87,-2.55) circle (1.1pt);
\fill (1.02,-1.02) circle (1.1pt);
\fill (-0.34,-0.34) circle (1.1pt);
\fill (-0.17,-1.87) circle (1.1pt);
\fill (1.02,-2.04) circle (1.1pt);
\fill (-0.34,-1.36) circle (1.1pt);
\fill (2.04,-2.72) circle (1.1pt);
\fill (0.68,-2.04) circle (1.1pt);
\fill (-0.51,-0.51) circle (1.1pt);
\fill (0.00,-0.34) circle (1.1pt);
\fill (1.53,-1.87) circle (1.1pt);
\fill (0.00,-1.36) circle (1.1pt);
\fill (0.17,-0.51) circle (1.1pt);
\fill (0.34,-2.38) circle (1.1pt);
\fill (1.19,-1.19) circle (1.1pt);
\fill (-0.34,-0.68) circle (1.1pt);
\fill (-0.17,-0.17) circle (1.1pt);
\fill (-0.68,-2.72) circle (1.1pt);
\fill (1.19,-2.21) circle (1.1pt);
\fill (0.51,-0.51) circle (1.1pt);
\fill (0.68,-2.38) circle (1.1pt);
\fill (-0.51,-0.85) circle (1.1pt);
\fill (0.85,-2.55) circle (1.1pt);
\fill (1.36,-2.38) circle (1.1pt);
\fill (1.53,-2.21) circle (1.1pt);
\fill (0.00,-1.70) circle (1.1pt);
\fill (0.17,-0.85) circle (1.1pt);
\fill (0.34,-2.72) circle (1.1pt);
\fill (1.19,-1.53) circle (1.1pt);
\fill (-0.68,-1.02) circle (1.1pt);
\fill (-0.17,-0.51) circle (1.1pt);
\fill (1.02,-2.72) circle (1.1pt);
\fill (1.19,-2.55) circle (1.1pt);
\fill (0.51,-0.85) circle (1.1pt);
\fill (0.68,-0.68) circle (1.1pt);
\fill (0.68,-2.72) circle (1.1pt);
\fill (0.85,-1.87) circle (1.1pt);
\fill (-0.51,-1.19) circle (1.1pt);
\fill (1.53,-1.53) circle (1.1pt);
\fill (0.17,-0.17) circle (1.1pt);
\fill (1.36,-2.72) circle (1.1pt);
\fill (1.53,-2.55) circle (1.1pt);
\fill (0.00,-2.04) circle (1.1pt);
\fill (1.70,-1.70) circle (1.1pt);
\fill (0.17,-1.19) circle (1.1pt);
\fill (0.34,-1.02) circle (1.1pt);
\fill (1.19,-1.87) circle (1.1pt);
\fill (-0.17,-0.85) circle (1.1pt);\draw[-{Stealth}] (-3.2,0) -- (-3.2,-2.7) node[below,font=\footnotesize] {time};
\node[font=\footnotesize] at (0,0.3) {seed};
\end{tikzpicture}
\caption{A directed-percolation cluster in one spatial dimension (horizontal) grown from a
single seed, for bond probability $p=0.66$ (here $p$ is a probability, not a prime),
slightly above $p_c\simeq0.6447$. Open bonds are drawn as lines;
active sites as dots; grey dots are inactive sites. Bonds can only be followed downward,
i.e.\ forward in time. In the neural reading, the cluster is an avalanche started by one
spike.}
\label{fig:DP}
\end{figure}

\paragraph*{The lattice model.}
In this appendix, $p$ denotes a probability, not a prime number, and $N$ denotes the number
of spatial dimensions of the lattice (the time direction is not counted); it plays the role
of the dimension $N$ of the neuron space in the main text. Place sites on a lattice in which one direction plays the role of \emph{time}
(Fig.~\ref{fig:DP}). Each site is connected to two sites in the next time step, by bonds
that are open with probability $p$ and closed with probability $1-p$, independently. A
site at time $t+1$ is \emph{active} if at least one open bond connects it to an active site
at time $t$. Starting from a single active site (the \emph{seed}), the active sites form a
\emph{cluster} that spreads forward in time. Because bonds can only be followed forward in
time, the percolation is \emph{directed}.

\paragraph*{Neural and machine-learning readings.}
For a neuroscientist, time steps are successive rounds of firing. Bonds are synapses
that transmit a spike with probability $p$, and a cluster started by one spike is a
\emph{neuronal avalanche}. For a machine-learning reader, each time step is a layer of a
deep feedforward network with random binary connections and ``OR'' units: a unit fires
if at least one of its inputs fires. The question ``does a signal injected at the input
reach arbitrarily deep layers?'' is exactly the question of whether a DP cluster survives.

\paragraph*{The phase transition.}
If $p$ is small, every cluster dies after a finite time: the empty configuration is an
absorbing state, and the system is in the \emph{absorbing phase}. If $p$ is large, a
cluster survives forever with positive probability and occupies a finite fraction of the
sites: this is the \emph{active phase}. The two phases are separated by a critical value
$p_c$. For bond DP in one spatial dimension, $p_c\simeq0.6447$~\cite{Hinrichsen2000}. The
value of $p_c$ depends on the details of the lattice. The behavior near $p_c$ does not.

\paragraph*{The contact process.}
A continuous-time version is the \emph{contact process}. Each active site becomes
inactive at rate $1$, and it activates each of its $2N$ nearest neighbors, if inactive, at rate
$\lambda_{\rm cp}/(2N)$ (on $\mathbb Z^N$).
It has a critical value $\lambda_{\rm cp,c}\simeq3.2978$ in one dimension, and the same critical
exponents as DP~\cite{Hinrichsen2000}. The branching process of Appendix~\ref{app:primer}
and the effective spike model are further members of the same family: decay plays the role
of closed bonds, branching the role of open bonds, and saturation the role of the rule
``a site is active if \emph{at least one} input is active'' (two inputs do not make a site
twice as active).

\subsection{What is measured}\label{app:DPobs}

The critical behavior is described by a small number of quantities. Two initial
conditions are natural.
\begin{itemize}
\item \emph{Fully active start.} The \emph{density} $\rho(t)$ of active sites decays at
  $p_c$ as $\rho(t)\propto t^{-\delta}$. In the active phase it saturates at
  $\rho_{\rm st}\propto(p-p_c)^{\beta}$.
\item \emph{Single seed.} The \emph{survival probability} $P(t)$ that the cluster is still
  alive at time $t$ decays as $P(t)\propto t^{-\delta'}$. The mean number of active sites
  grows as $\mathcal N(t)\propto t^{\vartheta}$, and the mean squared spreading distance
  from the seed grows as $R^2(t)\propto t^{2/z}$.
\end{itemize}
Away from $p_c$, correlations decay over a spatial length $\xi_\perp\propto|p-p_c|^{-\nu_\perp}$
and a time $\xi_\parallel\propto|p-p_c|^{-\nu_\parallel}$. The exponents are related by
\begin{equation}
\delta=\frac{\beta}{\nu_\parallel},\quad \delta'=\frac{\beta'}{\nu_\parallel},\quad
z=\frac{\nu_\parallel}{\nu_\perp},\quad
\vartheta=\frac{N\nu_\perp-\beta-\beta'}{\nu_\parallel},
\label{eq:app_scaling}
\end{equation}
the last one (``hyperscaling'') holding below the upper critical dimension. In DP,
$\beta=\beta'$ (Sec.~\ref{app:RFT}), so there are only three independent exponents,
e.g.\ $\beta$, $\nu_\perp$, $\nu_\parallel$. Their values are listed in
Appendix~\ref{app:universality}. Avalanche statistics follow from the same exponents; for
instance the avalanche-size exponent is
$\tau_S=1+\beta'/(\nu_\parallel+N\nu_\perp-\beta)$ (Sec.~\ref{sec:avalanches}).

\subsection{From DP to a field theory}\label{app:DPfield}

\paragraph*{A coarse-grained equation.}
On scales much larger than the lattice spacing, the state of the system is described by a
coarse-grained density $\rho(x,t)\ge0$. Janssen and Grassberger argued that its dynamics
is captured by the Langevin equation~\cite{Janssen1981,Grassberger1982}
\begin{align}
&\partial_t\rho(x,t)=a\rho(x,t)-b\rho(x,t)^2+D\nabla^2\rho(x,t)+\eta(x,t),\nonumber\\
&\langle\eta(x,t)\eta(x',t')\rangle=Q\,\rho(x,t)\,\delta(x-x')\,\delta(t-t'),
\label{eq:app_DPlangevin}
\end{align}
where $Q>0$ is a constant. The noise is uncorrelated in space and time, and its variance at
$(x,t)$ is proportional to the local density $\rho(x,t)$; the product $\rho(x,t)\,\delta(x-x')$
means that the variance is evaluated at the point where the noise acts (with the It\^o
convention in time).
Each term has a simple meaning: $a\rho$ is the net growth rate (branching minus decay),
$-b\rho^2$ is saturation, $D\nabla^2\rho$ is spreading to neighbors, and $\eta$ is the
noise from the randomness of individual events. The noise variance is proportional to
$\rho$, so it vanishes in the absorbing state $\rho=0$. This is the defining feature of the
class. The mean-field version ($\eta=0$) is a rate equation of the same form as
Eq.~\eqref{eq:app_rate}.

\paragraph*{The action.}
Applying the MSRJD construction of Appendix~\ref{app:action} to
Eq.~\eqref{eq:app_DPlangevin} gives the action
\begin{equation}
S=\int_0^\infty\!dt\!\int\!d^Nx\,\Big\{\tilde\rho\big(\partial_t-a-D\nabla^2\big)\rho+b\,\tilde\rho\rho^2
-\frac Q2\,\tilde\rho^{\,2}\rho\Big\}.
\label{eq:app_DPaction}
\end{equation}
Rescaling the fields as $\rho=c\,\psi$, $\tilde\rho=\tpsi/c$ with $c=\sqrt{Q/(2b)}$
makes the two cubic couplings equal. With $\mu=-a$ and $g=\sqrt{2bQ}$ one obtains
\begin{align}
S_{\rm RFT}=\int_0^\infty\!dt\!\int\!d^Nx\,\Big\{&\tpsi\big(\partial_t+\mu-D\nabla^2\big)\psi\nonumber\\
&+\frac g2\big(\tpsi\psi^2-\tpsi^{\,2}\psi\big)\Big\}.
\label{eq:app_RFT2}
\end{align}
This is Reggeon field theory, Eq.~\eqref{eq:app_RFT}. The Doi--Peliti action of the
branching process~\eqref{eq:app_S} leads to the same result after the same rescaling (the
quartic term $\lambda\tphi^2\varphi^2$ being irrelevant). So does the B-C action, whose
rescaling is Eq.~(58) of Ref.~\cite{BuiceCowan2007}. The \emph{signs} matter. The saturation
term $+\tfrac g2\tpsi\psi^2$ must enter the action with the sign shown, so that the
mean-field equation $\partial_t\psi=-\mu\psi-\tfrac g2\psi^2+D\nabla^2\psi$ saturates. In
the effective spike model this sign is guaranteed by $f''(0)<0$.

\paragraph*{MSRJD and Doi--Peliti.}
The two constructions lead to actions of the same form but start from different places.
The MSRJD construction~\cite{MSR1973,Janssen1976,DeDominicis1976} starts from a Langevin
equation for a continuous field, here the coarse-grained density $\rho$, with a postulated
Gaussian noise. The response field $\tilde\rho$ enforces the equation of motion, and
averaging over the noise produces the term $-\tfrac Q2\tilde\rho^{\,2}\rho$; the action is
therefore exactly quadratic in $\tilde\rho$, and $\rho$ is the density itself. The
Doi--Peliti construction~\cite{Doi1976a,Peliti1985} (Appendix~\ref{app:doi}) starts from the
master equation of the discrete particle numbers and involves no approximation. Its field
$\varphi$ is the eigenvalue of the annihilation operator, not the density: $\langle\varphi\rangle$
is the mean number, but higher moments of $\varphi$ are factorial moments of the counts
(Appendix~\ref{app:action}), and the action may contain any power of $\tphi$, such as the term
$\lambda\tphi^2\varphi^2$ of Eq.~\eqref{eq:app_S}. Truncating the Doi--Peliti action at second
order in $\tphi$ gives an action of MSRJD form, i.e.\ a Langevin equation (Appendix~\ref{app:action},
``Connection with stochastic differential equations''); this Langevin reading is formal,
since its noise covariance can be negative. Near the critical point the two descriptions
agree, because the higher powers of $\tphi$ are irrelevant and both actions reduce to
Eq.~\eqref{eq:app_RFT2}. The B-C theory uses the Doi--Peliti construction, so it is an exact
rewriting of the effective spike model before any truncation; the Langevin
equation~\eqref{eq:app_DPlangevin} of Janssen and Grassberger is a coarse-grained,
phenomenological description of the same universality class.

\paragraph*{Why ``Reggeon''?}
The same action appeared in the 1970s in high-energy physics, in Gribov's description of
hadron scattering at high energies in terms of ``Reggeons''. Grassberger and Sundermeyer,
and Cardy and Sugar, showed that this theory is equivalent to DP~\cite{GrassbergerSundermeyer1978,CardySugar1980}.
Only the name survives from that context.

\subsection{Rapidity reversal}\label{app:RFT}

The action~\eqref{eq:app_RFT2} is invariant under
\begin{equation}
\psi(x,t)\to-\tpsi(x,-t),\qquad \tpsi(x,t)\to-\psi(x,-t),
\label{eq:app_rapidity}
\end{equation}
which exchanges the density field and the response field and reverses time. It is
called \emph{rapidity reversal}, and it is an emergent symmetry: the microscopic model need
not have it. One consequence is that the density decay from a fully active start and the
survival probability from a single seed have the same exponent, $\beta=\beta'$ and
$\delta=\delta'$. In DP these are exactly the same statistics: on the lattice, the
probability that a seed survives to time $t$ equals the probability that a given site is
active at time $t$ when starting from a fully active configuration. In the neural reading,
the exponent governing how long an avalanche lasts equals the one governing how fast
activity decays after global stimulation.

\subsection{Mean-field theory}\label{app:DPmf}

Setting $\tpsi=0$ in the equations of motion of Eq.~\eqref{eq:app_RFT2} gives
$\partial_t\psi=-\mu\psi-\tfrac g2\psi^2+D\nabla^2\psi$. Its homogeneous stationary
solutions are $\psi=0$ and, for $\mu<0$, $\psi=2|\mu|/g$. The density therefore vanishes
linearly at the transition: $\beta=1$. The linear response is
$e^{-(\mu+Dq^2)t}$, which relaxes at rate $\mu$ and spreads diffusively. Hence
$\nu_\parallel=1$, $\nu_\perp=\tfrac12$ and $z=2$. At criticality, activity started from a
single seed is a critical branching process. Its survival probability decays as $1/t$, and
its avalanche sizes and durations are distributed as $S^{-3/2}$ and $T^{-2}$
(Appendix~\ref{app:universality}).

\subsection{Beyond mean field: the $\varepsilon$ expansion}\label{app:DPeps}

Power counting (Appendix~\ref{app:universality}) shows that the coupling $g$ has
dimension $2-N/2$. Fluctuations are therefore negligible above $N_c=4$ and change the
exponents below it. Near four dimensions the RG can be carried out perturbatively in
$\varepsilon=4-N$. The one-loop results are~\cite{Janssen1981,Hinrichsen2000,Tauber2014}
\begin{equation}
\beta=1-\frac\varepsilon6,\quad \nu_\perp=\frac12+\frac\varepsilon{16},\quad
\nu_\parallel=1+\frac\varepsilon{12},\quad z=2-\frac\varepsilon{12}.
\label{eq:app_eps}
\end{equation}
They reproduce the trend of the numerical values in $N=1,2,3$ (Appendix~\ref{app:universality}).
These coefficients agree with Eqs.~(208) and (210) and Table~2 of Ref.~\cite{Tauber2007},
where $1/\nu_\perp=2-\varepsilon/4$, $z=2-\varepsilon/12$ and $\beta=1-\varepsilon/6$, with
$\nu_\parallel=z\nu_\perp=1+\varepsilon/12$ to this order. They also agree with the
$d=4-\varepsilon$ column of Table~2 of Ref.~\cite{Hinrichsen2000}.

\paragraph*{Long-range spreading.}
If activity spreads by long jumps whose length distribution decays as $r^{-N-\sigma}$
(``L\'evy flights''), the operator $-D\nabla^2$ is replaced by a fractional operator with
symbol $|q|^\sigma$. Power counting then gives $N_c=2\sigma$. In the Euclidean setting the
diffusive term takes over again for $\sigma\ge2$~\cite{Janssen1999}. In the $p$-adic
theory of this paper there is no local diffusive term, the kinetic operator is the nonlocal diffusion operator ($W$-operator)
of Sec.~\ref{sec:Wop}, and every $\sigma>0$ gives a distinct class with $N_c=2\sigma$
(Sec.~\ref{sec:RG}).

\subsection{From the effective spike model to Reggeon field theory}\label{app:DPchecklist}

The following summary connects the main text with this appendix.
\begin{enumerate}
\item \emph{Absorbing state:} $f(0)=0$ with $I=0$ (Sec.~\ref{sec:split}).
\item \emph{Branching:} $f'(0)>0$; the linear term $f'(0)\,\tphi\,w\star\varphi$.
\item \emph{Saturation:} $f''(0)<0$; it fixes the sign of $\tfrac g2\tpsi\psi^2$.
\item \emph{Distance to criticality:} $\mu=\alpha-f'(0)\hat w(0)$.
\item \emph{Spreading operator:} $-D\nabla^2$ in the Euclidean theory; the $W$-operator,
  with symbol $\simeq\kappa_W\norm{q}^\sigma$, in the $p$-adic theory.
\item \emph{Upper critical dimension:} $N_c=4$ (Euclidean), $N_c=2\sigma$ ($p$-adic).
\end{enumerate}

\paragraph*{Experimental status.}
Although DP is the generic class for absorbing-state transitions, clean experimental
realizations are rare. A well-known example is the transition between turbulent states in
liquid-crystal convection~\cite{Takeuchi2007}. Neuronal avalanches, with size and duration
exponents close to $3/2$ and $2$~\cite{BeggsPlenz2003}, are consistent with the mean-field
DP (critical branching) values. This is the regime predicted by B-C for highly connected
cortical networks, and by the $p$-adic theory whenever $\sigma\le N/2$.

\section{Mean-field exponents of the $p$-adic theory}\label{app:MF}

In this appendix we derive the mean-field exponents~\eqref{eq:MF} from the
action~\eqref{eq:RFT}. Three of them ($\beta$, $\beta'$, $\nu_\parallel$) come from the
dynamics of the total activity alone and are the same as in B-C. The other two
($\nu_\perp$, $z$) come from how the relaxation rate depends on the length scale; this is
where $\sigma$ enters. Throughout, $\mu$ is the distance to criticality: the network is
subcritical (absorbing phase) for $\mu>0$, critical at $\mu=0$, and supercritical (active
phase) for $\mu<0$.

\subsection{Definitions}
Near the critical point the characteristic quantities behave as powers of $|\mu|$, and the
exponents are defined by these powers:
\begin{itemize}
\item $\beta$: the stationary density of activity, starting from a fully active network,
  $\rho\propto(-\mu)^{\beta}$ for $\mu<0$ [Eq.~\eqref{eq:betas}];
\item $\beta'$: the probability that the activity started by one spike never dies out,
  $P_\infty\propto(-\mu)^{\beta'}$ for $\mu<0$ [Eq.~\eqref{eq:betas}];
\item $\nu_\parallel$: the correlation time, $\xi_\parallel\propto|\mu|^{-\nu_\parallel}$;
\item $\nu_\perp$: the correlation length, $\xi_\perp\propto|\mu|^{-\nu_\perp}$;
\item $z$: the dynamic exponent, $\xi_\parallel\propto\xi_\perp^{\,z}$, i.e.\
  $z=\nu_\parallel/\nu_\perp$.
\end{itemize}
These are the standard exponents of directed percolation [Eq.~(60) of
Ref.~\cite{BuiceCowan2007}].

\subsection{The mean-field equation}
Mean-field theory is the saddle point of the action~\eqref{eq:RFT}. The equation
$\delta S_{\rm eff}^W/\delta\tpsi=0$ reads
\begin{equation*}
(\partial_t+\mu+\boldsymbol W)\psi+\frac g2\psi^2-g\tpsi\psi=0 .
\end{equation*}
At the saddle point the response field vanishes, $\tpsi=0$, by the same mechanism as
$\tilde a=0$ in Sec.~\ref{sec:meanfield} (conservation of probability). The mean activity
$\rho(x,t)=\langle\psi(x,t)\rangle$ therefore obeys
\begin{equation}
\partial_t\rho=-\mu\rho-\boldsymbol W\rho-\frac g2\rho^2 .
\label{eq:app_MFeq}
\end{equation}
This is the hierarchical Wilson--Cowan equation near the critical point, written for the
rescaled field: decay minus branching ($-\mu\rho$), spreading through the hierarchy
($-\boldsymbol W\rho$) and saturation ($-\tfrac g2\rho^2$).

\subsection{$\beta=1$}
For a homogeneous state $\boldsymbol W\rho=0$, since the symbol of $\boldsymbol W$
vanishes at $\xi=0$. Equation~\eqref{eq:app_MFeq} becomes
$d\rho/dt=-\rho(\mu+\tfrac g2\rho)$. Its stationary solutions are $\rho=0$ (the absorbing
state) and
\begin{equation*}
\rho=-\frac{2\mu}{g}=\frac{2|\mu|}{g}\qquad(\mu<0).
\end{equation*}
For $\mu>0$ only $\rho=0$ is admissible ($\rho\ge0$), and it is stable. For $\mu<0$ the
state $\rho=0$ is unstable, since $d\rho/dt\simeq|\mu|\rho>0$ for small $\rho$, and the
activity settles at $\rho=2|\mu|/g$. The density grows linearly with the distance to
criticality, so $\beta=1$.

\subsection{$\beta'=1$}
The survival probability is a property of a single avalanche. In mean-field theory an
avalanche is a branching process (Appendix~\ref{app:primer}): each active unit produces a
new one at rate $b$ and disappears at rate $d$, and $\mu\propto d-b$. For binary branching
in continuous time, let $q$ be the probability that the activity started by one unit
eventually dies out. The first event is a birth with probability $b/(b+d)$, after which
two independent units are present, or a death with probability $d/(b+d)$. Hence
\begin{equation*}
q=\frac{d}{b+d}+\frac{b}{b+d}\,q^2,\qquad\text{i.e.}\qquad bq^2-(b+d)q+d=0,
\end{equation*}
whose roots are $q=1$ and $q=d/b$. The extinction probability is the smallest root in
$[0,1]$, so the probability that the activity never dies out is
$P_\infty=1-d/b=(b-d)/b$ for $b>d$, and $P_\infty=0$ for $b\le d$. Near criticality
$b-d\propto|\mu|$, so $P_\infty\propto|\mu|$ and $\beta'=1$. The same result follows from
the action: the saddle-point equation $\delta S_{\rm eff}^W/\delta\psi=0$ has a stationary
homogeneous solution with $|\tpsi|=2|\mu|/g$, and rapidity reversal~\eqref{eq:rapidity},
which exchanges $\psi$ and $-\tpsi$, gives $\beta'=\beta$.

\subsection{$\nu_\parallel=1$}\label{app:MFnupar}
Linearize Eq.~\eqref{eq:app_MFeq} about its stable stationary state and expand the
deviation in Fourier modes (equivalently, in the wavelets $\psi_{\gamma nj}$). For $\mu>0$
the stable state is $\rho=0$, and each mode evolves independently,
$\hat\rho(\xi,t)=\hat\rho(\xi,0)\,e^{-\lambda(\xi)t}$, with $\lambda(\xi)=\mu+A_w(\xi)$. For
$\mu<0$ the stable state is $\rho^*=2|\mu|/g$; writing $\rho=\rho^*+\delta\rho$, the linear
term of Eq.~\eqref{eq:app_MFeq} is $-(\mu+g\rho^*)\delta\rho=-|\mu|\,\delta\rho$, so the
deviations relax with $\lambda(\xi)=|\mu|+A_w(\xi)$. On both sides of the transition
\begin{equation*}
\lambda(\xi)=|\mu|+A_w(\xi).
\end{equation*}
The slowest mode is the homogeneous one, for which $A_w(0)=0$; it relaxes on the time scale
$\xi_\parallel=1/|\mu|$. Hence $\nu_\parallel=1$, with the same amplitude on both sides.

\subsection{$\nu_\perp=1/\sigma$}
The relaxation rate has two parts: $\mu$, the same for all modes, and
$A_w(\xi)\simeq\kappa_W\norm{\xi}^{\sigma}$, which is larger for modes supported on smaller
balls. Modes with $\kappa_W\norm{\xi}^{\sigma}\ll|\mu|$, supported on large balls, relax at
essentially the same rate as the homogeneous mode: on these scales fluctuations are
correlated. Modes with $\kappa_W\norm{\xi}^{\sigma}\gg|\mu|$, supported on small balls,
relax much faster, and fluctuations die out before they spread. The two regimes are
separated by the scale at which $\kappa_W\norm{\xi}^{\sigma}\approx|\mu|$, and the
corresponding length is the correlation length
\begin{equation*}
\xi_\perp=\Big(\frac{\kappa_W}{|\mu|}\Big)^{1/\sigma},
\end{equation*}
so $\nu_\perp=1/\sigma$. In $\Qp^N$ the norm takes only the values $p^k$, so this scale
is a level $k^*=\log_p\xi_\perp$ of the hierarchy, and $\xi_\perp$ is defined up to
a factor $p$; this is the length $\ell_W$ of the Ginzburg criterion (Sec.~\ref{sec:ginzburg}),
whose discreteness produces log-periodic corrections without changing the exponent. In the
Euclidean theory $A_w(q)\simeq Dq^2$, and the same argument gives
$\xi_\perp=(D/|\mu|)^{1/2}$, i.e.\ $\nu_\perp=\tfrac12$; this is B-C's diffusion length
[Eqs.~(52)--(53) of Ref.~\cite{BuiceCowan2007}].

\subsection{$z=\sigma$}
At $\mu=0$ the only rate left is $A_w(\xi)\simeq\kappa_W\norm{\xi}^{\sigma}$. A fluctuation
of size $L$, supported on a ball of radius $L$ ($\norm{\xi}\approx1/L$), relaxes in a time
$T\approx L^{\sigma}/\kappa_W$. Time scales as (length)$^{\sigma}$, so $z=\sigma$;
equivalently, $z=\nu_\parallel/\nu_\perp=\sigma$. This is the dynamic exponent of the power
counting in Sec.~\ref{sec:RG}. In the Euclidean theory $T\approx L^2/D$ (diffusion), and
$z=2$.

\subsection{Ginzburg estimate}\label{app:MFginzburg}
We derive Eq.~\eqref{eq:ginzburg_heur}. By the correspondence between
Eqs.~\eqref{eq:app_DPlangevin} and~\eqref{eq:app_RFT2} (Appendix~\ref{app:DPfield}), the
action~\eqref{eq:RFT} is the response-field action of the Langevin equation
\begin{align*}
&\partial_t\psi=-\mu\psi-\boldsymbol W\psi-\frac g2\psi^2+\eta,\\
&\langle\eta(x,t)\eta(x',t')\rangle=g\,\psi(x,t)\,\delta(x-x')\delta(t-t'):
\end{align*}
the term $-\tfrac g2\tpsi^{\,2}\psi$ of Eq.~\eqref{eq:RFT} is the noise, with variance
proportional to the activity (multiplicative noise), and the term $\tfrac g2\tpsi\psi^2$ is
the saturation. Without noise this is Eq.~\eqref{eq:app_MFeq}.

Take $\mu<0$, so that mean-field theory predicts the homogeneous density
$\rho^*=2|\mu|/g$, and ask how large the fluctuations about $\rho^*$ are. On scales smaller
than $\xi_\perp$ the activity is correlated, so we average over a ball $B$ of radius
$\xi_\perp$, of volume $V=\xi_\perp^{\,N}$:
$\rho_B(t)=V^{-1}\int_B\psi(x,t)\,d^Nx$. Modes supported on balls of radius at least
$\xi_\perp$ have $A_w(\xi)\lesssim|\mu|$, so, up to factors of order one, $\boldsymbol W$
does not change the relaxation of $\rho_B$, and the deviation $\delta\rho=\rho_B-\rho^*$
obeys, by the linearization of Appendix~\ref{app:MFnupar},
\begin{equation*}
\frac{d\,\delta\rho}{dt}=-|\mu|\,\delta\rho+\eta_B,\qquad
\eta_B=\frac1V\int_B\eta\,d^Nx .
\end{equation*}
Since $\eta$ is uncorrelated in space,
$\langle\eta_B(t)\eta_B(t')\rangle=(g\rho^*/V)\,\delta(t-t')$. This is an
Ornstein--Uhlenbeck process: noise of intensity $g\rho^*/V$ is accumulated during the
relaxation time $\xi_\parallel=1/|\mu|$, and the stationary variance is
\begin{equation*}
\langle\delta\rho^2\rangle=\frac{g\rho^*/V}{2|\mu|}=\frac{g\rho^*\,\xi_\parallel}{2V}.
\end{equation*}
Dividing by $\rho^{*2}$ and using $\rho^*=2|\mu|/g$, $\xi_\parallel=1/|\mu|$ and
$V=\xi_\perp^{\,N}=(\kappa_W/|\mu|)^{N/\sigma}$ gives
\begin{equation*}
\frac{\langle\delta\rho^2\rangle}{\rho^{*2}}=\frac{g\,\xi_\parallel}{2\rho^*\,\xi_\perp^{\,N}}
=\frac{g^2}{4\,\kappa_W^{N/\sigma}}\;|\mu|^{N/\sigma-2},
\end{equation*}
which is Eq.~\eqref{eq:ginzburg_heur} with its prefactor. Mean-field theory is consistent
when this ratio is small. As $\mu\to0$ it tends to zero if $N>2\sigma$ and diverges if
$N<2\sigma$; at $N=2\sigma$ it is a constant of order $g^2/\kappa_W^2$, the marginal case.
The estimate treats the ball $B$ as well mixed and drops factors of order one; it locates
the upper critical dimension but not the exponents below it. The same condition follows
from the one-loop integral of Sec.~\ref{sec:ginzburg} and from power counting.

\subsection{Comparison and validity}
The exponents of the total activity ($\beta$, $\beta'$, $\nu_\parallel$) are those of B-C,
because they come from the equation $d\rho/dt=-\mu\rho-\tfrac g2\rho^2$, which does not see
the geometry. The spatial exponents ($\nu_\perp$, $z$) come from the kinetic operator,
$Dq^2$ in B-C and $\kappa_W\norm{\xi}^{\sigma}$ here; the two sets coincide for $\sigma=2$.
Mean-field exponents are exact only when the cubic coupling is irrelevant, i.e.\ for
$N>N_c=2\sigma$ (with logarithmic corrections at $N=2\sigma$); for $N<2\sigma$ loop
corrections change all five exponents. The mean-field avalanche exponents (sizes
distributed as $S^{-3/2}$, durations as $T^{-2}$) come from the same branching process as
$\beta'$ and do not depend on $\sigma$; the simulations of Sec.~\ref{sec:simulations} test
when they hold.

\section{Basic facts on $p$-adic analysis}\label{app:padic}

In this appendix we fix the notation and collect the results of $p$-adic analysis used
in the paper. For detailed expositions see
Refs.~\cite{VVZ1994,AKS2010,Taibleson1975,ZunigaLNM2016,ZunigaTextbook2025}.

\subsection{The field of $p$-adic numbers}

Throughout, $p$ denotes a fixed prime number. The field of $p$-adic numbers $\Qp$ is the
completion of the field of rational numbers $\mathbb Q$ with respect to the $p$-adic norm
\begin{equation}
|x|_p=
\begin{cases}
0 & \text{if }x=0,\\
p^{-\gamma} & \text{if }x=p^{\gamma}\,\dfrac ab,
\end{cases}
\end{equation}
where $a$ and $b$ are integers coprime with $p$. The integer $\gamma=\mathrm{ord}(x)$, with
$\mathrm{ord}(0)=+\infty$, is called the $p$-adic order of $x$.

Any $p$-adic number $x\ne0$ has a unique expansion
\begin{equation}
x=p^{\mathrm{ord}(x)}\sum_{j=0}^{\infty}x_jp^j,\qquad x_j\in\{0,1,\dots,p-1\},\quad x_0\ne0 ,
\label{eq:padic_expansion}
\end{equation}
and $|x|_p=p^{-\mathrm{ord}(x)}$. The \emph{fractional part} of $x\in\Qp$ is the rational
number
\begin{equation}
\{x\}_p=
\begin{cases}
0 & \text{if }x=0\text{ or }\mathrm{ord}(x)\ge0,\\[2pt]
p^{\mathrm{ord}(x)}\displaystyle\sum_{j=0}^{-\mathrm{ord}(x)-1}x_jp^j & \text{if }\mathrm{ord}(x)<0 .
\end{cases}
\end{equation}
The $p$-adic integers are the numbers with $|x|_p\le1$, i.e.\ those whose expansion has no
negative powers of $p$; they form the ring $\Zp$. The successive digits of the
expansion~\eqref{eq:padic_expansion} select a path in a rooted tree in which every vertex
has $p$ children. This is the origin of the correspondence between $\Zp$ and the leaves of
an infinite rooted tree used in Sec.~\ref{sec:geometry}. This description holds for $\Zp$
and, more generally, for any ball. It does not hold for $\Qp$: the expansion of a general
$p$-adic number starts at an arbitrary negative power of $p$, the balls $p^{-l}\Zp$ grow
without bound, and there is no root. The points of $\Qp$ are organized in a tree-like
structure, which is not a rooted tree.

The norm is extended to $\Qp^N$ by
\begin{equation}
\norm{x}=\max_{1\le i\le N}|x_i|_p,\qquad x=(x_1,\dots,x_N)\in\Qp^N .
\end{equation}
With $\mathrm{ord}(x)=\min_i\mathrm{ord}(x_i)$ one has $\norm{x}=p^{-\mathrm{ord}(x)}$, so
the norm takes only the values $p^k$, $k\in\mathbb Z$, and $0$. It satisfies the
\emph{ultrametric inequality}
\begin{equation}
\norm{x+y}\le\max\{\norm{x},\norm{y}\},
\end{equation}
which is stronger than the triangle inequality, and $(\Qp^N,\norm{\cdot})$ is a complete
ultrametric space.

\subsection{Balls, spheres and topology}

For $r\in\mathbb Z$ and $a\in\Qp^N$, the ball and the sphere of radius $p^r$ with center
$a$ are
\begin{align}
B_r^N(a)&=\{x\in\Qp^N:\norm{x-a}\le p^r\}=a+p^{-r}\Zp^N,\\
S_r^N(a)&=\{x\in\Qp^N:\norm{x-a}=p^r\}\nonumber\\
&=B_r^N(a)\smallsetminus B_{r-1}^N(a),
\end{align}
and we write $B_r^N=B_r^N(0)$, $S_r^N=S_r^N(0)$. In particular $B_0^N=\Zp^N$ is the unit ball. The
ultrametric inequality has the following consequences, which have no Euclidean
counterpart. Every point of a ball is a center of it. Two balls are either disjoint or
one is contained in the other. Each ball of radius $p^r$ is the disjoint union of $p^N$
balls of radius $p^{r-1}$. The balls therefore form a tree under inclusion, and the
distance $\norm{x-y}$ is the radius of the smallest ball that contains $x$ and $y$.

Balls and spheres are both open and closed, and they are compact. The space
$\Qp^N$ is locally compact and totally disconnected: its only connected subsets are the
empty set and the points. We denote by $\Omega(p^{-r}\norm{x-a})$ the characteristic
function of the ball $B_r^N(a)$; here $\Omega$ is the characteristic function of the
interval $[0,1]$.

\subsection{The Haar measure}

Since $(\Qp^N,+)$ is a locally compact topological group, it has a Haar measure $d^Nx$,
which is invariant under translations, $d^N(x+a)=d^Nx$. It is unique once normalized by
$\int_{\Zp^N}d^Nx=1$. With this normalization,
\begin{align}
\mathrm{vol}\big(B_r^N(a)\big)&=p^{rN}, &
\mathrm{vol}\big(S_r^N(a)\big)&=(1-p^{-N})\,p^{rN},
\end{align}
and $d^N(cx)=|c|_p^N\,d^Nx$ for $c\in\Qp\smallsetminus\{0\}$. These formulas give
Eq.~\eqref{eq:radial} for the integral of a radial function.

\subsection{Test functions and $L^\rho$ spaces}

A function $\varphi:\Qp^N\to\mathbb C$ is \emph{locally constant} if for every $x$
there is an integer $l(x)$ such that
\begin{equation}
\varphi(x+x')=\varphi(x)\quad\text{for all }x'\in B_{l(x)}^N .
\label{eq:loc_const}
\end{equation}
A \emph{test function} (or Bruhat--Schwartz function) is a locally constant function
with compact support. Every test function is a finite linear combination of
characteristic functions of balls. The space of test functions is denoted by
$\mathcal D(\Qp^N)$. For a test function, the integer $l(x)$ in Eq.~\eqref{eq:loc_const}
can be chosen independently of $x$; the largest such integer is its parameter of
constancy.

For integers $m\le l$, we denote by $\mathcal D^{l}_{m}(\Qp^N)$ the space of test functions
supported in the ball $B_{-m}^N=p^m\Zp^N$ and with parameter of constancy $\ge-l$. A
function in $\mathcal D^{l}_{m}(\Qp^N)$ is constant on the cosets of $p^l\Zp^N$ in
$p^m\Zp^N$, i.e.\ on the $p^{N(l-m)}$ balls of radius $p^{-l}$ contained in $B_{-m}^N$, and
it is determined by its values on them. Therefore $\mathcal D^{l}_{m}(\Qp^N)$ has finite
dimension $p^{N(l-m)}$. The space $\mathcal D^l$ of Sec.~\ref{sec:geometry}, which describes
the finite network $G_l^N$, is $\mathcal D^{l}_{0}(\Qp^N)$ in this notation: functions on
$\Zp^N$ that are constant on the balls of radius $p^{-l}$; its dimension is $p^{Nl}$.

For $1\le\rho<\infty$, $L^\rho(\Qp^N)$ is the space of functions $g$ with
\begin{equation}
\Vert g\Vert_\rho=\Big(\int_{\Qp^N}|g(x)|^\rho\,d^Nx\Big)^{1/\rho}<\infty .
\end{equation}
We use $\rho=1$ and $\rho=2$. The test functions are dense in $L^\rho(\Qp^N)$.

\subsection{The Fourier transform}

Set $\chi_p(y)=\exp(2\pi i\{y\}_p)$ for $y\in\Qp$. The map $\chi_p$ is an additive character
of $\Qp$: it is continuous, $|\chi_p(y)|=1$, and
$\chi_p(y+y')=\chi_p(y)\chi_p(y')$. It is trivial on $\Zp$. For
$\xi,x\in\Qp^N$ set $\xi\cdot x=\sum_{j=1}^N\xi_jx_j$. The Fourier transform of
$\varphi\in\mathcal D(\Qp^N)$ is
\begin{equation}
\hat\varphi(\xi)=(\mathcal F\varphi)(\xi)=\int_{\Qp^N}\chi_p(\xi\cdot x)\,\varphi(x)\,d^Nx .
\end{equation}
It is a linear isomorphism of $\mathcal D(\Qp^N)$ onto itself, and
\begin{equation}
\big(\mathcal F(\mathcal F\varphi)\big)(\xi)=\varphi(-\xi),
\label{eq:FF}
\end{equation}
which is the inversion formula of Sec.~\ref{sec:split}. The Fourier transform maps
$\mathcal D^{l}_{m}$ onto $\mathcal D^{-m}_{-l}$: a function that is constant on balls
of radius $p^l$ has a transform supported in $\norm{\xi}\le p^{-l}$. For the finite network $G_l^N$, whose
fields are constant on balls of radius $p^{-l}$, this gives the momentum cutoff
$\norm{\xi}\le p^{l}$. The basic example is Eq.~\eqref{eq:ballFT}, the transform of
the characteristic function of a ball.

The Fourier transform extends to a unitary operator on $L^2(\Qp^N)$,
$\Vert\mathcal Fg\Vert_2=\Vert g\Vert_2$, and Eq.~\eqref{eq:FF} holds in $L^2$. For
$g\in L^1(\Qp^N)$ the integral defining $\hat g$ converges, and $\hat g$ is bounded and
continuous. For $w,g\in L^1(\Qp^N)$ the convolution
\begin{equation}
(w\star g)(x)=\int_{\Qp^N}w(x-y)\,g(y)\,d^Ny
\end{equation}
is in $L^1(\Qp^N)$, with $\Vert w\star g\Vert_1\le\Vert w\Vert_1\Vert g\Vert_1$ and
$\widehat{w\star g}=\hat w\,\hat g$.

\subsection{Distributions}

A distribution is a linear functional on $\mathcal D(\Qp^N)$; the space of distributions
is denoted by $\mathcal D'(\Qp^N)$. In contrast with the Euclidean case, no continuity
condition is needed: every linear functional on $\mathcal D(\Qp^N)$ is continuous. We
write $(T,\varphi)$ for the value of $T$ on $\varphi$. A sequence $T_j$ converges to $T$
if $(T_j,\varphi)\to(T,\varphi)$ for every test function $\varphi$. Every locally
integrable function $g$ defines a distribution by
\begin{equation}
(g,\varphi)=\int_{\Qp^N}g(x)\,\varphi(x)\,d^Nx .
\end{equation}
The Dirac distribution is $(\delta,\varphi)=\varphi(0)$. It is the limit of the
functions $\delta_k(x)=p^{Nk}\,\Omega(p^k\norm{x})$ as $k\to\infty$, which are the
normalized characteristic functions of the balls $B_{-k}^N$. The function $\delta_l$ is
the regularized delta function of the finite network $G_l^N$ (Sec.~\ref{sec:split}).

The Fourier transform of a distribution $T$ is defined by
\begin{equation}
(\mathcal F[T],\varphi)=(T,\mathcal F[\varphi])\quad\text{for all }\varphi\in\mathcal D(\Qp^N).
\end{equation}
It is a linear isomorphism of $\mathcal D'(\Qp^N)$ onto itself, and it agrees with the
previous definitions on $L^1$ and $L^2$. In particular $\mathcal F[\delta]=1$ and
$\mathcal F[1]=\delta$. A bounded function such as $e^{t f'(0)\hat w(\xi)}$, which does not
decay at infinity, therefore has an inverse Fourier transform in $\mathcal D'(\Qp^N)$ but
not as a function. This is the sense of the propagator formula~\eqref{eq:Delta0x}.

The convolution of a distribution $T$ with a test function $\psi$ is the locally
constant function $(T\star\psi)(y)=(T(x),\psi(y-x))$; in particular
$\delta\star\psi=\psi$. Whenever the convolution of two distributions $F$ and $G$
exists, for instance if one of them has compact support or if both are in $L^1$, one has
\begin{equation}
\mathcal F[F\star G]=\mathcal F[F]\cdot\mathcal F[G],
\end{equation}
see Ref.~\cite{VVZ1994}, Sec.~7.

\subsection{Explicit constants in the heat-kernel estimate}\label{app:heatkernel}

Here we derive explicit admissible constants in Eq.~\eqref{eq:Z_estimate}.
Reference~\cite{BGPW2014} proves Eq.~\eqref{eq:Z_estimate} with constants that depend on the
doubling constant of the spectral distribution function (their Theorem~2.14 and
Lemmas~2.19--2.21), from which explicit values can also be extracted;
Ref.~\cite{RodriguezZuniga2008} proves the upper bound. Here we give a direct derivation
from the series~\eqref{eq:Z_series} below.

The calculation matters for the paper for two reasons. First, Eq.~\eqref{eq:Z_estimate} is
the basis of several statements of Sec.~\ref{sec:phase}: the spectral dimension
$d_s=2N/\sigma$, the power-law tail of the propagator in hierarchical distance, and the
spreading of activity in the active phase. With explicit constants these statements become
inequalities valid for all $t>0$ and all $x$, with constants that depend only on $p$, $N$ and
$\sigma$ and not on $\kappa_W$, so that the coupling enters only through the combination
$\kappa_Wt$. Second, the argument works directly with the positive series~\eqref{eq:Z_series}
and does not use the abstract theory of isotropic semigroups. It is therefore the natural
starting point for the open problem of Remark~\ref{rem:Kt_open}, the heat kernel of the
$W$-operator of the B-C theory, whose symbol is bounded and to which Theorem~5.17 of
Ref.~\cite{BGPW2014} does not apply.

Set
$s=\kappa_Wt>0$. The heat kernel of $\kappa_W\Dsig$ is
\begin{equation}
Z^{(\sigma)}_t(x)=\int_{\Qp^N}e^{-s\norm{\xi}^{\sigma}}\chi_p(\xi\cdot x)\,d^N\xi ,
\qquad x\in\Qp^N .
\end{equation}
Its integrand is radial, so $Z^{(\sigma)}_t$ is given by the well-known formula for the
Fourier transform of a radial function~\cite{VVZ1994}, Lemma~\ref{lem:radialFT} of
Appendix~\ref{app:integrals}. With $a_k=e^{-sp^{k\sigma}}$ and $\norm{x}=p^m$,
\begin{equation*}
Z^{(\sigma)}_t(x)=(1-p^{-N})\sum_{k\le-m}p^{kN}a_k-p^{-mN}a_{1-m}.
\end{equation*}
It is convenient to regroup the terms. Equivalently, decompose $\Qp^N\smallsetminus\{0\}$ into
the spheres $S_k^N=\{\norm{\xi}=p^k\}$; by Eq.~\eqref{eq:ballFT},
$\int_{S_k^N}\chi_p(\xi\cdot x)\,d^N\xi=b_k-b_{k-1}$ with $b_k=p^{kN}\Omega(p^k\norm{x})$, so that
$Z^{(\sigma)}_t(x)=\sum_ka_k(b_k-b_{k-1})$. A \emph{summation by parts}, the discrete analogue
of integration by parts, moves the difference from $b$ to $a$:
$\sum_ka_k(b_k-b_{k-1})=\sum_kb_k(a_k-a_{k+1})$, since the boundary terms vanish ($b_k\to0$ as
$k\to-\infty$ and $a_k\to0$ as $k\to+\infty$). This gives the positive series
\begin{equation}
Z^{(\sigma)}_t(x)=\sum_{k\in\mathbb Z}p^{kN}\,\Omega(p^k\norm{x})\,
\big(e^{-sp^{k\sigma}}-e^{-sp^{(k+1)\sigma}}\big),
\label{eq:Z_series}
\end{equation}
valid for all $x\in\Qp^N$, including $x=0$. For $\norm{x}=p^m$ the sum runs over $k\le-m$.
Each term equals
$T_k=p^{kN}e^{-y_k}\big(1-e^{-(p^\sigma-1)y_k}\big)$, with $y_k=sp^{k\sigma}$.
Let $F(s,x)=s\big(s^{1/\sigma}+\norm{x}\big)^{-N-\sigma}$, so that $F_t(x)=F(\kappa_Wt,x)$. Since
$\max\{s^{1/\sigma},\norm{x}\}\le s^{1/\sigma}+\norm{x}\le2\max\{s^{1/\sigma},\norm{x}\}$,
\begin{equation}
2^{-N-\sigma}M(s,x)\le F(s,x)\le M(s,x),
\label{eq:F_M}
\end{equation}
where $M(s,x)=\min\big\{s\norm{x}^{-N-\sigma},\,s^{-N/\sigma}\big\}$.

\emph{Upper bound.} Since $1-e^{-z}\le z$, $T_k\le(p^\sigma-1)\,s\,p^{k(N+\sigma)}e^{-y_k}$.
Dropping $e^{-y_k}$ and summing the geometric series over $k\le-m$ gives
$Z^{(\sigma)}_t(x)\le C_1\,s\norm{x}^{-N-\sigma}$, with
$C_1=(p^\sigma-1)/(1-p^{-N-\sigma})$. Alternatively, write $p^{kN}=(y_k/s)^{N/\sigma}$, so that
$T_k\le(p^\sigma-1)s^{-N/\sigma}g(\ln y_k)$, with $g(u)=e^{\beta u}e^{-e^u}$ and
$\beta=1+N/\sigma$. The points $\ln y_k$ are equally spaced with step $\sigma\ln p$, and $g$
is unimodal on $\mathbb R$, with maximum $(\beta/e)^\beta$ and
$\int_{\mathbb R}g(u)\,du=\Gamma(\beta)$. Hence
$\sum_kg(\ln y_k)\le(\beta/e)^\beta+\Gamma(\beta)/(\sigma\ln p)$, and
$Z^{(\sigma)}_t(x)\le C_2\,s^{-N/\sigma}$, with
$C_2=(p^\sigma-1)\big[(\beta/e)^\beta+\Gamma(\beta)/(\sigma\ln p)\big]$.
Together with Eq.~\eqref{eq:F_M}, $Z^{(\sigma)}_t(x)\le\max\{C_1,C_2\}M(s,x)\le
2^{N+\sigma}\max\{C_1,C_2\}F(s,x)$.

\emph{Lower bound.} Each $T_k$ is nonnegative, so $Z^{(\sigma)}_t(x)$ is bounded below by
any single term. We use $1-e^{-az}\ge(1-e^{-a})z$ for $z\in[0,1]$ and $a>0$, which
follows from concavity. If $s\le\norm{x}^\sigma$, take $k=-m$, so $y_{-m}=s\norm{x}^{-\sigma}\le1$
and
$Z^{(\sigma)}_t(x)\ge T_{-m}\ge e^{-1}\big(1-e^{-(p^\sigma-1)}\big)s\norm{x}^{-N-\sigma}$.
If $s>\norm{x}^\sigma$ (in particular if $x=0$), take $k=-\lceil\sigma^{-1}\log_ps\rceil$.
Then $p^{-\sigma}<y_k\le1$ and $p^k\le s^{-1/\sigma}<\norm{x}^{-1}$, so the term is present
in Eq.~\eqref{eq:Z_series}, and $p^{kN}=(y_k/s)^{N/\sigma}>p^{-N}s^{-N/\sigma}$. Hence
$Z^{(\sigma)}_t(x)\ge T_k\ge e^{-1}p^{-N}\big(1-e^{-(1-p^{-\sigma})}\big)s^{-N/\sigma}$.
The second constant is the smaller one, and by Eq.~\eqref{eq:F_M}, $M\ge F$.

Then, we have established the following result.

\begin{proposition}\label{prop:cC}
The estimate
\begin{equation*}
c\,F_t(x)\le Z^{(\sigma)}_t(x)\le C\,F_t(x)
\end{equation*}
(Eq.~\eqref{eq:Z_estimate}) holds for all $t>0$ and $x\in\Qp^N$ with
\begin{equation}
\begin{aligned}
c&=e^{-1}p^{-N}\big(1-e^{-(1-p^{-\sigma})}\big),\\
C&=2^{N+\sigma}(p^\sigma-1)\\
&\quad\times\max\Big\{\frac{1}{1-p^{-N-\sigma}},\,
\Big(\frac{\beta}{e}\Big)^{\beta}+\frac{\Gamma(\beta)}{\sigma\ln p}\Big\},
\end{aligned}
\label{eq:cC_explicit}
\end{equation}
where $\beta=1+N/\sigma$.
\end{proposition}

The constants depend only on $p$, $N$, and $\sigma$, and not on $\kappa_W$. They are
admissible, not sharp. The sharp constants are the infimum and the supremum of
$Z^{(\sigma)}_t(x)/F(s,x)$. This ratio is invariant under the dilation
$(s,x)\mapsto(p^\sigma s,p^{-1}x)$, because both $Z^{(\sigma)}$ and $F$ acquire the factor
$p^{-N}$. Hence it suffices to compute the ratio for $\norm{x}\in\{0,1\}$, as a function of
$s>0$. For example, for $p=2$, $N=1$, $\sigma=1/2$, a numerical evaluation of
Eq.~\eqref{eq:Z_series} gives a ratio between about $0.53$ and $1.49$, whereas
Eq.~\eqref{eq:cC_explicit} gives $c\approx0.047$ and $C\approx8.3$.

\section{Regional networks}\label{app:regional}

In this appendix we prove Lemmas~\ref{lem:regional} and~\ref{lem:tail}, and extend Sec.~\ref{sec:master} to the
regional networks of Sec.~\ref{sec:Wop}, Eqs.~\eqref{eq:K_union}--\eqref{eq:M}. In
Ref.~\cite{ZG2022Ultrametric}, $N=1$ and the regions are called basins.
The proof of Lemma~\ref{lem:regional} occupies Appendices~\ref{app:ultra}--\ref{app:spectrum}:
Appendix~\ref{app:ultra} collects two facts used below, part 2 is proved in
Appendix~\ref{app:reduction}, and parts 1 and 3 in Appendix~\ref{app:spectrum}. The proof
of Lemma~\ref{lem:tail} is given in Appendix~\ref{app:prooftail}.

\subsection{Two ultrametric facts}\label{app:ultra}
As in Sec.~\ref{sec:Wop}, $B_a=B_{l_a}^N(a)=a+p^{-l_a}\Zp^N$ denotes the ball of radius
$p^{l_a}$ with center $a$ (region $a$), not a ball centered at the origin. (i) If $a\neq b$, then
$\norm{x-y}=\norm{a-b}=:d_{ab}$ for all $x\in B_a$, $y\in B_b$, by the ultrametric inequality,
since the balls are disjoint. (ii) For $x\in B_a$,
$B_a-x=B_{l_a}^N$, so $\int_{B_a}w_a(\norm{x-y})\,d^Ny=\omega_a$ does not depend on
$x\in B_a$. By (i), a translation-invariant radial kernel $w$ restricted to $K$ defines a
regional network with $w_a=w$ and $c_{ab}=w(d_{ab})$. With $\alpha_a=\alpha$ and $f_a=f$,
Eq.~\eqref{eq:lin_K} is then Eq.~\eqref{eq:linearized} for a network whose neurons occupy
$K$; it is the restriction of Eq.~\eqref{eq:linearized} to $K$ when
$K+\operatorname{supp}w\subset K$.

\subsection{Reduction to a system}\label{app:reduction}
In this section, we prove part 2 of Lemma~\ref{lem:regional}. Write $u_a=u\,1_{B_a}$. Eq.~\eqref{eq:lin_K} becomes the system
\begin{align}
\partial_tu_a(x,t)=&-\alpha_a\,u_a(x,t)\nonumber\\
&+f_a'(0)\!\int_{B_a}w_a(\norm{x-y})\,u_a(y,t)\,d^Ny\nonumber\\
&+f_a'(0)\sum_{b\neq a}c_{ab}\,U_b(t)
\label{eq:system}
\end{align}
for $x\in B_a$, where $U_b(t)=\int_{B_b}u(y,t)\,d^Ny$ is the total activity of region $b$
(Lemma~\ref{lem:regional}). The integral describes branching inside region $a$; the sum describes
branching from region $b$ into region $a$, which depends on region $b$ only through its
total activity $U_b$. Integrating over $B_a$ and using (ii), with
$\int_{B_a}w_a(\norm{x-y})\,d^Nx=\omega_a$, gives
\begin{align*}
\frac{dU_a}{dt}&=\big(-\alpha_a+f_a'(0)\,\omega_a\big)U_a
+f_a'(0)\,\mathrm{vol}(B_a)\sum_{b\neq a}c_{ab}\,U_b\\
&=\sum_bM_{ab}U_b ,
\end{align*}
where $M$ is the matrix of Eq.~\eqref{eq:M}: $M_{aa}=-\alpha_a+f_a'(0)\,\omega_a$ and
$M_{ab}=f_a'(0)\,\mathrm{vol}(B_a)\,c_{ab}$ for $a\neq b$. This is part 2 of
Lemma~\ref{lem:regional}. Activity is
transferred from region $b$ to region $a$ at rate $f_a'(0)\,\mathrm{vol}(B_a)\,c_{ab}$. At
the level of regions the dynamics is a finite continuous-time Markov chain (with creation
and killing of activity), and each state of the chain is replaced by a ball carrying the
radial kernel $w_a$. This is the structure of the transition networks of
Ref.~\cite{ZG2022Ultrametric}.

\subsection{Spectrum and criticality}\label{app:spectrum}
In this section, we prove parts 1 and 3 of Lemma~\ref{lem:regional}. Recall that the neuron
space is $K=\bigsqcup_{a\in A}B_a$, the disjoint union of the finitely many balls
$B_a=B_{l_a}^N(a)$ [Eq.~\eqref{eq:K_union}]. The space $L^2(K)$ is spanned by the indicator functions $1_{B_a}$ and by the wavelets
$\psi_{\gamma nj}$ [Eq.~\eqref{eq:wavelets}] supported in some $B_a$, i.e.\ with
$\gamma\le l_a$. A wavelet supported in $B_a$ has mean zero, so the sum in
Eq.~\eqref{eq:system} annihilates it, and so do the equations for the other regions. By
Eq.~\eqref{eq:wavelet_eig}, applied to the kernel $w_a1_{B_{l_a}^N}$, it is an
eigenfunction with eigenvalue
\begin{equation*}
-\alpha_a+f_a'(0)\,\hat w_a\big(p^{1-\gamma}\big)\le-\alpha_a+f_a'(0)\,\omega_a=M_{aa}.
\end{equation*}
This is part 1. On the span of the $1_{B_a}$, the right-hand side of Eq.~\eqref{eq:lin_K}
acts by the matrix $\tilde M=D^{-1}MD$, with $D=\mathrm{diag}(\mathrm{vol}(B_a))$. The
off-diagonal entries of $M$ are nonnegative (a \emph{Metzler} matrix), so for $c>0$ large
enough $M+cI$ is a nonnegative matrix, to which the Perron--Frobenius theorem
applies~\cite{HornJohnson2013,Seneta2006}. Hence $s(M)=s(\tilde M)$ is a real eigenvalue with a
nonnegative eigenvector (the Perron eigenvector), and $s(M)\ge M_{aa}$ for every $a$, since the
Perron root of a nonnegative matrix is not smaller than its diagonal entries. Hence the leading growth rate of
Eq.~\eqref{eq:lin_K} is $s(M)$, which proves part 3. This completes the proof of
Lemma~\ref{lem:regional}.

\subsection{Markov property}
For $x\in B_a$, define the total input weight received by a neuron of region $a$,
\begin{equation*}
\hat w_{(a)}:=\int_Kw(x,y)\,d^Ny=\omega_a+\sum_{b\neq a}\mathrm{vol}(B_b)\,c_{ab} .
\end{equation*}
By fact~(ii) of Appendix~\ref{app:ultra}, it does not depend on $x\in B_a$. It is the value at
$\xi=0$ of the Fourier transform, in $y$, of $w(x,\cdot)1_K$; for a single ball and a
translation-invariant kernel it reduces to $\hat w(0)$. In general $\hat w_{(a)}$ differs from
$\hat w_a(0)=\omega_a$, which counts only the input from inside region $a$. Since
$\omega_a\ge0$ and $c_{ab}\ge0$, the condition $\hat w_{(a)}=0$ holds if and only if
$\omega_a=0$ and $c_{ab}=0$ for every $b\neq a$, i.e.\ $w_a\equiv0$ on $B_{l_a}^N$ ($w_a$ is
continuous and nonnegative) and region $a$ receives no input at all: its neurons are not
excited by one another or by the other regions. By Eq.~\eqref{eq:lin_K} their activity then
simply decays at rate $\alpha_a$, and, since $I=0$, they do not fire spontaneously. Such a
region is trivially subcritical and can be removed from the network. We therefore assume
$\hat w_{(a)}>0$ for every $a$; this is also needed to define the rates $j(y\,|\,x)$ below.
Set
$\mu_a=\alpha_a-f_a'(0)\,\hat w_{(a)}$. Then Eq.~\eqref{eq:lin_K} is the master
equation~\eqref{eq:masterK} on $K$ with
\begin{align*}
j(x\,|\,y)&=f_a'(0)\,w(x,y),\\
j(y\,|\,x)&=\frac{\alpha_a}{\hat w_{(a)}}\,w(x,y)\qquad(x\in B_a).
\end{align*}
Using the covering~\eqref{eq:K_union}, Hypothesis~1 of Ref.~\cite{ZG2022Ultrametric},
$j(x\,|\,y)\le j(y\,|\,x)$ for all $x,y\in K$, reduces to the list of conditions
\begin{equation}
\mu_a\ge0\qquad\text{for every }a\in A .
\label{eq:mu_a}
\end{equation}
Under these conditions, Theorem~3.1 of Ref.~\cite{ZG2022Ultrametric} gives a Feller
semigroup on $C(K,\mathbb R)$ and a Markov process with state space $K$. We rewrite
Eq.~\eqref{eq:lin_K} as $\partial_tu(x,t)=Ju(x,t)$, where $J$ is the linear operator on its
right-hand side. Then $J1_K=-\mu_a$ on $B_a$: the killing
rate is constant on each region, and the process is conservative if and only if
$\mu_a=0$ for every $a$. Conditions~\eqref{eq:mu_a} are also necessary, since a Feller
semigroup requires $J1_K\le0$.

The row sums of $\tilde M$ are $-\mu_a$, so by the row-sum bounds for the Perron root of a
nonnegative matrix (applied to $\tilde M+cI$; see Ref.~\cite{HornJohnson2013}, Chap.~8),
\begin{equation*}
-\max_a\mu_a\le s(M)\le-\min_a\mu_a .
\end{equation*}
Consequently: under conditions~\eqref{eq:mu_a} the dynamics is not supercritical; if
$\mu_a=0$ for every $a$, then $s(M)=0$, i.e.\ \emph{a conservative dynamics is critical}.
The converse holds when all $\mu_a$ are equal, in particular for a single ball containing
the support of a translation-invariant kernel, where $\hat w_{(a)}=\hat w(0)$ and
$\mu_a=\mu$ (Sec.~\ref{sec:master}). If the $\mu_a$ are not all equal and $M$ is
irreducible, then at $s(M)=0$ one has $\min_a\mu_a<0<\max_a\mu_a$, because for an
irreducible matrix the row-sum bounds are strict unless all row sums are
equal~\cite{HornJohnson2013,Seneta2006}; the critical dynamics
is then a Markov process only after the change of normalization given by the Perron
eigenvector of $\tilde M$. When $c_{ab}=c_{ba}$, $\alpha_a=\alpha$ and $f_a=f$ (for
instance, for a translation-invariant kernel), the column sums of $M$ are also $-\mu_b$,
so $\frac{d}{dt}\sum_aU_a=-\sum_b\mu_bU_b$, and a conservative process conserves the total
activity.

The Markov process given by Theorem~3.1 of Ref.~\cite{ZG2022Ultrametric} describes the linear (mean) dynamics. By the
standard many-to-one formula for branching processes, the mean activity equals the
transition density of a single walker, the ancestral lineage of the branching process,
weighted by its survival. The process adds no elementary events to
Eq.~\eqref{eq:app_reactions}: its jumps are branching events, and its killing is the
excess of decay over branching.

Finally, if every $w_a$ is supported in $\{\norm{x}\le p^{r}\}$, every region has radius
at least $p^{r}$, and the regional network comes from a translation-invariant kernel,
then $c_{ab}=w(d_{ab})=0$ for $a\neq b$, $M$ is diagonal, and the regions are independent
networks, each of the kind discussed before Lemma~\ref{lem:regional}.

\subsection{Proof of Lemma~\ref{lem:tail}}\label{app:prooftail}

\begin{proof}
For $\xi=0$ both sides vanish, so let $\norm{\xi}=p^m$ with $m\le-r$. Since $w\in L^1$,
\begin{equation*}
\hat w(0)-\hat w(\xi)=\int_{\Qp^N}w(\norm{x})\,\big[1-\chi_p(\xi\cdot x)\big]\,d^Nx .
\end{equation*}
\emph{Step 1: only the tail contributes.} If $\norm{x}\le\norm{\xi}^{-1}=p^{-m}$, then
$|\xi\cdot x|_p\le\norm{\xi}\,\norm{x}\le1$, so $\xi\cdot x\in\Zp$, $\{\xi\cdot x\}_p=0$ and
$\chi_p(\xi\cdot x)=1$: the bracket vanishes on the ball $B_{-m}^N$. Since $p^{-m}\ge p^r$,
this ball contains the region $\norm{x}\le p^r$, where $w$ is arbitrary. On the complement,
$\norm{x}>p^{-m}\ge p^r$, so $w(\norm{x})=C_w\norm{x}^{-N-\sigma}$, and
\begin{equation*}
\hat w(0)-\hat w(\xi)=C_w\sum_{k\ge1-m}p^{-k(N+\sigma)}
\int_{S_k^N}\big[1-\chi_p(\xi\cdot x)\big]\,d^Nx ,
\end{equation*}
where we decomposed $\{\norm{x}>p^{-m}\}$ into the spheres $S_k^N$, $k\ge1-m$, on which
$\norm{x}=p^k$ is constant.

\emph{Step 2: the sphere integrals.} By Eq.~\eqref{eq:ballFT},
$\int_{S_k^N}\chi_p(\xi\cdot x)\,d^Nx=p^{kN}\Omega(p^{k+m})-p^{(k-1)N}\Omega(p^{k-1+m})$, and
$\mathrm{vol}(S_k^N)=(1-p^{-N})p^{kN}$. For $k=1-m$ the character integral equals
$-p^{-mN}$, and for $k\ge2-m$ it vanishes. Hence
\begin{align*}
\int_{S_{1-m}^N}\big[1-\chi_p(\xi\cdot x)\big]\,d^Nx&=(1-p^{-N})p^{(1-m)N}+p^{-mN}\\
&=p^{(1-m)N},\\
\int_{S_k^N}\big[1-\chi_p(\xi\cdot x)\big]\,d^Nx&=(1-p^{-N})p^{kN},
\end{align*}
the second for $k\ge2-m$.

\emph{Step 3: summation.} Therefore
\begin{align*}
\frac{\hat w(0)-\hat w(\xi)}{C_w}
&=p^{-(1-m)\sigma}+(1-p^{-N})\sum_{k\ge2-m}p^{-k\sigma}\\
&=p^{(m-1)\sigma}\Big[1+\frac{(1-p^{-N})\,p^{-\sigma}}{1-p^{-\sigma}}\Big]\\
&=p^{(m-1)\sigma}\,\frac{1-p^{-N-\sigma}}{1-p^{-\sigma}}\\
&=\frac{1-p^{-N-\sigma}}{p^{\sigma}-1}\,p^{m\sigma},
\end{align*}
where the geometric series converges because $\sigma>0$. Since $p^{m\sigma}=\norm{\xi}^\sigma$,
this is Eq.~\eqref{eq:tail}. Finally, $A_w(\xi)=f'(0)\big(\hat w(0)-\hat w(\xi)\big)$ by
Eq.~\eqref{eq:Wsymbol}, which gives $A_w(\xi)=\kappa_W\norm{\xi}^\sigma$ with
$\kappa_W=f'(0)c_\sigma$; this is the symbol of $\kappa_W\Dsig$, Eq.~\eqref{eq:Taibleson}.
\end{proof}

\section{Evaluation of momentum integrals}\label{app:integrals}

In this appendix we state precisely which momentum integrals over $\Qp^N$ can be reduced to
one-dimensional sums, and prove it. The general claim that ``all momentum integrals are
evaluated by Eq.~\eqref{eq:radial}'' is \emph{not} made. Eq.~\eqref{eq:radial} applies to
radial integrands. A one-loop integral with a nonzero external momentum is not radial in
the loop momentum, but it still reduces to single sums (Lemma~\ref{lem:oneloop}).
Integrals with several independent external momenta, or with several loops, require
repeated use of these lemmas. We do not claim a general closed form for them.

Throughout, $S_k^N=\{\xi\in\Qp^N:\norm{\xi}=p^k\}$ and $B_k^N=\{\norm{\xi}\le p^k\}$, so
$\mathrm{vol}(B_k^N)=p^{kN}$ and $\mathrm{vol}(S_k^N)=(1-p^{-N})p^{kN}$.

\begin{lemma}[Radial integrals]\label{lem:radial}
Let $F:p^{\mathbb Z}\to\mathbb C$ with $\sum_kp^{kN}|F(p^k)|<\infty$. Then
$\int_{\Qp^N}F(\norm{\xi})\,d^N\xi=(1-p^{-N})\sum_{k\in\mathbb Z}p^{kN}F(p^k)$.
\end{lemma}
\begin{proof}
$\Qp^N\smallsetminus\{0\}$ is the disjoint union of the spheres $S_k^N$, on which
$F(\norm{\xi})=F(p^k)$ is constant. Since $S_k^N=B_k^N\smallsetminus B_{k-1}^N$,
$\mathrm{vol}(S_k^N)=p^{kN}-p^{(k-1)N}$. Absolute convergence justifies summing term by term.
\end{proof}

\begin{lemma}[Fourier transform of radial functions]\label{lem:radialFT}
If $g\in L^1(\Qp^N)$ is radial, then $\hat g$ is radial. Explicitly, for $\norm{\xi}=p^m$,
\begin{equation}
\hat g(\xi)=(1-p^{-N})\sum_{k\le-m}p^{kN}g(p^k)-p^{-mN}\,g(p^{1-m}).
\label{eq:radialFT}
\end{equation}
\end{lemma}
\begin{proof}
By Eq.~\eqref{eq:ballFT},
\begin{equation*}
\int_{S_k^N}\chi_p(\xi\cdot x)\,d^Nx=p^{kN}\Omega(p^{k+m})-p^{(k-1)N}\Omega(p^{k-1+m}).
\end{equation*}
This equals $(1-p^{-N})p^{kN}$ if $k\le-m$, equals $-p^{-mN}$ if $k=1-m$, and vanishes if
$k\ge2-m$. Sum over $k$; the result depends on $\xi$ only through $m$.
\end{proof}

Equation~\eqref{eq:radialFT} is the formula used in Proposition~\ref{prop:levels}.

\begin{lemma}[One loop with an external momentum]\label{lem:oneloop}
Let $F,G:p^{\mathbb Z}\to\mathbb C$ be such that the integrals below converge absolutely,
and let $\eta\in\Qp^N$ with $\norm{\eta}=p^m$. Then
\begin{align}
&\int_{\Qp^N}F(\norm{\xi})\,G(\norm{\eta-\xi})\,d^N\xi\nonumber\\
&\;=G(p^m)\,(1-p^{-N})\sum_{k<m}p^{kN}F(p^k)\nonumber\\
&\;+(1-p^{-N})\sum_{k>m}p^{kN}F(p^k)\,G(p^k)\nonumber\\
&\;+F(p^m)\Big[\int_{B_m^N}G(\norm{u})\,d^Nu-p^{(m-1)N}G(p^m)\Big],
\label{eq:oneloop}
\end{align}
where $\int_{B_m^N}G(\norm u)\,d^Nu=(1-p^{-N})\sum_{k\le m}p^{kN}G(p^k)$ by
Lemma~\ref{lem:radial}. In particular the left-hand side depends on $\eta$ only through
$\norm{\eta}$.
\end{lemma}
\begin{proof}
Split $\Qp^N$ into $\{\norm{\xi}<p^m\}$, $\{\norm{\xi}>p^m\}$ and $S_m^N$. By the strong
triangle inequality, $\norm{\eta-\xi}=\max(\norm{\eta},\norm{\xi})$ whenever
$\norm{\xi}\neq\norm{\eta}$. On the first region $\norm{\eta-\xi}=p^m$, and on the second
$\norm{\eta-\xi}=\norm{\xi}$. Lemma~\ref{lem:radial} gives the first two lines. On $S_m^N$,
$F(\norm{\xi})=F(p^m)$, and we substitute $u=\eta-\xi$. The condition $\norm{\xi}=p^m$
becomes $\norm{\eta-u}=p^m$, i.e.\ $u\in B_m^N\smallsetminus(\eta+B_{m-1}^N)$. On the removed
ball $\eta+B_{m-1}^N$ we have $\norm{u}=\norm{\eta}=p^m$, and its volume is $p^{(m-1)N}$. This
gives the third line.
\end{proof}

Lemma~\ref{lem:oneloop} extends to integrands that are arbitrary functions of the two
radii. The proof is a slight variation of that of Lemma~\ref{lem:oneloop}; the result is
used in the calculation of loop integrals with a radial integrand, in particular the
two-loop integrals below.

\begin{lemma}[One loop with a general radial integrand]\label{lem:oneloopH}
Let $H:p^{\mathbb Z}\times p^{\mathbb Z}\to\mathbb C$ be such that the integral below
converges absolutely, and let $\xi\in\Qp^N$ with $\norm{\xi}=p^m$. Then
\begin{align}
&\int_{\Qp^N}H\big(\norm{\zeta},\norm{\xi-\zeta}\big)\,d^N\zeta\nonumber\\
&\;=(1-p^{-N})\sum_{k<m}p^{kN}H(p^k,p^m)\nonumber\\
&\;+(1-p^{-N})\sum_{k>m}p^{kN}H(p^k,p^k)\nonumber\\
&\;+(1-p^{-N})\sum_{j\le m}p^{jN}H(p^m,p^j)\nonumber\\
&\;-p^{(m-1)N}H(p^m,p^m).
\label{eq:oneloopH}
\end{align}
In particular, the left-hand side is a radial function of $\xi$.
\end{lemma}
\begin{proof}
As in the proof of Lemma~\ref{lem:oneloop}, split $\Qp^N$ into $\{\norm{\zeta}<p^m\}$,
$\{\norm{\zeta}>p^m\}$ and $S_m^N$. On the first region $\norm{\xi-\zeta}=p^m$, and on the
second $\norm{\xi-\zeta}=\norm{\zeta}$. On $S_m^N$ put $u=\xi-\zeta$, which runs over
$B_m^N\smallsetminus(\xi+B_{m-1}^N)$; on the removed ball $\norm{u}=p^m$, and its volume is
$p^{(m-1)N}$.
\end{proof}

Lemma~\ref{lem:oneloop} is the case $H(a,b)=F(a)G(b)$.

Lemmas~\ref{lem:radial}, \ref{lem:radialFT}, and~\ref{lem:oneloop} are well known; see
Refs.~\cite{VVZ1994,ZunigaTextbook2025}. We include them here to provide a complete,
self-contained presentation.

\begin{lemma}[Iterated convolutions]\label{lem:convpowers}
Let $w\ge0$ be radial and integrable, with
$w_0=\int_{\Qp^N}w(\norm{x})\,d^Nx$, and radially nonincreasing:
$w(r)\ge w(r')$ for $r<r'$. Then, for all $n\ge1$ and $x\neq0$,
\begin{align}
\text{(i)}\quad&w^{\star n}(x)\le(2n-1)\,w_0^{\,n-1}\,w(\norm{x}),\nonumber\\
\text{(ii)}\quad&\lim_{\norm{x}\to\infty}\frac{w^{\star n}(x)}{w(\norm{x})}=n\,w_0^{\,n-1},
\label{eq:convpowers}
\end{align}
the limit holding whenever $w(r)>0$ for all large $r$.
\end{lemma}
\begin{proof}
Induction on $n$; the case $n=1$ is trivial. Let $G=w^{\star(n-1)}\ge0$, which is radial by
Lemma~\ref{lem:oneloop} and has $\int_{\Qp^N}G(u)\,d^Nu=w_0^{n-1}$. Fix $\norm{x}=p^m$
and split $w^{\star n}(x)=\int_{\Qp^N}w(\norm{y})\,G(x-y)\,d^Ny$ as in the proof of
Lemma~\ref{lem:oneloop}.
(a) On $\norm{y}<p^m$, $G(x-y)=G(x)$, and this region contributes
$G(x)\int_{\norm{y}<p^m}w(\norm{y})\,d^Ny\le G(x)\,w_0$.
(b) On $\norm{y}>p^m$, $\norm{x-y}=\norm{y}$ and $w(\norm y)\le w(p^m)$, so this region contributes
at most $w(p^m)\int_{\norm{u}>p^m}G(u)\,d^Nu$.
(c) On $\norm{y}=p^m$, $w(\norm y)=w(p^m)$, and this region contributes
$w(p^m)\big[\int_{B_m^N}G(u)\,d^Nu-p^{(m-1)N}G(x)\big]\le w(p^m)\int_{B_m^N}G(u)\,d^Nu$.
Adding the three and using the induction hypothesis
$G(x)\le(2n-3)w_0^{n-2}w(p^m)$ gives $w^{\star n}(x)\le\big[(2n-3)+2\big]w_0^{n-1}w(p^m)$.
For the limit, divide by $w(p^m)$ and let $m\to\infty$. By induction, term (a) tends to
$(n-1)w_0^{n-2}\cdot w_0$. Term (b) is at most $\int_{\norm{u}>p^m}G(u)\,d^Nu\to0$. Term (c)
tends to $\int_{\Qp^N}G(u)\,d^Nu=w_0^{n-1}$. Indeed $\int_{B_m^N}G(u)\,d^Nu\to w_0^{n-1}$,
and the subtracted term
satisfies
\begin{align*}
p^{(m-1)N}G(x)&\le(2n-3)w_0^{n-2}\,p^{(m-1)N}w(p^m)\\
&=(2n-3)w_0^{n-2}\,\frac{W_m}{p^N-1}\to0,
\end{align*}
with $W_m$ as in Eq.~\eqref{eq:Wk}. Hence the limit is $n\,w_0^{\,n-1}$.
\end{proof}

\begin{proposition}[The zero-loop propagator]\label{prop:propagator}
Let $w\ge0$ be radial and integrable, and set $c=f'(0)\ge0$.
\begin{enumerate}
\item In $\mathcal D'(\Qp^N)$,
$\mathcal F^{-1}_{\xi\to x}\big(e^{-\lambda_0(\xi)t}\big)=e^{-\alpha t}\big[\delta+K_t\big]$
with $K_t=\sum_{n\ge1}(ct)^nw^{\star n}/n!\in L^1(\Qp^N)$.
\item If in addition $\hat w\in L^1(\Qp^N)$, then $K_t$ is continuous and, for
$\norm{x}=p^k$,
\begin{equation}
K_t(x)=(1-p^{-N})\sum_{j\le-k}p^{jN}g_t(p^j)-p^{-kN}g_t(p^{1-k}),
\label{eq:Kt_explicit}
\end{equation}
where $g_t(\norm{\xi})=e^{ct\hat w(\xi)}-1$. If $\hat w$ is supported in
$\norm{\xi}\le p^{L}$, the sum is finite ($-L\le j\le-k$).
\item If $w$ is radially nonincreasing and positive at large distances, then
$K_t(x)\simeq ct\,e^{ctw_0}\,w(\norm{x})$ as $\norm{x}\to\infty$, and hence
$\Delta(x,t)\simeq\theta(t)\,ct\,e^{-\lambda_0(0)t}\,w(\norm{x})$, with
$\lambda_0(0)=\alpha-cw_0$.
\end{enumerate}
\end{proposition}
\begin{proof}
(1) $e^{-\lambda_0(\xi)t}=e^{-\alpha t}e^{ct\hat w(\xi)}$, and
$e^{ct\hat w}=1+\sum_{n\ge1}(ct)^n\hat w^{\,n}/n!$ with the series converging uniformly,
since $|\hat w|\le w_0$. The inverse Fourier transform of $1$ is $\delta$. Since
$\hat w^{\,n}=\widehat{w^{\star n}}$ and $\Vert w^{\star n}\Vert_1\le w_0^{\,n}$, the series
$\sum_{n\ge1}(ct)^nw^{\star n}/n!$ converges in $L^1$, and its Fourier transform is
$e^{ct\hat w}-1$. (2) $|g_t|\le ct\,|\hat w|\,e^{ctw_0}$, so $g_t\in L^1$ and
$K_t=\mathcal F^{-1}g_t$ is continuous. Since $g_t$ is radial and real,
$\mathcal F^{-1}g_t=\mathcal Fg_t$, and Eq.~\eqref{eq:Kt_explicit} is
Eq.~\eqref{eq:radialFT} of Lemma~\ref{lem:radialFT} applied to $g_t$. (3) By
Lemma~\ref{lem:convpowers}, $(ct)^nw^{\star n}(x)/(n!\,w(\norm x))$ is bounded by
$(ct)^n(2n-1)w_0^{n-1}/n!$, which is summable, and tends to $(ct)^nn\,w_0^{n-1}/n!$.
Dominated convergence gives
$K_t(x)/w(\norm x)\to\sum_{n\ge1}(ct)^nw_0^{n-1}/(n-1)!=ct\,e^{ctw_0}$.
\end{proof}

\begin{remark}\label{rem:what_compact}
Let $w\in L^1(\Qp^N)$ be radial and constant on $B_r^N$. Then $w(x+x')=w(x)$ for all
$x\in\Qp^N$ and all $x'\in B_r^N$: for $\norm{x}\le p^r$ both points lie in $B_r^N$, and for
$\norm{x}>p^r$ the ultrametric inequality gives $\norm{x+x'}=\norm{x}$. By the translation
invariance of the Haar measure, $\hat w(\xi)=\chi_p(\xi\cdot x')\,\hat w(\xi)$, i.e.\
$\hat w(\xi)\big(1-\chi_p(\xi\cdot x')\big)=0$ for all $x'\in B_r^N$. If $\hat w(\xi)\neq0$,
then $\chi_p(\xi\cdot x')=1$ for all $x'\in B_r^N=p^{-r}\Zp^N$, i.e.\ $\xi\cdot x'\in\Zp$
for all such $x'$, which holds if and only if $\norm{\xi}\le p^{-r}$. Hence $\hat w$ is
supported in the ball $\norm{\xi}\le p^{-r}$. If $w\neq0$, then $\hat w\not\equiv0$, since
the Fourier transform is injective on $L^1(\Qp^N)$ and $w$, being locally constant, is
continuous; thus the support of $\hat w$ is a nonempty subset of this ball. In particular
$\hat w$ has compact support, and,
being bounded, it is in $L^1(\Qp^N)$, so part~2 of Proposition~\ref{prop:propagator}
applies. See Ref.~\cite{ZunigaTextbook2025}, Theorem~1.161.
\end{remark}

\paragraph*{Proof of Lemma~\ref{lem:I2}.}
(1) Since $w$ is radial and real,
$A_w(\xi)=f'(0)\int_{\Qp^N}w(\norm{x})\,\big[1-\cos\big(2\pi\{\xi\cdot x\}_p\big)\big]\,d^Nx$, and the
bracket is non-negative. Suppose $A_w(\xi)=0$ for some $\xi\neq0$. Then
$\xi\cdot x\in\Zp$ for almost every $x$ in the support of $w$. This support is a union of
spheres. If it met a sphere of radius $p^{j}>\norm{\xi}^{-1}$, some $x$ on that sphere
would satisfy $|\xi\cdot x|_p>1$. Hence $w$ is supported in the ball of radius
$\norm{\xi}^{-1}$, and then $\hat w(0)-\hat w(\xi')=0$ for every
$\norm{\xi'}\le\norm{\xi}$, which contradicts Eq.~\eqref{eq:Wasym}.

(2) and (3). Split the sum~\eqref{eq:I2sum} at $k=k_0$, where $k_0\le-r$ is chosen so
that $|\varepsilon(\xi)|\le\frac12$ for $\norm{\xi}\le p^{k_0}$.

\emph{Terms with $k_0<k\le\Lambda$.} $A_w$ is constant on spheres, positive by (1), and
tends to $f'(0)\hat w(0)>0$ as $\norm{\xi}\to\infty$. Hence
$\inf_{\norm{\xi}>p^{k_0}}A_w>0$, and these finitely many terms are bounded uniformly in
$\mu\ge0$, by a constant times $p^{\Lambda N}$. Relative to $\mu^{N/\sigma-2}$, they give
the error $O(\mu^{2-N/\sigma})$.

\emph{Terms with $k\le k_0$.} Here $A_w(p^k)\ge\frac12\kappa_Wp^{k\sigma}$, and
\begin{equation*}
\Big|\frac{1}{(\mu+A_w(p^k))^2}-\frac{1}{(\mu+\kappa_Wp^{k\sigma})^2}\Big|
\le C'\,\frac{p^{k\delta}\,\kappa_Wp^{k\sigma}}{\big(\mu+\frac12\kappa_Wp^{k\sigma}\big)^3}.
\end{equation*}
The sum over $k\le k_0$ of $p^{kN}$ times the right-hand side is
$O\big(\mu^{(N+\delta)/\sigma-2}\big)$ if $N+\delta<2\sigma$, $O(|\ln\mu|)$ if
$N+\delta=2\sigma$, and $O(1)$ otherwise. This is the first error term.

\emph{The pure power law.} At $\mu=0$, the sum
$\sum_{k\le k_0}p^{kN}/(\kappa_Wp^{k\sigma})^2=\kappa_W^{-2}\sum_{k\le k_0}p^{k(N-2\sigma)}$
converges if and only if $N>2\sigma$, which, with the bounds above, proves (2). For
$N<2\sigma$, the sum of $p^{kN}/(\mu+\kappa_Wp^{k\sigma})^2$ over $k>k_0$ is bounded by
$\kappa_W^{-2}\sum_{k>k_0}p^{k(N-2\sigma)}<\infty$, so we may sum over all $k\in\mathbb Z$
at the cost of another $O(1)$. Finally, with $\mu=\kappa_Wp^{\sigma s}$,
\begin{equation*}
\sum_{k\in\mathbb Z}\frac{p^{kN}}{(\mu+\kappa_Wp^{k\sigma})^2}
=\kappa_W^{-N/\sigma}\,\mu^{N/\sigma-2}\,\Phi(s),
\end{equation*}
which gives Eq.~\eqref{eq:I2}. If only $\varepsilon\to0$ is known, dominated convergence
applied to the rescaled sum gives the error $o(1)$. \hfill$\square$

\paragraph*{Fourier series of $\Phi$.}
Write $\Phi(s)=\sum_kF(k-s)$, with $F(u)=p^{uN}/(1+p^{\sigma u})^2$. By the Poisson
summation formula, $\Phi(s)=\sum_n\hat F(n)e^{-2\pi ins}$, with
$\hat F(n)=\int_{\mathbb R}F(u)e^{-2\pi inu}\,du$. The substitution $t=p^{\sigma u}$ turns
this into a Beta integral,
$\hat F(n)=(\sigma\ln p)^{-1}\int_0^\infty t^{a-i\omega_n-1}(1+t)^{-2}\,dt
=(\sigma\ln p)^{-1}\,\Gamma(a-i\omega_n)\,\Gamma(2-a+i\omega_n)$, which is
Eq.~\eqref{eq:Phi_Fourier}.

\paragraph*{Two-loop integrals.}
We now write the two-loop self-energy of the action~\eqref{eq:RFT}. The propagator is
$\theta(t)e^{-\lambda(\xi)t}$, with $\lambda(\xi)=\mu+A_w(\xi)$. All lines run forward in
time, and integrating over the times between successive vertices gives, for each interval
in which the lines $\xi_1,\dots,\xi_r$ are present, a factor
$\big(-i\omega+\lambda(\xi_1)+\dots+\lambda(\xi_r)\big)^{-1}$, where $\eta$ and $\omega$ are
the external momentum and frequency. A two-loop self-energy has two splitting and two
merging vertices, and there are two topologies. In the nested one, a one-loop self-energy is
inserted on an internal line:
$\eta\to\xi+(\eta-\xi)$, $\xi\to\zeta+(\xi-\zeta)$, $\zeta+(\xi-\zeta)\to\xi$,
$\xi+(\eta-\xi)\to\eta$. In the overlapping one, a vertex is corrected:
$\eta\to\xi+(\eta-\xi)$, $\xi\to\zeta+(\xi-\zeta)$, $(\xi-\zeta)+(\eta-\xi)\to\eta-\zeta$,
$\zeta+(\eta-\zeta)\to\eta$. With
\begin{align*}
D_1&=-i\omega+\lambda(\xi)+\lambda(\eta-\xi),\\
D_2&=-i\omega+\lambda(\zeta)+\lambda(\xi-\zeta)+\lambda(\eta-\xi),\\
D_3&=-i\omega+\lambda(\zeta)+\lambda(\eta-\zeta),
\end{align*}
the two contributions are
\begin{align}
\Sigma_A(\eta,\omega)&=c_A\,g^4\iint\frac{d^N\xi\,d^N\zeta}{D_1^{\,2}\,D_2},\nonumber\\
\Sigma_B(\eta,\omega)&=c_B\,g^4\iint\frac{d^N\xi\,d^N\zeta}{D_1\,D_2\,D_3},
\label{eq:twoloop}
\end{align}
where both integrals are over $\{\norm{\xi}\le p^{\Lambda},\,\norm{\zeta}\le p^{\Lambda}\}$, and
$c_A$, $c_B$ are combinatorial factors.

\begin{proposition}[Two loops at zero external momentum]\label{prop:twoloop}
Write $\lambda_k=\lambda(p^k)$. Then
\begin{align}
\Sigma_A(0,0)&=\frac{c_Ag^4}{4}\,(1-p^{-N})\sum_{m\le\Lambda}\frac{p^{mN}K_m}{\lambda_m^2},\nonumber\\
\Sigma_B(0,0)&=\frac{c_Bg^4}{4}\,(1-p^{-N})\sum_{m\le\Lambda}\frac{p^{mN}L_m}{\lambda_m},
\label{eq:twoloop_sums}
\end{align}
where
\begin{align*}
K_m={}&2(1-p^{-N})\sum_{k<m}\frac{p^{kN}}{\lambda_k+2\lambda_m}
+\frac{(1-2p^{-N})\,p^{mN}}{3\lambda_m}\\
&+(1-p^{-N})\sum_{m<k\le\Lambda}\frac{p^{kN}}{2\lambda_k+\lambda_m},\\
L_m={}&(1-p^{-N})\sum_{k<m}\frac{p^{kN}}{\lambda_k(\lambda_k+2\lambda_m)}\\
&+(1-p^{-N})\sum_{m<k\le\Lambda}\frac{p^{kN}}{\lambda_k(2\lambda_k+\lambda_m)}\\
&+\frac{1-p^{-N}}{\lambda_m}\sum_{k<m}\frac{p^{kN}}{\lambda_k+2\lambda_m}
+\frac{(1-2p^{-N})\,p^{mN}}{3\lambda_m^2}.
\end{align*}
\end{proposition}
\begin{proof}
At $\eta=0$, $\omega=0$, since $\lambda$ is radial, $D_1=2\lambda(\xi)$,
$D_3=2\lambda(\zeta)$ and $D_2=\lambda(\zeta)+\lambda(\xi-\zeta)+\lambda(\xi)$. The inner
integrals over $\zeta$ are of the form~\eqref{eq:oneloopH}, with
$H(a,b)=\big(\lambda(a)+\lambda(b)+\lambda_m\big)^{-1}$ for $\Sigma_A$ and
$H(a,b)=\lambda(a)^{-1}\big(\lambda(a)+\lambda(b)+\lambda_m\big)^{-1}$ for $\Sigma_B$.
Lemma~\ref{lem:oneloopH} gives $K_m$ and $L_m$; in $K_m$ the terms with $k<m$ from the
first and third sums of Eq.~\eqref{eq:oneloopH} coincide, and the terms with $k=m$ combine
into $\big[(1-p^{-N})p^{mN}-p^{(m-1)N}\big]/(3\lambda_m)=(1-2p^{-N})p^{mN}/(3\lambda_m)$. The
remaining integral over $\xi$ is radial, and Lemma~\ref{lem:radial} applies.
\end{proof}

For $p=2$ and $N=1$ the factor $1-2p^{-N}$ vanishes: then no $\zeta$ on the sphere of radius
$p^m$ satisfies $\norm{\xi-\zeta}=p^m$. With $\lambda_k=\mu+\kappa_Wp^{k\sigma}$, both sums
in Eq.~\eqref{eq:twoloop_sums} scale as $g^4\mu^{2N/\sigma-3}$ times a function periodic in
$s$. Their singular part is finite at $\mu=0$ if $N>3\sigma/2$; like $J(0)$ in
Sec.~\ref{sec:ginzburg}, it only shifts the critical point. The $\mu$-derivative behaves as
$g^4\mu^{2N/\sigma-4}=\big(g^2\mu^{N/\sigma-2}\big)^2$, the square of the one-loop Ginzburg
ratio, consistently with $N_c=2\sigma$.

\paragraph*{Scope.}
Lemma~\ref{lem:radial} covers the one-loop integrals at zero external momentum used in the
Ginzburg criterion (Sec.~\ref{sec:ginzburg}) and the spectral sums of
Sec.~\ref{sec:phase}. Lemma~\ref{lem:oneloop} covers one-loop self-energy and vertex
integrals with a single external momentum, when the propagators are radial functions of
the momentum, for example $\big(\mu+A_w(\xi)\big)^{-1}$ after the
frequency integral. Lemma~\ref{lem:oneloopH} and Proposition~\ref{prop:twoloop} reduce
the two-loop self-energy at zero external momentum to explicit double sums. One-loop
integrals with two independent external momenta $\eta_1$, $\eta_2$, and two-loop
integrals at nonzero external momentum, require one more step: the regions of integration
are fixed by comparing $\norm{\xi}$ with the sides of the ultrametric (isosceles) triangle
with vertices $0$, $\eta_1$, $-\eta_2$.



\begin{thebibliography}{99}

\bibitem{BullmoreSporns2009}
E. Bullmore and O. Sporns, Nat. Rev. Neurosci. \textbf{10}, 186 (2009).

\bibitem{HilgetagGoulas2020}
C.~C. Hilgetag and A. Goulas, Phil. Trans. R. Soc. B \textbf{375}, 20190319 (2020).

\bibitem{BeggsPlenz2003}
J.~M. Beggs and D. Plenz, J. Neurosci. \textbf{23}, 11167 (2003).

\bibitem{BornholdtRohl2003}
S. Bornholdt and T. R\"ohl, Phys. Rev. E \textbf{67}, 066118 (2003).

\bibitem{Schoenholz2017}
S.~S. Schoenholz, J. Gilmer, S. Ganguli, and J. Sohl-Dickstein, in \emph{International
Conference on Learning Representations} (2017).

\bibitem{VockMeisel2025}
S. Vock and C. Meisel, \emph{Critical dynamics governs deep learning}, arXiv:2507.08527.

\bibitem{Sporns2006}
O. Sporns, BioSystems \textbf{85}, 55 (2006).

\bibitem{CostaSporns2005}
L.~da F. Costa and O. Sporns, Eur. Phys. J. B \textbf{48}, 555 (2005).

\bibitem{SpornsTononiKotter2005}
O. Sporns, G. Tononi, and R. K\"otter, PLoS Comput. Biol. \textbf{1}, e42 (2005).

\bibitem{WattsStrogatz1998}
D.~J. Watts and S.~H. Strogatz, Nature (London) \textbf{393}, 440 (1998).

\bibitem{BassettBullmore2006}
D.~S. Bassett and E. Bullmore, Neuroscientist \textbf{12}, 512 (2006).

\bibitem{ZZ2023WC}
W.~A. Z\'u\~niga-Galindo and B.~A. Zambrano-Luna, Entropy \textbf{25}, 949 (2023).

\bibitem{BuiceCowan2007}
M.~A. Buice and J.~D. Cowan, Phys. Rev. E \textbf{75}, 051919 (2007).

\bibitem{Doi1976a}
M. Doi, J. Phys. A \textbf{9}, 1465 (1976); \textbf{9}, 1479 (1976).

\bibitem{Peliti1985}
L. Peliti, J. Phys. (France) \textbf{46}, 1469 (1985).

\bibitem{Janssen1981}
H.~K. Janssen, Z. Phys. B \textbf{42}, 151 (1981).

\bibitem{Grassberger1982}
P. Grassberger, Z. Phys. B \textbf{47}, 365 (1982).

\bibitem{CardySugar1980}
J.~L. Cardy and R.~L. Sugar, J. Phys. A \textbf{13}, L423 (1980).

\bibitem{Hinrichsen2000}
H. Hinrichsen, Adv. Phys. \textbf{49}, 815 (2000).

\bibitem{HeliasDahmen2020}
M. Helias and D. Dahmen, \emph{Statistical Field Theory for Neural Networks}, Lecture
Notes in Physics Vol.~970 (Springer, Cham, 2020).

\bibitem{Halverson2021}
J. Halverson, A. Maiti, and K. Stoner, Mach. Learn.: Sci. Technol. \textbf{2}, 035002
(2021).

\bibitem{GrosvenorJefferson2022}
K.~T. Grosvenor and R. Jefferson, SciPost Phys. \textbf{12}, 081 (2022).

\bibitem{ZG2023}
W.~A. Z\'u\~niga-Galindo, Physica A \textbf{612}, 128492 (2023).

\bibitem{ZG2024}
W.~A. Z\'u\~niga-Galindo, J. Phys. A \textbf{58}, 125201 (2025); arXiv:2410.02206.

\bibitem{BuiceCowan2009}
M.~A. Buice and J.~D. Cowan, Prog. Biophys. Mol. Biol. \textbf{99}, 53 (2009).

\bibitem{RobertsYaida2022}
D.~A. Roberts and S. Yaida, \emph{The Principles of Deep Learning Theory} (Cambridge
University Press, Cambridge, 2022).

\bibitem{Segadlo2022}
K. Segadlo, B. Epping, A. van Meegen, D. Dahmen, M. Kr\"amer, and M. Helias, J. Stat.
Mech. (2022) 103401.

\bibitem{ZG2026}
W.~A. Z\'u\~niga-Galindo, \emph{Critical Organization of Deep Neural Networks, and
$p$-Adic Statistical Field Theories}, arXiv:2601.19070.

\bibitem{RTV1986}
R. Rammal, G. Toulouse, and M.~A. Virasoro, Rev. Mod. Phys. \textbf{58}, 765 (1986).

\bibitem{KKZ2018}
A. Khrennikov, S. Kozyrev, and W.~A. Z\'u\~niga-Galindo, \emph{Ultrametric Equations and
its Applications}, Encyclopedia of Mathematics and its Applications Vol.~168 (Cambridge
University Press, Cambridge, 2018).

\bibitem{Mezard1984}
M. M\'ezard, G. Parisi, N. Sourlas, G. Toulouse, and M. Virasoro, J. Phys. (Paris)
\textbf{45}, 843 (1984).

\bibitem{Frauenfelder1991}
H. Frauenfelder, S.~G. Sligar, and P.~G. Wolynes, Science \textbf{254}, 1598 (1991).

\bibitem{VVZ1994}
V.~S. Vladimirov, I.~V. Volovich, and E.~I. Zelenov, \emph{$p$-Adic Analysis and
Mathematical Physics} (World Scientific, Singapore, 1994).

\bibitem{Kochubei2001}
A.~N. Kochubei, \emph{Pseudo-Differential Equations and Stochastics over
Non-Archimedean Fields} (Marcel Dekker, New York, 2001).

\bibitem{AKS2010}
S. Albeverio, A.~Yu. Khrennikov, and V.~M. Shelkovich, \emph{Theory of $p$-adic
Distributions: Linear and Nonlinear Models} (Cambridge University Press, 2010).

\bibitem{ZunigaLNM2016}
W.~A. Z\'u\~niga-Galindo, \emph{Pseudodifferential Equations over Non-Archimedean
Spaces}, Lecture Notes in Mathematics Vol.~2174 (Springer, Cham, 2016).

\bibitem{ZunigaTextbook2025}
W.~A. Z\'u\~niga-Galindo, \emph{$p$-Adic Analysis: Stochastic Processes and
Pseudo-Differential Equations} (De Gruyter, Berlin, 2025).

\bibitem{ParisiSourlas2000}
G. Parisi and N. Sourlas, Eur. Phys. J. B \textbf{14}, 535 (2000).

\bibitem{ABKO2002}
V.~A. Avetisov, A.~H. Bikulov, S.~V. Kozyrev, and V.~A. Osipov, J. Phys. A
\textbf{35}, 177 (2002).

\bibitem{ZHZ2023}
W.~A. Z\'u\~niga-Galindo, C. He, and B.~A. Zambrano-Luna, Prog. Theor. Exp. Phys.
\textbf{2023}, 063A01 (2023).

\bibitem{ZG2024ATMP}
W.~A. Z\'u\~niga-Galindo, Adv. Theor. Math. Phys. \textbf{28}, 679 (2024).

\bibitem{ZZ2023CNN}
B.~A. Zambrano-Luna and W.~A. Z\'u\~niga-Galindo, $p$-adic cellular neural networks,
J. Nonlinear Math. Phys. \textbf{30}, 34--70 (2023), doi:10.1007/s44198-022-00071-8.

\bibitem{ZG2022Ultrametric}
W.~A. Z\'u\~niga-Galindo, Physica A \textbf{597}, 127221 (2022); arXiv:2201.06452.

\bibitem{Janssen1999}
H.~K. Janssen, K. Oerding, F. van Wijland, and H.~J. Hilhorst, Eur. Phys. J. B \textbf{7}, 137 (1999).

\bibitem{ZGshort}
W.~A. Z\'u\~niga-Galindo, Hierarchical connectivity sets the upper critical dimension of
neural avalanches. In preparation; available at arXiv.org.

\bibitem{Kozyrev2002}
S.~V. Kozyrev, Izv. Math. \textbf{66}, 367 (2002).

\bibitem{ZG2022}
W.~A. Z\'u\~niga-Galindo, Physica A \textbf{602}, 127648 (2022).

\bibitem{ZG2022RMP}
W.~A. Z\'u\~niga-Galindo, Rev. Math. Phys. \textbf{34}, 2250022 (2022).

\bibitem{HinrichsenHoward1999}
H. Hinrichsen and M. Howard, Eur. Phys. J. B \textbf{7}, 635 (1999).

\bibitem{Stein1970}
E.~M. Stein, \emph{Singular Integrals and Differentiability Properties of Functions}
(Princeton University Press, Princeton, 1970).

\bibitem{Kwasnicki2017}
M. Kwa\'snicki, Fract. Calc. Appl. Anal. \textbf{20}, 7 (2017).

\bibitem{Tauber2014}
U.~C. T\"auber, \emph{Critical Dynamics: A Field Theory Approach to Equilibrium and
Non-Equilibrium Scaling Behavior} (Cambridge University Press, Cambridge, 2014).

\bibitem{Dyson1969}
F.~J. Dyson, Commun. Math. Phys. \textbf{12}, 91 (1969).

\bibitem{BleherSinai1973}
P.~M. Bleher and Ya.~G. Sinai, Commun. Math. Phys. \textbf{33}, 23 (1973).

\bibitem{LernerMissarov1989}
\'E.~Yu. Lerner and M.~D. Missarov, Theor. Math. Phys. \textbf{78}, 177 (1989).

\bibitem{AbdesselamChandraGuadagni2013}
A. Abdesselam, A. Chandra, and G. Guadagni, Rigorous quantum field theory functional
integrals over the $p$-adics I: anomalous dimensions, arXiv:1302.5971.

\bibitem{Gubser2017}
S.~S. Gubser, C. Jepsen, S. Parikh, and B. Trundy, J. High Energy Phys. \textbf{11} (2017) 107.

\bibitem{ErbinLahocheSamary2021}
H. Erbin, V. Lahoche, and D.~O. Samary, arXiv:2108.01403.

\bibitem{BuiceCowanChow2010}
M.~A. Buice, J.~D. Cowan, and C.~C. Chow, Neural Comput. \textbf{22}, 377 (2010).

\bibitem{Munoz1999}
M.~A. Mu\~noz, R. Dickman, A. Vespignani, and S. Zapperi, Phys. Rev. E \textbf{59},
6175 (1999).

\bibitem{Gillespie1976}
D.~T. Gillespie, J. Comput. Phys. \textbf{22}, 403 (1976).

\bibitem{LewisShedler1979}
P.~A.~W. Lewis and G.~S. Shedler, Nav. Res. Logist. Q. \textbf{26}, 403 (1979).

\bibitem{GrassbergerTorre1979}
P. Grassberger and A. de la Torre, Ann. Phys. (N.Y.) \textbf{122}, 373 (1979).

\bibitem{Clauset2009}
A. Clauset, C.~R. Shalizi, and M.~E.~J. Newman, SIAM Rev. \textbf{51}, 661 (2009).

\bibitem{KovalMeesterTrapman2012}
V. Koval, R. Meester, and P. Trapman, Electron. J. Probab. \textbf{17}, 57 (2012).

\bibitem{BenjaminiBerger2001}
I. Benjamini and N. Berger, Random Struct. Algorithms \textbf{19}, 102 (2001).

\bibitem{Biskup2004}
M. Biskup, Ann. Probab. \textbf{32}, 2938 (2004).

\bibitem{BGPW2014}
A. Bendikov, A. Grigor'yan, Ch. Pittet, and W. Woess, Russ. Math. Surveys \textbf{69},
589 (2014) [arXiv:1304.6271].

\bibitem{RodriguezZuniga2008}
J.~J. Rodr\'{\i}guez-Vega and W.~A. Z\'u\~niga-Galindo, Pacific J. Math. \textbf{237},
327 (2008).

\bibitem{CabreRoquejoffre2013}
X. Cabr\'e and J.-M. Roquejoffre, Commun. Math. Phys. \textbf{320}, 679 (2013).

\bibitem{OrnsteinZernike1914}
L.~S. Ornstein and F. Zernike, Accidental deviations of density and opalescence at the
critical point of a single substance, Proc. Akad. Sci. Amsterdam \textbf{17}, 793 (1914).

\bibitem{vanKampen2007}
N.~G. van Kampen, \emph{Stochastic Processes in Physics and Chemistry}, 3rd ed.
(North-Holland, Amsterdam, 2007).

\bibitem{PenttonenBuzsaki2003}
M. Penttonen and G. Buzs\'aki, Thalamus Relat. Syst. \textbf{2}, 145 (2003).

\bibitem{AbbottNelson2000}
L.~F. Abbott and S.~B. Nelson, Synaptic plasticity: taming the beast, Nat. Neurosci.
\textbf{3}, 1178 (2000).

\bibitem{Levina2007}
A. Levina, J.~M. Herrmann, and T. Geisel, Nat. Phys. \textbf{3}, 857 (2007).

\bibitem{MorettiMunoz2013}
P. Moretti and M.~A. Mu\~noz, Nat. Commun. \textbf{4}, 2521 (2013).

\bibitem{Cardy1996notes}
J. Cardy, \emph{Renormalisation group approach to reaction-diffusion problems},
arXiv:cond-mat/9607163 (1996).

\bibitem{TauberHowardVollmayrLee2005}
U.~C. T\"auber, M. Howard, and B.~P. Vollmayr-Lee, J. Phys. A \textbf{38}, R79 (2005).

\bibitem{Harris1963}
T.~E. Harris, \emph{The Theory of Branching Processes} (Springer, Berlin, 1963).

\bibitem{Poole2016}
B. Poole, S. Lahiri, M. Raghu, J. Sohl-Dickstein, and S. Ganguli, in \emph{Advances in
Neural Information Processing Systems 29} (2016).

\bibitem{MSR1973}
P.~C. Martin, E.~D. Siggia, and H.~A. Rose, Phys. Rev. A \textbf{8}, 423 (1973).

\bibitem{Janssen1976}
H.~K. Janssen, Z. Phys. B \textbf{23}, 377 (1976).

\bibitem{DeDominicis1976}
C. De Dominicis, J. Phys. (Paris) Colloq. \textbf{37}, C1-247 (1976).

\bibitem{GrassbergerSundermeyer1978}
P. Grassberger and K. Sundermeyer, Phys. Lett. B \textbf{77}, 220 (1978).

\bibitem{Tauber2007}
U.~C. T\"auber, Lect. Notes Phys. \textbf{716}, 295 (2007) [arXiv:cond-mat/0511743].

\bibitem{Takeuchi2007}
K.~A. Takeuchi, M. Kuroda, H. Chat\'e, and M. Sano, Phys. Rev. Lett. \textbf{99}, 234503 (2007).

\bibitem{Taibleson1975}
M.~H. Taibleson, \emph{Fourier Analysis on Local Fields} (Princeton University Press,
Princeton, 1975).

\bibitem{HornJohnson2013}
R.~A. Horn and C.~R. Johnson, \emph{Matrix Analysis}, 2nd ed. (Cambridge University
Press, Cambridge, 2013).

\bibitem{Seneta2006}
E. Seneta, \emph{Non-negative Matrices and Markov Chains}, revised printing (Springer,
New York, 2006).

\end{thebibliography}
\end{document}